\documentclass[12pt,a4paper]{report}
\let\openright=\clearpage

\usepackage[utf8]{inputenc}
\usepackage[T1]{fontenc}
\usepackage[english]{babel}
\usepackage{lmodern}
\usepackage{graphicx}
\usepackage{amsmath}
\usepackage{amssymb}
\usepackage{mathtools}

\usepackage[pdftex,unicode]{hyperref}   
\hypersetup{pdftitle=Theoretical description of nuclear collective excitations}
\hypersetup{pdfauthor=Anton Repko}

\makeatletter
\def\@makechapterhead#1{
  {\parindent \z@ \raggedright \normalfont
   \huge\bfseries \thechapter. #1
   \par\nobreak
   \vskip 20\p@
}}
\def\@makeschapterhead#1{
  {\parindent \z@ \raggedright \normalfont
   \huge\bfseries #1
   \par\nobreak
   \vskip 20\p@
}}
\makeatother

\def\chapwithtoc#1{
\chapter*{#1}
\addcontentsline{toc}{chapter}{#1}
}

\begin{document}

\lefthyphenmin=2
\righthyphenmin=2


\pagestyle{empty}
\begin{center}

\large

Charles University in Prague

\medskip

Faculty of Mathematics and Physics

\vfill

{\bf\Large DOCTORAL THESIS}

\vfill

\centerline{\mbox{\includegraphics[width=60mm]{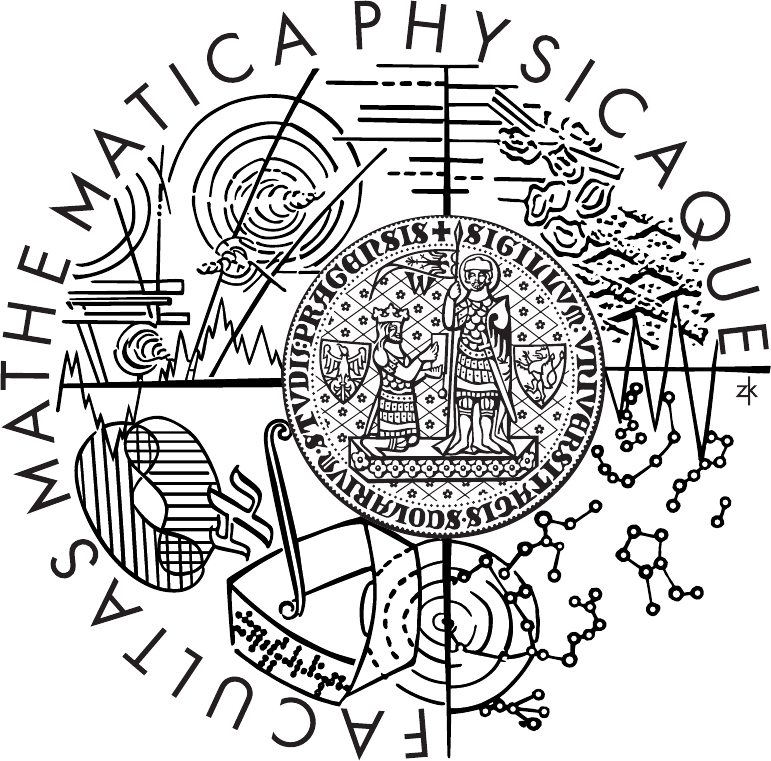}}}

\vfill
\vspace{5mm}

{\LARGE Anton Repko}

\vspace{15mm}

{\LARGE\bfseries Theoretical description of nuclear collective excitations}

\vfill

Institute of Particle and Nuclear Physics

\vfill

\begin{tabular}{rl}

Supervisor of the doctoral thesis: & prof. RNDr. Jan Kvasil, DrSc. \\
\noalign{\vspace{2mm}}
Study programme: & Physics \\
\noalign{\vspace{2mm}}
Specialization: & Nuclear Physics \\
\end{tabular}

\vfill

Prague 2015

\end{center}

\newpage


\openright

\noindent
I would like to thank my supervisor prof.~Jan Kvasil for patience during the explanation and discussions of the field of nuclear theory, and for various helpful suggestions. I am grateful also to Valentin Nesterenko for encouragement, for hosting me during my two stays in Dubna, and for pointing out perspective research directions.

\newpage


\vglue 0pt plus 1fill

\noindent
I declare that I carried out this doctoral thesis independently, and only with the cited
sources, literature and other professional sources.

\medskip\noindent
I understand that my work relates to the rights and obligations under the Act No.
121/2000 Coll., the Copyright Act, as amended, in particular the fact that the Charles
University in Prague has the right to conclude a license agreement on the use of this
work as a school work pursuant to Section 60 paragraph 1 of the Copyright Act.

\vspace{10mm}

\hbox{\hbox to 0.8\hsize{%
In Prague December 11$^\textrm{th}$ 2015;\rule{2mm}{0mm} \begin{footnotesize}3$^\textrm{rd}$ revision May 29$^\textrm{th}$ 2020\end{footnotesize}.
\hss}\hbox to 0.2\hsize{%
Anton Repko
\hss}}

\vspace{20mm}
\newpage


\vbox to 0.5\vsize{
\setlength\parindent{0mm}
\setlength\parskip{4mm}

\begin{small}
Název práce:
Teoretický popis kolektivních excitací jader

Autor:
Anton Repko

Katedra:  
Ústav částicové a jaderné fyziky

Vedoucí disertační práce:
prof. RNDr. Jan Kvasil, DrSc., ÚČJF MFF UK

Abstrakt:
Teorie funkcionálu hustoty je preferovaná mikroskopická metoda pro výpočet vlastností jader napříč celou tabulkou nuklidů. Vedle vlastností základního stavu, které se počítají Hartreeho-Fockovou metodou, vzbuzené stavy jader se dají popsat pomocí metody Random Phase Approximation (RPA). Hlavním cílem předkládané práce je podat formalismus RPA metody pro sféricky symetrická jádra, s použitím technik skládání momentu hybnosti. Probírají se také různá pomocná témata, jako Hartreeho-Fockova teorie, Coulombův integrál, těžišťové korekce a párování. Metoda RPA je odvozená rovněž pro axiálně deformovaná jádra. Odvozené vzorce byly zabudovány do počítačových programů a použity pro výpočet některých fyzikálních výsledků. Po zevrubném prozkoumání výpočtů z hlediska numerické přesnosti byla probrána tyto témata: toroidální povaha nízko-ležící části E1 rezonance (,,pygmy``), gigantické rezonance různé multipolarity v deformovaném jádře $^{154}$Sm a magnetické dipólové (M1) přechody v deformovaném $^{50}$Cr.

Klíčová slova:
Random Phase Approximation, Skyrme funkcionál, gigantické rezonance v jádrech, redukované maticové elementy
\end{small}

\vss}\nobreak\vbox to 0.49\vsize{
\setlength\parindent{0mm}
\setlength\parskip{4mm}

\begin{small}
Title:
Theoretical description of nuclear collective excitations

Author:
Anton Repko

Department:
Institute of Particle and Nuclear Physics

Supervisor:
prof. RNDr. Jan Kvasil, DrSc., IPNP Charles University in Prague

Abstract:
Density functional theory is a preferred microscopic method for calculation of nuclear properties over the whole nuclear chart. Besides ground-state properties, which are calculated by Hartree-Fock theory, nuclear excitations can be described by means of Random Phase Approximation (RPA). The main objective of the present work is to give the RPA formalism for spherically symmetric nuclei, using the techniques of angular-momentum coupling. Various auxiliary topics, such as Hartree-Fock theory, Coulomb integral, center-of-mass corrections and pairing, are treated as well. RPA method is derived also for axially deformed nuclei. The derived formulae are then implemented in the computer code and utilized for calculation of some physical results. After thorough investigation of the precision aspects of the calculation, the following topics are treated as examples: toroidal nature of the low-energy (pygmy) part of the E1 resonance, giant resonances of various multipolarities in deformed nucleus $^{154}$Sm, and magnetic dipole (M1) transitions in deformed $^{50}$Cr.

\end{small}

\vss}

\newpage


\openright
\pagestyle{plain}
\setcounter{page}{1}
\tableofcontents

\chapter{Introduction}
Microscopic quantum-theoretical description of nuclei (in terms of their ground-state and excited-state properties, transition probabilities and nuclear reactions) is a difficult task, due to poor knowledge of the nuclear interaction (as compared to electronic systems) and various other obstacles. Although the main features of the nucleon-nucleon interaction can be deduced from the scattering data, their straightforward application for a product wavefunction (Slater determinant) of the Hartree-Fock method, which is numerically the simplest approach to a quantum many-body problem, runs into the problem of strongly repulsive short-range part of the $N$-$N$ interaction. This obstacle can be circumvented by utilizing renormalized in-medium interaction, thus giving rise to \emph{Br\"uckner-Hartree-Fock} method \cite{Ring1980}. However, such approach did not give satisfactory results, also due to the need of three-body interactions, which are difficult to measure. The BHF method has currently attracted revived attention \cite{Ring2015}, due to the fact that its relativistic version \cite{Muther1988} does not need the three-body interaction.

An alternative approach is to diagonalize the Hamiltonian in the full configuration space of many-body wavefunctions, constructed from a given single-particle basis, and the resulting method is called \emph{large-scale shell model} \cite{Caurier2005}. Because the numerical cost grows exponentially with the basis, the shell model either has to use strongly truncated model space with phenomenological corrections to the interaction (usable for $A < 100$), or restrict itself to very light nuclei (up to $^{12}$C), giving rise to \emph{ab-initio} no-core shell model. The convergence and reach of the no-core shell model can be somewhat improved either by softening of the interaction (based on chiral forces) by similarity renormalization group \cite{Barett2013}, or by group-theoretical preselection of the basis in the symmetry-adapted no-core shell model \cite{Dytrych2013}.

Shell model is not suitable for heavier nuclei, which are therefore most often treated by the \emph{density functionals}. These were at first inspired by the Br\"uckner-Hartree-Fock method, so they are mean-field methods, based on a product wavefunction determined in an iterative way, but the interaction is phenomenological and no longer derived from the bare $N$-$N$ data. Energy density functional in nuclear physics is then a self-consistent microscopic approach to calculate nuclear properties and structure over the whole periodic table (except the lightest nuclei) \cite{Bender2003}. The method is analogous to Kohn-Sham density functional theory (DFT) used in electronic systems. Three types of functionals are frequently used nowadays: non-relativistic Skyrme functional \cite{Skyrme1959,Vautherin1972,Reinhard2011} with zero-range two-body and density dependent interaction, finite-range Gogny force \cite{Gogny1980,Gogny2009} and relativistic (covariant) mean-field \cite{Walecka1986,Vretenar2005,Niksic2014}. Typical approach employs Hartree-Fock-Bogoliubov or HF+BCS calculation scheme to obtain ground state and single-(quasi)particle wavefunctions and energies. These results are then utilized to fit the parameters of the functional to experimental data, thus obtaining various parametrizations suitable for specific aims, such as: calculation of mass-table, charge radii, fission barriers, spin-orbit splitting and giant resonances. Mean-field calculation can be extended by taking a superposition of more Slater determinants and by restoration of broken symmetries (particle number, angular momentum), leading to the \emph{generator coordinate method}, which is suitable for description of shape coexistence and low-energy excited states (including rotational) \cite{Rodriguez2010,Yao2011}.

Random Phase Approximation (RPA) is a textbook standard \cite{Ring1980} to calculate one-phonon excitations of the nucleus, suitable also for the mean-field functionals. In practice, it is a method widely utilized for calculation of giant dipole resonances and other strength functions (giant monopole, quadrupole and M1 spin-flip resonances). Increasing computing power has enabled to employ fully self-consistent residual interaction derived from the same density functional as the underlying ground state. While the spherical nuclei can be treated directly (by matrix diagonalization) \cite{Reinhard1992,Terasaki2005,Colo2013}, axially deformed nuclei still pose certain difficulties due to large matrix dimensions \cite{Terasaki2010,Yoshida2013}. Our group developed a separable RPA (SRPA) approach for Skyrme functional \cite{Nesterenko2002,Nesterenko2006}, which greatly reduces the computational cost for deformed nuclei by utilizing separable residual interaction, entirely derived from the underlying functional by means of multi-dimensional linear response theory.

Skyrme RPA is used in our group mainly in its separable form and assuming the axial symmetry. Therefore, the primary aim of my work was a derivation and implementation of RPA in the spherical symmetry. To clarify the remaining issues, I developed also the full RPA in axial symmetry and a spherical Hartree-Fock for closed-shell nuclei.

The present work gives a derivation of convenient formalism for rotationally-invariant treatment of spherical Skyrme RPA (both full and separable). Both time-even and time-odd terms of Skyrme functional are employed, so the method is suitable for various electric and magnetic multipolarities. Then, the corresponding computer codes were constructed. Programs \texttt{sph\_qrpa} and \texttt{sph\_srpa} take wavefunctions from Reinhard's \texttt{haforpa}, which is a grid-based Skyrme HF+BCS code. Due to a restricted model space in \texttt{haforpa} (22--23 major shells), I wrote also a Skyrme Hartree-Fock code (without pairing) based on the spherical-harmonic-oscillator (SHO) basis, which allows to extend the model space to over 100 major shells. Subsequent RPA then leads to almost complete elimination of the spurious center-of-mass contribution in E1 transitions.

Detailed expressions for matrix elements, applicable to full RPA, were also derived for axial symmetry. RPA code \texttt{skyax\_qrpa} was written to deal with wavefunctions of Reinhard's \texttt{skyax}, a Skyrme HF+BCS code for deformed nuclei working with a cylindrical coordinate grid. Due to large computational demands, special care was taken to vectorize and parallelize the code, to make it suitable for routine calculations on the available multi-processor workstations (with 12 CPU cores and more than 32 GB of RAM).

The new codes were first tuned with respect to the basis parameters and the size of the configuration space, with the aim of consistent RPA results. Then, full and separable RPA are compared to get a set of the most efficient input operators. Selected nuclear properties were then calculated and compared to the experimental data, such as giant electric dipolar (E1) resonance (GDR) with its low-energy ``pygmy'' part, isoscalar giant monopolar (E0) resonance (GMR), M1 and E2 strength functions in spherical and deformed nuclei. The importance of spin and tensor terms of Skyrme functional is demonstrated for M1 and toroidal E1 resonances. Since the strength functions are calculated in the long-wave approximation, a comparison with exact transition operator is presented as well. Finally, a toroidal nature of the low-energy E1 (pygmy) transitions is demonstrated, as was also published in our recent papers \cite{Repko2013,Reinhard2014}.

The thesis is organized as follows. First, I give a detailed treatment of various terms of the nuclear density functional in chapter \emph{\ref{ch_theory} Theoretical formalism}.
\begin{equation}
\label{full_hamil}
\mathcal{H} = \mathcal{H}_\mathrm{kin} + \mathcal{H}_\mathrm{Sk} + \mathcal{H}_\mathrm{coul} + \mathcal{H}_\mathrm{xc} + \mathcal{H}_\mathrm{pair} + \mathcal{H}_\mathrm{c.m.}
\end{equation}
Kinetic and direct Coulomb terms are
\begin{align}
\mathcal{H}_\mathrm{kin} &= \int\mathrm{d}^3r \bigg(\frac{\hbar^2}{2m_p}\tau_p(\vec{r}) + \frac{\hbar^2}{2m_n}\tau_n(\vec{r})\bigg), \\
\mathcal{H}_\mathrm{coul} &= \frac{1}{2}\frac{e^2}{4\pi\epsilon_0}
\iint\mathrm{d}^3r_1\mathrm{d}^3r_2
\frac{\rho_p(\vec{r}_1)\rho_p(\vec{r}_2)}{|\vec{r}_1-\vec{r}_2|},
\end{align}
where the densities ($\rho,\,\tau$) will be defined in (\ref{Jd_gs}). Skyrme functional $\mathcal{H}_\mathrm{Sk}$, including its implementation in RPA, is treated for spherical symmetry in section \ref{sec_skyr_sph} and for axial symmetry in section \ref{sec_skyr_ax}. Its derivation from the two-body interaction is given in appendix \ref{app_skyr-dft}. Direct Coulomb interaction and numerical integration in general are discussed in section \ref{sec_coul}, and the exchange Coulomb interaction is taken in Slater approximation \cite{Slater1951}:
\begin{equation}
\label{xc}
\mathcal{H}_\mathrm{xc} = -\frac{3}{4}\bigg(\frac{3}{\pi}\bigg)^{\!1/3}
\frac{e^2}{4\pi\epsilon_0}\int\mathrm{d}^3 r \rho_p^{4/3}(\vec{r})
\end{equation}
Pairing interaction $\mathcal{H}_\mathrm{pair}$ is given in section \ref{sec_pair}, and finally, the subtraction of center-of-mass energy is described in section \ref{sec_kin-cm}.  The computer programs and their tuning are discussed in chapter \emph{\ref{ch_num} Numerical codes} and the physical results of the calculations, mainly in terms of strength functions and transition currents, are given in chapter \emph{\ref{ch_results} Physical results}. SRPA formalism adapted to spherical symmetry is given in appendix \ref{app_SRPA}.

\chapter{Theoretical formalism}\label{ch_theory}
This chapter gives a detailed account of the calculation of Skyrme Hartree-Fock and RPA (i.e., nuclear ground state and small-amplitude excitations) \cite{Ring1980} by means of single-particle (s.p.) wavefunctions decomposed by assuming rotational symmetry -- either spherical or axial (cylindrical). Besides Skyrme functional, it was necessary to treat also the Coulomb interaction, pairing interaction, transition operators, and kinetic center-of-mass term \cite{Reinhard2011}. The derived formulae were implemented in the computer programs as mentioned in the introduction, with the exception of Coulomb integral in cartesian coordinates, which is given only as a kind-of toy-model. More specifically, the programs included: spherical closed-shell HF in SHO basis, spherical full and separable RPA in SHO basis and on the radial grid, and axial full RPA on the 2D grid -- the results of these calculations are given in chapters \ref{ch_num} and \ref{ch_results}. Separable RPA, which is a numerically efficient method based on the linear response theory \cite{Nesterenko2002,Nesterenko2006}, is treated only in appendix \ref{app_SRPA}, to avoid unnecessary details in this chapter.

Since the primary aim was to derive the spherical RPA, the formalism given below is optimized in this direction. Particular attention was given also to the precise evaluation of the Coulomb integral by means of Euler-Maclaurin corrections, and to the evaluation of kinetic center-of-mass term for HF and RPA. Both of these topics seem to have little coverage in the literature on nuclear density functionals.

Notation of Clebsch-Gordan coefficients and most of the formulae used in the derivation are taken from the book of Varshalovich \cite{Varshalovich1988}. Detailed derivation of the utilized formulae can be found also in my notes about special functions in quantum mechanics \cite{Repko-specf_qm} (in Slovak).

Before coming to the theory itself, a few preliminary comments are given here in order to clarify the further utilization of a bra-ket notation. When applying single-particle expressions to a many-body system, it is necessary to distinguish whether the bra-ket formulation of matrix elements is understood as in the antisymmetrized many-body system, described by the Slater determinants (or equivalently by creation and annihilation operators; $P$ is a permutation of indices)
\begin{subequations}
\label{slater}
\begin{align}
{\langle\vec{r}_1,\vec{r}_2,\ldots\vec{r}_n|1,2,3,\ldots n\rangle}_\mathrm{Slater}
&= \frac{1}{\sqrt{n!}}\sum_P \mathrm{sign}(P)
\psi_{P(1)}(\vec{r}_1)\psi_{P(2)}(\vec{r}_2)\ldots\psi_{P(n)}(\vec{r}_n) \\
\Leftrightarrow\quad
{|1,2,3,\ldots n\rangle}_\mathrm{Slater} &= \hat{a}_1^+\hat{a}_2^+\ldots\hat{a}_n^+|\rangle
\end{align}
\end{subequations}
or they are meant only as a shortcut for non-symmetrized integral
\begin{equation}
\label{nonsym}
{\langle\alpha\beta|\hat{V}|\gamma\delta\rangle}_\mathrm{nonsym} =
\int \psi_\alpha^\dagger(\vec{r}_1) \psi_\beta^\dagger(\vec{r}_2)
\hat{V}(\vec{r}_1,\vec{r}_2)
\psi_\gamma(\vec{r}_1) \psi_\delta(\vec{r}_2)\,\mathrm{d}^3 r_1\mathrm{d}^3 r_2
\end{equation}
In most cases below, the bra-ket notation is meant as (\ref{nonsym}), with the exception of sections \emph{\ref{sec_fullrpa} Full RPA} and \emph{\ref{sec_pair} Pairing}, where many-body Slater states (\ref{slater}) or their linear combinations are used. Many-body matrix element is presumed also in the following shortcut for commutators, which is evaluated in Hartree-Fock (or HF+BCS) ground state:
\begin{equation}
\langle[\hat{A},\hat{B}]\rangle \equiv
{\langle\mathrm{HF}|\hat{A}\hat{B}-\hat{B}\hat{A}|\mathrm{HF}\rangle}_\mathrm{Slater}
\end{equation}
Conversion between Slater and non-symmetrized two-body matrix element is
\begin{equation}
\label{V_2ph}
\langle\alpha\beta|\frac{1}{2}\sum_{i,j}^N\hat{V}(\vec{r}_i,\vec{r}_j)
{|\gamma\delta\rangle}_\mathrm{Slater}
= {\langle\alpha\beta|\hat{V}(\vec{r}_1,\vec{r}_2)|\gamma\delta\rangle}_\mathrm{nonsym}
- {\langle\alpha\beta|\hat{V}(\vec{r}_1,\vec{r}_2)|\delta\gamma\rangle}_\mathrm{nonsym}
\end{equation}
on the condition that all s.p.~states $\alpha,\beta,\gamma,\delta$ are different (with zero overlap) and $\hat{V}(\vec{r}_i,\vec{r}_j) = \hat{V}(\vec{r}_j,\vec{r}_i)$. Notation $|\alpha\beta\rangle_\mathrm{Slater}$ can mean either a two-particle state ($N=2$) or a many-particle state ($N\geq2$), where the undisclosed states are the same as in $|\gamma\delta\rangle_\mathrm{Slater}$. When the matrix element is calculated between the same many-body Slater states, as is the case of Hartree-Fock total energy, the result is the following (with sums running over the occupied single-particle states):
\begin{align}
\langle\mathrm{HF}|\sum_i\hat{T}(\vec{r}_i)+{}&\frac{1}{2}\sum_{i,j}\hat{V}(\vec{r}_i,\vec{r}_j){|\mathrm{HF}\rangle}_\mathrm{Slater} = \nonumber\\
\label{V_HF}
&= \sum_\gamma{\langle\gamma|\hat{T}(\vec{r})|\gamma\rangle}_\mathrm{nonsym}
+\frac{1}{2}\sum_{\alpha\beta}{\langle\alpha\beta|\hat{V}(\vec{r}_1,\vec{r}_2)|\alpha\beta\rangle}_\mathrm{nonsym} \nonumber\\
&\hspace{85pt}{}-\frac{1}{2}\sum_{q=p,n}\sum_{\alpha\beta\in q}
{\langle\alpha\beta|\hat{V}(\vec{r}_1,\vec{r}_2)|\beta\alpha\rangle}_\mathrm{nonsym}
\end{align}
Prescriptions (\ref{V_2ph}) and (\ref{V_HF}) can be unified by means of creation and annihilation operators:
\begin{equation}
\frac{1}{2}\sum_{i,j}^N\hat{V}(\vec{r}_i,\vec{r}_j) = \frac{1}{2}
\sum_{\alpha\beta\gamma\delta}
{\langle\alpha\beta|\hat{V}(\vec{r}_1,\vec{r}_2)|\gamma\delta\rangle}_\mathrm{nonsym}
\hat{a}^+_\alpha\hat{a}^+_\beta\hat{a}_\delta^{\phantom{|}}\hat{a}_\gamma^{\phantom{|}}
\end{equation}

\section{Skyrme interaction and density functional}\label{sec_skyrme}
Skyrme interaction is a phenomenological approach to nuclear potential, which includes spatial derivatives in addition to the local densities. Its definition usually starts with a two-body density-dependent interaction \cite{Vautherin1972}
\begin{align}
\hat{V}_\mathrm{Sk}(\vec{r}_1,\vec{r}_2) & =
t_0(1+x_0\hat{P}_\sigma)\delta(\vec{r}_1-\vec{r}_2)
-\frac{1}{8}t_1(1+x_1\hat{P}_\sigma) \nonumber\\
&\qquad\qquad\qquad{}\times \big[
(\overleftarrow{\nabla}_1-\overleftarrow{\nabla}_2)^2\delta(\vec{r}_1-\vec{r}_2) + \delta(\vec{r}_1-\vec{r}_2)(\overrightarrow{\nabla}_1-\overrightarrow{\nabla}_2)^2
\big] \nonumber\\
& \quad{}+\frac{1}{4}t_2(1+x_2\hat{P}_\sigma)(\overleftarrow{\nabla}_1-\overleftarrow{\nabla}_2) \cdot \delta(\vec{r}_1-\vec{r}_2)(\overrightarrow{\nabla}_1-\overrightarrow{\nabla}_2)
\nonumber\\
& \quad{}+\frac{1}{6}t_3(1+x_3\hat{P}_\sigma)\delta(\vec{r}_1-\vec{r}_2) \rho^\alpha\Big(\frac{\vec{r}_1+\vec{r}_2}{2}\Big) \nonumber\\
\label{V_skyrme}
& \quad {}+\frac{\mathrm{i}}{4}t_4
(\vec{\sigma}_1+\vec{\sigma}_2)\cdot\big[(\overleftarrow{\nabla}_1-\overleftarrow{\nabla}_2) \times \delta(\vec{r}_1-\vec{r}_2)(\overrightarrow{\nabla}_1-\overrightarrow{\nabla}_2)\big]
\end{align}
with parameters $t_0,t_1,t_2,t_3,t_4,x_0,x_1,x_2,x_3,\alpha$ and a spin-exchange operator
\begin{equation}
\label{P_sigma}
\hat{P}_\sigma = \frac{1}{2}(1+\vec{\sigma}_1\cdot\vec{\sigma}_2) =
\frac{1+\sigma_{1z}\sigma_{2z}}{2} + \sigma_{1+}\sigma_{2-} + \sigma_{1-}\sigma_{2+}, \quad
\sigma_\pm = \frac{\sigma_x\pm\sigma_y}{2}.
\end{equation}
Since it is a zero-range interaction, the solution of a many-body problem by Hartree-Fock can be equivalently reformulated as a density functional theory \cite{Vautherin1972,Reinhard1992} (given in detail in appendix \ref{app_skyr-dft}), and the complete density functional is
\begin{small}
\begin{align}
\mathcal{H}_\mathrm{Sk} & = \frac{1}{2}\sum_{\alpha\beta} \langle \alpha\beta|\hat{V}_\mathrm{Sk}|\alpha\beta\rangle -
\frac{1}{2}\sum_{q=p,n}\sum_{\alpha\beta\in q} \langle \alpha\beta|\hat{V}_\mathrm{Sk}|\beta\alpha\rangle \nonumber\\
& = \int\mathrm{d}^3 r\bigg\{ \frac{b_0}{2}\rho^2 - \frac{b_0'}{2}\sum_q\rho_q^2
+b_1(\rho\tau\!-\!\vec{j}^{\,2}) - b_1'\sum_q(\rho_q\tau_q\!-\!\vec{j}_q^{\,2})
+\frac{b_2}{2}(\vec{\nabla}\rho)^2 - \frac{b_2'}{2}\sum_q(\vec{\nabla}\rho_q)^2 \nonumber\\
&\qquad {}+\tilde{b}_1\Big(\vec{s}\cdot\vec{T} - \sum_{ij} \mathcal{J}_{ij}^2\Big)
+\tilde{b}_1'\sum_q\Big(\vec{s}_q\cdot\vec{T}_q-\sum_{ij} \mathcal{J}_{q;ij}^2\Big)
+\frac{b_3}{3}\rho^{\alpha+2}-\frac{b_3'}{3}\rho^\alpha\sum_q\rho_q^2 \nonumber\\
&\qquad {}-b_4\big[\rho\vec{\nabla}\cdot\vec{\mathcal{J}}
+\vec{s}\cdot(\vec{\nabla}\times\vec{j})\big]
-b_4'\sum_q\big[\rho_q\vec{\nabla}\cdot\vec{\mathcal{J}}_q+\vec{s}_q\cdot (\vec{\nabla}\times\vec{j}_q)\big] \nonumber\\
\label{Skyrme_DFT}
&\qquad {}+\frac{\tilde{b}_0}{2}\vec{s}^2-\frac{\tilde{b}_0'}{2}\sum_q\vec{s}_q^{\,2}
+\frac{\tilde{b}_2}{2}\sum_{ij}(\nabla_i s_j)^2-\frac{\tilde{b}_2'}{2}\sum_q \sum_{ij}(\nabla_i s_j)_q^2
+\frac{\tilde{b}_3}{3}\rho^\alpha\vec{s}^{\,2}-\frac{\tilde{b}_3'}{3}\rho^\alpha\sum_q \vec{s}_q^{\,2} \bigg\}
\end{align}
\end{small} \\[-6pt]
where the last line contains the spin terms, which are usually omitted. However, they have quite important contribution for magnetic excitations, as will be shown in section \ref{sec_spin-tens}, so I am using them in all calculations. Parameters $b_j$ depend on the parameters $t_j,x_j$ from (\ref{V_skyrme}):
\begin{align}
b_0 &= \tfrac{t_0(2+x_0)}{2}, \quad b_0' = \tfrac{t_0(1+2x_0)}{2}, \quad
\tilde{b}_0 = \tfrac{t_0 x_0}{2}, \quad \tilde{b}_0' = \tfrac{t_0}{2}, \nonumber\\
b_1 &= \tfrac{t_1(2+x_1)+t_2(2+x_2)}{8}, \quad
b_1' = \tfrac{t_1(1+2x_1)-t_2(1+2x_2)}{8}, \quad
\tilde{b}_1 = \tfrac{t_1 x_1 + t_2 x_2}{8}, \quad
\tilde{b}_1' = \tfrac{-t_1+t_2}{8}, \nonumber\\
b_2 &= \tfrac{3t_1(2+x_1)-t_2(2+x_2)}{16}, \quad
b_2' = \tfrac{3t_1(1+2x_1)+t_2(1+2x_2)}{16}, \quad
\tilde{b}_2 = \tfrac{3t_1 x_1 - t_2 x_2}{16}, \quad
\tilde{b}_2' = \tfrac{3t_1 + t_2}{16}, \nonumber\\
b_3 &= \tfrac{t_3(2+x_3)}{8}, \quad b_3' = \tfrac{t_3(1+2x_3)}{8}, \quad
\tilde{b}_3 = \tfrac{t_3 x_3}{8}, \quad \tilde{b}_3' = \tfrac{t_3}{8}, \quad
b_4 = b_4' = \tfrac{t_4}{2}
\end{align}
Most Skyrme parametrizations set explicitly $\tilde{b}_1=\tilde{b}_1'=0$ and this fact is denoted here as exclusion of the ``tensor term'' (not to be confused with spin-tensor term utilized in the shell model). There are also parametrizations fitted with the tensor term included, e.g. SGII \cite{SGII}, SLy7 \cite{SLy6}, SkT6 \cite{SkT6}.

The ground state densities (denoted in general as $J_d(\vec{r}) = \langle\hat{J}_d(\vec{r})\rangle$) are defined:
\begin{align}
\rho_q(\vec{r}) &= \sum_{\alpha\in q}v_\alpha^2 \psi_\alpha^\dagger(\vec{r})\psi_\alpha^{\phantom{|}}(\vec{r}),\quad
\rho(\vec{r}) = \sum_{q=p,n}\rho_q(\vec{r}),\quad
\tau(\vec{r}) = \sum_{\alpha}v_\alpha^2 [\vec{\nabla}\psi_\alpha(\vec{r})]^\dagger
\cdot[\vec{\nabla}\psi_\alpha(\vec{r})], \nonumber\\
\mathcal{J}_{jk}(\vec{r}) &= \frac{\mathrm{i}}{2}\sum_{\alpha}v_\alpha^2
\big\{[\partial_j\sigma_k\psi_\alpha(\vec{r})]^\dagger\psi_\alpha(\vec{r}) -
\psi_\alpha^\dagger(\vec{r})[\partial_j\sigma_k\psi_\alpha(\vec{r})]\big\},
\quad \mathcal{J}_i(\vec{r}) = \sum_{ijk} \varepsilon_{ijk}\mathcal{J}_{jk}(\vec{r}), \nonumber\\
\vec{\mathcal{J}}(\vec{r}) &= \frac{\mathrm{i}}{2}\sum_{\alpha}v_\alpha^2
\big\{[(\vec{\nabla}\times\vec{\sigma})\psi_\alpha(\vec{r})]^\dagger
\psi_\alpha(\vec{r}) - \psi_\alpha^\dagger(\vec{r})[(\vec{\nabla}\times\vec{\sigma})\psi_\alpha(\vec{r})]\big\}, \nonumber\\
\vec{j}(\vec{r}) &= \frac{\mathrm{i}}{2}\sum_{\alpha}v_\alpha^2
\big\{[\vec{\nabla}\psi_\alpha(\vec{r})]^\dagger\psi_\alpha(\vec{r}) -
\psi_\alpha^\dagger(\vec{r})[\vec{\nabla}\psi_\alpha(\vec{r})]\big\},\quad
\vec{s}(\vec{r}) = \sum_{\alpha} v_\alpha^2
\psi_\alpha^\dagger(\vec{r})\vec{\sigma}\psi_\alpha^{\phantom{|}}(\vec{r}), \nonumber\\
\label{Jd_gs}
\vec{T}(\vec{r}) &= \sum_{\alpha} v_\alpha^2 \sum_j
[\partial_j\psi_\alpha(\vec{r})]^\dagger\vec{\sigma}[\partial_j\psi_\alpha(\vec{r})]
\end{align}
Densities $\rho,\tau,\mathcal{J}$ are time-even and currents $\vec{j},\vec{s},\vec{T}$ are time-odd, and $v_\alpha^2$ represents occupation probability, defined later in (\ref{bogoliubov}). Time-odd currents are zero in the ground state ($0^+$) of the even-even nuclei. The operators corresponding to the densities and currents are
\begin{small}
\begin{align}
\textrm{density:}\quad & \hat{\rho}(\vec{r}_0) = \delta(\vec{r}-\vec{r}_0), \qquad
\textrm{kinetic energy:} \quad \hat{\tau}(\vec{r}_0) =
\overleftarrow{\nabla}\cdot\delta(\vec{r}-\vec{r}_0)\overrightarrow{\nabla}, \nonumber\\
\textrm{spin-orbital:}\quad & \hat{\mathcal{J}}_{jk}(\vec{r}_0) = \tfrac{\mathrm{i}}{2}
\big[\overleftarrow{\nabla}_{\!j}\sigma_k\delta(\vec{r}-\vec{r}_0)
-\delta(\vec{r}-\vec{r}_0)\overrightarrow{\nabla}_{\!j}\sigma_k\big], \nonumber\\
\textrm{vector spin-orbital:}\quad & \hat{\vec{\mathcal{J}}}(\vec{r}_0) = \tfrac{\mathrm{i}}{2} \big[\overleftarrow{\nabla}\times\vec{\sigma}\delta(\vec{r}-\vec{r}_0)
-\delta(\vec{r}-\vec{r}_0)\overrightarrow{\nabla}\times\vec{\sigma}\big],\quad
\hat{\mathcal{J}}_i = \textstyle\sum_{ijk}\varepsilon_{ijk}\mathcal{J}_{jk}, \nonumber\\
\textrm{current:}\quad & \hat{\vec{j}}(\vec{r}_0) = \tfrac{\mathrm{i}}{2}
\big[\overleftarrow{\nabla}\delta(\vec{r}-\vec{r}_0)
-\delta(\vec{r}-\vec{r}_0)\overrightarrow{\nabla}\big],\quad
\textrm{spin:}\quad \hat{\vec{s}}(\vec{r}_0) = \vec{\sigma}\delta(\vec{r}-\vec{r}_0), \nonumber\\
\label{Jd_op}
\textrm{kinetic energy-spin:}\quad & \hat{T}_j(\vec{r}_0) =
\overleftarrow{\nabla}\cdot\sigma_j\delta(\vec{r}-\vec{r}_0)\overrightarrow{\nabla}
\end{align}
\end{small} \\[-6pt]
and they are understood as single-particle operators in many-body system; more explicit notation would be, e.g.
\begin{equation*}
\hat{\rho}_q(\vec{r}_0) = \sum_{i\in q}\delta(\vec{r}_i-\vec{r}_0)
\end{equation*}

Spin-orbital current $\mathcal{J}_{jk}$ and current $\nabla_j s_k$ have two indices, so they can be interpreted as spherical tensor operators and then decomposed into scalar, vector and (symmetric) rank-2 tensor part, using orthogonality of Clebsch-Gordan coefficients, with components corresponding to angular quantum numbers 0, 1 and 2 (i.e., total number of components is $1+3+5=9=3^2$).
\begin{small}
\begin{equation}
\sum_{i,j=x,y,z} \mathcal{J}_{ij}^2 = \sum_{\mu,\nu}^{{-1},0,1} (-1)^{\mu+\nu} \mathcal{J}_{\mu\nu} \mathcal{J}_{-\mu,-\nu}
\end{equation}
\begin{equation}
\mathcal{J}_{\tilde{\mu}\tilde{\nu}} = C_{1\tilde{\mu}1\tilde{\nu}}^{00}{[\mathcal{J}_{\mu\otimes\nu}]}_0
+ \sum_{M=-1}^1 C_{1\tilde{\mu}1\tilde{\nu}}^{1M}{[\mathcal{J}_{\mu\otimes\nu}]}_{1M}
+ \sum_{M=-2}^2 C_{1\tilde{\mu}1\tilde{\nu}}^{2M}{[\mathcal{J}_{\mu\otimes\nu}]}_{2M}
\end{equation}
\begin{align}
\,{[\mathcal{J}_{\mu\otimes\nu}]}_0 &= \sum_{\mu\nu}C_{1\mu1\nu}^{00}\mathcal{J}_{\mu\nu} =
-\sum_{\mu=-1}^1 \frac{(-1)^\mu}{\sqrt{3}} \mathcal{J}_{\mu,-\mu} =
-\frac{1}{\sqrt{3}}\sum_i \mathcal{J}_{ii} = -\frac{1}{\sqrt{3}}\mathcal{J}_s \nonumber\\[-2pt]
\,{[\mathcal{J}_{\mu\otimes\nu}]}_{1M} &= \sum_{\mu\nu}C_{1\mu1\nu}^{1M}\mathcal{J}_{\mu\nu} =
\frac{\mathrm{i}}{\sqrt{2}} \Big[\sum_{ij} \varepsilon_{ijk}\mathcal{J}_{ij}\Big]_{k\rightarrow M} =
\frac{\mathrm{i}}{\sqrt{2}} {[\vec{\mathcal{J}}]}_M \\
\,{[\mathcal{J}_{\mu\otimes\nu}]}_{2M} &= \sum_{\mu\nu}C_{1\mu1\nu}^{2M}\mathcal{J}_{\mu\nu} =
\mathcal{J}_{tM} = {[\boldsymbol{\mathcal{J}}_{\!t}]}_{M} \nonumber
\end{align}
\begin{equation}
\label{Jsq-decom}
\sum_{ij} \mathcal{J}_{ij}^2 = \frac{1}{3}\Big(\sum_i \mathcal{J}_{ii}\Big)^2 +
\frac{1}{2}\vec{\mathcal{J}}^2 + \sum_{m=-2}^2 (-1)^m
{[\boldsymbol{\mathcal{J}}_{\!t}^{\vphantom{*}}]}_m {[\boldsymbol{\mathcal{J}}_{\!t}^{\vphantom{*}}]}_{-m} =
\frac{1}{3}\mathcal{J}_s^2 + \frac{1}{2}\vec{\mathcal{J}}^2 + \boldsymbol{\mathcal{J}}_{\!t}^2
\end{equation}
\end{small} \\[-6pt]
Decomposition of vector spin-orbital current in the convention of tensor operators is (see \cite[(1.2.28)]{Varshalovich1988} for vector product formula)
\begin{equation}
{[\vec{\mathcal{J}}]}_\mathrm{sph} = \begin{pmatrix} (-\mathcal{J}_x-\mathrm{i}\mathcal{J}_y)/\sqrt{2} \\
\mathcal{J}_z \\ (\mathcal{J}_x-\mathrm{i}\mathcal{J}_y)/\sqrt{2} \end{pmatrix} = \begin{pmatrix}
\mathrm{i}(-\mathcal{J}_{10} + \mathcal{J}_{01}) \\ \mathrm{i}(-\mathcal{J}_{1,-1} + \mathcal{J}_{-1,1}) \\
\mathrm{i}(\mathcal{J}_{-1,0} - \mathcal{J}_{0,-1}) \end{pmatrix}
\end{equation}
The tensor part is then
\begin{small}
\begin{align}
{[\boldsymbol{\mathcal{J}}_{\!t}]}_{\pm2} &= \mathcal{J}_{\pm1,\pm1}
= \frac{\mathcal{J}_{xx} \pm \mathrm{i}\mathcal{J}_{xy} \pm \mathrm{i}\mathcal{J}_{yx} - \mathcal{J}_{yy}}{2} \nonumber\\[2pt]
{[\boldsymbol{\mathcal{J}}_{\!t}]}_{\pm1} &= \frac{\mathcal{J}_{\pm1,0}+\mathcal{J}_{0,\pm1}}{\sqrt{2}}
= \frac{\mp\mathcal{J}_{xz} - \mathrm{i}\mathcal{J}_{yz} \mp \mathcal{J}_{zx} - \mathrm{i}\mathcal{J}_{zy}}{2} \\
{[\boldsymbol{\mathcal{J}}_{\!t}]}_0 &= \frac{\mathcal{J}_{1,-1} + 2\mathcal{J}_{00} + \mathcal{J}_{-1,1}}{\sqrt{6}} =
\frac{-\mathcal{J}_{xx} - \mathcal{J}_{yy} + 2\mathcal{J}_{zz}}{\sqrt{6}} \nonumber
\end{align}
\end{small} \\[-6pt]
To check the decomposition (\ref{Jsq-decom}), I can substitute above expressions into it:
\begin{align*}
\mathcal{J}_s^2 &= \mathcal{J}_{xx}^2 + \mathcal{J}_{yy}^2 + \mathcal{J}_{zz}^2 
+2(\mathcal{J}_{xx}\mathcal{J}_{yy} + \mathcal{J}_{yy}\mathcal{J}_{zz} + \mathcal{J}_{zz}\mathcal{J}_{xx}) \qquad\qquad \times\frac{1}{3} \\
\vec{\mathcal{J}}^2 &= \mathcal{J}_{yz}^2 + \mathcal{J}_{zy}^2 + \mathcal{J}_{zx}^2 + \mathcal{J}_{xz}^2 + \mathcal{J}_{xy}^2 + \mathcal{J}_{yx}^2
-2(\mathcal{J}_{yz}\mathcal{J}_{zy} + \mathcal{J}_{zx}\mathcal{J}_{xz} + \mathcal{J}_{xy}\mathcal{J}_{yx}) \quad \times\frac{1}{2} \\
\boldsymbol{\mathcal{J}}_{\!t}^2 &= \frac{2}{3}(\mathcal{J}_{xx}^2 + \mathcal{J}_{yy}^2 + \mathcal{J}_{zz}^2
-\mathcal{J}_{xx}\mathcal{J}_{yy} - \mathcal{J}_{yy}\mathcal{J}_{zz} - \mathcal{J}_{zz}\mathcal{J}_{xx}) + \frac{1}{2}(\mathcal{J}_{xy}+\mathcal{J}_{yx})^2 \\
&\quad{}+\frac{1}{2}[(\mathcal{J}_{zx}+\mathcal{J}_{xz})^2 + (\mathcal{J}_{yz}+\mathcal{J}_{zy})^2]
\end{align*}

\section{Skyrme RPA in the spherically symmetric case}\label{sec_skyr_sph}
The complete treatment of various terms of Skyrme density functional, and the residual interaction derived from it, is given below for spherical symmetry. Some of these concepts are valid also for the axial symmetry, so the corresponding section will be accordingly shorter.
\subsection{Notation for one-body matrix elements}\label{sec_notation}
Spherical decomposition of a single-particle wavefunction (spin 1/2) is
\begin{equation}
\langle\vec{r}|\alpha\rangle = \psi_\alpha(\vec{r}) =
R_\alpha(r)\,\Omega_{j_\alpha m_\alpha}^{l_\alpha}(\vartheta,\varphi) =
R_\alpha(r) \sum_{\nu \xi}
C_{l_\alpha,\nu,\frac{1}{2},\xi}^{j_\alpha,m_\alpha}
Y_{l_\alpha\nu}^{\phantom{|}}(\vartheta,\varphi)\,
\chi_{\xi}^{\phantom{|}}
\end{equation}
with $\Omega_{j_\alpha m_\alpha}^{l_\alpha}$ denoting spin-orbitals and $\chi_{\xi}^{\phantom{|}}$ are spinors. Greek letters will be used for labeling of single-particle and single-quasiparticle states and should not be later confused with creation and annihilation operators for quasiparticles which are always denoted by a hat (i.e.~$\hat{\alpha}_\alpha,\,\hat{\alpha}_\beta^+$).

Time reversal of a nucleon wavefunction is defined as
\begin{equation}
\label{t_inv}
\psi_{\bar{\alpha}}(\vec{r}) = \hat{T}\psi_\alpha(\vec{r}) =
\mathrm{i}\sigma_y\psi_\alpha^*(\vec{r}) =
(-1)^{l_\alpha+j_\alpha+m_\alpha}\psi_{-\alpha}(\vec{r}),\quad
\hat{T}|\bar{\alpha}\rangle = -|\alpha\rangle
\end{equation}
where ${-}\alpha\equiv\{n_\alpha,j_\alpha,l_\alpha,-m_\alpha\}$. Time-parity of an operator $\hat{A}$, denoted as $\gamma_T^A$, is defined by the relation
\begin{equation}
\hat{T}^{-1}\hat{A}\hat{T} = \gamma_T^A\hat{A}^\dagger, \quad
\gamma_T^A = \Big\{
\begin{array}{l} +1:\quad\textrm{time-even} \\ -1:\quad\textrm{time-odd}\end{array}
\end{equation}
Single-particle matrix elements then satisfy
\begin{subequations}
\label{time-inv}
\begin{align}
\langle\bar{\alpha}|\hat{A}|\bar{\beta}\rangle &=
\langle\bar{\alpha}|\hat{A}\hat{T}|\beta\rangle =
\langle\alpha|\hat{T}^{-1}\hat{A}\hat{T}|\beta\rangle^* = 
\gamma_T^A\langle\beta|\hat{A}|\alpha\rangle \\
\langle\alpha|\hat{A}|\bar{\beta}\rangle &=
\langle\alpha|\hat{A}\hat{T}|\beta\rangle =
-\langle\bar{\alpha}|\hat{T}^{-1}\hat{A}\hat{T}|\beta\rangle^* = 
-\gamma_T^A\langle\beta|\hat{A}|\bar{\alpha}\rangle
\end{align}
\end{subequations}

To account for pairing (treated in more detail in section \emph{\ref{sec_pair} Pairing interaction}), quasiparticles are introduced by Bogoliubov transformation \cite[p.~234]{Ring1980}
\begin{equation}
\label{bogoliubov}
\begin{array}{ll}
\hat{a}_\beta^+ = u_\beta^{\phantom{|}} \hat{\alpha}_\beta^+
+ v_\beta^{\phantom{|}} \hat{\alpha}_{\bar{\beta}}^{\phantom{|}},\quad &
\{\hat{\alpha}_\alpha^{\phantom{|}},\hat{\alpha}_\beta^+\} =
\delta_{\alpha\beta},\\
\hat{a}_{\bar{\beta}}^+ = u_\beta^{\phantom{|}} \hat{\alpha}_{\bar{\beta}}^+
- v_\beta^{\phantom{|}} \hat{\alpha}_\beta^{\phantom{|}},\quad &
\{\hat{\alpha}_\alpha^{\phantom{|}},\hat{\alpha}_\beta^{\phantom{|}}\} = 0
\end{array}
\end{equation}
with real positive coefficients $u_\beta,v_\beta$ satisfying $u_\beta^2+v_\beta^2 = 1$. States $\alpha,\beta,\ldots$ are obtained from Skyrme Hartree-Fock iteration, which is appended by a solution of BCS equations to obtain $u_\beta,v_\beta$. The HF+BCS groud state of an even-even nucleus is then a vacuum with respect to quasiparticle annihilation operators $\hat{\alpha}_\beta$. One-body operator can be expressed as
\begin{align}
\hat{A} = \sum_{\alpha\beta}
\langle\alpha|\hat{A}|\beta\rangle\hat{a}_\alpha^+\hat{a}_\beta^{\phantom{+}}
= \sum_{\alpha\beta} &\langle\alpha|\hat{A}|\beta\rangle
\big(u_\alpha^{\phantom{|}}v_\beta^{\phantom{|}}
\hat{\alpha}_\alpha^+\hat{\alpha}_{\bar{\beta}}^+ + v_\alpha^{\phantom{|}}u_\beta^{\phantom{|}} \hat{\alpha}_{\bar{\alpha}}^{\phantom{|}}\hat{\alpha}_\beta^{\phantom{|}}
\nonumber\\[-8pt]
 &\qquad\ \ 
{}+u_\alpha^{\phantom{|}}u_\beta^{\phantom{|}}\hat{\alpha}_\alpha^+\hat{\alpha}_\beta^{\phantom{|}}
-v_\alpha^{\phantom{|}}v_\beta^{\phantom{|}}\hat{\alpha}_{\bar{\beta}}^+\hat{\alpha}_{\bar{\alpha}}^{\phantom{|}} + v_\alpha^2\delta_{\alpha\beta}\big)
\end{align}
In the RPA, the operators are evaluated only in the commutators in the ground state, $\langle[\hat{A},\hat{B}]\rangle$, and here, the non-zero contributions come only from $\hat{\alpha}_{\alpha}^{\phantom{|}}\hat{\alpha}_\beta^{\phantom{|}},\hat{\alpha}_{\alpha}^+\hat{\alpha}_\beta^+$. So I will drop all other terms and symmetrize according to (\ref{time-inv}) to obtain
\begin{subequations}
\label{2qp_operator}
\begin{align}
\hat{A} &= \frac{1}{2}\sum_{\alpha\beta}u_{\alpha\beta}^{(\gamma_T^A)}
\langle\alpha|\hat{A}|\bar{\beta}\rangle
(-\hat{\alpha}_\alpha^+\hat{\alpha}_\beta^+ + \gamma_T^A
\hat{\alpha}_{\bar{\alpha}}^{\phantom{*}}\hat{\alpha}_{\bar{\beta}}^{\phantom{*}}) \\
&= \frac{1}{2}\sum_{\alpha\beta}u_{\alpha\beta}^{(\gamma_T^A)}
\langle\bar{\alpha}|\hat{A}|\beta\rangle
(\hat{\alpha}_{\bar{\alpha}}^+\hat{\alpha}_{\bar{\beta}}^+ - \gamma_T^A \hat{\alpha}_\alpha^{\phantom{|}}\hat{\alpha}_\beta^{\phantom{|}})
\end{align}
\end{subequations}
where the pairing factors were abbreviated as
\begin{equation}
\label{u_ab}
u_{\alpha\beta}^{(+)} = u_\alpha^{\phantom{|}}v_\beta^{\phantom{|}}+ v_\alpha^{\phantom{|}} u_\beta^{\phantom{|}}, \qquad
u_{\alpha\beta}^{(-)} = u_\alpha^{\phantom{|}}v_\beta^{\phantom{|}}- v_\alpha^{\phantom{|}} u_\beta^{\phantom{|}}
\end{equation}
The sums are not restricted with respect to double counting, so the diagonal matrix elements are treated correctly. Ordering of the pairs $\alpha\beta$ with properly included diagonal matrix elements will be discussed at (\ref{order2qp}).

Vector hermitian operators can be rewritten as tensor operators of rank 1
\begin{equation}
\hat{A}_1 = (-\hat{A}_x-\mathrm{i}\hat{A}_y)/\sqrt{2},\quad
\hat{A}_0 = \hat{A}_z,\quad
\hat{A}_{-1} = (\hat{A}_x-\mathrm{i}\hat{A}_y)/\sqrt{2}
\end{equation}
I therefore define hermiticity of the tensor operator by a condition
\begin{equation}
\label{hermit}
\hat{A}_m^\dagger = (-1)^m\hat{A}_{-m}
\end{equation}
The same rule applies also to higher-rank tensor operators. Besides scalars and vectors, I will use rank-2 tensors (denoted by boldface, $\boldsymbol{A}$). When the rank is not specified, I will use upright bold symbols ($\mathbf{A}$). By the term ``rank'', I refer to the ``spin'' part of an operator (i.e., to its multi-component nature; but its meaning is closer to a photon spin, and not a nucleon spin).

Since the density and current operators depend on position, their angular part will be decomposed by orbital angular momentum ($L$) and total angular momentum ($J$ or $\lambda$) in terms of scalar ($Y_{LM}$), vector ($\vec{Y}_{JM}^L$) and tensor ($\boldsymbol{Y}_{\!JM}^L$) spherical harmonics, whose decomposition in terms of Clebsch-Gordan coefficients and tensor-operator-like components (denoted by $[\ ]_m$) is in general
\begin{align}
\label{sph_vectors}
\mathbf{Y}_{JM}^L(\vartheta,\varphi) &=
\sum_{m\mu}C_{Lms\mu}^{JM} Y_{Lm}\mathbf{e}_\mu =
\sum_{m=-s}^s (-1)^m \big[\mathbf{Y}_{JM}^L\big]_m\mathbf{e}_{-m} \\
&= (-1)^{J+L+M+s}\mathbf{Y}_{J,-M}^{L*}(\vartheta,\varphi)
\end{align}
where I choose $\vec{e}_0 = \vec{e}_z$ and $\mathbf{e}_0 = (2\vec{e}_z\vec{e}_z-\vec{e}_x\vec{e}_x-\vec{e}_y\vec{e}_y)/\sqrt{6}$; $s$ denotes the rank (0: scalar, 1: vector, 2: tensor), and $L\in\{J-s,\ldots J+s\}$. 

When $\hat{A}$ is a tensor operator with multipolarity $\lambda,\mu$, then, according to Wigner-Eckart theorem, I can factorize a Clebsch-Gordan coefficient from $\langle\alpha|\hat{A}|\beta\rangle$, and obtain a reduced matrix element, which will be denoted by $A_{\alpha\beta}$ (including the pairing factor), instead of bra-ket, not to cause confusion in many-body quasiparticle formalism.
\begin{equation}
\label{A_rme}
u_{\alpha\beta}^{(\gamma_T^A)}\langle\alpha|\hat{A}|\bar{\beta}\rangle =
\frac{(-1)^{l_\beta+j_\beta+m_\beta}A_{\alpha\beta}}{\sqrt{2j_\alpha+1}}
C_{j_\beta,-m_\beta,\lambda,\mu}^{j_\alpha,m_\alpha} =
\frac{(-1)^{l_\beta}A_{\alpha\beta}}{\sqrt{2\lambda+1}}
C_{j_\alpha m_\alpha j_\beta m_\beta}^{\lambda\mu}
\end{equation}
\begin{subequations}
\begin{align}
\hat{A} & = \frac{1}{2}\sum_{\alpha\beta}
\frac{(-1)^{l_\beta}A_{\alpha\beta}}{\sqrt{2\lambda+1}}
C_{j_\alpha m_\alpha j_\beta m_\beta}^{\lambda\mu}
(-\hat{\alpha}_\alpha^+\hat{\alpha}_\beta^+ + \gamma_T^A
\hat{\alpha}_{\bar{\alpha}}^{\phantom{*}}
\hat{\alpha}_{\bar{\beta}}^{\phantom{*}}) \\
& = \frac{1}{2}\sum_{\alpha\beta}
\frac{(-1)^{l_\alpha+\lambda+\mu+1}A_{\alpha\beta}}{\sqrt{2\lambda+1}}
C_{j_\alpha m_\alpha j_\beta m_\beta}^{\lambda,-\mu}
(\hat{\alpha}_{\bar{\alpha}}^+\hat{\alpha}_{\bar{\beta}}^+ - \gamma_T^A \hat{\alpha}_\alpha^{\phantom{|}}\hat{\alpha}_\beta^{\phantom{|}})
\end{align}
\end{subequations}
The commutator in the ground state then evaluates as
\begin{equation}
\label{comm}
\langle[\hat{A},\hat{B}]\rangle = \frac{1}{2}\sum_{\alpha\beta}
\frac{(-1)^{l_\alpha+l_\beta+\lambda+\mu}}{2\lambda+1}
(\gamma_T^A-\gamma_T^B)A_{\alpha\beta}B_{\alpha\beta}
\end{equation}
where the sum does not run over $m_\alpha,m_\beta$ anymore, and the operators are supposed to have the same $\lambda$ and opposite $\mu$.

The formalism of reduced matrix elements needs to be generalized to density and current operators (\ref{Jd_op}), which are position-dependent, in contrast with usual tensor operators (\ref{A_rme}). The outcome will be first demonstrated for ordinary density, by using (7.2.40) in \cite{Varshalovich1988}:
\begin{align}
\langle\alpha|\hat{\rho}(\vec{r})|\bar{\beta}\rangle &= R_\alpha(r) R_\beta(r)
(-1)^{l_\beta+j_\beta+m_\beta}
\Omega_{j_\alpha m_\alpha}^{l_\alpha\dagger}(\vartheta,\varphi)
\Omega_{j_\beta,-m_\beta}^{l_\beta}(\vartheta,\varphi)  \nonumber\\
\Omega_{j_\alpha m_\alpha}^{l_\alpha\dagger}
\Omega_{j_\beta,-m_\beta}^{l_\beta} &=
\sum_L (-1)^{j_\alpha+m_\alpha+j_\beta+L+\frac{1}{2}}
\sqrt{\tfrac{(2j_\alpha+1)(2j_\beta+1)(2l_\alpha+1)(2l_\beta+1)}{4\pi(2L+1)}}
\begin{Bmatrix} l_\alpha & l_\beta & L \\ j_\beta & j_\alpha & \frac{1}{2} \end{Bmatrix}
\nonumber\\[-5pt]
&\qquad\qquad{}\times C_{l_\alpha 0 l_\beta 0}^{L 0}
C_{j_\alpha,-m_\alpha,j_\beta,-m_\beta}^{L,m_\beta-m_\alpha}
Y_{L,-m_\beta-m_\alpha} \nonumber\\
&= \sum_L (-1)^{j_\beta+\frac{1}{2}}
\sqrt{\tfrac{(2j_\alpha+1)(2j_\beta+1)(2l_\alpha+1)(2l_\beta+1)}{4\pi(2j_\alpha+1)}}
\begin{Bmatrix} l_\alpha & l_\beta & L \\ j_\beta & j_\alpha & \frac{1}{2} \end{Bmatrix}
\nonumber\\[-5pt]
\label{rho_rme_example}
&\qquad\qquad{}\times C_{l_\alpha 0 l_\beta 0}^{L 0}
C_{j_\beta,-m_\beta,L,m_\beta+m_\alpha}^{j_\alpha,m_\alpha}
Y^*_{L,m_\beta+m_\alpha}(\vartheta,\varphi)
\end{align}
As can be seen, besides Clebsch-Gordan coefficient and numerical factors, there is a radial-dependent function and the complex-conjugated spherical harmonics (appearance of $Y^*$ can be understood as coming from the multipolar decomposition of the delta function to $\delta(r)Y(\hat{r})Y^*(\hat{r})$). In the generalization of the Wigner-Eckart theorem (\ref{A_rme}), I will absorb the radial dependence into the reduced matrix element, which will be denoted like $\rho_{\alpha\beta}^L(r)$. In general (see appendix \ref{app_Jab}), the multipolar expansion of the density and current operators, $\hat{\mathbf{J}}_d(\vec{r})$ (\ref{Jd_op}), contains spherical harmonics in its scalar, vector or tensor form: $\mathbf{Y}_{J,M}^{L*}(\vartheta,\varphi)$. I then define a reduced matrix element $J_{d;\alpha\beta}^{JL}(r)$ as
\begin{align}
u_{\alpha\beta}^{(\gamma_T^d)}\langle\alpha|\hat{\mathbf{J}}_d|\bar{\beta}\rangle
& = \sum_{LJ} \frac{(-1)^{l_\beta+j_\beta+m_\beta}
J_{d;\alpha\beta}^{JL}(r)}{\sqrt{2j_\alpha+1}}
C_{j_\beta,-m_\beta,J,m_\alpha+m_\beta}^{j_\alpha,m_\alpha}
\mathbf{Y}_{J,m_\alpha+m_\beta}^{L*}(\vartheta,\varphi) \nonumber\\
\label{dens_rme}
& = \sum_{LJ} J_{d;\alpha\beta}^{JL}(r)\frac{(-1)^{l_\beta}}{\sqrt{2J+1}}
C_{j_\alpha m_\alpha j_\beta m_\beta}^{J,m_\alpha+m_\beta}
\mathbf{Y}_{J,m_\alpha+m_\beta}^{L*}(\vartheta,\varphi)
\end{align}
The operators are then expressed in terms of quasiparticles (\ref{2qp_operator})
\begin{subequations}
\label{rme}
\begin{align}
\!\!\hat{\mathbf{J}}_d(\vec{r}) & = \tfrac{1}{2}\!\!\!\sum_{\alpha\beta LJM}\!\!\!
J_{d;\alpha\beta}^{JL}(r)\frac{(-1)^{l_\beta}}{\sqrt{2J+1}}
C_{j_\alpha m_\alpha j_\beta m_\beta}^{JM} \mathbf{Y}_{JM}^{L*}(\vartheta,\varphi)
(-\hat{\alpha}_{\alpha}^+ \hat{\alpha}_{\beta}^+ + \gamma_T^d
\hat{\alpha}_{\bar{\alpha}} \hat{\alpha}_{\bar{\beta}}) \\
& = \tfrac{1}{2}\!\!\!\sum_{\alpha\beta LJM}\!\!\!
J_{d;\alpha\beta}^{JL}(r)\frac{(-1)^{l_\alpha+L+s+1}}{\sqrt{2J+1}}
C_{j_\alpha m_\alpha j_\beta m_\beta}^{JM} \mathbf{Y}_{JM}^L(\hat{r})
(\hat{\alpha}_{\bar{\alpha}}^+ \hat{\alpha}_{\bar{\beta}}^+ - \gamma_T^d
\hat{\alpha}_{\alpha}^{\phantom{*}} \hat{\alpha}_{\beta}^{\phantom{*}})
\end{align}
\end{subequations}
All the density and current operators are hermitian and their reduced matrix elements satisfy
\begin{equation}
\label{rme_hermit}
J_{d;\alpha\beta}^{JL}(r) = \gamma_T^{d}(-1)^{l_\alpha+l_\beta+L+s}
J_{d;\alpha\beta}^{JL*}(r) = (-1)^{l_\alpha+l_\beta+j_\alpha+j_\beta+J+1}
J_{d;\beta\alpha}^{JL}(r)
\end{equation}

\subsection{Reduced matrix elements of densities and currents}\label{sec_rme}
To simplify the expressions for reduced matrix elements, it is convenient to absorb certain numerical factors to the radial wavefunctions, e.g.~factor $\sqrt{(2j+1)(2l+1)}$ in ordinary density (\ref{rho_rme_example}). Other densities and currents will employ also derivative operators and the following shorthand notation of radial wavefunctions turns out to be convenient
\begin{subequations}
\label{Rpm}
\begin{align}
R_{\alpha}^{(0)} &\equiv \sqrt{(2j_\alpha+1)(2l_\alpha+1)}\, R_\alpha(r) \phantom{\bigg(} \\
R_{\alpha}^{(+)} &\equiv -\sqrt{(2j_\alpha+1)(l_\alpha+1)(2l_\alpha+3)}\,\bigg( \frac{\mathrm{d}R_\alpha(r)}{\mathrm{d}r} - \frac{l_\alpha}{r} R_\alpha(r)\bigg) \\
R_{\alpha}^{(-)} &\equiv \sqrt{(2j_\alpha+1)l_\alpha(2l_\alpha-1)}\,\bigg( \frac{\mathrm{d}R_\alpha(r)}{\mathrm{d}r} + \frac{l_\alpha+1}{r} R_\alpha(r)\bigg)
\end{align}
\end{subequations}
and a shifted angular momentum will be denoted by
\begin{equation}
l_\alpha^+ = l_\alpha+1,\quad l_\alpha^- = l_\alpha-1
\end{equation}
An example of the derivation of vector spin-orbital current $\mathcal{J}_{\alpha\beta}^{JL}(r)$ is given in appendix \ref{app_Jab}, which illustrates main steps involved in the remaining densities/currents.

Precise differentiation of the wavefunctions in (\ref{Rpm}), which are defined on an equidistant grid (with spacing $\Delta$, going from $-n\Delta$ to $n\Delta$), is achieved through their discrete Fourier transformation.
\begin{align}
R(r) &= \sum_{k=-n}^{n-1}\tilde{R}_k\mathrm{e}^{\mathrm{i}\pi kr/n\Delta}
= \frac{1}{2n}\sum_{k=-n}^{n-1}\sum_{j=-n}^{n-1}
\mathrm{e}^{\mathrm{i}\pi k(r/\Delta-j)/n} R(j\Delta)
\nonumber\\
\frac{\mathrm{d}R(r)}{\mathrm{d}r}\bigg|_{r=m\Delta} &=
\frac{1}{2n}\sum_{k=-n+1}^{n-1}\sum_{j=-n}^{n-1}
\frac{\mathrm{i}\pi k}{n\Delta}\mathrm{e}^{\mathrm{i}\pi k(m-j)/n} R(j\Delta)
\nonumber\\
&= \sum_{j=-n}^{n-1}\bigg(\frac{-\pi}{n^2\Delta}\sum_{k=1}^{n-1}
k\sin\frac{\pi k(m-j)}{n}\bigg) R(j\Delta)
\end{align}
In practice, the convolution matrix (in large parentheses) is calculated in advance for two cases, even and odd $R(r)$, and then applied to functions $R_\alpha(r)$. Alternatively, expressions (\ref{Rpm}) can be calculated analytically, if the radial wavefunctions are expressed in the basis of spherical harmonic oscillator (see later (\ref{Rpm_sho})).

The reduced matrix elements (\ref{dens_rme},\ref{rme}) of quantities used in Skyrme functional are listed below. I will later complement the r.m.e.~by index $q\in\{p,n\}$ (e.g.~$\rho_{q;\alpha\beta}^{L}(r)$, where $\alpha\beta\in q$).
\begin{footnotesize}
\begin{subequations}
\begin{align}
\rho_{\alpha\beta}^L(r) & =
u_{\alpha\beta}^{(+)}R_\alpha^{(0)}(r) R_\beta^{(0)}(r)
\frac{(-1)^{j_\beta+\frac{1}{2}}}{\sqrt{4\pi}}
\begin{Bmatrix} 
l_\alpha & \!l_\beta\! & L \\ j_\beta & \!j_\alpha\! & \frac{1}{2} \end{Bmatrix}
C_{l_\alpha 0 l_\beta 0}^{L 0} \\
\tau_{\alpha\beta}^L(r) & = u_{\alpha\beta}^{(+)}
\bigg[ \sum_{ss'}^{\pm\pm} R_\alpha^{(s)}(r) R_\beta^{(s')}(r) \begin{Bmatrix}
l_\alpha^s & l_\beta^{s'} & L \\ l_\beta & l_\alpha & 1 \end{Bmatrix}
C_{l_\alpha^s 0 l_\beta^{s'} 0}^{L 0} \bigg]
\frac{(-1)^{j_\beta-\frac{1}{2}}}{\sqrt{4\pi}}
\begin{Bmatrix} l_\alpha & l_\beta & L \\ j_\beta & j_\alpha & \frac{1}{2} \end{Bmatrix} \\
\label{spin-orb_me}
\mathcal{J}_{\alpha\beta}^{JL}(r) & =
\frac{1}{2}u_{\alpha\beta}^{(+)}\bigg[\sum_{ss'}^{0\pm,\pm0}
\mathcal{A}_{\alpha\beta LJ}^{\vec{\mathcal{J}},ss'} R_\alpha^{(s)}(r) R_\beta^{(s')}(r) \bigg]
(-1)^{j_\beta+\frac{1}{2}}
\sqrt{\frac{2J+1}{4\pi}} \\
& \mathcal{A}_{\alpha\beta LJ}^{\vec{\mathcal{J}},0\pm} =
C_{l_\alpha 0 l_\beta^\pm 0}^{L 0}
\bigg[
\begin{Bmatrix} l_\alpha & \!l_\beta\! & J \\ j_\beta & \!j_\alpha\! & \frac{1}{2} \end{Bmatrix}
\begin{Bmatrix} l_\alpha & \!l_\beta\! & J \\ 1 & \!L\! & l_\beta^\pm \end{Bmatrix}
-\frac{2\sqrt{3}}{\sqrt{2j_\beta+1}}
\begin{Bmatrix} j_\alpha & j_\beta & J \\ l_\alpha & l_\beta^\pm & L \\ \frac{1}{2} & \!\frac{1}{2}\! & 1 \end{Bmatrix}
\bigg] \nonumber\\
& \mathcal{A}_{\alpha\beta LJ}^{\vec{\mathcal{J}},\pm0} =
(-1)^{J+L+1} C_{l_\alpha^\pm 0 l_\beta 0}^{L 0}
\bigg[
\begin{Bmatrix} l_\alpha & \!l_\beta\! & J \\ j_\beta & \!j_\alpha\! & \frac{1}{2} \end{Bmatrix}
\begin{Bmatrix} l_\beta & \!l_\alpha\! & J \\ 1 & \!L\! & l_\alpha^\pm \end{Bmatrix}
-\frac{2\sqrt{3}}{\sqrt{2j_\alpha+1}}
\begin{Bmatrix} j_\beta & \!j_\alpha\! & J \\ l_\beta & \!l_\alpha^\pm\! & L \\ \frac{1}{2} & \!\frac{1}{2}\! & 1 \end{Bmatrix}
\bigg] \nonumber\\
\label{spin-orb-s_me}
\mathcal{J}_{s;\alpha\beta}^L(r) & = \mathrm{i}
u_{\alpha\beta}^{(+)}\frac{(-1)^{j_\beta+\frac{1}{2}}}{\sqrt{8\pi}}\Big[
\frac{{\mp}R_\alpha^{(0)}R_\beta^{(\pm)}\!\!\!}{\sqrt{2j_\beta+1}}
C_{l_\alpha 0 l_\beta^\pm 0}^{L0} \begin{Bmatrix}
l_\alpha & l_\beta^\pm & L \\ j_\beta & j_\alpha & \frac{1}{2} \end{Bmatrix}
\pm \frac{R_\alpha^{(\pm)}R_\beta^{(0)}}{\sqrt{2j_\alpha+1}}
C_{l_\alpha^\pm 0 l_\beta 0}^{L0} \begin{Bmatrix}
l_\alpha^\pm & l_\beta & L \\ j_\beta & j_\alpha & \frac{1}{2} \end{Bmatrix}
\Big] \\
\label{spin-orb-t_me}
\mathcal{J}_{t;\alpha\beta}^{JL}(r) & = \mathrm{i} u_{\alpha\beta}^{(+)}
\bigg[\sum_{ss'}^{0\pm,\pm0} \mathcal{A}_{\alpha\beta LJ}^{\boldsymbol{\mathcal{J}},ss'}
R_\alpha^{(s)}(r)R_\beta^{(s')}(r) \bigg]
(-1)^{j_\beta+\frac{1}{2}} \sqrt{\frac{5(2J+1)}{6\pi}} \\
& \mathcal{A}_{\alpha\beta LJ}^{\boldsymbol{\mathcal{J}},0\pm} =
(j_\beta-l_\beta)
{\textstyle \sqrt{\frac{(4j_\beta-2l_\beta+1\pm1)[2\mp2(j_\beta-l_\beta)]}
{(2l_\beta+1)(2j_\beta+1)}} }
\begin{Bmatrix} j_\alpha & j_\beta & J \\ l_\alpha & l_\beta^\pm & L \\ \frac{1}{2} & \frac{3}{2} & 2 \end{Bmatrix} C_{l_\alpha 0 l_\beta^\pm 0}^{L0} \nonumber\\
& \mathcal{A}_{\alpha\beta LJ}^{\boldsymbol{\mathcal{J}},\pm0} =
(-1)^{J+L+1}(j_\alpha-l_\alpha)
{\textstyle \sqrt{\frac{(4j_\alpha-2l_\alpha+1\pm1)[2\mp2(j_\alpha-l_\alpha)]}
{(2l_\alpha+1)(2j_\alpha+1)}} }
\begin{Bmatrix} j_\beta & j_\alpha & J \\ l_\beta & l_\alpha^\pm & L \\ \frac{1}{2} & \frac{3}{2} & 2 \end{Bmatrix} C_{l_\alpha^\pm 0 l_\beta 0}^{L0} \nonumber \\
\!\!\!\!\!\!\!(\nabla\cdot \mathcal{J})_{\alpha\beta}^L(r) & = \tau_{\alpha\beta}^L(r) -
u_{\alpha\beta}^{(+)}\frac{2(-1)^{j_\alpha+L+\frac{1}{2}}
R_\alpha^{(\pm)}(r) R_\beta^{(\pm)}(r)
}{\sqrt{4\pi(2j_\alpha+1)(2j_\beta+1)}}
\begin{Bmatrix} l_\alpha^\pm & \!l_\beta^\pm\! & L \\ j_\beta & \!j_\alpha\! & \frac{1}{2} \end{Bmatrix}
C_{l_\alpha^\pm 0 l_\beta^\pm 0}^{L 0} \bigg|_{\pm:\,j=l\pm\frac{1}{2}}\! \\
\label{j_me}
j_{\alpha\beta}^{JL}(r) & =
\frac{\mathrm{i}}{2}u_{\alpha\beta}^{(-)}
\bigg[ \sum_{ss'}^{0\pm,\pm0}
\mathcal{A}_{\alpha\beta LJ}^{\vec{j},ss'} R_\alpha^{(s)}(r) R_\beta^{(s')}(r) \bigg]
(-1)^{j_\beta-\frac{1}{2}}
\sqrt{\frac{2J+1}{4\pi}}
\begin{Bmatrix} l_\alpha & \!l_\beta\! & J \\ j_\beta & \!j_\alpha\! & \frac{1}{2} \end{Bmatrix} \\
& \mathcal{A}_{\alpha\beta LJ}^{\vec{j},0\pm} =
\begin{Bmatrix} l_\alpha & \!l_\beta\! & J \\ 1 & \!L\! & l_\beta^\pm \end{Bmatrix}
C_{l_\alpha 0 l_\beta^\pm 0}^{L 0},
\qquad
\mathcal{A}_{\alpha\beta LJ}^{\vec{j},\pm0} = (-1)^{L+J}
\begin{Bmatrix} l_\beta & \!l_\alpha\! & J \\ 1 & \!L\! & l_\alpha^\pm \end{Bmatrix}
C_{l_\alpha^\pm 0 l_\beta 0}^{L 0} \nonumber\\
\!\!\!\!\!\!\!\!(\nabla\times j)_{\alpha\beta}^{JL}(r) & =
u_{\alpha\beta}^{(-)}\bigg[ \sum_{ss'}^{\pm\pm}
R_\alpha^{(s)}R_\beta^{(s')}
\begin{Bmatrix} l_\alpha & \!l_\beta\! & J \\ l_\alpha^{s} & \!l_\beta^{s'}\! & L \\ 1 & \!1\! & 1 \end{Bmatrix} C_{l_\alpha^s 0 l_\beta^{s'} 0}^{L0} \bigg]
(-1)^{j_\beta+L+J-\frac{1}{2}}\sqrt{\frac{3(2J+1)}{2\pi}}
\begin{Bmatrix} l_\alpha & \!l_\beta\! & J \\ j_\beta & \!j_\alpha\! & \frac{1}{2} \end{Bmatrix} \\
s_{\alpha\beta}^{JL}(r) & =
u_{\alpha\beta}^{(-)}R_\alpha^{(0)}(r) R_\beta^{(0)}(r)
(-1)^{l_\beta} \sqrt{\frac{3(2J+1)}{2\pi}}\,
\begin{Bmatrix} j_\alpha & \!j_\beta\! & J \\ l_\alpha & \!l_\beta\! & L \\ \frac{1}{2} & \!\frac{1}{2}\! & 1 \end{Bmatrix}
C_{l_\alpha 0 l_\beta 0}^{L 0} \\
T_{\alpha\beta}^{JL}(r) & =
u_{\alpha\beta}^{(-)}\bigg[\sum_{ss'}^{\pm\pm}
R_\alpha^{(s)} R_\beta^{(s')} \begin{Bmatrix}
l_\alpha^s & l_\beta^{s'} & L \\ l_\beta & l_\alpha & 1 \end{Bmatrix}
C_{l_\alpha^s 0 l_\beta^{s'} 0}^{L 0} \bigg]
(-1)^{l_\beta+1} \sqrt{\frac{3(2J+1)}{2\pi}}\,
\begin{Bmatrix} j_\alpha & \!j_\beta\! & J \\ l_\alpha & \!l_\beta\! & L \\ \frac{1}{2} & \!\frac{1}{2}\! & 1 \end{Bmatrix}
\end{align}
\end{subequations}
\end{footnotesize}

I am interested in electric and magnetic transitions of multipolarity $\lambda$, so the relevant matrix elements follow the selection rules
\begin{equation}
\label{EM_sel_rules}
J=\lambda,\quad(-1)^{l_\alpha+l_\beta+\lambda}=\Big\{\begin{array}{l}
{+}1:\quad\textrm{electric} \\ {-}1:\quad\textrm{magnetic} \end{array}
\end{equation}
These selection rules together with (\ref{rme_hermit}) lead to the conditions on non-zero $L$-components as listed in the Table \ref{tab-L}.
\begin{table}[t]
\caption{Selection rules on $L\protect\phantom{p}\!\!\!$ in r.m.e.~of densities and currents.}\label{tab-L}
\centering
\begin{tabular}{|c||c|c|c|c|c|}
\hline
\rule{0pt}{13pt}$(-1)^{l_\alpha+l_\beta+\lambda}$ & $\rho$ & $\tau$ & $\mathcal{J},\vec{\mathcal{J}},\boldsymbol{\mathcal{J}}$ & $\vec{j}$ & $\vec{s},\,\vec{T},\,\vec{\nabla}\times\vec{j}$ \\
\hline
\hline
$+1$ & $L=\lambda$ & $L=\lambda$ & $L=\lambda\pm1$ & $L=\lambda\pm1$ & $L=\lambda$ \\
\hline
$-1$ & 0 & 0 & $L=\lambda,\lambda\pm2$ & $L=\lambda$ & $L=\lambda\pm1$ \\
\hline
\end{tabular}
\end{table}

Reduced matrix elements of $\vec{\nabla}\rho$ and $(\nabla s)$ are not given here, since they are simply related to $\vec{j}$ and $\mathcal{J}$ (\ref{Jd_op}) and differ only in the relative sign and the imaginary constant.
\begin{equation}
(\vec{\nabla}\rho)(\vec{r}_0) =
\overleftarrow{\nabla}_{\!(\vec{r})}\delta(\vec{r}-\vec{r}_0)
+\delta(\vec{r}-\vec{r}_0)\overrightarrow{\nabla}_{\!(\vec{r})}
\end{equation}
The differentiation in the definitions above does not spoil the hermiticity of the corresponding operators, because the resulting operators can be given equivalently as commutators, e.g.
\begin{equation}
\label{diff_op}
(\vec{\nabla}\rho)(\vec{r}) = -\sum_j[\vec{\nabla}_j,\hat{\rho}(\vec{r})] = -\mathrm{i}\sum_j[\hat{\vec{p}}_j,\hat{\rho}(\vec{r})]/\hbar
\end{equation}
where $j$ labels the particles.

Most of the $9j$ symbols given above do not have to be calculated explicitly, since their product with $C_{l_\alpha 0 l_\beta 0}^{L 0}$ can be expressed by Clebsch-Gordan coefficients, e.g.~$C_{j_\alpha,-\frac{1}{2},j_\beta,\frac{1}{2}}^{J,0}$, see \cite[eq.~10.9.10--12]{Varshalovich1988}.

\subsection{Hartree-Fock in the basis of spherical harmonic oscillator}
Solution of the Hartree-Fock (in its density-functional form) corresponds to a variation of the full Hamiltonian $\mathcal{H}$ with respect to densities $J_d(\vec{r})$ to obtain single-particle Hamiltonian $\hat{h}$:
\begin{equation}
\hat{h} = \int\mathrm{d}\vec{r}\,\sum_d\frac{\delta\mathcal{H}}{\delta J_d(\vec{r})}
\hat{J}_d(\vec{r})
\end{equation}
Ground state densities, which are contained in $\frac{\delta\mathcal{H}}{\delta J_d}$, are non-zero in spherical even-even nuclei only in their monopole component ($J=0$) and for time-even case. They can be calculated from the reduced matrix elements of the previous section, re-evaluating (\ref{dens_rme}) without $u_{\alpha\beta}^{(\gamma_T^d)}$, assuming $j_a = j_b = j$, $l_a=l_b=l$, $m_a=m_b=-m_\beta$, or, more precisely, $|\bar{\beta}\rangle\mapsto|\overline{-b}\rangle = (-1)^{l_b+j_b-m_b}|b\rangle$.
\begin{equation}
\label{HF_me}
\langle a|\hat{\mathbf{J}}_d|b\rangle =
\frac{J_{d;ab}^{0L\mathrm{(HF)}}(r)}{\sqrt{2j+1}} \mathbf{Y}_{00}^{L*}
\end{equation}
Terms with $J>0$ cancel in the summation over $m$ during the calculation of ground-state densities. $L$ is irrelevant for scalar densities, but is fixed as $L=1$ for vector densities and $L=2$ for tensor densities (due to triangular inequality in the coupling of orbital and spin angular momentum). Index (HF) and latin letters emphasize that indices $a,\,b$ correspond to the basis of spherical harmonic oscillator, instead of HF basis, so the factor $u_{ab}^{(+)}$ is absent here (instead, factors $v^2$ will be included later)
\begin{subequations}
\begin{align}
\rho_{ab}^{0\mathrm{(HF)}}(r) &= \frac{1}{\sqrt{4\pi(2j+1)}}
\frac{\displaystyle R_a^{(0)} R_b^{(0)}}{2l+1} = \sqrt{\frac{2j+1}{4\pi}}\,R_a(r) R_b(r) \\
\tau_{ab}^{0\mathrm{(HF)}}(r) &= \frac{1}{\sqrt{4\pi(2j+1)}}\frac{1}{2l+1}
\bigg(\frac{\displaystyle R_a^{(+)}R_b^{(+)}}{2l^+ + 1}+\frac{\displaystyle R_a^{(-)}R_b^{(-)}}{2l^- + 1}\bigg) \\
\!\!(\nabla\!\cdot\!\mathcal{J})_{ab}^{0\mathrm{(HF)}}(r) &= \tau_{ab}^{0\mathrm{(HF)}}(r)
- \frac{1}{\sqrt{4\pi(2j+1)}}\frac{\displaystyle 2R_a^{(\pm)}R_b^{(\pm)}}{(2j+1)(2l^\pm+1)}
\bigg|_{\pm:\,j=l\pm\frac{1}{2}} \\
(\nabla\rho)_{ab}^{01\mathrm{(HF)}}(r) &= \frac{1}{\sqrt{4\pi(2j+1)(2l+1)}}
\frac{1}{2l+1} \bigg[
\sqrt{\frac{l+1}{2l^++1}}\big(R_a^{(+)}R_b^{(0)}+R_a^{(0)}R_b^{(+)}\big) \nonumber\\
&\qquad\quad\qquad\qquad\qquad{}-\sqrt{\frac{l}{2l^-+1}}
\big(R_a^{(-)}R_b^{(0)}+R_a^{(0)}R_b^{(-)}\big) \bigg] \\
\mathcal{J}_{ab}^{01\mathrm{(HF)}}(r) &= \frac{(\nabla\rho)_{ab}^{01\mathrm{(HF)}}(r)}{2}
\mp \frac{\displaystyle R_a^{(\pm)}R_b^{(0)}+R_a^{(0)}R_b^{(\pm)}}{\sqrt{8\pi(2l+1)(2l^\pm+1)}\,(2j+1)} \bigg|_{\pm:\,j=l\pm\frac{1}{2}}
\end{align}
\end{subequations}
where $l^\pm = l\pm1$. Scalar and tensor spin-orbital currents are zero due to Clebsch-Gordan coefficient $C_{l_\alpha 0 l_\alpha^\pm 0}^L$ in (\ref{spin-orb-s_me}), (\ref{spin-orb-t_me}) with $L=0$ or $2$, respectively.

The basis of spherical harmonic oscillator (SHO) is defined by oscillator length $b$ (not to be confused with the w.f.~labels above), orbital angular momentum $l$ and radial quantum number $\nu\in\{0,1,2,\ldots\}$.
\begin{equation}
\label{SHO}
\psi_{\nu l m_l}^\mathrm{SHO}(r) = R_{\nu l}(r)Y_{lm_l}(\vartheta,\varphi),\quad
E_{\nu l}^\mathrm{SHO} = \hbar\omega\bigg(2\nu+l+\frac{3}{2}\bigg),\quad
b = \sqrt{\frac{\hbar}{m\omega}},
\end{equation}
Radial part of s.p.~HF matrix elements is evaluated directly in SHO basis, and the derivatives in the definition of $R^{(\pm)}$ (\ref{Rpm}) can be calculated analytically
\begin{align*}
{-}\tfrac{\mathrm{d}R_{\nu l}(r)}{\mathrm{d}r}+\tfrac{l}{r}R_{\nu l}(r) &=
\tfrac{1}{b}\big[\sqrt{\nu+l+3/2}\,R_{\nu,l+1}(r)+\sqrt{\nu}\,R_{\nu-1,l+1}(r)\big]\\
\tfrac{\mathrm{d}R_{\nu l}(r)}{\mathrm{d}r}+\tfrac{l+1}{r}R_{\nu l}(r) &=
\tfrac{1}{b}\big[\sqrt{\nu+l+1/2}\,R_{\nu,l-1}(r)+\sqrt{\nu+1}\,R_{\nu+1,l-1}(r)\big]
\end{align*}
so the expressions for $R^{(\pm)}$ are
\begin{small}
\begin{subequations}
\label{Rpm_sho}
\begin{align}
R_{\nu l}^{(0)} &= \sqrt{(2j+1)(2l+1)}\,R_{\nu l}(r) \\
\!R_{\nu l}^{(+)} &= \sqrt{(2j+1)(l+1)(2l^++1)}\,\tfrac{1}{b}
\Big[\sqrt{\nu+l+3/2}\,R_{\nu,l+1}(r)+\sqrt{\nu}\,R_{\nu-1,l+1}(r)\Big] \\
\!R_{\nu l}^{(-)} &= \sqrt{(2j+1)l(2l^-+1)}\,\tfrac{1}{b}
\Big[\sqrt{\nu+l+1/2}\,R_{\nu,l-1}(r)+\sqrt{\nu+1}\,R_{\nu+1,l-1}(r)\Big]
\end{align}
\end{subequations}
\end{small}

Kinetic energy can be evaluated in a similar way, and the only non-zero matrix elements are
\begin{subequations}
\label{HF_kinetic}
\begin{align}
\langle\nu-1,l|\nabla^2|\nu,l\rangle &= {-}\sqrt{\nu(\nu+l+1/2)}/b^2 \\
\langle\nu,l|\nabla^2|\nu,l\rangle &= {-}\big(2\nu+l+\tfrac{3}{2}\big)/b^2 \\
\langle\nu+1,l|\nabla^2|\nu,l\rangle &= {-}\sqrt{(\nu+1)(\nu+l+3/2)}/b^2
\end{align}
\end{subequations}

\newpage Skyrme HF calculation then proceeds by a straightforward iterative way:
\begin{enumerate}
\item HF wavefunctions $R_\alpha(r)$ are evaluated on the radial grid from the orthogonal matrices $U_{a\alpha\phantom{b}}^{(j,l)}$ in each subspace of $j$ and $l$ (index $a$ is essentially equivalent to $\nu$ in spherical-harmonic-oscillator basis).
\item Ground state densities are calculated from $R_\alpha(r)$, taking into acount pairing factors $v_\alpha^2$ (given by the separate BCS step, or taken as $0/1$ according to occupancy). Densities from the previous iteration are admixed to the new densities (by 50\%) to stabilize the convergence. Coulomb potential is calculated by folding with $1/|\vec{r}_1-\vec{r_2}|$ according to section \ref{sec_sph_coul}. The total energy can be calculated here as well.
\item Matrix elements of single-particle HF Hamiltonian are calculated by radial integration of a product of the ground-state densities and the matrix elements of densities (\ref{HF_me}) in SHO basis. Kinetic term (\ref{HF_kinetic}) is shown separately from $\mathcal{H}$ in the formula below. Moreover, it is possible to include center-of-mass correction for the kinetic energy (see section \ref{sec_kin-cm}), if correction-before-variation is needed -- this option requires also the calculation of density matrix $D_{ab}^{(j,l)} = \sum_\alpha v_\alpha^2 U_{a\alpha\phantom{b}}^{(j,l)} U_{b\alpha}^{(j,l)}$, which is $(2j+1)$-times degenerated in quantum number $m$.
\item Diagonalization of the single-particle HF Hamiltonian $\hat{h}$ to get single-particle energies and matrices $U_{a\alpha}^{(j,l)}$ (with eigenvectors in columns).
\end{enumerate}
\begin{align*}
&R_\alpha(r)=\sum_{a}U_{a\alpha}^{(j,l)}R_a(r) \quad\Rightarrow\quad
J_d(r) = \sum_{j,l} \sum_{\alpha\in(j,l)} (2j+1)v_\alpha^2
\frac{J_{d;\alpha\alpha}^{0L\mathrm{(HF)}}(r)}{\sqrt{2j+1}} \\
&\qquad\qquad\Rightarrow\quad
\langle a|\hat{h}|b\rangle = -\frac{\hbar^2}{2m_q}\langle a|\nabla^2|b\rangle
+ \int\mathrm{d}^3 r \sum_d \frac{\delta\mathcal{H}}{\delta J_d(r)}
\frac{J_{d;ab}^{0L\mathrm{(HF)}}(r)}{\sqrt{2j+1}}
\end{align*}
The iterations are repeated until the relative difference in the total energy becomes lower than $10^{-14}$ (it takes from 50 iteration for Ca up to 90 iterations for Pb). Then, four iterations are done without admixing previous densities.

\subsection{Full RPA}\label{sec_fullrpa}
Excitations of a given multipolarity will be treated as RPA phonons. One-phonon state is denoted as $|\nu\rangle$, with energy $E_\nu = \hbar\omega_\nu$ above ground state, and was created by action of operator $\hat{C}_\nu^+$ on the RPA ground state $|\textrm{RPA}\rangle$.
\begin{equation}
\hat{C}_\nu^+|\textrm{RPA}\rangle=|\nu\rangle,\quad
\hat{C}_\nu^{\phantom{|}}|\textrm{RPA}\rangle=0
\end{equation}
Operator $\hat{C}_\nu^+$ is a two-quasiparticle ($2qp$) operator defined by real coefficients $c_{\alpha\beta}^{(\nu\pm)}$
\begin{equation}
\label{phonon_sph}
\hat{C}_\nu^+ = \frac{1}{2}\sum_{\alpha\beta}
C_{j_\alpha m_\alpha j_\beta m_\beta}^{\lambda_\nu\mu_\nu}\Big(
c_{\alpha\beta}^{(\nu-)}\hat{\alpha}_{\alpha}^+\hat{\alpha}_{\beta}^+ +
c_{\alpha\beta}^{(\nu+)}\hat{\alpha}_{\bar{\alpha}}^{\phantom{*}}
\hat{\alpha}_{\bar{\beta}}^{\phantom{*}} \Big)
\end{equation}
(in the following, I will drop the index $\nu$ in $\lambda_\nu,\:\mu_\nu$), its normalization is given by
\begin{equation}
\label{RPA_norm}
\langle[\hat{C}_\nu^{\phantom{|}},\hat{C}_{\nu'}^+]\rangle = \delta_{\nu\nu'} \quad\Rightarrow\quad
\frac{1}{2}\sum_{\alpha\beta} \Big(\big|c_{\alpha\beta}^{(\nu-)}\big|^2
- \big|c_{\alpha\beta}^{(\nu+)}\big|^2 \Big) = 1
\end{equation}
and it satisfies the RPA equation
\begin{equation}
\label{RPA_eq}
{[\hat{H},\hat{C}_\nu^+]}_{2qp} = E_\nu\hat{C}_\nu^+
\end{equation}
where the index $2qp$ means that I take only the two-quasiparticle portion of the commutator (after normal ordering). Although all commutators should be evaluated in the RPA ground state, I evaluate them in the HF+BCS ground state (i.e., I am using quasi-boson approximation), which is a common practice, as the contribution of $4qp$ and higher correlations in the ground state to the expectation value of commutators is assumed to be low \cite{Ring1980}.

The Hamiltonian is taken as a sum of mean-field part (HF+BCS) and the second functional derivative of the energy density functional (\ref{full_hamil}).
\begin{equation}
\hat{H} = \hat{H}_0 + \hat{V}_\mathrm{res} =
\sum_\gamma\varepsilon_\gamma\hat{\alpha}_\gamma^+\hat{\alpha}_\gamma^{\phantom{|}}
+ \frac{1}{2}\sum_{dd'}\int\!\!\!\int\mathrm{d}^3 r\,\mathrm{d}^3 r'
\frac{\delta^2\mathcal{H}}{\delta J_d(\vec{r})\delta J_{d'}({\vec{r}\,}')}
:\!\hat{J}_{d}(\vec{r})\hat{J}_{d'}({\vec{r}\,}')\!:
\end{equation}
Left hand side of (\ref{RPA_eq}) is then evaluated as (with $\varepsilon_{\alpha\beta} = \varepsilon_\alpha + \varepsilon_\beta$)
\begin{align}
{[\hat{H}_0,\hat{C}_\nu^+]} & = \frac{1}{2}\sum_{\alpha\beta}
\varepsilon_{\alpha\beta} C_{j_\alpha m_\alpha j_\beta m_\beta}^{\lambda \mu}
\Big( c_{\alpha\beta}^{(\nu-)} \hat{\alpha}_\alpha^+ \hat{\alpha}_\beta^+
- c_{\alpha\beta}^{(\nu+)} \hat{\alpha}_{\bar{\alpha}}^{\phantom{*}} \hat{\alpha}_{\bar{\beta}}^{\phantom{*}} \Big) \\
{[\hat{V}_\mathrm{res},\hat{C}_\nu^+]}_{2qp} & = \sum_{dd'}
\int\!\!\!\int\mathrm{d}^3 r\,\mathrm{d}^3 r'
\frac{\delta^2\mathcal{H}}{\delta J_d(\vec{r})\delta J_{d'}({\vec{r}\,}')}
\langle[\hat{J}_d(\vec{r}),\hat{C}_\nu^+]\rangle\hat{J}_{d'}({\vec{r}\,}') \\
& = \sum_{dd'}\gamma_T^d\,
\frac{1}{4}\sum_{\alpha\beta\gamma\delta L}
\int_0^\infty r^2\mathrm{d}r \frac{\delta^2\mathcal{H}}{\delta J_d\delta J_{d'}}
J_{d;\alpha\beta}^{\lambda L*}(r)J_{d';\gamma\delta}^{\lambda L}(r)
\frac{(-1)^{l_\beta+l_\delta+1}}{2\lambda+1}
C_{j_\gamma m_\gamma j_\delta m_\delta}^{\lambda\mu} \nonumber\\
& \qquad\times
\Big(\gamma_T^d c_{\alpha\beta}^{(\nu-)}+c_{\alpha\beta}^{(\nu+)}\Big)
(-\hat{\alpha}_{\gamma}^+ \hat{\alpha}_{\delta}^+ + \gamma_T^{d'}
\hat{\alpha}_{\bar{\gamma}}^{\phantom{*}} \hat{\alpha}_{\bar{\delta}}^{\phantom{*}})
\end{align}

At this point, I will remove duplicate $2qp$ pairs. To do it consistently, I will rescale diagonal pairing factors
\begin{equation}
\label{order2qp}
\frac{1}{2}\sum_{\alpha\beta}\mapsto\sum_{\alpha\geq\beta},\quad
u_{\alpha\alpha}^{(+)} = \sqrt{2}\,u_\alpha v_\alpha\quad\textrm{(instead of $2u_\alpha v_\alpha$)}
\end{equation}
and $c_{\alpha\alpha}^{(\nu\pm)}$ will be rescaled automatically. Diagonal matrix elements contribute only to electric transitions with $\lambda$ even. Then, comparison of coefficients at $\hat{\alpha}_{\gamma}^+ \hat{\alpha}_{\delta}^+$ and $\hat{\alpha}_{\bar{\gamma}}^{\phantom{*}} \hat{\alpha}_{\bar{\delta}}^{\phantom{*}}$ in (\ref{RPA_eq}) leads to
\begin{small}
\begin{subequations}
\begin{align}
(E_\nu-\varepsilon_{\gamma\delta})c_{\gamma\delta}^{(\nu-)} & =
\sum_{dd'L} \sum_{\alpha\geq\beta} \int_0^\infty
\frac{\delta^2\mathcal{H}}{\delta J_d\delta J_{d'}}
J_{d;\alpha\beta}^{\lambda L*}(r)J_{d';\gamma\delta}^{\lambda L}(r)
r^2\mathrm{d}r \frac{(-1)^{l_\beta+l_\delta}}{2\lambda+1}
\Big(c_{\alpha\beta}^{(\nu-)}+\gamma_T^d c_{\alpha\beta}^{(\nu+)}\Big) \\
(E_\nu+\varepsilon_{\gamma\delta})c_{\gamma\delta}^{(\nu+)} & =
-\sum_{dd'L} \sum_{\alpha\geq\beta} \int_0^\infty
\frac{\delta^2\mathcal{H}}{\delta J_d\delta J_{d'}}
J_{d;\alpha\beta}^{\lambda L*}(r)J_{d';\gamma\delta}^{\lambda L}(r)
r^2\mathrm{d}r \frac{(-1)^{l_\beta+l_\delta}}{2\lambda+1}
\Big(\gamma_T^d c_{\alpha\beta}^{(\nu-)}+c_{\alpha\beta}^{(\nu+)}\Big)
\end{align}
\end{subequations}
\end{small}\\[-6pt]
and these equations can be expressed in a compact matrix form
\begin{equation}
\label{fullRPA_eq}
\begin{pmatrix} A & B \\ B & A \end{pmatrix} \binom{c^{(\nu-)}}{c^{(\nu+)}} =
\begin{pmatrix} E_\nu & 0 \\ 0 & -E_\nu \end{pmatrix}
\binom{c^{(\nu-)}}{c^{(\nu+)}}
\end{equation}
where the real matrices $A,\,B$ in the ordered $2qp$ basis ($p\equiv\alpha\beta,\,p'\equiv\gamma\delta$) are
\begin{subequations}
\begin{align}
\label{fullRPA_A}
A_{pp'} & = \delta_{pp'}\varepsilon_p + \sum_{dd'L}\frac{(-1)^{l_\beta+l_\delta}}{2\lambda+1}
\int_0^\infty \frac{\delta^2\mathcal{H}}{\delta J_d\delta J_{d'}}
J_{d;p}^{\lambda L}(r)J_{d';p'}^{\lambda L*}(r)r^2\mathrm{d}r \\
\label{fullRPA_B}
B_{pp'} & = \sum_{dd'L} \gamma_T^d \frac{(-1)^{l_\beta+l_\delta}}{2\lambda+1}
\int_0^\infty \frac{\delta^2\mathcal{H}}{\delta J_d\delta J_{d'}}
J_{d;p}^{\lambda L}(r)J_{d';p'}^{\lambda L*}(r)
r^2\mathrm{d}r
\end{align}
\end{subequations}
Expression $\frac{\delta^2\mathcal{H}}{\delta J_d\delta J_{d'}}$ is symbolical, and includes integration of the delta function, yielding ${\vec{r}\,}' = \vec{r}$. The exchange Coulomb interaction can be treated by Slater approximation as a density functional (\ref{xc})
\begin{equation}
\frac{\delta^2\mathcal{H}_\mathrm{xc}}{\delta\rho_p\delta\rho_p}
= \frac{-1}{\sqrt[3\,]{9\pi}} \frac{e^2}{4\pi\epsilon_0} \rho_{0p}^{-2/3}(r)
\end{equation}
where $\rho_{0p}(r)$ is the ground-state proton density. However, the direct Coulomb interaction gives rise to a double integral instead (see also corrections in (\ref{coul_EM2}))
\begin{align}
\!\!\int_0^\infty\frac{\delta^2\mathcal{H}}{\delta J_d\delta J_{d'}}
J_{d;\alpha\beta}^{\lambda L*}(r)J_{d';\gamma\delta}^{\lambda L}(r)
r^2\mathrm{d}r \quad\mapsto \nonumber\\
\frac{e^2}{4\pi\epsilon_0} \frac{4\pi}{2\lambda+1}
\int_0^\infty\! r^2\mathrm{d}r \int_0^\infty\! r'^2\mathrm{d}r' 
&\rho_{\alpha\beta}^\lambda(r)\rho_{\gamma\delta}^\lambda(r')
\times \bigg\{\!\!\begin{array}{l}
r^\lambda / r'^{\lambda+1} \ \ (r<r') \phantom{\big|}\\
r'^\lambda / r^{\lambda+1} \ \ (r\geq r')\phantom{\big|} \end{array}
\end{align}

Matrix equation (\ref{fullRPA_eq}) can be reduced to a diagonalization of a symmetric matrix of half dimension. I define
\begin{equation}\label{half_RPA}
\!\!\begin{array}{l} x_p = c_p^{(\nu-)} + c_p^{(\nu+)}\phantom{\big|} \\
y_p = c_p^{(\nu-)} - c_p^{(\nu+)}, \end{array} \ \ 
c_p^{(\nu-)} = \tfrac{x_p+y_p}{2}, \ \ 
c_p^{(\nu+)} = \tfrac{x_p-y_p}{2}, \ \ 
\begin{array}{l} Q = A + B \\
P = A - B = CC^T \end{array}
\end{equation}
where the lower-triangular matrix $C$ was defined as a square root of $P$. Equation (\ref{fullRPA_eq}) then turns into
\begin{equation}
Q\vec{x}=E_\nu\vec{y}, \quad P\vec{y}=E_\nu\vec{x}
\end{equation}
and the eigenvalue problem can be formulated in terms of a symmetric matrix $C^T Q C$ with eigenvalues $E_\nu^2$ and eigenvectors $\vec{R}_\nu$.
\begin{equation}
\label{CQC}
\vec{x} = C\vec{R}_\nu,\quad C^T\vec{y}=E_\nu\vec{R}_\nu,\quad
C^T Q C\vec{R}_\nu = E_\nu^2\vec{R}_\nu
\end{equation}
Normalization condition (\ref{RPA_norm}) then becomes
\begin{equation}
\vec{x}\cdot\vec{y} = 1,\quad E_\nu = E_\nu\vec{x}\cdot\vec{y} =
\vec{x}\cdot Q\vec{x} = \vec{R}_\nu\cdot C^T Q C\vec{R}_\nu\quad\rightarrow\quad
\vec{R}_\nu^2 = 1/E_\nu
\end{equation}

\subsection{Transition operators}
After calculation of the RPA states, yielding $E_\nu$ and $c_{\alpha\beta}^{(\nu\pm)}$, I am interested in the matrix elements of electric and magnetic transition operators and in the transition densities and currents.
\begin{equation}
\label{trans_me}
\langle\nu|\hat{M}_{\lambda\mu}|\textrm{RPA}\rangle =
\langle[\hat{C}_\nu^{\phantom{|}},\hat{M}_{\lambda\mu}]\rangle =
\sum_{\alpha\geq\beta} \frac{(-1)^{l_\beta+1}}{\sqrt{2\lambda+1}}
M_{\lambda;\alpha\beta}
\Big( c_{\alpha\beta}^{(\nu-)} + \gamma_T^M c_{\alpha\beta}^{(\nu+)} \Big)^{\!*}
\end{equation}
\begin{align}
\label{trans_rho}
\!\delta\rho_{q;\nu}(\vec{r}) & =
\langle[\hat{C}_\nu^{\phantom{|}},\hat{\rho}_{q}(\vec{r})]\rangle =
\sum_{\alpha\geq\beta} \frac{(-1)^{l_\beta+1}}{\sqrt{2\lambda+1}}
\rho_{q;\alpha\beta}^\lambda(r)
\big( c_{\alpha\beta}^{(\nu-)} + c_{\alpha\beta}^{(\nu+)} \big)^{\!*}
Y_{\lambda\mu}^*(\vartheta,\varphi) \\
\label{trans_cur}
\!\delta\vec{j}_{q;\nu}(\vec{r}) & =
\langle[\hat{C}_\nu^{\phantom{|}},\hat{\vec{j}}_{q}(\vec{r})]\rangle =
\sum_L\sum_{\alpha\geq\beta} \frac{(-1)^{l_\beta+1}}{\sqrt{2\lambda+1}}
j_{q;\alpha\beta}^{\lambda L}(r)
\big( c_{\alpha\beta}^{(\nu-)} - c_{\alpha\beta}^{(\nu+)} \big)^{\!*}
\vec{Y}_{\lambda\mu}^{L*}(\vartheta,\varphi)\!
\end{align}
Besides electric ($\gamma_T^{\mathrm{E}\lambda} = 1$) and magnetic ($\gamma_T^{\mathrm{M}\lambda} = -1$) operators in long-wave approximation ($k \equiv E_\nu/\hbar c \ll 1/r$), I will use also electric vortical, toroidal and compression operators \cite{Kvasil2011}
\begin{small}
\begin{subequations}
\label{tran}
\begin{align}
\label{M_E}
\hat{M}_{\lambda\mu}^\mathrm{E} & = \sum_i \hat{M}^{\mathrm{E}}_{\lambda \mu}(\vec{r}_i)
= e\sum_q z_q\sum_{i\in q}
\Big( r^\lambda Y_{\lambda\mu}(\vartheta,\varphi) \Big)_i \\
\label{M_M}
\hat{M}_{\lambda\mu}^\mathrm{M} & = \frac{\mu_N}{c}\sqrt{\lambda(2\lambda+1)}\,
\sum_q \sum_{i\in q} \bigg(
r^{\lambda-1}\vec{Y}_{\lambda\mu}^{\lambda-1}(\vartheta,\varphi)
\cdot\bigg[ \frac{g_q}{2}\vec{\sigma}
+ \frac{2 z_q}{\lambda+1} \hat{\vec{l}}\,\bigg]
\bigg)_i \\
\hat{M}_{\mathrm{vor};\lambda\mu}^\mathrm{E} & = \frac{-\mathrm{i}/c}{2\lambda+3}
\sqrt{\frac{2\lambda+1}{\lambda+1}}\int\!\mathrm{d}^3 r\,
\hat{\vec{j}}_\mathrm{nuc}(\vec{r})r^{\lambda+1}
\vec{Y}_{\lambda\mu}^{\lambda+1}(\vartheta,\varphi) =
\hat{M}_{\mathrm{tor};\lambda\mu}^\mathrm{E} + \hat{M}_{\mathrm{com};\lambda\mu}^E \\
\label{M_tor}
\hat{M}_{\mathrm{tor};\lambda\mu}^\mathrm{E} & =
\frac{-1}{2c(2\lambda+3)}\sqrt{\frac{\lambda}{\lambda+1}}
\int\mathrm{d}^3 r\,\hat{\vec{j}}_\mathrm{nuc}(\vec{r})\cdot\vec{\nabla}\times
\big[r^{\lambda+2}\vec{Y}_{\lambda\mu}^\lambda(\vartheta,\varphi)\big] \\
\label{M_com}
\hat{M}_{\mathrm{com};\lambda\mu}^\mathrm{E} & =
\frac{\mathrm{i}}{2c(2\lambda+3)}
\int\mathrm{d}^3 r\,\hat{\vec{j}}_\mathrm{nuc}(\vec{r})\cdot
\vec{\nabla}\big[r^{\lambda+2} Y_{\lambda\mu}(\vartheta,\varphi)\big]\quad\ 
\big(\approx -k\hat{M}_{\mathrm{com};\lambda\mu}^{\mathrm{E}\,\prime}\big) \\
\hat{M}_{\mathrm{com};\lambda\mu}^{\mathrm{E}\,\prime} & =
\sum_i \hat{M}_{\mathrm{com};\lambda\mu}^{\mathrm{E}\,\prime}(\vec{r}_i) =
\frac{e}{2(2\lambda+3)} 
\sum_q z_q \sum_{i\in q} \Big( r^{\lambda+2} Y_{\lambda\mu}(\vartheta,\varphi) \Big)_i
\end{align}
\end{subequations}
\end{small}\\[-6pt]
where $z_q$ are effective charges of the nucleons, $g_{q}$ are spin g-factors (these are reduced by a quenching factor 0.7), $\hat{\vec{l}} = -\mathrm{i}\vec{r}\times\vec{\nabla}$, and $\vec{j}_\mathrm{nuc}$ is a nuclear current composed of convective and magnetization part
\begin{equation}
\label{j_nuc}
\hat{\vec{j}}_\mathrm{nuc}(\vec{r}) = \frac{e\hbar}{m_p}\sum_{q=p,n}\sum_{i\in q}
\Big[ z_q\hat{\vec{j}}_i(\vec{r}) + \frac{1}{4} g_q\vec{\nabla}_{\!(\vec{r})}\times\hat{\vec{s}}_i(\vec{r})\Big]
\end{equation}
where (convective) current and spin one-body operators are the same as in Skyrme functional (\ref{Jd_op}):
\[ \hat{\vec{j}}(\vec{r}_0) = \tfrac{\mathrm{i}}{2}
\big[\overleftarrow{\nabla}\delta(\vec{r}-\vec{r}_0)
-\delta(\vec{r}-\vec{r}_0)\overrightarrow{\nabla}\big],\qquad
\hat{\vec{s}}(\vec{r}_0) = \vec{\sigma}\delta(\vec{r}-\vec{r}_0) \]
Formula (\ref{j_nuc}) can be derived by the non-relativistic reduction of Dirac current $\vec{j} = ec\psi^\dagger\vec{\alpha}\psi$, and by replacing electron-like factor $g=2$ by generic $g_q$. Reduced matrix element of the orbital-angular-momentum-like operator
\begin{equation}
\hat{l}(\vec{r}) = \sum_j\delta(\vec{r}_j-\vec{r})\hat{l}_j
= -\mathrm{i}\sum_j\delta(\vec{r}_j-\vec{r})\vec{r}_j\times\vec{\nabla}_j
\end{equation}
involved in $\hat{M}_{\lambda\mu}^\mathrm{M}$ (\ref{M_M}) is evaluated as
\begin{align}
l_{\alpha\beta}^{JL}(r) &=
u_{\alpha\beta}^{(-)}R_\alpha^{(0)}(r)R_\beta^{(0)}(r)\frac{(-1)^{j_\beta+\frac{1}{2}}}{\sqrt{4\pi}}
\sqrt{(2\lambda+1)(2l_\beta+1)(l_\beta+1)l_\beta} \nonumber\\
&\qquad\qquad{}\times\begin{Bmatrix} L & J & 1 \\ l_\beta & l_\beta & l_\alpha \end{Bmatrix}
\begin{Bmatrix} l_\alpha & l_\beta & J \\ j_\beta & j_\alpha & \frac{1}{2} \end{Bmatrix}
C_{l_\alpha 0 l_\beta 0}^{L 0}
\end{align}
usually with $J=\lambda$ and $L=\lambda-1$.

Operators (\ref{tran}) can be derived by the long-wave approximation ($kr\ll 1$) of the exact transition operators \cite{Greiner1996}, using $\vec{\nabla}\cdot\delta\vec{j} = -\partial_t\delta\rho = -\mathrm{i}kc\delta\rho$.
\begin{subequations}\label{exactM}
\begin{align}
\hat{M}_{\lambda\mu}^\mathrm{exactE} & =
-\frac{(2\lambda+1)!!}{ck^{\lambda+1}} \sqrt{\frac{\lambda}{\lambda+1}}
\int\mathrm{d}^3 r\,\hat{\vec{j}}_\mathrm{nuc}(\vec{r})\cdot\vec{\nabla}\times
\big[j_\lambda(kr)\vec{Y}_{\lambda\mu}^\lambda(\vartheta,\varphi)\big] \\
&\approx \hat{M}_{\lambda\mu}^\mathrm{E} - k\hat{M}_{\mathrm{tor};\lambda\mu}^\mathrm{E}
+ \ldots \nonumber\\
\hat{M}_{\lambda\mu}^\mathrm{exactM} & =
-\mathrm{i}\frac{(2\lambda+1)!!}{ck^\lambda} \sqrt{\frac{\lambda}{\lambda+1}}
\int\mathrm{d}^3 r\,\hat{\vec{j}}_\mathrm{nuc}(\vec{r})\cdot
\big[j_\lambda(kr)\vec{Y}_{\lambda\mu}^\lambda(\vartheta,\varphi)\big] \\
&\approx \hat{M}_{\lambda\mu}^\mathrm{M} + \ldots \nonumber
\end{align}
\end{subequations}
where $j_\lambda(kr)$ is the spherical Bessel function and $k = E_\nu/\hbar c$.
\begin{equation}
j_\lambda(kr) = \sum_{n=0}^\infty
\frac{(-1)^n (kr)^{\lambda+2n}}{\displaystyle2^n n!(2\lambda+2n+1)!!}
= \frac{(kr)^\lambda}{(2\lambda+1)!!}\bigg(1
-\frac{(kr)^2}{2(2\lambda+3)}+O[(kr)^4]\bigg)
\end{equation}
Quantity $k$ is not a constant: it depends on the particular transition, and also changes sign under hermitian conjugation. For this reason, the electric operators containing an odd power of $k$ (including $j_\lambda(kr)$) are time-even, despite the time-odd nature of the current $\hat{\vec{j}}_\mathrm{nuc}$ (please notice that $\hat{M}_{\mathrm{tor};\lambda\mu}^\mathrm{E}$ in our definition is (strictly speaking) time-odd and non-hermitian, because it was stripped of $k$).

The constant involved in magnetic and toroidal/compression transition operators is
\begin{equation}
\frac{\mu_N}{ec} = \frac{\hbar}{2m_p c} = 0.10515445\ \mathrm{fm}
\end{equation}
and the elementary charge $e$ (as a symbolical parameter without a specific unit system) is usually excluded from the numerical evaluation. Then the matrix element is said to be in units $[e.\mathrm{fm}^\lambda]$ (or $[e.\mathrm{fm}^{\lambda+1}]$ for $\hat{M}_{\mathrm{vtc};\lambda\mu}^\mathrm{E}$, or $[e.\mathrm{fm}^{\lambda+2}]$ for $\hat{M}_{\mathrm{com};\lambda\mu}^{\mathrm{E}\,\prime}$). Magnetic transitions are often enumerated excluding the whole $\mu_N/c$ factor, and are then reported as being in units $[\mu_N.\mathrm{fm}^{\lambda-1}]$ (because $\mu_N/c$ in SI units is equivalent to $\mu_N$ in cgs units).

Gamma absorption cross section is related to the transition probability
\begin{equation}
B(\lambda\mu,0\rightarrow\nu) = \big|\langle\nu|\hat{M}_{\lambda\mu}|\textrm{RPA}\rangle\big|^2
= \big|\langle[\hat{C}_\nu^{\phantom{|}},\hat{M}_{\lambda\mu}]\rangle\big|^2
\end{equation}
by the formula \cite{VeselyPhD} (assuming the exact transition operators (\ref{exactM})):
\begin{align}
\label{cross_sec}
\sigma_\gamma(E) = \frac{8\pi^3\alpha}{e^2}\sum_\nu\sum_{\lambda\mu}
\frac{E_\nu^{2\lambda-1}}{(\hbar c)^{2\lambda-2}} \frac{\lambda+1}{\lambda[(2\lambda+1)!!]^2}
&\big[B(\mathrm{E}\lambda\mu,0\rightarrow\nu)+B(\mathrm{M}\lambda\mu,0\rightarrow\nu)\big] \nonumber\\[-8pt]
&\quad{}\times\delta_\Delta(E_\nu-E)
\end{align}
where the Lorentz function
\begin{equation}
\delta_\Delta(E_\nu-E) = \frac{\Delta}{2\pi[(E_\nu-E)^2+(\Delta/2)^2]}
\end{equation}
accounts for a finite half-life, but in practice, other effects are included by choosing a larger width $\Delta$ (such as finite experimental energy resolution, inability to calculate fragmentation of the states due to complex configurations etc.). The observed absorption cross-section is mostly dominated by long-wave isovector E1 transitions, so the larger multipolarities (and also monopole and isoscalar transitions) can be measured only indirectly, for example by electron or alpha scattering. The individual states are usually not distinguishable, and the distribution of the transition probability is depicted by means of a \emph{strength function}
\begin{equation}
\label{sf}
S_n(\mathrm{E/M}\lambda\mu; E) = \sum_\nu E_\nu^n
B(\mathrm{E/M}\lambda\mu,0\rightarrow\nu)\delta_\Delta(E_\nu-E)
\end{equation}
where $n$ is usually 0 or 1. Value $n=0$ is assumed in the case of omitted index.

The isoscalar toroidal and compression E1 transitions are very sensitive to the spurious center-of-mass motion, which can be subtracted by a correction $r^3\mapsto r^3-\frac{5}{3}r{\langle r^2\rangle}_0$ \cite{Kvasil2011}.
\begin{subequations}\label{E1vtccm}
\begin{align}
\hat{M}_{\mathrm{tor};1\mu}^{\mathrm{E},\Delta T=0} &= \frac{-1}{10\sqrt{2}\,c}
\int\mathrm{d}^3 r\,\hat{\vec{j}}_\mathrm{nuc}(\vec{r})\cdot\vec{\nabla}\times
\Big[\Big(r^3-\frac{5}{3}r{\langle r^2\rangle}_0\Big)
\vec{Y}_{1\mu}^1(\vartheta,\varphi)\Big] \nonumber\\
\label{M_torE1cm}
&= \frac{-\mathrm{i}}{2\sqrt{3}\,c} \int\mathrm{d}^3 r\,
\hat{\vec{j}}_\mathrm{nuc}(\vec{r})\cdot
\Big[\big(r^2-{\langle r^2\rangle}_0\big)\vec{Y}_{1\mu}^0
+\frac{\sqrt{2}}{5} r^2\vec{Y}_{1\mu}^2\Big] \\
\hat{M}_{\mathrm{com};1\mu}^{\mathrm{E},\Delta T=0} &= \frac{\mathrm{i}}{10c}
\int\mathrm{d}^3 r\,\hat{\vec{j}}_\mathrm{nuc}(\vec{r})\cdot\vec{\nabla}
\Big[\Big(r^3-\frac{5}{3}r{\langle r^2\rangle}_0\Big)
Y_{1\mu}(\vartheta,\varphi)\Big] \nonumber\\
\label{M_comE1cm}
&= \frac{\mathrm{i}}{2\sqrt{3}\,c} \int\mathrm{d}^3 r\,
\hat{\vec{j}}_\mathrm{nuc}(\vec{r})\cdot
\Big[\big(r^2-{\langle r^2\rangle}_0\big)\vec{Y}_{1\mu}^0
-\frac{2\sqrt{2}}{5} r^2\vec{Y}_{1\mu}^2\Big] \\
\label{M_com2E1cm}
\hat{M}_{\mathrm{com'};1\mu}^{\mathrm{E},\Delta T=0} &= \frac{e}{10} \sum_{i} \big[
\big(r^3-\tfrac{5}{3}r{\langle r^2\rangle}_0\big) Y_{1\mu}(\vartheta,\varphi)\big]_i
\end{align}
\end{subequations}
Center-of-mass correction essentially integrates and removes the contribution of homogeneous motion of the whole nucleus, since $\vec{Y}_{1\mu}^0 = \vec{e}_\mu/\sqrt{4\pi}$. Below is a derivation, suitable also for non-isoscalar transitions (with a c.m.~velocity $\vec{v}_\nu^\mathrm{\,c.m.}$ and a ground-state density $\rho_p(\vec{r})+\rho_n(\vec{r})$). For simplicity, I am taking $m_p=m_n$.
\begin{align*}
\vec{j}_{q;\nu}^\mathrm{\,c.m.}(\vec{r}) &=
\frac{\rho_q(\vec{r})m_q\vec{v}_\nu^\mathrm{\,c.m.}}{\hbar}
= \frac{\rho_q(\vec{r})}{A}\int\delta\vec{j}_\nu(\vec{r}_1)\,\mathrm{d}^3 r_1 \\
&= \frac{\rho_q(r)}{A}\vec{e}_{\mu}^{\,*}\sum_{\alpha\geq\beta}
\frac{(-1)^{l_\beta+1}}{\sqrt{3}}
\big( c_{\alpha\beta}^{(\nu-)} - c_{\alpha\beta}^{(\nu+)} \big)^{\!*}
\int j_{q;\alpha\beta}^{10}(r_1)\sqrt{4\pi}\,r_1^2\mathrm{d}r_1 \\
\delta\vec{j}_{q;\nu}^\textrm{\,corrected}(\vec{r}) &=
\delta\vec{j}_{q;\nu}(\vec{r}) - \vec{j}_{q;\nu}^\mathrm{\,c.m.}(\vec{r})
\end{align*}
After rearrangement of the integrals in the transition matrix element, the convective current (its lower component) and the density in the transition operator need to be substituted by
\begin{subequations}\label{cmc-generic}
\begin{align}
z_q\vec{j}_i(\vec{r})\cdot r^2\vec{Y}_{1\mu}^0(\vartheta,\varphi)\quad &\mapsto\quad
\vec{j}_i(\vec{r})\cdot
\big(z_q r^2 - {\langle r^2 \rangle}_t\big)\vec{Y}_{1\mu}^0(\vartheta,\varphi) \\
z_q \big[r^3 Y_{1\mu}(\vartheta,\varphi)\big]_i\quad &\mapsto\quad
\big[\big(z_q r^3-\tfrac{5}{3}r{\langle r^2\rangle}_t\big)
Y_{1\mu}(\vartheta,\varphi)\big]_i\\
&\textrm{with }{\langle r^2\rangle}_t = \int
\frac{z_p\rho_p(r)+z_n\rho_n(r)}{A} 4\pi r^4\mathrm{d}r \nonumber
\end{align}
\end{subequations}
It is not necessary to apply these corrections, if the spurious mode is sufficiently well separated (e.g.~by employing a large SHO basis), but then the spurious state has to be excluded from the calculation of the strength function.

The accuracy of the calculation for electric transitions can be checked by evaluation of the energy-weighted sum rule $m_1$ (EWSR), which relates certain commutators in the ground state to transition probabilities:
\begin{equation}
m_1 = \frac{1}{2}\sum_\mu\langle\mathrm{HF}|[\hat{M}_{\lambda\mu}^\dagger,
[\hat{H},\hat{M}_{\lambda\mu}]|\mathrm{HF}\rangle = \sum_\mu\sum_\nu E_\nu B(\mathrm{E}\lambda\mu,0\rightarrow\nu)
\end{equation}
In spherical symmetry, the transition probability doesn't depend on $\mu$
\begin{equation}
m_1\mathrm{(RPA)} = (2\lambda+1)\sum_\nu E_\nu
\big|\langle\nu|\hat{M}_{\lambda\mu}|\textrm{RPA}\rangle\big|^2
\end{equation}
and the ground-state estimate is
\begin{equation}
\label{EWSR-wf}
m_1 = (1 + \mathcal{K})\frac{\hbar^2}{2m}\sum_\mu\int
[\vec{\nabla},\hat{M}_{\lambda\mu}^\dagger]\cdot[\vec{\nabla},\hat{M}_{\lambda\mu}]
\rho(\vec{r}) \mathrm{d}^3 r
\end{equation}
where $\mathcal{K} = 0$ for isoscalar transitions, and for isovector case it is necessary to include non-zero enhancement factor $\mathcal{K}$ acting as a reduced effective mass \cite{Lipparini1989}:
\begin{equation}
\mathcal{K} =\frac{8mb_1}{\hbar^2}\frac{\int[\vec{\nabla},\hat{M}]^2
\rho_n(\vec{r})\rho_p(\vec{r})\mathrm{d}^3r}{\textstyle\int[\vec{\nabla},\hat{M}]^2
\rho(\vec{r})\mathrm{d}^3r}
\end{equation}
Commutator $[\vec{\nabla},\hat{M}_{\lambda\mu}]$ leads to a simple function for long-wave and time-even compression transitions
\begin{subequations}
\begin{align}
\vec{\nabla}r^\lambda Y_{\lambda\mu} &= \sqrt{\lambda(2\lambda+1)}\,
r^{\lambda-1}\vec{Y}_{\lambda\mu}^{\lambda-1} \\
\vec{\nabla}r^{\lambda+2} Y_{\lambda\mu} &= \frac{r^{\lambda+1}}{\sqrt{2\lambda+1}}
\big[(2\lambda+3)\sqrt{\lambda}\,\vec{Y}_{\lambda\mu}^{\lambda-1}
- 2\sqrt{\lambda+1}\,\vec{Y}_{\lambda\mu}^{\lambda+1}\big]
\end{align}
\end{subequations}
Isoscalar E1 compressional transition ($z_p=z_n=1$) with center-of-mass correction (\ref{M_com2E1cm}) gives
\begin{equation}
m_1\big[\hat{M}=\tfrac{1}{2}\big(r^3-\tfrac{5}{3}r{\langle r^2\rangle}_0\big) Y_{1\mu}\big]
= \frac{\hbar^2}{2m} \frac{3A}{16\pi} \bigg( 11\langle r^4 \rangle
- \frac{25}{3} \langle r^2 \rangle^2 \bigg)
\end{equation}

\section{Skyrme RPA in the axially deformed case}\label{sec_skyr_ax}
Full RPA was derived also for the axial symmetry, and the corresponding formalism is given below. Some of the concepts are similar to the spherical case (such as pairing factors, transition operators), so the reader is referred to the previous sections.

Cylindrical coordinates are
\begin{equation}
\varrho=\sqrt{x^2+y^2},\ z,\ \varphi=\mathrm{arctg}\,\frac{y}{x};\qquad x=\varrho\cos\varphi,\ y=\varrho\sin\varphi
\end{equation}
Calculations in axially deformed nuclei don't conserve total angular momentum, nevertheless, they conserve its $z$-projection and parity, so it is convenient to preserve part of the formalism from the spherical symmetry, namely the convention of $m$-components in vector and tensor operators, and the rule (\ref{hermit}) for their hermitian conjugation:
\begin{equation}
\label{hermit-copy}
\hat{A}_m^\dagger = (-1)^m\hat{A}_{-m}
\end{equation}
The operators of differentiation and the spin matrices are then
\begin{equation}
\begin{array}{rlrl}
\nabla_{+1}\!\!\!\! &=
\bigg({-}\dfrac{\partial}{\partial x}-\mathrm{i}\dfrac{\partial}{\partial y}\bigg)
= {-}\dfrac{\mathrm{e}^{\mathrm{i}\varphi}}{\sqrt{2}}
\bigg( \dfrac{\partial}{\partial\varrho} + \dfrac{\mathrm{i}}{\varrho}\dfrac{\partial}{\partial\varphi} \bigg)\quad\ 
& \sigma_{+1}\!\!\!\! &= \begin{pmatrix} 0 & \!{-}\sqrt{2}\, \\ 0 & 0 \end{pmatrix} \\[9pt]
\nabla_0\!\! &= \dfrac{\partial}{\partial z}\qquad
& \sigma_0\!\! &= \begin{pmatrix} 1 & 0 \\ 0 & {-1} \end{pmatrix} \\[9pt]
\nabla_{-1}\!\!\!\! &= \bigg(\dfrac{\partial}{\partial x}-\mathrm{i}\dfrac{\partial}{\partial y}\bigg)
= \dfrac{\mathrm{e}^{-\mathrm{i}\varphi}}{\sqrt{2}}
\bigg( \dfrac{\partial}{\partial\varrho}
 - \dfrac{\mathrm{i}}{\varrho}\dfrac{\partial}{\partial\varphi} \bigg)\qquad
& \sigma_{-1}\!\!\!\! &= \begin{pmatrix} 0 & 0 \\ \sqrt{2} & 0 \end{pmatrix}
\end{array}
\end{equation}
Single-particle wavefunction (and its time-reversal conjugate) is expressed as a spinor (with $m_\alpha^\pm = m_\alpha \pm \frac{1}{2}$)
\begin{equation}
\psi_\alpha(\vec{r}) =
\binom{R_{\alpha\uparrow}(\varrho,z)\,\mathrm{e}^{\mathrm{i}m_\alpha^-\varphi}}
{R_{\alpha\downarrow}(\varrho,z)\,\mathrm{e}^{\mathrm{i}m_\alpha^+\varphi}}, \quad
\psi_{\bar{\alpha}}(\vec{r}) = \binom{R_{\alpha\downarrow}(\varrho,z)\,\mathrm{e}^{-\mathrm{i}m_\alpha^+\varphi}}
{-R_{\alpha\uparrow}(\varrho,z)\,\mathrm{e}^{-\mathrm{i}m_\alpha^-\varphi}},
\end{equation}
and the radial parts of its derivatives will be denoted by a shorthand notation similar to (\ref{Rpm})
\begin{align}
\nabla_{+1}\psi_\alpha &= {-}\frac{\mathrm{e}^{\mathrm{i}\varphi}}{\sqrt{2}}
\binom{(\partial_\varrho R_{\alpha\uparrow} - m_\alpha^- R_{\alpha\uparrow}/\varrho)\,\mathrm{e}^{\mathrm{i}m_\alpha^-\varphi}}
{(\partial_\varrho R_{\alpha\downarrow} - m_\alpha^+ R_{\alpha\downarrow}/\varrho)\,\mathrm{e}^{\mathrm{i}m_\alpha^+\varphi}}
\equiv \mathrm{e}^{\mathrm{i}\varphi}
\begin{pmatrix} R_{\alpha\uparrow}^{(+)}\mathrm{e}^{\mathrm{i}m_\alpha^-\varphi} \\
R_{\alpha\downarrow}^{(+)}\mathrm{e}^{\mathrm{i}m_\alpha^+\varphi}
\end{pmatrix} \nonumber\\
\nabla_0\psi_\alpha &=
\binom{\partial_z R_{\alpha\uparrow}\mathrm{e}^{\mathrm{i}m_\alpha^-\varphi}}
{\partial_z R_{\alpha\downarrow}\mathrm{e}^{\mathrm{i}m_\alpha^+\varphi}}
\equiv \begin{pmatrix} R_{\alpha\uparrow}^{(0)}\mathrm{e}^{\mathrm{i}m_\alpha^-\varphi} \\
R_{\alpha\downarrow}^{(0)}\mathrm{e}^{\mathrm{i}m_\alpha^+\varphi} \end{pmatrix} \\
\nabla_{-1}\psi_\alpha &= \frac{\mathrm{e}^{-\mathrm{i}\varphi}}{\sqrt{2}}
\binom{(\partial_\varrho R_{\alpha\uparrow} + m_\alpha^- R_{\alpha\uparrow}/\varrho)\,\mathrm{e}^{\mathrm{i}m_\alpha^-\varphi}}
{(\partial_\varrho R_{\alpha\downarrow} + m_\alpha^+ R_{\alpha\downarrow}/\varrho)\,\mathrm{e}^{\mathrm{i}m_\alpha^+\varphi}}
\equiv \mathrm{e}^{-\mathrm{i}\varphi}
\begin{pmatrix} R_{\alpha\uparrow}^{(-)}\mathrm{e}^{\mathrm{i}m_\alpha^-\varphi} \\
R_{\alpha\downarrow}^{(-)}\mathrm{e}^{\mathrm{i}m_\alpha^+\varphi} \end{pmatrix} \nonumber
\end{align}
Let's emphasize that the index $(\pm)$ in the axial case stands for a shift in $m$, whereas in the spherical case, there was a shift in $l$.

Radial functions $R_{\alpha\uparrow\!\downarrow}(\varrho,z),\,R_{\alpha\uparrow\!\downarrow}^{(\pm)}$ are real, and their spinor-wise products will be denoted by a dot to keep the expressions simple:
\begin{equation}
R_\alpha \cdot R_\beta \ \equiv\ R_{\alpha\uparrow}(\varrho,z) R_{\beta\uparrow}(\varrho,z)
+ R_{\alpha\downarrow}(\varrho,z) R_{\beta\downarrow}(\varrho,z)
\end{equation}
Vector currents will be decomposed in the style of tensor operators of rank 1. Vector product in the expression for spin-orbital current leads to (for vector product in $m$-scheme see \cite[(1.2.28)]{Varshalovich1988})
\begin{equation}
(\vec{\nabla}\times\vec{\sigma})\psi_\alpha = \left\{\begin{array}{rl}
{+}1: & \mathrm{i}\,\mathrm{e}^{\mathrm{i}\varphi}
\begin{pmatrix} \big({-}R_{\alpha\uparrow}^{(+)} -\sqrt{2}\,R_{\alpha\downarrow}^{(0)}\big)
\mathrm{e}^{\mathrm{i}m_\alpha^-\varphi} \\
R_{\alpha\downarrow}^{(+)}\mathrm{e}^{\mathrm{i}m_\alpha^+\varphi} \end{pmatrix} \\
0: & \mathrm{i}\,\begin{pmatrix}
{-}\sqrt{2}\,R_{\alpha\downarrow}^{(-)}\mathrm{e}^{\mathrm{i}m_\alpha^-\varphi} \\
{-}\sqrt{2}\,R_{\alpha\uparrow}^{(+)}\mathrm{e}^{\mathrm{i}m_\alpha^+\varphi}
\end{pmatrix} \\
{-}1: & \mathrm{i}\,\mathrm{e}^{-\mathrm{i}\varphi}
\begin{pmatrix} R_{\alpha\uparrow}^{(-)}\mathrm{e}^{\mathrm{i}m_\alpha^-\varphi} \\
\big({-}R_{\alpha\downarrow}^{(-)} -\sqrt{2}\,R_{\alpha\uparrow}^{(0)}\big)
\mathrm{e}^{\mathrm{i}m_\alpha^+\varphi} \end{pmatrix} \end{array} \right.
\end{equation}

Matrix elements of densities and currents are then
\begin{subequations}
\label{axial_me}
\begin{align}
\langle\alpha|\hat{\rho}|\beta\rangle &= R_\alpha \cdot R_\beta
\,\mathrm{e}^{\mathrm{i}(m_\beta-m_\alpha)\varphi} \\
\langle\alpha|\hat{\tau}|\beta\rangle &= \big( R_\alpha^{(0)} \cdot R_\beta^{(0)}
+ R_\alpha^{(+)} \cdot R_\beta^{(+)} + R_\alpha^{(-)} \cdot R_\beta^{(-)} \big)
\,\mathrm{e}^{\mathrm{i}(m_\beta-m_\alpha)\varphi}
\end{align}
Factor $\mathrm{e}^{\mathrm{i}(m_\beta-m_\alpha)\varphi}$ will be omitted in the following expressions.
\begin{small}
\begin{align}
\langle\alpha|\vec{\mathcal{J}}|\beta\rangle &= \left\{\!\! \begin{array}{rl}
{+}1:\!& \tfrac{1}{2}\mathrm{e}^{\mathrm{i}\varphi}\big[
\big(R_{\alpha\downarrow}^{(-)} + \sqrt{2}\,R_{\alpha\uparrow}^{(0)}\big)R_{\beta\downarrow}
- R_{\alpha\uparrow}^{(-)} R_{\beta\uparrow} \\
&\qquad\qquad{}-R_{\alpha\uparrow}
\big(R_{\beta\uparrow}^{(+)}+\sqrt{2}\,R_{\beta\downarrow}^{(0)}\big)
+ R_{\alpha\downarrow} R_{\beta\downarrow}^{(+)} \big] \\
0:\!& \tfrac{1}{2}\big[{-}\sqrt{2}\,\big( R_{\alpha\downarrow}^{(-)} R_{\beta\uparrow} + R_{\alpha\uparrow}^{(+)}R_{\beta\downarrow}
+ R_{\alpha\uparrow}R_{\beta\downarrow}^{(-)} + R_{\alpha\downarrow}R_{\beta\uparrow}^{(+)} \big) \big] \\
{-}1:\!& \tfrac{1}{2}\mathrm{e}^{-\mathrm{i}\varphi}\big[
\big( R_{\alpha\uparrow}^{(+)} + \sqrt{2}\,R_{\alpha\downarrow}^{(0)} \big) R_{\beta\uparrow}
- R_{\alpha\downarrow}^{(+)} R_{\beta\downarrow} \\
&\qquad\qquad{}-R_{\alpha\downarrow}
\big( R_{\beta\downarrow}^{(-)}+\sqrt{2}\,R_{\beta\uparrow}^{(0)} \big)
+ R_{\alpha\uparrow} R_{\beta\uparrow}^{(-)} \big]
\end{array} \right. \\
\langle\alpha|\vec{j}|\beta\rangle &= \left\{\!\! \begin{array}{rl}
{+}1:\!& \tfrac{\mathrm{i}}{2}\mathrm{e}^{\mathrm{i}\varphi}
\big( {-}R_\alpha^{(-)}\cdot R_\beta - R_\alpha\cdot R_\beta^{(+)} \big) \\
0:\!& \tfrac{\mathrm{i}}{2}
\big( R_\alpha^{(0)}\cdot R_\beta - R_\alpha\cdot R_\beta^{(0)} \big) \\
{-}1:\!& \tfrac{\mathrm{i}}{2}\mathrm{e}^{-\mathrm{i}\varphi}
\big( {-}R_\alpha^{(+)}\cdot R_\beta - R_\alpha\cdot R_\beta^{(-)} \big) \end{array} \right. \\
\quad \langle\alpha|\vec{s}|\beta\rangle &= \Bigg\{\!\! \begin{array}{rl}
{+}1:\!& \mathrm{e}^{\mathrm{i}\varphi}\big({-}\sqrt{2}\,R_{\alpha\uparrow} R_{\beta\downarrow}\big) \\
0:\!& R_{\alpha\uparrow} R_{\beta\uparrow} - R_{\alpha\downarrow} R_{\beta\downarrow} \\
{-}1:\!& \mathrm{e}^{-\mathrm{i}\varphi}\big(\sqrt{2}\,R_{\alpha\downarrow} R_{\beta\uparrow}\big)
\end{array}\!\! \\
\langle\alpha|\vec{T}|\beta\rangle &= \left\{\!\! \begin{array}{rl}
{+}1:\!& \mathrm{e}^{\mathrm{i}\varphi}(-\sqrt{2}\,)
\big[ R_{\alpha\uparrow}^{(0)} R_{\beta\downarrow}^{(0)}
+ R_{\alpha\uparrow}^{(+)} R_{\beta\downarrow}^{(+)}
+ R_{\alpha\uparrow}^{(-)} R_{\beta\downarrow}^{(-)} \big] \\
0:\!& R_{\alpha\uparrow}^{(0)} R_{\beta\uparrow}^{(0)} - R_{\alpha\downarrow}^{(0)} R_{\beta\downarrow}^{(0)}
+ R_{\alpha\uparrow}^{(+)} R_{\beta\uparrow}^{(+)} \\
&\qquad\qquad{}- R_{\alpha\downarrow}^{(+)} R_{\beta\downarrow}^{(+)}
+ R_{\alpha\uparrow}^{(-)} R_{\beta\uparrow}^{(-)} - R_{\alpha\downarrow}^{(-)} R_{\beta\downarrow}^{(-)} \\
{-}1:\!& \mathrm{e}^{-\mathrm{i}\varphi}\sqrt{2}\,
\big[ R_{\alpha\downarrow}^{(0)} R_{\beta\uparrow}^{(0)}
+ R_{\alpha\downarrow}^{(+)} R_{\beta\uparrow}^{(+)}
+ R_{\alpha\downarrow}^{(-)} R_{\beta\uparrow}^{(-)} \big] \end{array} \right.\\
\langle\alpha|\vec{\nabla}\times\vec{j}|\beta\rangle
&= -\mathrm{i}\big(\vec{\nabla}\psi_\alpha\big)^\dagger\!\times\!\vec{\nabla}\psi_\beta
= \left\{\!\! \begin{array}{rl}
{+}1:\!& \mathrm{e}^{\mathrm{i}\varphi}\big( R_\alpha^{(-)} \cdot R_\beta^{(0)}
+ R_\alpha^{(0)} \cdot R_\beta^{(+)} \big) \\
0:\!& R_\alpha^{(-)} \cdot R_\beta^{(-)} - R_\alpha^{(+)} \cdot R_\beta^{(+)} \\
{-}1:\!& \mathrm{e}^{-\mathrm{i}\varphi}\big( {-}R_\alpha^{(+)} \cdot R_\beta^{(0)}
- R_\alpha^{(0)} \cdot R_\beta^{(-)} \big) \end{array}\right. \\
\langle\alpha|\vec{\nabla}\cdot\vec{\mathcal{J}}|\beta\rangle
&= {-}R_{\alpha\uparrow}^{(+)} R_{\beta\uparrow}^{(+)} + R_{\alpha\downarrow}^{(+)} R_{\beta\downarrow}^{(+)}
+ R_{\alpha\uparrow}^{(-)} R_{\beta\uparrow}^{(-)} - R_{\alpha\downarrow}^{(-)} R_{\beta\downarrow}^{(-)} \nonumber\\
& \quad{}-\sqrt{2}\,\big( R_{\alpha\uparrow}^{(0)} R_{\beta\downarrow}^{(-)} + R_{\alpha\downarrow}^{(-)} R_{\beta\uparrow}^{(0)}
+ R_{\alpha\downarrow}^{(0)} R_{\beta\uparrow}^{(+)} + R_{\alpha\uparrow}^{(+)} R_{\beta\downarrow}^{(0)} \big) \\
\langle\alpha|\mathcal{J}_s|\beta\rangle &= \tfrac{\mathrm{i}}{2}\big[
\big( R_{\alpha\uparrow}^{(0)} + \sqrt{2}\,R_{\alpha\downarrow}^{(-)} \big) R_{\beta\uparrow}
- \big( R_{\alpha\downarrow}^{(0)} + \sqrt{2}\,R_{\alpha\uparrow}^{(+)} \big) R_{\beta\downarrow} \nonumber\\
&\qquad {}- R_{\alpha\uparrow} \big( R_{\beta\uparrow}^{(0)} + \sqrt{2}\,R_{\beta\downarrow}^{(-)} \big)
+ R_{\alpha\downarrow} \big( R_{\beta\downarrow}^{(0)} + \sqrt{2}\,R_{\beta\uparrow}^{(+)} \big)\big] \\[5pt]
\langle\alpha|\mathcal{J}_t|\beta\rangle &= \left\{\!\! \begin{array}{rl}
{+}2:\!& \tfrac{\mathrm{i}}{\sqrt{2}} \mathrm{e}^{2\mathrm{i}\varphi}
\big( R_{\alpha\uparrow}^{(-)} R_{\beta\downarrow}
+ R_{\alpha\uparrow} R_{\beta\downarrow}^{(+)} \big) \\
{+}1:\!& \tfrac{\mathrm{i}}{2\sqrt{2}} \mathrm{e}^{\mathrm{i}\varphi}
\big[ {-}R_{\alpha\uparrow}^{(-)} R_{\beta\uparrow}
+ \big(R_{\alpha\downarrow}^{(-)}
 -\sqrt{2}\,R_{\alpha\uparrow}^{(0)} \big)R_{\beta\downarrow} \\
&\qquad{}- R_{\alpha\uparrow}\big(R_{\beta\uparrow}^{(+)}
 -\sqrt{2}\,R_{\beta\downarrow}^{(0)}\big)
+ R_{\alpha\downarrow}R_{\beta\downarrow}^{(+)} \big] \\
0:\!& \tfrac{\mathrm{i}}{2\sqrt{3}}\big[
\big(\sqrt{2}\,R_{\alpha\uparrow}^{(0)}-R_{\alpha\downarrow}^{(-)}\big)R_{\beta\uparrow}
-\big(\sqrt{2}\,R_{\alpha\downarrow}^{(0)}-R_{\alpha\uparrow}^{(+)}\big)R_{\beta\downarrow} \\
 & \qquad{}-R_{\alpha\uparrow}\big(\sqrt{2}\,R_{\beta\uparrow}^{(0)}-R_{\beta\downarrow}^{(-)}\big)
+R_{\alpha\downarrow}\big(\sqrt{2}\,R_{\beta\downarrow}^{(0)}-R_{\beta\uparrow}^{(+)}\big)
\big] \\
{-}1:\!& \tfrac{\mathrm{i}}{2\sqrt{2}} \mathrm{e}^{-\mathrm{i}\varphi}
\big[ {-}\big(R_{\alpha\uparrow}^{(+)}
 -\sqrt{2}\,R_{\alpha\downarrow}^{(0)} \big)R_{\beta\uparrow}
+ R_{\alpha\downarrow}^{(+)} R_{\beta\downarrow} \\
&\qquad{}- R_{\alpha\uparrow}R_{\beta\uparrow}^{(-)}
+ R_{\alpha\downarrow}\big(R_{\beta\downarrow}^{(-)}
 -\sqrt{2}\,R_{\beta\uparrow}^{(0)}\big) \big] \\
{-}2:\!& \tfrac{\mathrm{i}}{\sqrt{2}} \mathrm{e}^{-2\mathrm{i}\varphi}
\big( {-}R_{\alpha\downarrow}^{(+)} R_{\beta\uparrow}
- R_{\alpha\downarrow} R_{\beta\uparrow}^{(-)} \big)
\end{array} \right. \\
\hat{\vec{L}}\psi_\beta &= -\mathrm{i}(\vec{r}\times\vec{\nabla})\psi_\beta
= \left\{\!\! \begin{array}{rl}
{+}1:\!& \mathrm{e}^{\mathrm{i}\varphi}\big( \tfrac{\varrho}{\sqrt{2}}R_\beta^{(0)}
 + z R_\beta^{(+)} \big) \\
0:\!& \tfrac{\varrho}{\sqrt{2}}\big(R_\beta^{(+)}+R_\beta^{(-)}\big) \\
{-}1:\!& \mathrm{e}^{-\mathrm{i}\varphi}\big( \tfrac{\varrho}{\sqrt{2}}R_\beta^{(0)}
 - z R_\beta^{(-)} \big) \end{array} \right.
\end{align}
\end{small} \\[-6pt]
\end{subequations}

In the actual calculation, it is necessary to choose projection of angular momentum $\mu$ and parity $\pi$ (together denoted also as $K^\pi$, where $K=\mu$). Selection of the two-quasiparticle pairs is then restricted by $m_\alpha-m_\beta=\mu$. Transition operators have the form of
\begin{equation}
\hat{M}_{\lambda\mu} = \sum_i M_{\lambda\mu}(\varrho_i,z_i)\,
\mathrm{e}^{\mathrm{i}\mu\varphi_i}
\end{equation}
where $M_{\lambda\mu}(\varrho,z)$ contains a function (or even derivatives) not dependent on $\varphi$.

Single-particle operators (including densities and currents) can be expressed in terms of quasiparticles
\begin{align}
\hat{A} &= \frac{1}{2}\sum_{\alpha\beta}u_{\alpha\beta}^{(\gamma_T^A)}
\langle\alpha|\hat{A}|\beta\rangle
\big(\hat{\alpha}_\alpha^+\hat{\alpha}_{\bar{\beta}}^+ + \gamma_T^A
\hat{\alpha}_{\bar{\alpha}}^{\phantom{*}}\hat{\alpha}_{\beta}^{\phantom{|}}\big) \\
\label{axial_dens}
\hat{\mathbf{J}}_d(\vec{r}) &= \frac{1}{2}\sum_\mu\sum_{\alpha\beta\in\mu}
\mathbf{J}_{d;\alpha\beta}(\varrho,z)
\big(\hat{\alpha}_\alpha^+\hat{\alpha}_{\bar{\beta}}^+ + \gamma_T^d
\hat{\alpha}_{\bar{\alpha}}^{\phantom{*}}\hat{\alpha}_{\beta}^{\phantom{|}}\big)
\mathrm{e}^{-\mathrm{i}\mu\varphi} \\[-6pt]
&\qquad\qquad\qquad\qquad
\textrm{with the selection rule }\ m_\alpha-m_\beta=\mu \nonumber
\end{align}
Expression (\ref{axial_dens}) is defining the shorthand notation $\mathbf{J}_{d;\alpha\beta}(\varrho,z)$ for matrix elements of densities and currents which can have scalar, vector or tensor character (thus bold font). $\mathbf{J}_{d;\alpha\beta}(\varrho,z)$ is derived from (\ref{axial_me}) by adding a pairing factor and omitting $\mathrm{e}^{-\mathrm{i}\mu\varphi}$.

Commutators are evaluated in quasiparticle vacuum as
\begin{align}
\langle[\hat{A}^\dagger,\hat{B}]\rangle &= \frac{\gamma_T^A-\gamma_T^B}{2}
\sum_{\alpha\beta}u_{\beta\alpha}^{(\gamma_T^A)} u_{\alpha\beta}^{(\gamma_T^B)}
\langle\beta|\hat{A}^\dagger|\alpha\rangle\langle\alpha|\hat{B}|\beta\rangle \nonumber\\
&= \frac{1-\gamma_T^A\gamma_T^B}{2}\sum_{\alpha\beta}
u_{\alpha\beta}^{(\gamma_T^A)} u_{\alpha\beta}^{(\gamma_T^B)}
\langle\alpha|\hat{A}|\beta\rangle^*\,\langle\alpha|\hat{B}|\beta\rangle
\end{align}

RPA phonons can be defined as
\begin{equation}
\label{RPA_ax}
\hat{C}_\nu^+ = \frac{1}{2}\sum_{\alpha\beta}\big(
c_{\alpha\beta}^{(\nu-)} \hat{\alpha}_\alpha^+ \hat{\alpha}_{\bar{\beta}}^+
- c_{\alpha\beta}^{(\nu+)} \hat{\alpha}_{\bar{\alpha}} \hat{\alpha}_\beta \big)
\end{equation}
(factor $1/2$ is due to double counting of $\alpha\beta$ vs.~$\bar{\beta}\bar{\alpha}$) and their commutator with hermitian density/current operator is
\begin{subequations}
\label{comm2_ax}
\begin{align}
\langle[\hat{\mathbf{J}}_d(\vec{r}),\hat{C}_\nu^+]\rangle &= \frac{1}{2}\sum_{\alpha\beta}
u_{\alpha\beta}^{(\gamma_T^d)} \langle\alpha|\hat{\mathbf{J}}_d(\vec{r})|\beta\rangle^*\,
\big( c_{\alpha\beta}^{(\nu-)} + \gamma_T^d c_{\alpha\beta}^{(\nu+)} \big) \\
&= \frac{1}{2}\sum_{\alpha\beta}\mathbf{J}_{d;\alpha\beta}^\dagger(\varrho,z)
\big( c_{\alpha\beta}^{(\nu-)} + \gamma_T^d c_{\alpha\beta}^{(\nu+)} \big)\mathrm{e}^{\mathrm{i}\mu\varphi}
\end{align}
\end{subequations}
where hermitian conjugation is understood in the sense of (\ref{hermit-copy}) for vector/tensor components (see decomposition of matrix elements (\ref{axial_me})) and the factor $\mathrm{e}^{\mathrm{i}\mu\varphi}$ will get cancelled by $\mathrm{e}^{-\mathrm{i}\mu\varphi}$ from another $\hat{\mathbf{J}}_{d'}(\vec{r})$ in Skyrme interaction or in Coulomb integral (\ref{coul_ax}). RPA equations $[\hat{H},\hat{C}_\nu^+] = E_\nu\hat{C}_\nu^+$ are then
\begin{equation}
\label{fullRPA_eq-copy}
\begin{pmatrix} A & B \\ B & A \end{pmatrix} \binom{c^{(\nu-)}}{c^{(\nu+)}} =
\begin{pmatrix} E_\nu & 0 \\ 0 & -E_\nu \end{pmatrix}
\binom{c^{(\nu-)}}{c^{(\nu+)}}
\end{equation}
where the matrices $A$ and $B$ are
\begin{subequations}
\begin{align}
\label{axial_fullRPA_A}
A_{pp'} & = \delta_{pp'}\varepsilon_p + \sum_{dd'}
\iint\mathrm{d}\vec{r}_1\mathrm{d}\vec{r}_2
\frac{\delta^2\mathcal{H}}{\delta J_d(\vec{r}_1)\delta J_{d'}(\vec{r}_2)}
\mathbf{J}_{d;p}^\dagger(\varrho_1,z_1)\cdot\mathbf{J}_{d';p'}(\varrho_2,z_2) \\
\label{axial_fullRPA_B}
B_{pp'} & = \sum_{dd'} \gamma_T^d
\iint\mathrm{d}\vec{r}_1\mathrm{d}\vec{r}_2
\frac{\delta^2\mathcal{H}}{\delta J_d(\vec{r}_1)\delta J_{d'}(\vec{r}_2)}
\mathbf{J}_{d;p}^\dagger(\varrho_1,z_1)\cdot\mathbf{J}_{d';p'}(\varrho_2,z_2)
\end{align}
\end{subequations}
Index $p$ labels the $2qp$ pair (e.g.~$\alpha\beta$), satisfying $m_\alpha-m_\beta=\mu$ and the scalar product is understood in the spherical-tensor sense
\begin{equation}
\mathbf{A}^\dagger\cdot\mathbf{B} = \sum_\sigma (-1)^\sigma \big[\mathbf{A}^\dagger\big]_{-\sigma}\big[\mathbf{B}\big]_\sigma = \sum_\sigma \big[\mathbf{A}\big]_\sigma^*\,\big[\mathbf{B}\big]_\sigma
\end{equation}

Removal of the duplicate $2qp$ pairs (such as (\ref{order2qp}) in the spherical case) is done by omitting states $\alpha$ with $m_\alpha < 0$, but including pairs $\alpha\bar{\beta}$ with $m_\alpha+m_\beta=\mu$ and $\alpha>\beta$ in some ordering (and with omission of the Pauli-violating case $\alpha\bar{\alpha}$). The equivalent duplicates are then $\alpha\beta\leftrightarrow\bar{\beta}\bar{\alpha}$ and $\alpha\bar{\beta}\leftrightarrow\beta\bar{\alpha}$. The $2qp$ space for $\mu = 0$ splits into an independent electric and magnetic subspace with symmetric/antisymmetric combination of pairs $\alpha\beta\leftrightarrow\beta\alpha$, respectively: $(\alpha\beta\pm\beta\alpha)/\sqrt{2}$; then the diagonal pairs ($\alpha=\beta$) are present only in electric transitions with even multipolarity [revision 22.02.2018; papers from 2017 are already correct].

\section{Coulomb integral}\label{sec_coul}
In the following text, I will analyze the correct way of integration for direct two-body Coulomb interaction. The discussion deals also with accuracy of numerical integration in general, which is an important aspect of nuclear calculations, due to the rapid increase of computational cost in reduced symmetry (axial and triaxial nuclei). No further physical questions are treated in this section.

The calculation of Coulomb potential involves a problem of integrable singularity ($1/r$) during the evaluation of discretized integrals in axial and cartesian coordinates. Even the spherical case contains a kink for $r_1 = r_2$, which prevents from the accurate application of Gaussian quadrature. One possible solution employs Talmi-Moshinski transformation to center-of-mass coordinates \cite{Hassan1980}, which shifts the singularity to $r=0$, where it can be integrated easily (it is cancelled by $r^2$ in spherical Jacobian). However, this method is not suitable for DFT, since the calculation of coefficients becomes unfeasible for higher shells ($N\geq12$).

It turns out that Gaussian quadrature is not necessary, and very precise results can be obtained also with equidistant lattice, as follows from Euler-Maclaurin summation formula for a smooth function $f(x)$ \cite{Edwards1974}
\begin{equation}
\label{EMformula}
\sum_{n=M}^{N} f(n) \sim \int_M^N f(x)\mathrm{d}x + \frac{f(M)+f(N)}{2}
+ \sum_{j=1}^{\nu} \frac{B_{2j}}{(2j)!}\big[f^{(2j-1)}(N)-f^{(2j-1)}(M)\big]
\end{equation}
where $B_{2j}$ are Bernoulli numbers
\begin{equation}
B_2 = \frac{1}{6},\ \ B_4 = -\frac{1}{30},\ \ B_6 = \frac{1}{42},\ \
B_8 = -\frac{1}{30},\ldots\quad
\frac{x}{1-\mathrm{e}^{-x}} = \sum_{n=0}^{\infty} \frac{B_n}{n!} x^n
\end{equation}
The Euler-Maclaurin formula (further abbreviated as E-M) is an asymptotic series, which doesn't have to converge, and its error is similar to the last included term (which is usually small, since the growth begins only in high-order terms, which are difficult to calculate anyway). When the integration grid is sufficiently large, the harmonic oscillator wavefunction (including its derivatives) on the boundaries is negligible, so the error of integration by simple summation rapidly vanishes, provided the oscillation wavelength $\lambda$ is sufficiently larger than grid spacing $\Delta$. Nyquist limit is $\lambda < 2\Delta$, while the double precision accuracy can be reached already with $\lambda<4\Delta$ for harmonic oscillator basis. However, due to uncertainities arising from numerical differentiation and its use in E-M corrections in Coulomb integral (\ref{coul_EM_HF}), it is advisable to shift the limit to $\lambda<6\Delta$ in HF and $\lambda<8\Delta$ in RPA. Together with the appropriate integration boundary, it gives
\begin{equation}
\label{int_params}
\Delta \leq \frac{\pi b}{3\sqrt{2N}}\ \textrm{ for HF},\quad \Delta \leq \frac{\pi b}{4\sqrt{2N}}\ \textrm{ for RPA}, \quad r_\mathrm{max} \geq 1.3 b\sqrt{2N}
\end{equation}
where $b=\sqrt{\tfrac{\hbar}{m\omega}}$ is the oscillator length and $N=2\nu_\mathrm{max}+l$ is the number of major shells. These choices correspond to $2.5N$ integration points in HF for spherical symmetry, or $3.3N$ in RPA.

In fact, the methods like Simpson and Romberg integration take advantage of cancellation of the boundary terms in (\ref{EMformula}) by admixing sums with larger spacing ($2\Delta,\,4\Delta,$ etc.). Such approach is not suitable here, due to oscillatory character of the wavefunctions, which make the wider-spaced sums incorrect. It is much better to include E-M corrections directly, if needed.

\subsection{Spherical symmetry}\label{sec_sph_coul}
Coulomb interaction is usually taken into account by assuming point charge of proton. Numerical value of the interaction constant in nuclear units is
\begin{equation}
\frac{e^2}{4\pi\epsilon_0} = \alpha\hbar c = \frac{197.32697\ \mathrm{MeV.fm}}{137.035999}
= 1.4399645\ \mathrm{MeV.fm}.
\end{equation}
Spatial part of the interaction can be decomposed in spherical coordinates as \cite[(5.17.21)]{Varshalovich1988}
\begin{equation}
\frac{1}{|\vec{r}_1-\vec{r}_2|} = \sum_{lm}\frac{4\pi}{2l+1}Y_{lm}(\vartheta_1,\varphi_1)
Y_{lm}^*(\vartheta_2,\varphi_2)\cdot\bigg\{\begin{array}{ll}
r_1^l/r_2^{l+1} \ \ & \textrm{for } r_1\leq r_2 \\
r_2^l/r_1^{l+1} \ \ & \textrm{for } r_1\geq r_2 \end{array}
\end{equation}
The value of the integrand is then finite for all $r_1,\,r_2$, and has a kink in $r_1=r_2$. To get an acceptable accuracy of the result, evaluation of the Coulomb integral on equidistant grid needs a correction in $r_1=r_2$ coming from Euler-Maclaurin (E-M) series (\ref{EMformula}). Let's suppose that the grid spacing is $\Delta$ and the kink is located at $r=n\Delta$. Then, E-M series has the form:
\begin{align}
\int_0^{+\infty} f(r)\mathrm{d}r &= \Delta\bigg[\frac{f(0)}{2}+\sum_{m=1}^\infty f(m\Delta)\bigg]
+\frac{\Delta^2}{12}\big[f'(n\Delta^+)-f'(n\Delta^-)\big] \nonumber\\
&\quad{}-\frac{\Delta^4}{720}\big[f'''(n\Delta^+)-f'''(n\Delta^-)\big]
+\frac{\Delta^6}{30240}\big[f^{(5)}(n\Delta^+)-f^{(5)}(n\Delta^-)\big]-\ldots
\end{align}
There is no correction in $r_1=0$ or $r_2=0$ due to the presence of only even powers of $r$ in the integrand. The first case will be explained at (\ref{r1zeroEM}) and the second one is obvious.

The integral to be evaluated is
\begin{equation}
\label{coul_int}
\int_0^\infty \rho_1^{L*}(r_1)r_1^2\mathrm{d}r_1 \int_0^\infty \rho_2^L(r_2)\mathrm{d}r_2
\cdot \bigg\{\begin{array}{ll} r_2^{L+2}/r_1^{L+1} \quad &\textrm{for }r_2 \leq r_1 \\
r_1^L/r_2^{L-1} \quad &\textrm{for }r_2 \geq r_1 \end{array}
\end{equation}
where $\rho^L(r)$ is component of multipolarity $L$ (in the sense $\rho(\vec{r}) = \rho^L(r)Y_{LM}(\vartheta,\varphi)$), having a generic power expansion around $r=0$ like $\rho^L(r) = ar^L + br^{L+2} + cr^{L+4} \ldots$ Corrections are then applied to diagonal terms as follows:
\begin{align}
\int_0^\infty r_1^2\mathrm{d}r_1 &\int_0^\infty r_2^2\mathrm{d}r_2
\frac{\rho_1(\vec{r}_1)\rho_2(\vec{r}_2)}{|\vec{r}_1-\vec{r}_2|} = \frac{4\pi}{2L+1}
\sum_{n=1}^\infty n^2\Delta^3\rho_1^{L*}(n\Delta) \nonumber \\
&\!\!\!\!\!{}\times\bigg\{\sum_{m=1}^\infty \bigg[ \Delta^2\rho_2^L(m\Delta)\cdot\Big\{
\begin{array}{ll} m^{L+2}/n^{L+1} & \textrm{for }m\leq n \\
n^L/m^{L-1} & \textrm{for }m > n \end{array} \bigg] 
-\frac{\Delta^2}{12}(2L+1)\rho_2(n\Delta) \nonumber\\
\label{coul_EM2}
& {}+\frac{\Delta^4}{720}(2L+1)\bigg[ \frac{L(L+1)}{(n\Delta)^2}\rho_2(n\Delta) + \frac{6}{n\Delta}\rho'_2(n\Delta) + 3\rho''_2(n\Delta)\bigg]\bigg\}
\end{align}
The second E-M correction (last line of (\ref{coul_EM2})) contains derivatives and can be quantified by using the neighboring grid points as
\begin{equation}
\label{coul_EM2b}
\frac{\Delta^2}{720}(2L+1)\big\{ \tfrac{L(L+1)}{n^2}\rho_2(n\Delta)
+3\big[\tfrac{n+1}{n}\rho_2((n\!+\!1)\Delta) + \tfrac{n-1}{n}\rho_2((n\!-\!1)\Delta)-2\rho_2(n\Delta)\big] \big\}
\end{equation}

Let's return to the question of behavior of integral (\ref{coul_int}) over $r_2$ in the limit $r_1\to0$. It can be separated in two parts
\begin{equation}
\label{r1zeroEM}
r_1^L\int_0^\infty \frac{\rho_2^L(r_2)}{r_2^{L-1}}\mathrm{d}r_2 +
\frac{1}{r_1^{L+1}}\int_0^{r_1}\rho_2^L(r_2)\bigg(r_2^{L+2}-\frac{r_1^{2L+1}}{r_2^{L-1}}\bigg)\mathrm{d}r_2
\end{equation}
The first integral is a constant with respect to $r_1$, while the second part leads to a polynomial of the form $ar_1^{L+2}+br_1^{L+4}+\cdots$, which after multiplication with $\rho_1^L(r_1)$ gives zero correction in subsequent integration over $r_1$.

For the case $L=0$ (used in Hartree-Fock), I will give also the third-order E-M correction, so the diagonal term in summation (\ref{coul_EM2}) becomes
\begin{equation}
\label{coul_EM3}
\Delta^2 n \rho - \frac{\Delta^2}{12}\rho
+ \frac{\Delta^4}{240}\bigg(\frac{2\rho'}{n\Delta}+\rho''\bigg)
- \frac{\Delta^6}{6048}\bigg(\frac{4\rho'''}{n\Delta}+\rho^{(4)}\bigg)
\end{equation}
However, the approximations given previously correspond to
\begin{equation*}
\textrm{diag.(\ref{coul_EM2})+(\ref{coul_EM2b}), }L=0:\quad
\Delta^2 n \rho - \frac{\Delta^2}{12}\rho + \frac{\Delta^4}{240}\bigg(
\frac{2\rho'}{n\Delta} + \rho'' + \frac{\Delta\rho'''}{3n}
+ \frac{\Delta^2\rho^{(4)}}{12n} \bigg)
\end{equation*}
To correct the last terms of this series into the form of (\ref{coul_EM3}), it is necessary to subtract $31\Delta^6(4\rho'''/(n\Delta)+\rho^{(4)})/60480$. The derivatives can be estimated by
\begin{small}
\begin{align*}
2\Delta^3\rho'''(n\Delta) &= \rho((n\!+\!2)\Delta) - 2\rho((n\!+\!1)\Delta)
+ 2\rho((n\!-\!1)\Delta) - \rho((n\!-\!2)\Delta) + O(\Delta^5)\\
\Delta^4\rho^{(4)}(n\Delta) &= \rho((n\!+\!2)\Delta) \!-\! 4\rho((n\!+\!1)\Delta)
 \!+\! 6\rho(n\Delta) \!-\! 4\rho((n\!-\!1)\Delta) \!+\! \rho((n\!-\!2)\Delta) \!+\! O(\Delta^6)
\end{align*}
\end{small}\\[-8pt]
The diagonal term ($r_1=r_2=n\Delta$) for $L=0$, together with up to third-order E-M correction, then becomes
\begin{subequations}
\label{coul_EM_HF}
\begin{align}
\Delta^2 n\rho(n\Delta) &{}- \frac{\Delta^2}{12}\rho(n\Delta) \\
&{}+\frac{\Delta^2}{240}\bigg(\frac{n+1}{n}\rho((n\!+\!1)\Delta)
 - 2\rho(n\Delta) + \frac{n-1}{n}\rho((n\!-\!1)\Delta)\bigg) \\
&{}-\frac{31\Delta^2}{60480}\bigg( \frac{n+2}{n}\rho((n\!+\!2)\Delta)
 - 4\frac{n+1}{n}\rho((n\!+\!1)\Delta) + 6\rho(n\Delta) \nonumber\\
& \qquad\qquad{}- 4\frac{n-1}{n}\rho((n\!-\!1)\Delta) + \frac{n-2}{n}\rho((n\!-\!2)\Delta) \bigg)
\end{align}
\end{subequations}

\subsection{Cartesian coordinates}
Estimation of the Coulomb potential
\begin{equation}
V(x_0,y_0,z_0) = \iiint_{-\infty}^{+\infty} \frac{\rho(x,y,z)}{\sqrt{
(x-x_0)^2+(y-y_0)^2+(z-z_0)^2}} \mathrm{d}x\,\mathrm{d}y\,\mathrm{d}z
\end{equation}
on an equidistant coordinate grid runs into singularity at $\vec{r}=\vec{r}_0$, so the integral in its vicinity has to be evaluated analytically (in the following: $\vec{r}_0 = 0$). The border between integrated and summed function leads to E-M correction, which can be most easily estimated by inverse procedure -- cutting out the cube $C=({-}\Delta,\Delta)^3$ from the integral/sum
\begin{equation}
\label{coul_3D}
\iiint_{-\infty}^{+\infty} \!\!\frac{\rho(x,y,z)}{\sqrt{
x^2+y^2+z^2}} \mathrm{d}x\,\mathrm{d}y\,\mathrm{d}z =
\iiint_{-\infty}^{+\infty} \!f(x,y,z)\mathrm{d}x\,\mathrm{d}y\,\mathrm{d}z =
\sum_{j,k,l} f(j\Delta,k\Delta,l\Delta) \Delta^3
\end{equation}
where the accuracy of the integral estimation by summation is satisfied for functions vanishing at the integration limits (as discussed previously) -- this assumption holds for finite functions. In the case of Coulomb singularity, the central cube is evaluated by integral, instead of summation, by the E-M formula (\ref{EMformula}), generalized stepwise to three dimensions:
\begin{small}
\begin{equation*}
\bigg[\frac{f(-\Delta)+f(\Delta)}{2}+f(0)\bigg]\Delta = \int_{-\Delta}^\Delta f(x)\mathrm{d}x
+ \frac{\Delta^2}{12}\big[f'(\Delta)-f'(-\Delta)\big] + O(\Delta^4)
\end{equation*}
\begin{align*}
\Big[\tfrac{f(-\Delta,-\Delta)+f(-\Delta,\Delta)+f(\Delta,-\Delta)+f(\Delta,\Delta)}{4}
+ \tfrac{f(0,-\Delta)+f(0,\Delta)+f(-\Delta,0)+f(\Delta,0)}{2} + f(0,0)\Big]\Delta^2 = \\
= \iint_{-\Delta}^\Delta f(x,y)\mathrm{d}x\mathrm{d}y + \frac{\Delta^2}{12}\bigg\{
\int_{-\Delta}^\Delta \big[f'(\Delta,y)-f'(-\Delta,y)\big]\mathrm{d}y \\
{}+\int_{-\Delta}^\Delta \big[f'(x,\Delta)-f'(x,-\Delta)\big]\mathrm{d}x\bigg\}+O(\Delta^4)
\end{align*}
\begin{align}
\bigg\{\frac{1}{8}\sum_{s_1 s_2 s_3}^{\pm1} f(s_1\Delta,s_2\Delta,s_3\Delta)
+\frac{1}{4}\sum_{s_1 s_2}^{\pm1} \big[f(s_1\Delta,s_2\Delta,0)+f(s_1\Delta,0,s_2\Delta)
+f(0,s_1\Delta,s_2\Delta)\big] \nonumber\\
 {}+\frac{1}{2}\sum_{s=\pm1}\big[f(s\Delta,0,0)+f(0,s\Delta,0)+f(0,0,s\Delta)\big]
+ f(0,0,0)\bigg\}\Delta^3 = \nonumber\\
\label{int_cube}
= \iiint_{-\Delta}^\Delta f(x,y,z)\,\mathrm{d}x\,\mathrm{d}y\,\mathrm{d}z
+ \frac{\Delta^2}{12}\oint_{\partial C}\vec{\nabla}f\cdot\mathrm{d}\vec{S} + O(\Delta^4)
\end{align}
\end{small}\\[-6pt]
The double and triple integrals should now be estimated analytically. Let's emphasize at this point that the aim of this somewhat cumbersome workaround is to obtain an effective value of $f_0 = f(0,0,0)$ to be plugged into sum (\ref{coul_3D}) instead of the infinite value.

To calculate the integrals in the cube $C=({-}\Delta,\Delta)^3$, the density $\rho(\vec{r})$ will be approximated by Taylor expansion, where only even terms contribute to the integration:
\begin{align}
\rho(\vec{r}) &= \rho_0+\rho_x\tfrac{x^2}{2}+\rho_y\tfrac{y^2}{2}+\rho_z\tfrac{z^2}{2}
+\rho_{xy}\tfrac{x^2 y^2}{4}+\rho_{xz}\tfrac{x^2 z^2}{4}+\rho_{yz}\tfrac{y^2 z^2}{4} \nonumber\\
\label{rho_tayl4}
&\qquad\ {}+\rho_{x4}\tfrac{x^4}{24}+\rho_{y4}\tfrac{y^4}{24}+\rho_{z4}\tfrac{z^4}{24} +\textrm{odd} + O(r^6)
\end{align}
Following integrals will be needed, which can be derived using hyperbolic sine, \emph{per partes} with $f=(xf)'-xf'$, substitution $\sqrt{x^2+a^2}=x+t$ and other tricks.
\begin{subequations}
\begin{align}
\int\frac{\mathrm{d}x}{\sqrt{a^2+x^2}} &= \ln(x+\sqrt{a^2+x^2})-\ln a \\
\int\ln\frac{a+\sqrt{a^2+b^2+x^2}}{\sqrt{b^2+x^2}}\mathrm{d}x &=
x\ln\frac{a+\sqrt{a^2+b^2+x^2}}{\sqrt{b^2+x^2}}
+ a\ln\frac{x+\sqrt{a^2+b^2+x^2}}{\sqrt{a^2+b^2}} \nonumber\\
&\quad{}- b\,\mathrm{arctg}\frac{ax}{b\sqrt{a^2+b^2+x^2}} \\
\int x\,\mathrm{arctg}\frac{ab}{x\sqrt{a^2+b^2+x^2}}\mathrm{d}x &=
\frac{x^2}{2}\mathrm{arctg}\frac{ab}{x\sqrt{a^2+b^2+x^2}}
-\frac{a^2}{2}\mathrm{arctg}\frac{bx}{a\sqrt{a^2+b^2+x^2}} \nonumber\\
&\quad{}-\frac{b^2}{2}\mathrm{arctg}\frac{ax}{b\sqrt{a^2+b^2+x^2}}
+ab\ln\frac{x+\sqrt{a^2+b^2+x^2}}{\sqrt{a^2+b^2}}
\end{align}
\end{subequations}
So the basic three-dimensional integral over cube $C$ in (\ref{int_cube}) becomes
\begin{equation}
\iiint_{-\Delta}^\Delta \frac{\mathrm{d}x\,\mathrm{d}y\,\mathrm{d}z}{\sqrt{x^2+y^2+z^2}}
= \bigg(3\ln\frac{1+\sqrt{3}}{\sqrt{2}}-\frac{3}{2}\,\mathrm{arctg}\frac{1}{\sqrt{3}}\bigg)8\Delta^2 = \Delta^2(24\beta-2\pi)
\end{equation}
where I defined a useful constant
\begin{equation}
\beta = \ln\frac{1+\sqrt{3}}{\sqrt{2}} = 0.658478948
\end{equation}
A more general evaluation of (\ref{int_cube}) by assuming Taylor expansion (\ref{rho_tayl4}) then leads to
\begin{subequations}
\begin{align}
\int_C\frac{\rho(\vec{r})\,\mathrm{d}x\,\mathrm{d}y\,\mathrm{d}z}{\sqrt{x^2+y^2+z^2}}
&= 12\Delta^2\big(2\beta-\tfrac{\pi}{6}\big)\rho_0
+\tfrac{\Delta^4}{3}\big(\sqrt{3}+4\beta-\tfrac{\pi}{6}\big)(\rho_x+\rho_y+\rho_z) \nonumber\\
&\quad{}+\tfrac{\Delta^6}{5}\big(\sqrt{3}-2\beta+\tfrac{\pi}{6}\big)(\rho_{xy}\!+\!\rho_{xz}\!+\!\rho_{yz}) \nonumber\\
&\quad{}-\tfrac{\Delta^6}{90}\big(2\sqrt{3}-19\beta+7\tfrac{\pi}{6}\big)(\rho_{x4}\!+\!\rho_{y4}\!+\!\rho_{z4}) \\
\oint_{\partial C}\vec{\nabla}f(\vec{r})\cdot\mathrm{d}\vec{S} &=
-4\pi\rho_0 + 4\Delta^2\big(2\beta-\tfrac{\pi}{6}\big)(\rho_x\!+\!\rho_y\!+\!\rho_z) \nonumber\\
&\quad{}+2\Delta^4\big(\sqrt{3}-4\beta+3\tfrac{\pi}{6}\big)(\rho_{xy}\!+\!\rho_{xz}\!+\!\rho_{yz}) \nonumber\\
&\quad{}-\tfrac{\Delta^4}{3}\big(\sqrt{3}-12\beta+7\tfrac{\pi}{6}\big)(\rho_{x4}+\rho_{y4}+\rho_{z4})
\end{align}
\end{subequations}
Derivatives can be estimated from the neighboring points, defining convenient symbols $\rho_1\cdots\rho_4$:
\begin{subequations}
\begin{align}
\rho_1 &=
\sum_{s=\pm1}\big[\rho(s\Delta,0,0)+\rho(0,s\Delta,0)+\rho(0,0,s\Delta)\big]
- 6\rho(0,0,0) \nonumber\\
&= \Delta^2(\rho_x+\rho_y+\rho_z)
+ \tfrac{\Delta^4}{12}(\rho_{x4}+\rho_{y4}+\rho_{z4}) \\
\rho_2 &=
\sum_{s_1s_2}^{\pm1}\big[\rho(s_1\Delta,s_2\Delta,0)+\rho(s_1\Delta,0,s_2\Delta)
+\rho(0,s_1\Delta,s_2\Delta)\big] - 12\rho(0,0,0) \nonumber\\
&= 4\Delta^2(\rho_x+\rho_y+\rho_z)
+ \Delta^4(\rho_{xy}+\rho_{xz}+\rho_{yz})
+ \tfrac{\Delta^4}{3}(\rho_{x4}+\rho_{y4}+\rho_{z4})
\end{align}
\begin{align}
\rho_3 &= \sum_{s_1s_2s_3}^{\pm1} \rho(s_1\Delta,s_2\Delta,s_3\Delta)-8\rho(0,0,0) \nonumber\\
&= 4\Delta^4(\rho_x+\rho_y+\rho_z) + 2\Delta^4(\rho_{xy}+\rho_{xz}+\rho_{yz})
+\tfrac{\Delta^4}{3}(\rho_{x4}+\rho_{y4}+\rho_{z4}) \\
\rho_4 &= \sum_{s=\pm1}\big[\rho(2s\Delta,0,0)+\rho(0,2s\Delta,0)+\rho(0,0,2s\Delta)\big]
- 6\rho(0,0,0) \nonumber\\
&= 4\Delta^2(\rho_x+\rho_y+\rho_z)
+ \tfrac{4}{3}\Delta^4(\rho_{x4}+\rho_{y4}+\rho_{z4})
\end{align}
\end{subequations}
These relations can be easily inverted to get
\begin{subequations}
\begin{align}
\Delta^2(\rho_x+\rho_y+\rho_z) &= \tfrac{4}{3}\rho_1 - \tfrac{1}{12}\rho_4 \\
\Delta^4(\rho_{xy}+\rho_{xz}+\rho_{yz}) &= \rho_2 - 4\rho_1 \\
\Delta^4(\rho_{x4}+\rho_{y4}+\rho_{z4}) &= \rho_4 - 4\rho_1
\end{align}
\end{subequations}
Desired value of $f(0,0,0)$, according to (\ref{int_cube}), is then
\begin{subequations}
\begin{align}
f_0\Delta &\approx \big(24\beta-\tfrac{7}{3}\pi
-\tfrac{1}{\sqrt{3}}-\tfrac{3}{\sqrt{2}}-3\big)\rho_0
+ \Delta^2\big(\tfrac{\sqrt{3}}{6}+2\beta-\tfrac{\pi}{9}
-\tfrac{1}{\sqrt{2}}-\tfrac{1}{2}\big)(\rho_x+\rho_y+\rho_z) \nonumber\\
&\quad{}+\Delta^4\big(\tfrac{17}{60}\sqrt{3}-\tfrac{16}{15}\beta+\tfrac{7}{60}\pi
-\tfrac{1}{4\sqrt{2}}\big)(\rho_{xy}+\rho_{xz}+\rho_{yz}) \nonumber\\
&\quad{}-\Delta^4\big(\tfrac{23}{360}\sqrt{3}-\tfrac{49}{90}\beta+\tfrac{49}{1080}\pi
+\tfrac{\sqrt{2}+1}{24}\big)(\rho_{x4}+\rho_{y4}+\rho_{z4}) \\
&\approx 2.774441\rho_0 + 0.049460(\rho_x+\rho_y+\rho_z)
- 0.021887(\rho_{xy}+\rho_{xz}+\rho_{yz}) \nonumber\\
&\quad{}+ 0.004719(\rho_{x4}+\rho_{y4}+\rho_{z4}) \\
&\approx 2.774441\rho_0 + 0.134621\rho_1 - 0.021887\rho_2 + 0.000597\rho_4
\end{align}
\end{subequations}
By taking E-M corrections up to $\Delta^4$ in (\ref{int_cube}), the coefficients become
\begin{subequations}
\begin{align}
f_0\Delta &\approx 2.864251\rho_0 + 0.044052(\rho_x+\rho_y+\rho_z)\Delta^2
+ 0.003330(\rho_{xy}+\rho_{xz}+\rho_{yz})\Delta^4 \nonumber\\
&\quad{}- 0.013328(\rho_{x4}+\rho_{y4}+\rho_{z4})\Delta^4 \\
&\approx 2.864251\rho_0 + 0.098728\rho_1 + 0.003330\rho_2 - 0.016999\rho_4
\end{align}
\end{subequations}
As can be seen, taking higher orders of E-M corrections does not increase the order of integral convergence due to a divergent nature of the integrand. Nevertheless, the accurate value of needed coefficients can be obtained by empirical evaluation of the convergence of Coulomb integral for various charge distributions. Such approach gives
\begin{subequations}
\label{f0_exact}
\begin{align}
f_0\Delta &= 2.8372974794806\rho_0 + 0.04443271312(\rho_x+\rho_y+\rho_z)\Delta^2
\nonumber\\
&\quad{}+ 0.01962487(\rho_{xy}+\rho_{xz}+\rho_{yz})\Delta^4 \nonumber\\
&\quad{}- 0.00825759(\rho_{x4}+\rho_{y4}+\rho_{z4})\Delta^4 + O(\Delta^6) \\
&\approx 2.83729748\rho_0 + 0.01377450\rho_1 + 0.01962487\rho_2 - 0.01196032\rho_4
\end{align}
\end{subequations}
and the Coulomb integral then converges as $O(\Delta^8)$ -- assuming that the charge density vanishes near the integration boundary, or the E-M corrections up to third order are included there.

Since the calculation of Coulomb potential is a convolution, it would be natural to apply Fourier transformation during the process. Convolution with $1/r$ is then replaced by a multiplication of the frequency domain by $4\pi/k^2$ (derived by taking the limit $\mu\to0$ in $\mathrm{e}^{-\mu r}/r$, whose Fourier transformation is $4\pi/(\mu^2+k^2)$). Again, there is a singularity in $k=0$. In fact, the whole procedure -- the F.T.~of the density, multiplication by $4\pi/k^2$ and the inverse F.T.~-- can be expressed as an integral
\begin{equation}
\frac{4\pi}{(2\pi)^3}\iiint_{-\pi/\Delta}^{\pi/\Delta}
\frac{\mathrm{e}^{\mathrm{i}\vec{k}\cdot(\vec{r}_2-\vec{r}_1)}}{k^2}\mathrm{d}^3 k
= \frac{1}{2\pi\Delta}\iiint_{-1}^1
\frac{\cos(\pi n_x q_x)\cos(\pi n_y q_y)\cos(\pi n_z q_z)}{q_x^2+q_y^2+q_z^2}
\mathrm{d}^3 q
\end{equation}
where $\vec{r}_2-\vec{r}_1=(n_x,n_y,n_z)\Delta$. This integral should be evaluated in continuum limit, which corresponds to a shift of the periodic boundary to infinity. To get $O((\mathrm{d}q)^7)$ convergence, it is necessary to include up to third-order E-M correction on the boundary and to take the value of the central point as
\begin{align}
\label{coul_FT0}
&\frac{\cos(\pi n_x q_x)\cos(\pi n_y q_y)\cos(\pi n_z q_z)}{q_x^2+q_y^2+q_z^2}\bigg|_{q=0}
= \frac{8.91363291758515}{(\mathrm{d}q)^2} - \frac{\pi^2}{6}(n_x^2+n_y^2+n_z^2) \nonumber\\
&\qquad\qquad{}+ 0.610299(\mathrm{d}q)^2 [3(n_x^2 n_y^2 + n_x^2 n_z^2 + n_y^2 n_z^2)
-(n_x^4 + n_y^4 + n_z^4)] + O((\mathrm{d}q)^4)
\end{align}
Convolution array obtained by this method should include derivative corrections to all orders, as compared to (\ref{f0_exact}), which includes only up to fourth derivative. However, this method of Fourier-like array is probably not usable due to computational cost of calculating all $N^3$ coefficients, which have to be calculated accurately, and their integration time grows rapidly for large $n$. At least it provides a comparison with the convolution coefficients obtained by the previous methods, see Table \ref{tab_coul3D}.
\begin{table}
\caption{Convolution coefficients for integration of Coulomb interaction on cartesian grid according to naive method, Euler-Maclaurin estimation up to first and second order, exact numerical estimate with up to fourth derivative of $\rho$, and the central part of Fourier array.}\label{tab_coul3D}
\centering
\begin{tabular}{|c||c|c|c|c|c|}
\hline
$(\vec{r}_1-\vec{r}_2)/\Delta$ & $\Delta/r$ & E-M$\leq\!\!\Delta^2$
 & E-M$\leq\!\!\Delta^4$ & exact $\Delta^4$ & F.T. \\
\hline
(0,0,0) & $\infty$ & 2.2329 & 2.3342 & 2.590914 & 2.4427 \\
(1,0,0) & 1.0000 & 1.1346 & 1.0987 & 1.013775 & 1.0517 \\
(1,1,0) & 0.7071 & 0.6852 & 0.7104 & 0.726732 & 0.7268 \\
(1,1,1) & 0.5774 & 0.5774 & 0.5774 & 0.577350 & 0.5851 \\
(2,0,0) & 0.5000 & 0.5006 & 0.4830 & 0.488040 & 0.4740 \\
\hline
\end{tabular}
\end{table}

One can also use Fourier method directly: by appling direct and then inverse fast Fourier transformation (FFT) to the density, which should reduce computational cost from $O(N^6)$ to $O((N\log N)^3)$. However, there is a problem of the potential leaking from the periodic boundary (due to discretized momentum), and inability to apply the corrections beyond the first term in (\ref{coul_FT0}). Both difficulties may be solved by placing proper compensating charges on the boundary of the coordinate grid (e.g., employing a multipolar expansion of the nuclear charge distribution, where the main contribution comes from the first few terms \cite{Dobaczewski1997}).

\subsection{Axial symmetry}
In axial symmetry (using $m$-scheme and coordinates $\varrho=\sqrt{x^2+y^2}\,,\,z,\,\varphi$, see also section \ref{sec_skyr_ax}), the direct Coulomb integral is
\begin{small}
\begin{align}
\iint \frac{\rho_1^*(\varrho_1,z_1)\rho_2(\varrho_2,z_2)}{|\vec{r}_1-\vec{r}_2|}
&\exp(\mathrm{i}m_1\varphi_1-\mathrm{i}m_2\varphi_2)\mathrm{d}\vec{r}_1\mathrm{d}\vec{r}_2 = \Bigg| \begin{array}{c}
\mathrm{d}\vec{r}=\varrho\,\mathrm{d}\varrho\,\mathrm{d}z\,\mathrm{d}\varphi \\
\varphi = \varphi_1-\varphi_2 \\
\int \mathrm{e}^{\mathrm{i}(m_1-m_2)\varphi_2}\mathrm{d}\varphi_2 = 2\pi\delta_{m_1m_2}
\end{array} \Bigg| = \nonumber\\
&=2\pi\delta_{m_1m_2}
\int_0^\infty\rho_1^*(\varrho_1,z_1)\varrho_1\mathrm{d}\varrho_1\mathrm{d}z_1
\int_0^\infty\rho_2(\varrho_2,z_2)\varrho_2\mathrm{d}\varrho_2\mathrm{d}z_2 \nonumber\\
\label{coul_ax}
&\qquad\times\int_0^{2\pi}\frac{\exp(\mathrm{i}m\varphi)}{\sqrt{(z_1-z_2)^2+\varrho_1^2+\varrho_2^2
-2\varrho_1\varrho_2\cos\varphi}}\mathrm{d}\varphi
\end{align}
\end{small} \\[-6pt]
where at least the first E-M correction should be taken into account for $\varrho=0$ (as compared to spherical case, where it is zero), so the integral is evaluated like
\begin{equation}
\label{axial_int}
\int_0^\infty f(\varrho)\varrho\,\mathrm{d}\varrho = \Big[
\tfrac{\Delta}{12}f(0) + 1\Delta f(1\Delta) + 2\Delta f(2\Delta) + 3\Delta f(3\Delta)
+\ldots \Big]\Delta
\end{equation}

As can be seen, straightforward evaluation of (\ref{coul_ax}) gives an integral over $\varphi$
\begin{align}
\label{coul_ax_int}
\int_0^{2\pi} \frac{\exp(\mathrm{i}m\varphi)}{\sqrt{(z_1-z_2)^2
+\varrho_1^2+\varrho_2^2-2\varrho_1\varrho_2\cos\varphi}}
&= \frac{g_m\big(\tfrac{2\varrho_1\varrho_2}{(z_1-z_2)^2
+\varrho_1^2+\varrho_2^2}\big)}{\sqrt{(z_1-z_2)^2+\varrho_1^2+\varrho_2^2}} \\
\label{coul_ax_g}
\textrm{where }\ g_m(x) &= \int_{-\pi}^{\pi}
\frac{\cos(m\varphi)}{\sqrt{1-x\cos\varphi}} \mathrm{d}\varphi
\end{align}
which cannot be expressed in a closed form, but there is a Taylor expansion
\begin{equation}
\label{gm_taylor}
g_m(x) = -\mathrm{i}\oint\limits_{|z|=1} \frac{z^{m-1}\mathrm{d}z}{\sqrt{1-x(z+1/z)/2}}
= 2\pi\sum_{k=0}^{\infty} \frac{(4k+2m-1)!!}{k!(k+m)!}\bigg(\frac{x}{4}\bigg)^{m+2k}
\end{equation}
Function $g_m(x)$ has a logarithmic singularity in $x\to1^{-}$
\begin{equation}
g_m(1-t) = (O(t)-\sqrt{2})\ln t + O(1)
\end{equation}

It is usually suggested \cite{Dobaczewski2005} to reformulate the original integral by a Gaussian substitution, e.g.
\begin{equation}
\label{gaus_subs}
\frac{1}{|\vec{r}_1-\vec{r}_2|} = \frac{2}{\sqrt{\pi}}\int_0^\infty
\exp[-(\vec{r}_1-\vec{r}_2)^2 t^2]\mathrm{d}t
\end{equation}
The integral (\ref{coul_ax_int}) is then replaced by
\begin{small}
\begin{align}
\frac{2}{\sqrt{\pi}}&\int_0^{2\pi}\!\!\mathrm{d}\varphi\int_0^\infty\!\mathrm{d}t
\exp\{-[(z_1\!-\!z_2)^2\!+\!\varrho_1^2\!+\!\varrho_2^2\!
-\!2\varrho_1\varrho_2\cos\varphi]t^2
+\mathrm{i}m\varphi\}
= \Bigg|\!\! \begin{array}{c} u=\mathrm{e}^{\mathrm{i}\varphi} \\
\mathrm{d}u = \mathrm{i}\,\mathrm{e}^{\mathrm{i}\varphi}\mathrm{d}\varphi \\
2\cos\varphi = u + 1/u \end{array} \!\!\Bigg| = \nonumber\\
&= -\frac{2}{\sqrt{\pi}}\mathrm{i}\oint u^{m-1}\mathrm{d}u\int_0^\infty\mathrm{d}t
\exp\big\{{-}\big[(z_1-z_2)^2+\varrho_1^2+\varrho_2^2-\big(u+\tfrac{1}{u}\big)\big]t^2\big\} \nonumber \\
\label{coul_ax2}
&= \frac{2}{\sqrt{\pi}} 2\pi\int_0^\infty
\mathrm{e}^{-[(z_1-z_2)^2+\varrho_1^2+\varrho_2^2]t^2}
\sum_{n=-\infty}^{+\infty}I_n(2\rho_1\rho_2 t^2)u^n \nonumber\\
&= \frac{2}{\sqrt{\pi}} 2\pi\int_0^\infty
\mathrm{e}^{-[(z_1-z_2)^2+(\varrho_1-\varrho_2)^2]t^2}
\frac{I_m(2\rho_1\rho_2 t^2)}{\exp(2\rho_1\rho_2 t^2)}\mathrm{d}t
\end{align}
\end{small} \\[-6pt]
where the modified Bessel function \cite{Abramowitz1972} was used, which has an asymptotic behavior of $\exp(x)/\sqrt{x}$ (therefore, the computer libraries give it as $I_m(x)/\exp(x)$). Laurent series of modified Bessel function $I_n(x)$ can be derived from normal Bessel function $J_n(x)$ by evaluating it at imaginary axis ($z=\mathrm{i}x$)
\begin{align}
&\exp\bigg[\frac{z}{2}\bigg(t-\frac{1}{t}\bigg)\bigg]
= \sum_{n=-\infty}^{+\infty} J_n(z)\,t^n
\quad\Rightarrow\quad
\exp\bigg[\frac{x}{2}\bigg(\mathrm{i}t+\frac{1}{\mathrm{i}t}\bigg)\bigg] = \sum_{n=-\infty}^{+\infty} J_n(\mathrm{i}x)\,t^n \nonumber\\
&\qquad\qquad\qquad\Rightarrow\quad
\exp\bigg[\frac{x}{2}\bigg(u+\frac{1}{u}\bigg)\bigg] = \sum_{n=-\infty}^{+\infty} I_n(u)\,t^n
\end{align}
where
\begin{equation}
I_n(-x) = I_{-n}(x) = I_n(x) = (-\mathrm{i})^n J_n(\mathrm{i}x) =
\sum_{k=0}^\infty \frac{(x/2)^{n+2k}}{k!(n+k)!}
\end{equation}

The reformulated integral (\ref{coul_ax2}) is then rescaled to fit the interval $x\in(0,1)$ of the Gauss-Legendre quadrature:
\begin{equation}
\label{GL_subs}
t^2 = \frac{x^2}{1-x^2},\quad \mathrm{d}t = \frac{\mathrm{d}x}{(1-x^2)\sqrt{1-x^2}},\quad
t\in(0,\infty)\ \Rightarrow\ x\in(0,1)
\end{equation}
leading to
\begin{equation}
\frac{2}{\sqrt{\pi}} 2\pi\int_0^1
\exp\big\{{-}[(z_1-z_2)^2+(\varrho_1-\varrho_2)^2]\tfrac{x^2}{1-x^2}\big\}
\frac{I_m(2\rho_1\rho_2 \frac{x^2}{1-x^2})}{\exp(2\rho_1\rho_2 \frac{x^2}{1-x^2})}
\frac{\mathrm{d}x}{(1-x^2)\sqrt{1-x^2}}
\end{equation}
As can be seen, the reformulation of integral (\ref{coul_ax}) did not remove the singularity in $\varrho_1=\varrho_2,\,z_1=z_2$, and the result of integration remains finite only due to the finite number of integration points of the subsequent Gauss-Legendre quadrature (e.g.~20-point quadrature gives around two-fold overestimation). A correct removal of this singularity (i.e., its analytical integration) has to take into account the value of a finite grid spacing $\Delta$, as was demonstrated in the previous section for cartesian coordinates (and will be done also for axial case later in this section, see (\ref{coul_ax_0})).

However, the representation (\ref{coul_ax2}) enables to precisely evaluate the Coulomb integral in certain circumstances. Namely, by assuming a finite charge distribution of proton, here taken as $\sqrt{\langle r^2\rangle}=0.87\ \mathrm{fm}$ (which is larger than the usual grid spacing 0.4--0.7 fm). I will assume Gaussian distribution in the following treatment, instead of usually employed exponential distribution, since the physical properties should not depend very much on the type of the distribution \cite{Carroll2011}.
\begin{equation}
\label{proton_gaus}
\rho(\vec{r}) = \bigg(\frac{a}{\pi}\bigg)^{\!3/2}\mathrm{e}^{-a(\vec{r}-\vec{r}_0)^2},
\qquad \textrm{where }\ a = \frac{3}{2\langle r^2\rangle}
\end{equation}
Distribution (\ref{proton_gaus}) can be directly convoluted with Gaussian in (\ref{gaus_subs}). Convolution should be done twice (two smeared protons are interacting), nevertheless, the commutativity and associativity of convolutions simplifies the calculations to
\begin{equation*}
\bigg(\frac{a}{2\pi}\bigg)^{3/2} \int_{-\infty}^{+\infty}
\mathrm{e}^{-(\vec{r_1}-\vec{r})^2 t^2} \mathrm{e}^{-a(\vec{r}-\vec{r}_2)^2/2}
\mathrm{d}\vec{r} = \bigg(\frac{a}{a+2t^2}\bigg)^{3/2}
\exp\bigg[{-}\frac{at^2}{a+2t^2}(\vec{r}_1-\vec{r}_2)^2\bigg]
\end{equation*}
Integral (\ref{coul_ax2}) is then replaced by
\begin{small}
\begin{equation}
\frac{2}{\sqrt{\pi}} 2\pi\int_0^\infty \bigg(\frac{a}{a+2t^2}\bigg)^{\!3/2}\!\!
\exp\bigg\{{-}[(z_1\!-\!z_2)^2+(\varrho_1\!-\!\varrho_2)^2]\frac{at^2}{a+2t^2}\bigg\}
\frac{I_m(2\rho_1\rho_2 \frac{at^2}{a+2t^2})}{\exp(2\rho_1\rho_2 \frac{at^2}{a+2t^2})}\mathrm{d}t
\end{equation}
\end{small} \\[-10pt]
Application of the substitution (\ref{GL_subs}) gives
\begin{equation*}
\frac{a}{a+2t^2} = \frac{a(1-x^2)}{a+(2-a)x^2},\qquad
\frac{at^2}{a+2t^2} = \frac{ax^2}{a+(2-a)x^2},
\end{equation*}
leading to the integral
\begin{small}
\begin{equation}
\frac{2}{\sqrt{\pi}} 2\pi\int_0^1 \Big(\tfrac{a}{a+(2-a)x^2}\Big)^{\!3/2}\!\!
\exp\Big\{{-}[(z_1\!-\!z_2)^2\!+\!(\varrho_1\!-\!\varrho_2)^2]\tfrac{ax^2}{a+(2-a)x^2}\Big\}
\frac{I_m\big(2\rho_1\rho_2 \frac{ax^2}{a+(2-a)x^2}\big)}{\exp\big(2\rho_1\rho_2 \frac{ax^2}{a+(2-a)x^2}\big)}\mathrm{d}x
\end{equation}
\end{small} \\[-10pt]
which is finite and well defined for any $\varrho,\,z$.

It has to be noted that Skyrme functionals were usually fitted assuming point-like charges, and also Hartree-Fock calculation is done this way. So the usage of smeared charge in RPA can be considered as a violation of self-consistency, and is therefore disabled in the presented calculations (its usage almost doesn't change the results, only a slight downshift (ca.~0.1 MeV) of spurious state is observed).

Finally, it is also possible to employ empirical procedure similar to (\ref{f0_exact}), which gives the following replacement of the divergent point in (\ref{coul_ax_int}):
\begin{equation}
\label{coul_ax_0}
\frac{g_m\Big(\tfrac{2\varrho_1\varrho_2}{(z_1-z_2)^2+\varrho_1^2+\varrho_2^2}\Big)}
{\sqrt{(z_1-z_2)^2+\varrho_1^2+\varrho_2^2}} \bigg|_{z_1=z_2,\,\varrho_1=\varrho_2}
\!\!\!\!= \frac{1}{\varrho} \bigg[2\ln\frac{\varrho}{\Delta} + 6.779948935
- 4\sum_{n=1}^m \frac{1}{2n-1} \bigg] + O(\Delta^2)
\end{equation}
where $\Delta = \mathrm{d}z = \mathrm{d}\varrho$. For the point on the axis ($z_1=z_2$ and $\varrho_1=\varrho_2=0$, assuming $m=0$, otherwise the contribution is zero), the first term in (\ref{axial_int}) is replaced as
\begin{equation}
\frac{\Delta}{12} \frac{\rho_2(0,z_2) g_0(0)}{\sqrt{0^2+0^2+0^2}}
\quad\mapsto\quad 2.1770180559\,\rho_2(0,z_1)
\end{equation}
For all other points, the integral in the function $g_m(x)$ (\ref{coul_ax_g}) can be calculated as a simple sum of equidistantly sampled integrand, which converges rapidly (due to periodicity); or by Taylor series (\ref{gm_taylor}) for small $x$ and $m>0$, where the direct integration runs into numerical problems (subtraction of large numbers to get a small number). It is also advisable to use extended precision (long double) internally during the calculation of $g_m(x)$, to get an accurate result in double precision.

\section{Pairing interaction}\label{sec_pair}
Short-range part of the nuclear interaction gives rise to a superfluid phase transition in open-shell nuclei. This interaction gives rise to even-odd staggering of the nuclear masses and separation energies, and is therefore denoted as \emph{pairing}. Pairing was implemented on the BCS level, so that the HF+BCS ground state is
\begin{equation}
\label{BCS_gs}
|\mathrm{BCS}\rangle = \prod_{\beta}^{m_\beta>0}(u_\beta+v_\beta\hat{a}_\beta^+\hat{a}_{\bar{\beta}}^+)|0\rangle,
\qquad\textrm{where }u_\beta^2 + v_\beta^2 = 1.
\end{equation}
Since Skyrme interaction is assumed only in the $ph$ channel, pairing interaction is added separately and acts only in the $pp$ channel, either as a ``volume pairing'' ($\hat{V}_\mathrm{pair}$) or as a ``surface pairing'' ($\hat{V}'_\mathrm{pair}$):
\begin{subequations}
\label{V_pair}
\begin{align}
\label{V_pair1}
\hat{V}_\mathrm{pair} &= \sum_{q=p,n}\sum_{ij\in q}^{i<j} V_q\delta(\vec{r}_i-\vec{r}_j) \\
\label{V_pair2}
\hat{V}'_\mathrm{pair} &= \sum_{q=p,n}\sum_{ij\in q}^{i<j}
V_q\bigg(1-\frac{\rho(\vec{r}_i)}{\rho_0}\bigg)\delta(\vec{r}_i-\vec{r}_j)
\end{align}
\end{subequations}
The matrix element between two many-body states (Slater determinants) differing by two wavefunctions is then:
\begin{equation}
\label{pair_me}
\langle\alpha\beta|\hat{V}_\mathrm{pair}|\gamma\delta\rangle =
\iint \psi_\alpha^\dagger(\vec{r}_1)\psi_\beta^\dagger(\vec{r}_2)\hat{V}_\mathrm{pair}
\big[\psi_\gamma(\vec{r}_1)\psi_\delta(\vec{r}_2)
-\psi_\delta(\vec{r}_1)\psi_\gamma(\vec{r}_2)\big]\mathrm{d}\vec{r}_1\mathrm{d}\vec{r}_2
\end{equation}

For further evaluation of the matrix elements, I will explicitly separate the spin part ($\chi$) of the wavefunction:
\begin{align}
\psi_\alpha(\vec{r}) = \sum_s^{\pm1/2} \psi_{\alpha s}(\vec{r})\chi_s
&= R_\alpha(r)\sum_s^{\pm1/2}
C_{l_\alpha,m_\alpha-s,\frac{1}{2},s}^{j_\alpha,m_\alpha}
Y_{l_\alpha,m_\alpha-s}(\vartheta,\varphi)\chi_s, \\
&\qquad\qquad\textrm{where }\chi_{+1/2} = \binom{1}{0},\ \chi_{-1/2} = \binom{0}{1}
\nonumber
\end{align}
In the pairing channel, the wavefunctions are coupled to pairs $(\alpha\beta)$ and $(\gamma\delta)$, and the $\delta$-interaction doesn't depend on spin, so it useful to decompose the spin part of the 2-body wavefunction to triplet and singlet. The matrix element can be then decomposed schematically as
\begin{equation}
\sum_{s_1 s_2}^{\pm1/2} f_{s_1 s_2}^* g_{s_1 s_2} = \sum_{s_1 s_2}
\Big(\sum_{JM}C_{\frac{1}{2}s_1\frac{1}{2}s_2}^{JM} f_{JM}^*\Big)
\Big(\sum_{J'M'}C_{\frac{1}{2}s_1\frac{1}{2}s_2}^{J'M'} g_{J'M'}^{\phantom{*}}\Big)
= \sum_{JM} f_{JM}^*g_{JM}^{\phantom{*}}
\end{equation}
where $J,J'\in{0,1}$, and the symbols $f_{JM}$ and $g_{JM}^{\phantom{*}}$ were defined using orthogonality of Clebsch-Gordan coefficients:
\begin{equation}
f_{JM} = \sum_{s_1 s_2}C_{\frac{1}{2}s_1\frac{1}{2}s_2}^{JM} f_{s_1 s_2},\qquad
f_{s_1 s_2} = \sum_{JM}C_{\frac{1}{2}s_1\frac{1}{2}s_2}^{JM} f_{JM}
\end{equation}
\begin{equation}
f_{00} = \frac{f_{\uparrow\downarrow}-f_{\downarrow\uparrow}}{\sqrt{2}},\quad
f_{11} = f_{\uparrow\uparrow},\quad
f_{10} = \frac{f_{\uparrow\downarrow}+f_{\downarrow\uparrow}}{\sqrt{2}},\quad
f_{1,-1} = f_{\downarrow\downarrow}
\end{equation}
Evaluation of the pairing matrix element of $\delta$-force (\ref{V_pair1}) (and similarly for (\ref{V_pair2})) leads to the cancellation of the triplet component due to antisymmetrization.
\begin{align}
\langle\alpha\bar{\alpha}|\hat{V}_\mathrm{pair}|\beta\bar{\beta}\rangle
&= V_q\sum_{s_1s_2}\int\psi_{\alpha s_1}^*(\vec{r})\psi_{\bar{\alpha}s_2}^*(\vec{r})
\big[\psi_{\beta s_1}(\vec{r})\psi_{\bar{\beta}s_2}(\vec{r})-
\psi_{\bar{\beta}s_1}(\vec{r})\psi_{\beta s_2}(\vec{r})\big]\mathrm{d}^3 r
\nonumber\\
&= V_q\int(\psi_{\alpha\uparrow}^*\psi_{\bar{\alpha}\downarrow}^* -
\psi_{\alpha\downarrow}^*\psi_{\bar{\alpha}\uparrow}^*)
(\psi_{\beta\uparrow}\psi_{\bar{\beta}\downarrow}-\psi_{\beta\downarrow}\psi_{\bar{\beta}\uparrow})\,\mathrm{d}^3 r \quad (\alpha,\beta\in q)
\end{align}
I will denote one of the parentheses as $\sqrt{2}\,\big[\psi_{\beta}\psi_{\bar{\beta}}\big]_{00}$ and define the pairing density $\kappa(\vec{r})$, using $\psi_{\bar{\beta}} = \mathrm{i}\sigma_y\psi_\beta^*$ and Bogoliubov transformation \ref{bogoliubov} to quasiparticles:
\begin{align}
\label{pair_dens}
\hat{\kappa}(\vec{r}) &= -\sqrt{2}\sum_{\beta>0}\big[\psi_\beta(\vec{r})\psi_{\bar{\beta}}(\vec{r})\big]_{00}
(\hat{a}_\beta^+\hat{a}_{\bar{\beta}}^+ +\hat{a}_{\bar{\beta}}\hat{a}_\beta) \\
\label{pair_dens_qp}
&= \sum_{\beta>0} \psi_\beta^\dagger(\vec{r})\psi_\beta(\vec{r})
\big[2 u_\beta v_\beta(1-\hat{\alpha}_\beta^+\hat{\alpha}_\beta-\hat{\alpha}_{\bar{\beta}}^+\hat{\alpha}_{\bar{\beta}})
+ (u_\beta^2-v_\beta^2)
(\hat{\alpha}_\beta^+\hat{\alpha}_{\bar{\beta}}^+ + \hat{\alpha}_{\bar{\beta}}\hat{\alpha}_\beta) \big]
\end{align}
In spherical symmetry, pairing is applied only in the monopole part of the interaction, so the summation over $m_\beta$ leads to
\begin{equation}
\label{summ_pair}
\sum_{m_\beta>0}\psi_\beta^\dagger(\vec{r})\psi_\beta(\vec{r}) = \frac{2j_\beta+1}{8\pi} R_\beta^2(r)
\end{equation}

In the formalism of density functional theory, it is necessary to reformulate the two-body pairing interaction (\ref{pair_me}) to a functional of pairing density (\ref{pair_dens}). This is done by comparing the expectation value of $\hat{V}_\mathrm{pair}$ and $\hat{\kappa}_q$ in the BCS ground state (\ref{BCS_gs}).
\begin{align}
\langle\mathrm{BCS}|\hat{\kappa}_q(\vec{r})|\mathrm{BCS}\rangle &=
\sum_{\beta>0}^{\beta\in q} 2 u_\beta v_\beta
\psi_\beta^\dagger(\vec{r})\psi_\beta(\vec{r}) \stackrel{\mathrm{def}}{=} \kappa_q(\vec{r}) \\
\langle\mathrm{BCS}|\hat{V}_\mathrm{pair}|\mathrm{BCS}\rangle &=
\sum_{\beta>0}v_\beta^2
\langle\beta\bar{\beta}|\hat{V}_\mathrm{pair}|\beta\bar{\beta}\rangle
+ \sum_{\alpha,\beta>0}^{\alpha\neq\beta}u_\alpha v_\alpha u_\beta v_\beta
\langle\alpha\bar{\alpha}|\hat{V}_\mathrm{pair}|\beta\bar{\beta}\rangle \nonumber\\
&= \frac{1}{4}\sum_{q=p,n}V_q\int\kappa_q^2(\vec{r})\mathrm{d}\vec{r}
+ \sum_{\beta>0}v_\beta^4
\langle\beta\bar{\beta}|\hat{V}_\mathrm{pair}|\beta\bar{\beta}\rangle
\end{align}
The last term corresponds to an interaction in $ph$ channel and is therefore dropped (it is already included in the non-pairing part of the functional). Pairing part of density functional is therefore
\begin{subequations}
\begin{align}
\mathcal{H}_\mathrm{pair} &= \frac{1}{4}\sum_{q=p,n}V_q\int\kappa_q^2(\vec{r})\mathrm{d}\vec{r} \\
\mathcal{H}'_\mathrm{pair} &= \frac{1}{4}\sum_{q=p,n}
V_q\int\bigg(1-\frac{\rho(\vec{r})}{\rho_0}\bigg)\kappa_q^2(\vec{r})\mathrm{d}\vec{r}
\end{align}
\end{subequations}

Reduced matrix element of the pairing density for RPA in the spherical symmetry can be derived by rewriting the second part of (\ref{pair_dens_qp}).
\begin{align*}
\hat{\alpha}_\beta^+\hat{\alpha}_{\bar{\beta}}^+ + \hat{\alpha}_{\bar{\beta}}\hat{\alpha}_\beta &= (-1)^{l_\beta+j_\beta+m_\beta}
(\hat{\alpha}_\beta^+\hat{\alpha}_{-\beta}^+ - \hat{\alpha}_{\bar{\beta}}\hat{\alpha}_{\overline{-\beta}}) \\
&=
(-1)^{l_\beta}\sqrt{2j_\beta+1}\,C_{j_\beta,m_\beta,j_\beta,-m_\beta}^{0,0}
(-\hat{\alpha}_\beta^+\hat{\alpha}_{-\beta}^+ + \hat{\alpha}_{\bar{\beta}}\hat{\alpha}_{\overline{-\beta}})
\end{align*}
Comparison of this expression and (\ref{pair_dens_qp})+(\ref{summ_pair}) with (\ref{rme}) then leads to r.m.e.
\begin{equation}
\kappa_{\beta,-\beta}^{J=0}(r) = \sqrt{\frac{2j_\beta+1}{4\pi}}\,(u_\beta^2-v_\beta^2)R_\beta^2(r)
\end{equation}
which has to be further divided by $\sqrt{2}$ to provide correct treatment in the convention of omitted duplicate pairs (\ref{order2qp}).

In fact, $\delta$-interaction gives rise to a diverging pairing energy. This problem can be circumvented by using finite-range pairing interaction, as is done in Gogny force, or by applying a cutoff weight $f_\beta$ to the pairing density, as is usually done for Skyrme \cite{Bender2000}:
\begin{equation}
\label{pair_dens+f}
\hat{\kappa}(\vec{r}) = \sum_{\beta>0} f_\beta\psi_\beta^\dagger(\vec{r})\psi_\beta(\vec{r})
\big[2 u_\beta v_\beta(1-\hat{\alpha}_\beta^+\hat{\alpha}_\beta-\hat{\alpha}_{\bar{\beta}}^+\hat{\alpha}_{\bar{\beta}})
+ (u_\beta^2-v_\beta^2)
(\hat{\alpha}_\beta^+\hat{\alpha}_{\bar{\beta}}^+ + \hat{\alpha}_{\bar{\beta}}\hat{\alpha}_\beta) \big]
\end{equation}
\begin{equation}
f_\beta = \frac{1}{1+\exp[10(\epsilon_\beta-\lambda_q-\Delta E_q)/\Delta E_q]}
\end{equation}
Cutoff weight is meant to damp higher-lying levels, and the cutoff parameter $\Delta E_q$ (usually in the range of 5-9 MeV) is adjusted during the HF+BCS iterations, according to the actual level density, to yield
\begin{equation}
2\sum_{\beta\in q}^{m_\beta>0} f_\beta = N_q + 1.65N_q^{2/3} \qquad\textrm{($N_q$ is the particle number).}
\end{equation}
Pairing strengths $V_p,\,V_n$ (which are negative), obtained with this condition, are given in \cite{Reinhard1999} for SkM*, SkT6, SLy4, SkI1, SkI3, SkI4, SkP, SkO, and in \cite{Guo2007} for SLy6.

\section{Center-of-mass correction of the kinetic energy}\label{sec_kin-cm}
Many-body wavefunction in the form of Slater determinant does not guarantee that the center of mass is fixed in the center of coordinates. In fact, it has certain distribution around the center, and the expectation value of linear momentum is fluctuating as well. In this way the main-field theory breaks the translational symmetry, which can be approximately restored by subtraction of the center-of-mass kinetic energy from the total ground-state energy \cite{Reinhard2011}.
\begin{align}
\mathcal{H}_\mathrm{c.m.}
&= -\frac{1}{2M}{\langle\mathrm{HF}|\hat{P}_\mathrm{tot}^2|\mathrm{HF}\rangle}_\mathrm{Slater} \nonumber\\
\label{H_cm}
&= \frac{-1}{2(Zm_p + Nm_n)}\bigg(
\sum_{i}{\langle\mathrm{HF}|\hat{p}_i^2|\mathrm{HF}\rangle}_\mathrm{Slater} +
\sum_{i\neq j} {\langle\mathrm{HF}|\hat{\vec{p}}_i\cdot\hat{\vec{p}}_j|\mathrm{HF}\rangle}_\mathrm{Slater} \bigg)
\end{align}
The first term in (\ref{H_cm}) is similar to single-particle kinetic energy and can be included by rescaling of the nucleon mass (before or after variation). The second term looks like a two-body interaction, for which the direct term is zero in spherical symmetry (operator $\hat{p}$ shifts the angular momentum $l$ by $\pm1$ and changes the parity), and only the exchange term contributes.
\begin{subequations}
\begin{align}
\label{me_cm2a}
\sum_{j\neq k} \langle\mathrm{HF}|\hat{\vec{p}}_j\cdot\hat{\vec{p}}_k|\mathrm{HF}\rangle_\mathrm{Slater}
&= \hbar^2\sum_{\alpha\beta}v_\alpha^2 v_\beta^2
\langle\alpha\beta|\vec{\nabla}_1\cdot\vec{\nabla}_2|\beta\alpha\rangle \\
&= \hbar^2\sum_{\alpha\beta}v_\alpha^2 v_\beta^2\!\!\sum_\mu^{-1,0,1}\!\!(-1)^\mu
\langle\alpha|\nabla_\mu|\beta\rangle\langle\beta|\nabla_{-\mu}|\alpha\rangle
\end{align}
\end{subequations}
Matrix element of the derivative operator is evaluated in spherical symmetry according to \cite[(7.1.24)]{Varshalovich1988} and (\ref{Rpm}):
\begin{equation}
\label{deriv_me}
\langle\alpha|\nabla_\mu|\beta\rangle
= \frac{(-1)^{j_\beta+l_\beta-\frac{1}{2}}}{\sqrt{2l_\alpha+1}}
\bigg(\int R_\alpha(r)R_\beta^{(\pm)}(r)r^2\mathrm{d}r\bigg)
\begin{Bmatrix} j_\beta & j_\alpha & 1 \\ l_\alpha & l_\beta & \frac{1}{2} \end{Bmatrix}
C_{j_\beta,m_\beta,1,\mu}^{j_\alpha,m_\alpha+\mu}
\end{equation}
and similar expression is found for $\langle\beta|\nabla_{-\mu}|\alpha\rangle$ (with $R_\alpha^{(\mp)}$). In the following, I will assume that selection rules on $j_\alpha,j_\beta,l_\alpha,l_\beta$ are satified. Clebsch-Gordan coefficients are then eliminated by employing their symmetry \cite[(8.4.10)]{Varshalovich1988} and orthogonality \cite[(8.1.8)]{Varshalovich1988}, and including the summation over $m_\beta$.
\begin{align*}
\sum_{\mu,m_\beta}(-1)^\mu C_{j_\beta,m_\beta,1,\mu}^{j_\alpha,m_\alpha+\mu}
C_{j_\alpha,m_\alpha,1,-\mu}^{j_\beta,m_\beta-\mu}
&= (-1)^{j_\alpha-j_\beta}\sqrt{\frac{2j_\beta+1}{2j_\alpha+1}}
\sum_{\mu,m_\beta} C_{j_\beta,m_\beta,1,\mu}^{j_\alpha,m_\alpha+\mu}
C_{j_\beta,m_\beta,1,\mu}^{j_\alpha,m_\alpha+\mu} \\
&= (-1)^{j_\alpha-j_\beta}\sqrt{\frac{2j_\beta+1}{2j_\alpha+1}}
\end{align*}
Summation over $m_\alpha$ then gives additional factor $(2j_\alpha+1)$. The second radial integral will be modified by \emph{per partes}, taking into account the definition of $R_\alpha^{(\pm)}$ (\ref{Rpm}) and $l_\alpha = l_\beta\pm1$.
\[ \int R_\beta(r)R_\alpha^{(\mp)}(r)r^2\mathrm{d}r =
\sqrt{\frac{(2j_\alpha+1)(2l_\beta+1)}{(2j_\beta+1)(2l_\alpha+1)}}
\int R_\alpha(r)R_\beta^{(\pm)}(r)r^2\mathrm{d}r \]
Matrix element (\ref{me_cm2a}) is then
\begin{equation}
\label{me_cm2b}
\sum_{m_\alpha m_\beta}
\langle\alpha\beta|\vec{\nabla}_1\cdot\vec{\nabla}_2|\beta\alpha\rangle
= -\frac{2j_\alpha+1}{2l_\alpha+1}
\bigg(\int R_\alpha(r) R_\beta^{(\pm)}(r)r^2\mathrm{d}r\bigg)^{\!2}
\begin{Bmatrix} j_\beta & j_\alpha & 1 \\ l_\alpha & l_\beta &\frac{1}{2} \end{Bmatrix}^2
\end{equation}

Variation of $\psi_\alpha$ in Hartree-Fock style with general wavefunctions then gives non-local term in single-particle Hamiltonian (besides common local terms, such as kinetic single-particle term, Skyrme and direct Coulomb, collected in $\hat{h}_\mathrm{local}$)
\[ \varepsilon_\alpha R_\alpha(r) = \hat{h}_\mathrm{local}(r)R_\alpha(r)
+ const\cdot\sum_\beta v_\beta^2 R_\beta^{(\pm)}(r)
\int R_\beta^{(\pm)}(r')R_\alpha(r')r'^2\mathrm{d}r' \]
Numerical difficulties involved in the evaluation of exchange integral can be avoided with the basis of spherical harmonic oscillator. Integral (\ref{me_cm2b}) is then evaluated analytically in terms of density matrix $D_{\nu\nu'}^{(j,l)}$ (given in large square brackets).
\begin{align}
\sum_{\alpha\in(j,l,m)}\!\!\!v_\alpha^2 R_\alpha(r_1)R_\alpha(r_2) &=
\sum_{\nu,\nu'}\bigg[\sum_\alpha v_\alpha^2 U_{\nu\alpha}^{(j,l)} U_{\nu'\alpha}^{(j,l)}\bigg] R_{\nu l}(r_1) R_{\nu'l}(r_2) \nonumber\\
&= \sum_{\nu,\nu'} D_{\nu\nu'}^{(j,l)} R_{\nu l}(r_1) R_{\nu'l}(r_2)
\end{align}
Product of wavefunctions shifted in $l$ by differentiation are then evaluated using (\ref{Rpm_sho}).
\newpage
\begin{small}
\begin{subequations}
\begin{align}
\!\sum_{\alpha\in(j,l,m)}\!\!\!v_\alpha^2 R_\alpha^{(+)}(r_1)R_\alpha^{(+)}(r_2) &=
\tfrac{1}{b^2}(2j+1)(l+1)(2l+3)\sum_{\nu,\nu'}
\sum_\alpha v_\alpha^2 U_{\nu\alpha}^{(j,l)} U_{\nu'\alpha}^{(j,l)}
 \nonumber\\[-4pt]   &\qquad\qquad\cdot
\big[\sqrt{\nu+l+3/2}\,R_{\nu,l+1}(r_1) + \sqrt{\nu}\,R_{\nu-1,l+1}(r_1)\big]
 \nonumber\\    &\qquad\qquad\cdot
\big[\sqrt{\nu'+l+3/2}\,R_{\nu',l+1}(r_2) + \sqrt{\nu'}\,R_{\nu'-1,l+1}(r_2)\big]
 \nonumber\\
&= \tfrac{1}{b^2}(2j+1)(l+1)(2l+3)\sum_{\nu,\nu'} D_{\nu\nu'}^{(j,l{+})}
R_{\nu,l+1}(r_1) R_{\nu',l+1}(r_2) \\
\!\sum_{\alpha\in(j,l,m)}\!\!\!v_\alpha^2 R_\alpha^{(-)}(r_1)R_\alpha^{(-)}(r_2) &=
\tfrac{1}{b^2}(2j+1)l(2l-1)\sum_{\nu,\nu'}
\sum_\alpha v_\alpha^2 U_{\nu\alpha}^{(j,l)} U_{\nu'\alpha}^{(j,l)}
 \nonumber\\[-4pt]    &\qquad\qquad\cdot
\big[\sqrt{\nu+l+1/2}\,R_{\nu,l-1}(r_1) + \sqrt{\nu+1}\,R_{\nu+1,l-1}(r_1)\big]
 \nonumber\\     &\qquad\qquad\cdot
\big[\sqrt{\nu'+l+1/2}\,R_{\nu',l-1}(r_2) + \sqrt{\nu'+1}\,R_{\nu'+1,l-1}(r_2)\big]
 \nonumber\\
&= \tfrac{1}{b^2}(2j+1)l(2l-1)\sum_{\nu,\nu'} D_{\nu\nu'}^{(j,l{-})}
R_{\nu,l-1}(r_1) R_{\nu',l-1}(r_2)
\end{align}
\end{subequations}
\end{small} \\[-6pt]
where I defined modified density matrices $D_{\nu\nu'}^{(j,l{\pm})}$, which can be calculated easily from the standard density matrix $D_{\nu\nu'}^{(j,l)}$.
\begin{subequations}
\begin{align}
D_{\nu\nu'}^{(j,l{+})} &= \sqrt{\nu+l+3/2}
\big(\sqrt{\nu'+l+3/2}\,D_{\nu\nu'}^{(j,l)}
+ \sqrt{\nu'+1}\,D_{\nu,\nu'+1}^{(j,l)}\big) \nonumber\\
&\quad+\sqrt{\nu+1} \big(\sqrt{\nu'+l+3/2}\,D_{\nu+1,\nu'}^{(j,l)}
+ \sqrt{\nu'+1}\,D_{\nu+1,\nu'+1}^{(j,l)}\big) \\
D_{\nu\nu'}^{(j,l{-})} &= \sqrt{\nu+l+1/2}
\big(\sqrt{\nu'+l+1/2}\,D_{\nu\nu'}^{(j,l)}
+ \sqrt{\nu'}\,D_{\nu,\nu'-1}^{(j,l)}\big) \nonumber\\
&\quad+\sqrt{\nu} \big(\sqrt{\nu'+l+1/2}\,D_{\nu-1,\nu'}^{(j,l)}
+ \sqrt{\nu'}\,D_{\nu-1,\nu'-1}^{(j,l)}\big)
\end{align}
\end{subequations}
Matrix element (\ref{me_cm2b}) can be now summed within the corresponding spaces $(j_\alpha,l_\alpha),\,(j_\beta,l_\beta)$, using orthogonality of $R_{\nu l}(r)$ to get:
\begin{subequations}
\begin{align}
\!\!\!\sum_{\alpha\beta}^{l_\alpha=l_\beta+1}\!\!\!\! \langle\alpha\beta|\vec{\nabla}_1\!\cdot\!\vec{\nabla}_2|\beta\alpha\rangle
&= -\frac{(2j_\alpha+1)(2j_\beta+1)l_\alpha}{b^2}
\begin{Bmatrix} j_\beta & \!j_\alpha\! & 1 \\
l_\alpha & \!l_\alpha-1\! & \frac{1}{2} \end{Bmatrix}^2
\sum_{\nu\nu'} D_{\nu\nu'}^{(j_\alpha,l_\alpha)} D_{\nu\nu'}^{(j_\beta,l_\beta{+})} \\
\!\!\!\sum_{\alpha\beta}^{l_\alpha=l_\beta-1}\!\!\!\! \langle\alpha\beta|\vec{\nabla}_1\!\cdot\!\vec{\nabla}_2|\beta\alpha\rangle
&= -\frac{(2j_\alpha+1)(2j_\beta+1)l_\beta}{b^2}
\begin{Bmatrix} j_\beta & \!j_\alpha\! & 1 \\
l_\beta-1 & \!l_\beta\! & \frac{1}{2} \end{Bmatrix}^2
\sum_{\nu\nu'} D_{\nu\nu'}^{(j_\alpha,l_\alpha)} D_{\nu\nu'}^{(j_\beta,l_\beta{-})}
\end{align}
\end{subequations}
where $\sum_{\alpha\beta}$ runs also over $m$, while $\sum_{\nu\nu'}$ is understood for a fixed $m$, since $D_{\nu\nu'}$ is degenerate in $m$. Evaluation of $6j$ symbol according to \cite[tab.~9.1]{Varshalovich1988} gives
\[ \begin{Bmatrix} j' & j & 1 \\
l & l-1 & \frac{1}{2} \end{Bmatrix}^2 = \Bigg\{ \begin{array}{ll}
\frac{1}{2j(2j+1)^2(j+1)} & \textrm{for }j' = j = l-\frac{1}{2} \\[4pt]
\frac{1}{4jl} & \textrm{for }j' = j-1
\end{array} \]
leading to
\begin{subequations}
\begin{align}
\sum_{\alpha\beta}^{j_\alpha=j_\beta}\! v_\alpha^2 v_\beta^2\langle\alpha\beta|\vec{\nabla}_1\!\cdot\!\vec{\nabla}_2|\beta\alpha\rangle
&= \frac{-(2j_\alpha+1)}{b^2(2l_\alpha+1)(2l_\beta+1)}
\sum_{\nu\nu'} D_{\nu\nu'}^{(j_\alpha,l_\alpha)} D_{\nu\nu'}^{(j_\beta,l_\beta{\pm})} \\
\!\!\!\sum_{\alpha\beta}^{j_\alpha=j_\beta\pm1}\!\!\!\! v_\alpha^2 v_\beta^2\langle\alpha\beta|\vec{\nabla}_1\!\cdot\!\vec{\nabla}_2|\beta\alpha\rangle
&= -\frac{(2j_\alpha+1)(2j_\beta+1)}{2b^2(j_\alpha+j_\beta+1)}
\sum_{\nu\nu'} D_{\nu\nu'}^{(j_\alpha,l_\alpha)} D_{\nu\nu'}^{(j_\beta,l_\beta{\pm})}
\end{align}
\end{subequations}
\begin{equation}
\langle\nu,j,l|\hat{h}_\mathrm{c.m.ex}|\nu',j,l\rangle = \frac{\hbar^2}{Mb^2}
\sum_{\scriptstyle l'=l\pm1} \bigg[\frac{D_{\nu\nu'}^{(j,l'{\mp})}}{(2l+1)(2l'+1)}
+ \frac{(2j'+1)D_{\nu\nu'}^{(j\pm1,l'{\mp})}}{2(j+j'+1)}\bigg]
\end{equation}

In open-shell nuclei, the center-of-mass term contributes also in the pairing channel, according to (\ref{deriv_me}):
\begin{align}
\mathcal{H}_\mathrm{c.m.} &= \frac{\hbar^2}{2M} \frac{1}{4}\sum_{\alpha\beta}
u_\alpha^2 v_\alpha^2 u_\beta^2 v_\beta^2
{\langle\alpha\bar{\alpha}|2\vec{\nabla}_1\cdot\vec{\nabla}_2|\beta\bar{\beta}\rangle}_\mathrm{Slater} \nonumber\\
&= \frac{\hbar^2}{2M}\sum_{\alpha\beta}u_\alpha^2 v_\alpha^2 u_\beta^2 v_\beta^2
\sum_\mu \langle\bar{\alpha}|\nabla_{-\mu}|\bar{\beta}\rangle\langle\alpha|\nabla_\mu|\beta\rangle \nonumber\\
\label{cm_pair}
&= \frac{\hbar^2}{2M}\sum_{\alpha\beta}^{(j,l,\not m)}
u_\alpha^2 v_\alpha^2 u_\beta^2 v_\beta^2 \frac{2j_\alpha+1}{2l_\alpha+1}
\bigg(\int R_\alpha(r) R_\beta^{(\pm)}(r)r^2\mathrm{d}r\bigg)^{\!2}
\begin{Bmatrix} j_\beta & j_\alpha & 1 \\ l_\alpha & l_\beta &\frac{1}{2} \end{Bmatrix}^2
\end{align}
where the $1/4$ in the first line is due to summation over positive and negative $m$, the exchange term is absorbed to the direct term by
\[ \psi_{\bar{\beta}}(\vec{r}_1)\psi_\beta(\vec{r}_2) =
-\psi_{-\beta}^{\phantom{|}}(\vec{r}_1)\psi_{\overline{-\beta}}(\vec{r}_2), \]
and the summation in (\ref{cm_pair}) doesn't run over $m$, as it was already included like in (\ref{me_cm2b}).

It is possible also to include $\mathcal{H}_\mathrm{c.m.}$ in the residual interaction of RPA, which seems necessary for the self-consistency, when starting with a ground state calculated in variation-after-projection (VAP) approach. RPA already restores the symmetry to a certain degree, mainly limited by the size of the model space, and it can be expected that the includion of $\mathcal{H}_\mathrm{c.m.}$ will make the separation of the spurious motion even better. The derivation of the residual interaction from the two-body part of $\mathcal{H}_\mathrm{c.m.}$ is a bit cumbersome, as it requires to take into account both direct and exchange terms, which in the spherical symmetry require recoupling of the angular momenta.
\begin{equation}
\hat{V}_\mathrm{res}^\mathrm{(c.m.)} = \frac{\hbar^2}{2M} \sum_{\alpha\beta\gamma\delta} \sum_\mu (-1)^\mu
\langle\bar{\alpha}|\nabla_{-\mu}|\beta\rangle \langle\bar{\gamma}|\nabla_\mu|\delta\rangle
:\! \hat{a}_{\bar{\alpha}}^+\hat{a}_\beta^{\phantom{|}}
\hat{a}_{\bar{\gamma}}^+\hat{a}_\delta^{\phantom{|}} \!:
\end{equation}
Matrix element of the derivative operator is, according to (\ref{deriv_me}) and (\ref{t_inv}):
\begin{equation}
\langle\bar{\alpha}|\nabla_\mu|\beta\rangle =
\frac{(-1)^{\mu+j_\beta+\frac{1}{2}}}{\sqrt{3}\,(2l_\alpha+1)}
\bigg(\int R_\alpha^{(0)}R_\beta^{(\pm)}r^2\mathrm{d}r\bigg)
\begin{Bmatrix} l_\alpha & l_\beta & 1 \\ j_\beta & j_\alpha & \frac{1}{2} \end{Bmatrix}
C_{j_\alpha m_\alpha j_\beta m_\beta}^{1,-\mu}
= \langle\bar{\beta}|\nabla_\mu|\alpha\rangle
\end{equation}
The symmetry $(\alpha\leftrightarrow\beta)$ is then applied together with a transformation to quasiparticles (\ref{bogoliubov}):
\begin{align}
\hat{a}_{\bar{\alpha}}^+\hat{a}_\beta^{\phantom{|}} \mapsto
\tfrac{1}{2}\big(\hat{a}_{\bar{\alpha}}^+\hat{a}_\beta^{\phantom{|}} \!+\!
\hat{a}_{\bar{\beta}}^+\hat{a}_\alpha^{\phantom{|}}\big) &=
\tfrac{1}{2}\big[
\big(u_\alpha\hat{\alpha}_{\bar{\alpha}}^+ \!-\! v_\alpha\hat{\alpha}_\alpha^{\phantom{|}}\big)
\big(u_\beta\hat{\alpha}_\beta^{\phantom{|}} \!+\! v_\beta\hat{\alpha}_{\bar{\beta}}^+\big) \nonumber\\
&\qquad{}+\big(u_\beta\hat{\alpha}_{\bar{\beta}}^+ \!-\! v_\beta\hat{\alpha}_\beta^{\phantom{|}}\big)
\big(u_\alpha\hat{\alpha}_\beta^{\phantom{|}} \!+\! v_\alpha\hat{\alpha}_{\bar{\beta}}^+\big)
\big] \nonumber\\
&= \tfrac{1}{2}\big[u_{\alpha\beta}^{(p{-})}
\big(\hat{\alpha}_{\bar{\alpha}}^+\hat{\alpha}_\beta^{\phantom{|}}
+ \hat{\alpha}_{\bar{\beta}}^+\hat{\alpha}_\alpha^{\phantom{|}}\big)
+ u_{\alpha\beta}^{(-)}\big(\hat{\alpha}_{\bar{\alpha}}^+\hat{\alpha}_{\bar{\beta}}^+
+ \hat{\alpha}_\alpha^{\phantom{|}}\hat{\alpha}_\beta^{\phantom{|}}\big)\big]
\end{align}
where, besides already defined pairing factors (\ref{u_ab}), I introduced a corresponding factor for the particle-particle channel:
\begin{equation}
u_{\alpha\beta}^{(\pm)} = u_\alpha v_\beta \pm v_\alpha u_\beta, \qquad
u_{\alpha\beta}^{(p{\pm})} = u_\alpha u_\beta \mp v_\alpha v_\beta
\end{equation}
RPA phonons are (\ref{phonon_sph})
\begin{align}
\hat{C}_\nu^+ &= \frac{1}{2}\sum_{\alpha\beta}
C_{j_\alpha m_\alpha j_\beta m_\beta}^{\lambda \mu}\Big(
c_{\alpha\beta}^{(\nu-)}\hat{\alpha}_{\alpha}^+\hat{\alpha}_{\beta}^+ +
c_{\alpha\beta}^{(\nu+)}\hat{\alpha}_{\bar{\alpha}}^{\phantom{*}}
\hat{\alpha}_{\bar{\beta}}^{\phantom{*}} \Big) \nonumber\\
&= \frac{1}{2}\sum_{\alpha\beta}(-1)^{l_\alpha + l_\beta + \lambda + \mu}
C_{j_\alpha m_\alpha j_\beta m_\beta}^{\lambda,-\mu}\Big(
c_{\alpha\beta}^{(\nu-)}\hat{\alpha}_{\bar{\alpha}}^+\hat{\alpha}_{\bar{\beta}}^+ +
c_{\alpha\beta}^{(\nu+)}\hat{\alpha}_{\alpha}^{\phantom{|}}
\hat{\alpha}_{\beta}^{\phantom{|}} \Big)
\end{align}
Then, in the evaluation of commutator $[\hat{V}_\mathrm{res}^\mathrm{(c.m.)},\hat{C}_\nu^+]$, there are three types of terms:
\begin{itemize}
\item direct (active only in E1)
\[ u_{\alpha\beta}^{(-)}\big[\hat{\alpha}_{\bar{\alpha}}^+\hat{\alpha}_{\bar{\beta}}^+
+ \hat{\alpha}_\alpha^{\phantom{|}}\hat{\alpha}_\beta^{\phantom{|}},\hat{C}_\nu^+\big]
= u_{\alpha\beta}^{(-)}C_{j_\alpha m_\alpha j_\beta m_\beta}^{\lambda \mu}\big(
{-}c_{\alpha\beta}^{(\nu-)} + c_{\alpha\beta}^{(\nu+)} \big) \quad
(\times 2\textrm{ for }(\alpha\beta\leftrightarrow\gamma\delta)) \]
\item exchange normal (contributing negatively to $B$ matrix in RPA eq.~(\ref{fullRPA_eq}))
\begin{align*}
u_{\alpha\beta}^{(-)}u_{\gamma\delta}^{(-)}
&\big[-\hat{\alpha}_{\bar{\alpha}}^+\hat{\alpha}_{\bar{\gamma}}^+
\hspace{7pt}\underbracket[0.5pt]{\hspace{-8pt}\hat{\alpha}_{\bar{\beta}}^+\hat{\alpha}_{\bar{\delta}}^+\hspace{-4pt}}\hspace{2pt}
{}-{} \hat{\alpha}_\alpha^{\phantom{|}}\hat{\alpha}_\gamma^{\phantom{|}}
\hspace{7pt}\underbracket[0.5pt]{\hspace{-8pt}\hat{\alpha}_\beta^{\phantom{|}}\hat{\alpha}_\delta^{\phantom{|}}\hspace{-2pt}}\,,\hat{C}_\nu^+\big] = \\
&= u_{\alpha\beta}^{(-)}u_{\gamma\delta}^{(-)}C_{j_\beta m_\beta j_\delta m_\delta}^{\lambda\mu}
\Big(c_{\beta\delta}^{(\nu-)}\hat{\alpha}_\alpha^{\phantom{|}}\hat{\alpha}_\gamma^{\phantom{|}}
- c_{\beta\delta}^{(\nu+)}\hat{\alpha}_{\bar{\alpha}}^+\hat{\alpha}_{\bar{\gamma}}^+ \Big)
\qquad(\times 2)
\end{align*}
+ a similar term coupled as $(\alpha\delta)(\beta\gamma)$
\item exchange pairing (contributing to $A$ matrix)
\begin{align*}
u_{\alpha\beta}^{(p{-})}u_{\gamma\delta}^{(p{-})}
&\big[-\hat{\alpha}_{\bar{\alpha}}^+\hat{\alpha}_{\bar{\gamma}}^+
\hspace{7pt}\underbracket[0.5pt]{\hspace{-8pt}\hat{\alpha}_\beta^{\phantom{|}}\hat{\alpha}_\delta^{\phantom{|}}\hspace{-2pt}}
{}-{} \hat{\alpha}_\alpha^{\phantom{|}}\hat{\alpha}_\gamma^{\phantom{|}}
\hspace{7pt}\underbracket[0.5pt]{\hspace{-8pt}\hat{\alpha}_{\bar{\beta}}^+\hat{\alpha}_{\bar{\delta}}^+\hspace{-4pt}}\hspace{3pt},\hat{C}_\nu^+\big] = \\
&= u_{\alpha\beta}^{(p{-})}u_{\gamma\delta}^{(p{-})}C_{j_\beta m_\beta j_\delta m_\delta}^{\lambda\mu}
\Big(c_{\beta\delta}^{(\nu-)}\hat{\alpha}_{\bar{\alpha}}^+\hat{\alpha}_{\bar{\gamma}}^+ -
c_{\beta\delta}^{(\nu+)}\hat{\alpha}_\alpha^{\phantom{|}}\hat{\alpha}_\gamma^{\phantom{|}} \Big)
\qquad(\times 2)
\end{align*}
+ a similar term coupled as $(\alpha\delta)(\beta\gamma)$
\end{itemize}
Two exchange terms can be combined to provide time-even and time-odd contribution to the residual interaction:
\begin{subequations}
\begin{align}
-B:\ \quad u_{\alpha\beta}^{(-)}u_{\gamma\delta}^{(-)} &= u_\alpha v_\beta u_\gamma v_\delta
- u_\alpha v_\beta v_\gamma u_\delta - v_\alpha u_\beta u_\gamma v_\delta
+ v_\alpha u_\beta v_\gamma u_\delta \nonumber\\
A:\quad u_{\alpha\beta}^{(p{-})}u_{\gamma\delta}^{(p{-})} &= u_\alpha u_\beta u_\gamma u_\delta
+ u_\alpha u_\beta v_\gamma v_\delta + v_\alpha v_\beta u_\gamma u_\delta
+ v_\alpha v_\beta v_\gamma v_\delta \nonumber\\
V_\mathrm{even} = \frac{A+B}{2}:\quad &\frac{1}{2}\big(
u_{\alpha\gamma}^{(p{+})}u_{\beta\delta}^{(p{+})}
+ u_{\alpha\gamma}^{(+)}u_{\beta\delta}^{(+)}\big) \\
V_\mathrm{odd} = \frac{A-B}{2}:\quad &\frac{1}{2}\big(
u_{\alpha\gamma}^{(p{-})}u_{\beta\delta}^{(p{-})}
+ u_{\alpha\gamma}^{(-)}u_{\beta\delta}^{(-)}\big)
\end{align}
\end{subequations}

The direct term then contributes only to the time-odd part of $\hat{V}_\mathrm{res}$ in E1
\begin{align}
\label{V_pair-dir}
\!V_\mathrm{odd}:\ \frac{\hbar^2}{2M} \frac{1}{4}\sum_{\alpha\beta\gamma\delta}
&\frac{2u_{\alpha\beta}^{(-)}u_{\gamma\delta}^{(-)}(-1)^{j_\beta+j_\delta}}{3(2l_\alpha+1)(2l_\gamma+1)}
\bigg(\int R_\alpha^{(0)}R_\beta^{(\pm)}r^2\mathrm{d}r\bigg)
\bigg(\int R_\gamma^{(0)}R_\delta^{(\pm)}r^2\mathrm{d}r\bigg) \nonumber\\
& {\quad}\times
\begin{Bmatrix} l_\alpha & l_\beta & 1 \\ j_\beta & j_\alpha & \frac{1}{2} \end{Bmatrix}
\begin{Bmatrix} l_\gamma & l_\delta & 1 \\ j_\delta & j_\gamma & \frac{1}{2} \end{Bmatrix}
\end{align}
and the exchange term contributes to both time-even ($s={+}$) and time-odd ($s={-}$) part
\begin{small}
\begin{align}
\!\!V_\mathrm{even/odd}:\ &\frac{\hbar^2}{2M} \frac{1}{4}\sum_{\alpha\beta\gamma\delta}^{(\alpha\beta)(\gamma\delta)}
\bigg[\big(u_{\alpha\beta}^{(ps)}u_{\gamma\delta}^{(ps)}
+ u_{\alpha\beta}^{(s)}u_{\gamma\delta}^{(s)}\big)
\frac{(-1)^{l_\alpha+l_\beta+\lambda+j_\beta+j_\delta}
\delta_{l_\alpha l_\gamma^\pm} \delta_{l_\beta l_\delta^\pm}}{(2l_\alpha+1)(2l_\beta+1)}
\begin{Bmatrix} j_\alpha & j_\gamma & 1 \\ j_\delta & j_\beta & \lambda \end{Bmatrix}
\nonumber\\
&{\quad}\times\bigg(\int R_\alpha^{(0)}R_\gamma^{(\pm)}r^2\mathrm{d}r\bigg)
\bigg(\int R_\beta^{(0)}R_\delta^{(\pm)}r^2\mathrm{d}r\bigg)
\begin{Bmatrix} l_\alpha & l_\gamma & 1 \\ j_\gamma & j_\alpha & \frac{1}{2} \end{Bmatrix}
\begin{Bmatrix} l_\beta & l_\delta & 1 \\ j_\delta & j_\beta & \frac{1}{2} \end{Bmatrix}
\nonumber\\  \label{V_pair-ex}
&{}+\big(u_{\alpha\beta}^{(ps)}u_{\gamma\delta}^{(ps)}
\pm u_{\alpha\beta}^{(s)}u_{\gamma\delta}^{(s)}\big)
\frac{(-1)^{l_\alpha+l_\beta+j_\beta+j_\delta}
\delta_{l_\alpha l_\delta^\pm} \delta_{l_\beta l_\gamma^\pm}}{(2l_\alpha+1)(2l_\beta+1)}
\begin{Bmatrix} j_\alpha & j_\delta & 1 \\ j_\gamma & j_\beta & \lambda \end{Bmatrix} \nonumber\\
&{\quad}\times\bigg(\int R_\alpha^{(0)}R_\delta^{(\pm)}r^2\mathrm{d}r\bigg)
\bigg(\int R_\beta^{(0)}R_\gamma^{(\pm)}r^2\mathrm{d}r\bigg)
\begin{Bmatrix} l_\alpha & l_\delta & 1 \\ j_\delta & j_\alpha & \frac{1}{2} \end{Bmatrix}
\begin{Bmatrix} l_\beta & l_\gamma & 1 \\ j_\gamma & j_\beta & \frac{1}{2} \end{Bmatrix}
\bigg]
\end{align}
\end{small} \\[-6pt]
where corresponding substitutions (like $\beta\leftrightarrow\gamma$ etc.) were made to arrange the pairs in the residual interaction $V_{pp'}$ to $p=(\alpha\beta)$ and $p'=(\gamma\delta)$, which are assumed to satisfy the selection rules for the given multipolarity (besides selection rules like $\delta_{l_\alpha l_\gamma^\pm}$ and $\delta_{l_\beta l_\delta^\pm}$ which follow from the cross matrix elements of $\nabla$). Duplicate pairs can be now safely removed according to (\ref{order2qp}), since the matrix element (\ref{V_pair-ex}) is fully symmetrized.

The exchange kinetic c.m.~term was not implemented in axial HF nor in SRPA (which would be too much complicated). However, in both cases the direct term can be included alone in E1, providing somewhat similar effect to full approach of HF VAP + RPA with $\mathcal{H}_\mathrm{c.m.}$. This direct term is then expressed in terms of current density, more precisely by its $L=0$ component (independent on angle; $\vec{Y}_{1\mu}^0 = \vec{e}_\mu/\sqrt{4\pi}$):
\begin{subequations}\label{Hcm_dir}
\begin{align}
\mathcal{H}_\mathrm{c.m.}^\mathrm{dir} &= -\frac{\hbar^2}{2M}
\bigg(\int \vec{j}(\vec{r})\,\mathrm{d}^3 r\bigg)\cdot
\bigg(\int \vec{j}(\vec{r})\,\mathrm{d}^3 r\bigg) \\
&= -\frac{\hbar^2}{2M} 4\pi\sum_\mu (-1)^\mu
\bigg(\int \vec{j}(\vec{r})\cdot\vec{Y}_{1,\mu}^0 \,\mathrm{d}^3 r\bigg)
\bigg(\int \vec{j}(\vec{r})\cdot\vec{Y}_{1,-\mu}^0 \,\mathrm{d}^3 r\bigg)
\end{align}
In the spherical symmetry, the reduced-matrix-element formula is
\begin{equation}
V_\mathrm{odd}:\quad -\frac{\hbar^2}{2M} 8\pi \frac{1}{4}\sum_{\alpha\beta\gamma\delta}
\bigg(\int j_{\alpha\beta}^{10*}(r) r^2 \mathrm{d}r\bigg)
\bigg(\int j_{\gamma\delta}^{10}(r) r^2 \mathrm{d}r\bigg)
\end{equation}
and in the axial symmetry:
\begin{equation}
V_\mathrm{odd}:\quad -\frac{\hbar^2}{2M} 2 \frac{1}{4}\sum_{\alpha\beta\gamma\delta}
\bigg(\int \vec{j}_{\alpha\beta}^{\,\dagger}(\varrho,z) 2\pi\varrho\,\mathrm{d}\varrho\,\mathrm{d}z\bigg)
\cdot\bigg(\int \vec{j}_{\gamma\delta}(\varrho,z) 2\pi\varrho\,\mathrm{d}\varrho\,\mathrm{d}z\bigg)
\end{equation}
\end{subequations}
The response for SRPA (see the large parentheses in (\ref{XY_op})) is an ordinary vector, not a position-dependent quantity.

\chapter{Numerical codes}\label{ch_num}
The following computer programs dealing with Skyrme functional were developed and utilized in the calculations:
\begin{itemize}
\item spherical HF in SHO basis (\texttt{sph\_hf}) -- applicable only for closed-shell nuclei. The main parameters are oscillator length $b=\sqrt{\hbar/m\omega}$ (\ref{SHO}) and the basis size -- as a number of SHO major shells $N_\mathrm{HF}$ (understood as $N=2\nu_\mathrm{max}+l$, where $\nu$ is the radial quantum number).
\item spherical full RPA (\texttt{sph\_qrpa}) in SHO basis or with wavefunctions given on equidistant grid (provided by HF+BCS in Reinhard's \texttt{haforpa}). The main input parameters are the multipolarity and parity of the transition (and corresponding transition operator), and the number of major shells $N_\mathrm{RPA}$ (with the lowest energy) passed from HF to RPA.
\item spherical separable RPA (\texttt{sph\_srpa}) -- same as before, but taking also the input operators, which induce the separable form of the residual interaction.
\item axial full RPA (\texttt{skyax\_qrpa}), taking the single-particle HF+BCS basis from axial Hartree-Fock (\texttt{skyax\_hfb}, provided by Paul-Gerhard Reinhard). Separable axial RPA (\texttt{skyax\_me} and \texttt{sky\_srpa}) was provided by Wolfgang Kleinig. These programs will be utilized only in the next chapter. They were used with a fixed grid spacing of 0.4 fm (the smallest allowed value).
\end{itemize}
This chapter will give an analysis of various factors influencing the accuracy of calculation in the spherical symmetry for SHO and grid-based codes. These calculations were done on 2.5 GHz Intel i5 (Sandy Bridge) processor using single thread (with vectorization in the matrix algorithms), for which the computation times are given.

Most of the calculations below were done with SLy7 parametrization \cite{SLy6} of Skyrme functional, which contains both $\mathcal{J}^2$ (tensor) term and center-of-mass correction. The mass of proton and neutron are taken as equal with $\hbar^2/2m = 20.73553\ \mathrm{MeV.fm^2}$. Calculations with large spherical-harmonic-oscillator (SHO) basis were done for double-magic nuclei, due to absence of pairing in my Skyrme HF program. Parametrization SGII \cite{SGII}, which includes $\mathcal{J}^2$ term (and no c.m.c.; $\hbar^2/2m = 20.7525\ \mathrm{MeV.fm^2}$), was used for some calculations of magnetic transitions, because it was fitted on Gamow-Teller transitions (therefore, a better agreement with experiments on M1 is expected).

Strength functions $S_0(\mathrm{E})$ (\ref{sf}) will be given only for one component $\lambda\mu$, so the results should be multiplied with $2\lambda+1$ to get the total strength, except the plots of $\sigma_\gamma(\mathrm{E1})$ (\ref{cross_sec}) which already have the correct scaling.

\section{Effects of the basis parameters}
As will be shown below, utilization of the SHO basis has certain advantages. First, it allows to employ approximate restoration of the translational symmetry in HF by subtraction of the center-of-mass kinetic energy before variation (section \ref{sec_kin-cm}) at almost no cost. Second, it allows to push E1 spurious state to almost zero energy and reduce the amount of center-of-mass contribution to the time-even transition density of the remaining states. This section gives an analysis with the aim of proper choice of parameters of the basis, and its relation to the kinetic center-of-mass correction and to the separation of E1 spurious mode (i.e., the translational motion of the nucleus as a whole).

\begin{table}[t]
\caption{Ground state energy (by SLy7) and some of its contributions: single-particle kinetic energy, direct and exchange Coulomb energy, one- and two-body center-of-mass energy. Experimental data are from \cite{Wang2012}.}\label{tab_HFcm}
\begin{scriptsize}
\begin{tabular}{|c|c|c|c|c|c|c|c|c|c|c|}
\hline
\rule{0pt}{8pt}\!\!SLy7, ground\!\! & \multicolumn{2}{|c|}{$^{40}$Ca} & \multicolumn{2}{|c|}{$^{48}$Ca} & \multicolumn{2}{|c|}{$^{56}$Ni} & \multicolumn{2}{|c|}{$^{132}$Sn} & \multicolumn{2}{|c|}{$^{208}$Pb} \\
state [MeV] & VBP & VAP & VBP & VAP & VBP & VAP & VBP & VAP & VBP & VAP \\
\hline
\rule{0pt}{8pt}$T_\mathrm{s.p.}$ & \!652.06\! & \!656.91\! & \!840.40\! & \!845.50\! & \!1016.53\! & \!1021.44\! & \!2461.93\! & \!2466.02\! & \!3881.66\! & \!3885.26\! \\
$V_\textrm{coul-dir}$ & 79.66 & 79.92 & 78.53 & 78.71 & 143.25 & 143.54 & 359.65 & 359.83 & 826.81 & 827.04 \\
$V_\textrm{coul-ex}$ & -7.50 & -7.53 & -7.42 & -7.44 & -10.88 & -10.91 & -18.82 & -18.83 & -31.26 & -31.27 \\
$T_\mathrm{c.m.1}$ & -16.30 & -16.42 & -17.51 & -17.61 & -18.15 & -18.24 & -18.65 & -18.68 & -18.66 & -18.68 \\
$T_\mathrm{c.m.2}$ & 8.21 & 8.15 & 9.42 & 9.37 & 10.05 & 10.01 & 12.15 & 12.13 & 12.87 & 12.85 \\
$E_\mathrm{total}$ & \!\!-344.92\!\! & \!\!-345.01\!\! & \!\!-415.89\!\! & \!\!-415.97\!\! & \!\!-482.26\!\! & \!\!-482.32\!\! & \!\!-1102.85\!\! & \!\!-1102.88\!\! & \!\!-1636.84\!\! & \!\!-1636.85\!\! \\
\hline
\rule{0pt}{8pt}$E_\mathrm{exp}$ & \multicolumn{2}{|c|}{-342.052} & \multicolumn{2}{|c|}{-416.001} & \multicolumn{2}{|c|}{-483.994} & \multicolumn{2}{|c|}{-1102.84} & \multicolumn{2}{|c|}{-1636.43} \\
\hline
\end{tabular}
\end{scriptsize}
\end{table}
The center-of-mass correction in Hartree-Fock can be applied either after diagonalization, to get corrected total energy only (variation before projection, VBP), or already in the HF Hamiltonian (variation after projection, VAP). Comparison of both approaches is shown in Table \ref{tab_HFcm}, giving important contributions to the total energy. As can be seen, the effect of VBP/VAP on the total energy is below 0.1 MeV and decreases for heavier nuclei. For further RPA calculations, when $\mathcal{H}_\mathrm{c.m.}$ is not explicitly mentioned, I will use HF VBP approach with no $\mathcal{H}_\mathrm{c.m.}$ in RPA residual interaction.

\begin{table}[t]
\caption{Values of the oscillator length $b$ which lead to the minimum ground-state energy for the given size of the basis (number of major shells).}\label{tab_b-optim}
\centering
\begin{footnotesize}
\begin{tabular}{|c|c|c|c|c|c|c|c|c|c|c|}
\hline
$b_\mathrm{min}$ [fm] & \multicolumn{5}{|c|}{$\mathcal{H}_\mathrm{c.m.}$: variation before projection} & \multicolumn{5}{|c|}{$\mathcal{H}_\mathrm{c.m.}$: variation after projection} \\
\hline
\rule{0pt}{10pt}$N_\mathrm{HF}$ & $^{40}$Ca & $^{48}$Ca & $^{56}$Ni & $^{132}$Sn & $^{208}$Pb
 & $^{40}$Ca & $^{48}$Ca & $^{56}$Ni & $^{132}$Sn & $^{208}$Pb \\
\hline
30 & 1.577 & 1.603 & 1.614 & 1.770 & 1.933 & 1.577 & 1.603 & 1.614 & 1.771 & 1.933 \\
40 & 1.573 & 1.615 & 1.505 & 1.775 & 1.825 & 1.573 & 1.615 & 1.506 & 1.775 & 1.825 \\
60 & 1.550 & 1.535 & 1.502 & 1.686 & 1.808 & 1.550 & 1.538 & 1.504 & 1.686 & 1.808 \\
80 & 1.515 & 1.546 & 1.481 & 1.656 & 1.734 & 1.515 & 1.547 & 1.482 & 1.656 & 1.734 \\
100 & 1.469 & 1.515 & 1.467 & 1.638 & 1.697 & 1.471 & 1.516 & 1.468 & 1.639 & 1.697 \\
120 & 1.48 & 1.48 & 1.49 & 1.624 & 1.683 & 1.48 & 1.49 & 1.49 & 1.62 & 1.684 \\
\hline
\end{tabular}
\end{footnotesize}
\end{table}
\begin{figure}[t]
\includegraphics[width=\textwidth]{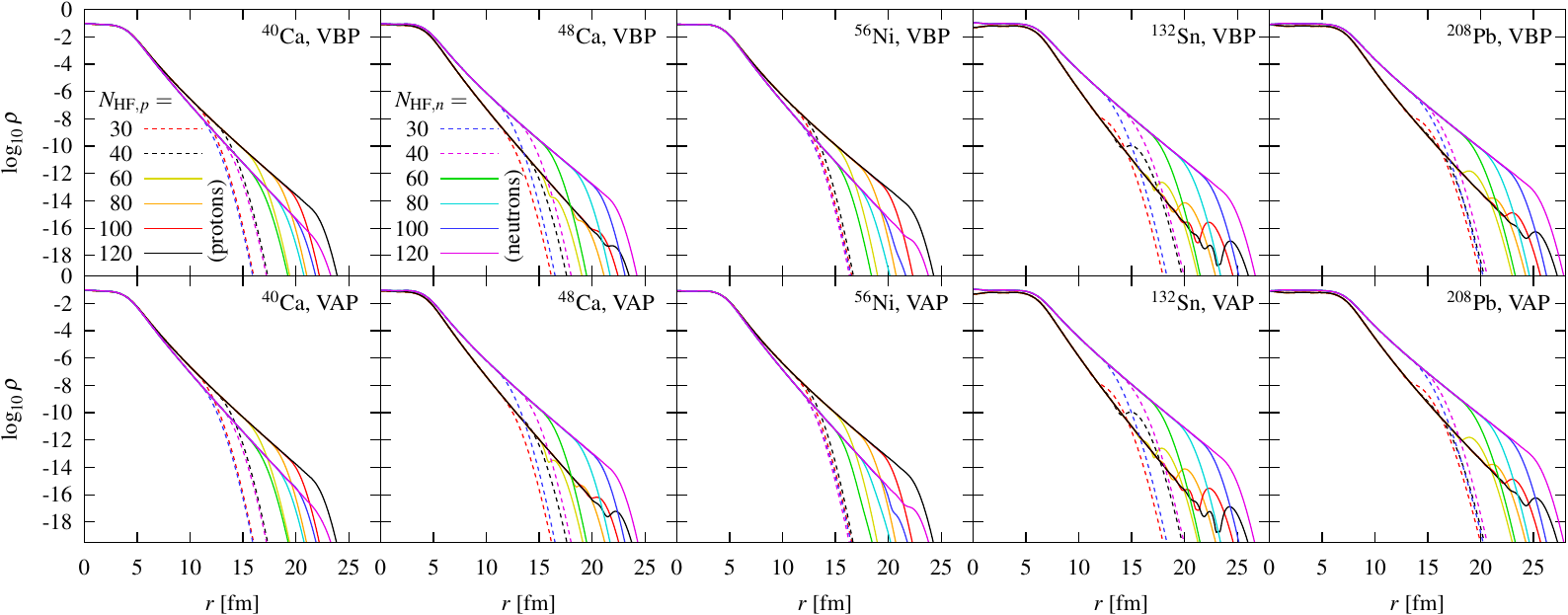}
\caption{Nucleon densities for the optimal parameters $b$ for the given size of the basis as listed in table \ref{tab_b-optim}.}\label{fig_bmin_dens}
\end{figure}
The calculation with the SHO basis has one free parameter -- oscillator length $b$ (\ref{SHO}) -- which takes the role of grid size from grid-based HF solvers. As a first estimate of $b$, I looked for the value which minimizes the ground state energy for the given basis size (Table \ref{tab_b-optim}, Fig.~\ref{fig_bmin_dens}). To exclude any possible bias due to a discrete integration grid, I employed the following integration parameters instead of (\ref{int_params}):
\begin{equation}
\Delta_\mathrm{grid} = 0.05\ \mathrm{fm},\qquad r_\mathrm{max} = 1.4 b\sqrt{2N}
\end{equation}
The ground state energy is converging rapidly with increasing basis. The upshift of the energy minimum in comparison to $N_\mathrm{HF}=120$ was from 1 keV (Ca) to 15 keV (Pb) for $N_\mathrm{HF}=30$, from 0.05 keV (Ca) to 2 keV (Pb) for $N_\mathrm{HF}=40$, and from 2 meV (Ca) to 60 meV (Pb) for $N_\mathrm{HF}=100$. Although such a large basis is certainly not needed for the evaluation of the ground-state energy, it becomes important in subsequent RPA step, where it helps to separate center-of-mass motion (in E1) and provides a sufficiently dense sampling of the continuum -- with energy step ca.~5 MeV per major shell in the range 50--100 MeV for s.p. excitation energy (grid based calculation with $R_\mathrm{box} = 3\cdot1.16 A^{1/3}$ led to 10--12 MeV / major shell for calcium and 5 MeV / major shell for $^{208}$Pb).

\begin{figure}[t]
\includegraphics[width=\textwidth]{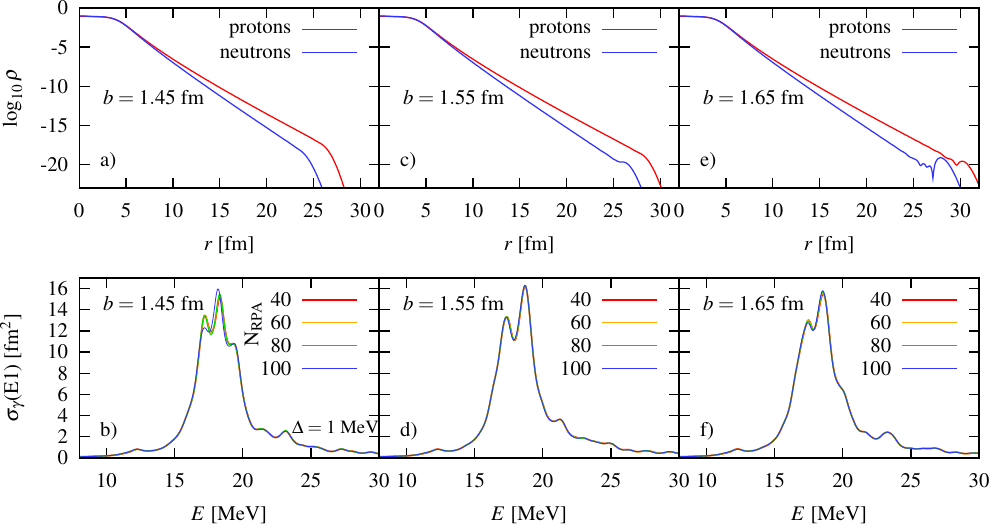}
\caption{Nucleon densities and isovector E1 strength functions (smoothing $\Delta=1\ \mathrm{MeV}$) for $^{40}$Ca for various oscillator lengths with $N_\mathrm{HF}(p,n) = 175,150$.}\label{fig_40Ca_b}
\end{figure}
In further calculations, parameter $b$ is chosen close to the minimum in energy at $N=120$, and the number of major shells is chosen separately for protons and neutrons in order to minimize the oscillations on the logarithmic plot of ground-state proton and neutron densities (Fig.~\ref{fig_40Ca_b}), and for the linear part of $\log_{10}\rho$ to reach certain reasonable level (-18 for Ca, -15.5 for Pb). It was found that this criterion leads also to consistent RPA results, i.e., that the strength function doesn't depend much on the number of major shells passed to RPA (assuming $N_\mathrm{RPA}\geq40$). This fact is demonstrated for $^{40}$Ca in Fig.~\ref{fig_40Ca_b} where the deviations in RPA results are found for the cases of $b$ shifted by $\pm0.1\ \mathrm{fm}$. As can be seen, the converged shape of the strength function depends somewhat on $b$ -- this effect is probably a consequence of particular discretization of the continuum (i.e., the nodal structure of the wavefunctions). The dependence of shape and convergence of the strength function on $b$ is not so pronounced for heavier nuclei. The choice of $b$ and $N_\mathrm{HF}$ (see Table \ref{tab_E1RPA}) deduced in this way for $^{40,48}$Ca and $^{208}$Pb will be used also in the following sections.

\begin{table}[t]
\caption{E1 RPA results (the spurious and the first excited state, which has mostly isoscalar nature), matrix dimension (number of $2qp$ pairs) and calculation times depending on the number of major shells passed from HF to RPA ($N_\mathrm{RPA}$), with corresponding maximum covered single-particle energy $E_\mathrm{wf}$. Isoscalar EWSR (relative to ground-state estimate (\ref{EWSR-wf})) is divided into spurious state and the sum of remaining states. Results labeles as ``grid'' are based on the \texttt{haforpa} program (VBP) with 22 proton and 23 neutron major shells. Missing results correspond to the calculations which failed during the square root of a non-positive-definite matrix $P$ (\ref{half_RPA}).}\label{tab_E1RPA}
\centering
\begin{scriptsize}
\begin{tabular}{|r|r|r|c|c c|r r|c c|}
\hline
\rule{0pt}{9pt}$\!N_\mathrm{RPA}\!\!$ & $E_\mathrm{wf}\ $ & \#\  & $t$
& \multicolumn{2}{|c|}{$E_\mathrm{spurious}$ [keV]}
& \multicolumn{2}{|c|}{$E_1\ (1^{-})$ [MeV]}
& \multicolumn{2}{|c|}{isoscalar EWSR fraction} \\
 & \![MeV]\! & $2qp$ & \![min]\! & VBP & VAP & VBP~ & VAP~ & VBP & VAP \\
\hline
\multicolumn{10}{|c|}{\rule{0pt}{10pt}$^{40}$Ca, $b=1.55$, $N_\mathrm{HF}(p,n) = 175,150$, $t_\mathrm{HF} = 24\ \mathrm{s}$} \\
\hline \rule{0pt}{9pt}
 20 &  26 &  260 & 0.15 & 2660 &  170 & 9.452 & 9.508
 & $1.001 + 10^{-2.8}$ & $\!10^{-2.4} + 10^{-4.6}$ \\
 40 & 103 &  560 & 0.22 & 1490 & 17.1 & 9.199 & 9.217
 & $1.000 + 10^{-3.7}$ & $\!10^{-3.9} + 10^{-4.7}$ \\
 60 & 230 &  860 & 0.39 &  532 & 1.28 & 8.773 & 8.623
 & $1.000 + 10^{-5.2}$ & $\!10^{-5.3} + 10^{-5.5}$ \\
 80 & 430 & 1160 & 0.70 & 82.5 & \!0.002i\! & 7.998 & 7.123
 & $1.000 + 10^{-8.0}$ & $\quad(-) + 10^{-8.0}$ \\
100 & 690 & 1460 & 1.17 & 21.5 &  --  & 7.473 & --
 & $1.000 + 10^{-9.1}$ & -- \\
\hline
\rule{0pt}{9pt}grid & 200 & 293 & \!0.008\! & 827 & 2.25 & 8.974 & 8.974
 & 1.000 + $10^{-4.4}$ & $\!10^{-5.1}$ + $10^{-5.2}$ \\
\hline \hline
\multicolumn{10}{|c|}{\rule{0pt}{10pt}$^{48}$Ca, $b=1.55$, $N_\mathrm{HF}(p,n) = 137,185$, $t_\mathrm{HF} = 30\ \mathrm{s}$} \\
\hline \rule{0pt}{9pt}
 20 &  21 &  283 & 0.19 & 2937 &  283 & \!*10.985 & \!\!\!*11.024
 & $1.002 + 10^{-2.8}$ & $\!10^{-2.1} + 10^{-4.0}$ \\
 40 &  95 &  613 & 0.28 & 1415 & 23.9 & \!*10.501 & \!\!\!*10.550
 & $1.000 + 10^{-3.6}$ & $\!10^{-3.6} + 10^{-4.4}$ \\
 60 & 220 &  943 & 0.50 &  438 & 1.42 & \!*10.134 & \!\!\!*10.012
 & $1.000 + 10^{-5.3}$ & $\!10^{-5.0} + 10^{-5.6}$ \\
 80 & 400 & 1273 & 0.90 & 88.4 & 0.020 & \!*9.581 & 9.130
 & $1.000 + 10^{-7.9}$ & $\!10^{-7.5} + 10^{-8.1}$ \\
100 & 650 & 1603 & 1.52 & 26.2 & 0.012 &  9.393 & 8.747
 & $1.000 + 10^{-9.1}$ & $\!10^{-9.5} + 10^{-9.0}$ \\
120 & 970 & 1933 & 2.43 & 5.38 &   --  &  9.165 & --
 & $\!1.000 + 10^{-10.6}\!\!\!$ & -- \\
\hline
\rule{0pt}{9pt}grid & 170 & 322 & 0.01 &  899 & 1.62 & 10.397 & 10.397
 & 1.000 + $10^{-4.2}$ & $\!10^{-5.5}$ + $10^{-5.1}$ \\
\hline \hline
\multicolumn{10}{|c|}{\rule{0pt}{10pt}$^{208}$Pb, $b=1.66$, $N_\mathrm{HF}(p,n) = 120,160$, $t_\mathrm{HF} = 40\ \mathrm{s}$} \\
\hline \rule{0pt}{9pt}
 20 &  18 &  743 & 0.45 & 2069 &  134 & 7.527 & 7.545
 & $0.999 + 10^{-2.8}$ & $\!10^{-2.4} + 10^{-4.2}$ \\
 40 &  94 & 1773 & 1.81 &  770 & 6.81 & 7.445 & 7.460
 & $1.000 + 10^{-4.2}$ & $\!10^{-4.1} + 10^{-5.4}$ \\
 60 & 220 & 2803 & 6.31 &  243 & 0.31 & 7.207 & 7.194
 & $1.000 + 10^{-5.8}$ & $\!10^{-5.8} + 10^{-6.2}$ \\
 80 & 420 & 3833 & 15.8 & 33.8 & 0.015 & 6.444 & 6.252
 & $1.000 + 10^{-8.2}$ & $\!10^{-7.4} + 10^{-8.1}$ \\
100 & 690 & 4863 & 32.3 & 7.81 & \!0.008i\! & 6.324 & 6.073
 & $1.000 + 10^{-10.7}\!\!$ & $\quad(-) + 10^{-10.8}\!\!$ \\
120 & 1060 & 5893 & 57.6 & 0.955 & \!0.023i\! & 6.316 & 6.059
 & $1.000 + 10^{-11.7}\!\!$ & $\quad(-) + 10^{-11.8}\!\!$ \\
\hline
\rule{0pt}{9pt}grid & 60 & 873 & 0.31 & 1131 & 17.0 & 7.537 & 7.537
 & $1.000 + 10^{-3.6}$ & $\!10^{-3.6}$ + $10^{-5.0}$ \\
\hline
\end{tabular}\\
\rule{0pt}{8pt}* Collective state, which has not yet the second lowest energy due to a small basis.\hspace{80pt}
\end{scriptsize}
\end{table}
Table \ref{tab_E1RPA} gives the results of long-wave isoscalar electric dipole RPA calculation ($z_p=z_n=1$) for the nuclei $^{40,\,48}$Ca and $^{208}$Pb, where the whole strength should be accumulated in the spurious state close to zero energy. $\mathcal{H}_\mathrm{c.m.}$ was either included (VAP) or not included (VBP) in the self-consistent interaction. The calculation time is very similar for VBP and VAP, when using SHO basis. It was found that the good separation of the spurious state in E1 in the more physically appropriate choice of HF VAP and RPA with $\mathcal{H}_\mathrm{c.m.}$ can be achieved also by using HF VBP and E1 RPA including only the direct term $\mathcal{H}_\mathrm{c.m.}^\mathrm{dir}$ (\ref{Hcm_dir}) -- this approach is suitable for SRPA (where the full $\mathcal{H}_\mathrm{c.m.}$ is very cumbersome to apply) and for axial nuclei (there, its application was not found to be much beneficial, apparently due to low precision of the HF results). However, such a trick doesn't offer much advantage (besides shifting $E_\mathrm{spurious}$ closer to zero) over a simple elimination of the E1 spurious contribution by the proper effective charges ($z_p=N/A,\,z_n=-Z/A$) or by the cmc term in E1 tor/com operators (\ref{cmc-generic}).

$\mathcal{H}_\mathrm{c.m.}^\mathrm{dir}$ was also used for the grid-based calculation starting with \texttt{haforpa} (given under VAP column in table \ref{tab_E1RPA}), because there the rigorous HF VAP approach leads to a crash of the full RPA calculation for calcium, while $^{208}$Pb succeeds only with a smaller basis (20+21 major shells), and then the results are even slightly worse than with a simple HF VBP + $\mathcal{H}_\mathrm{c.m.}^\mathrm{dir}$ approach.

\begin{figure}[!b]
\includegraphics[width=\textwidth]{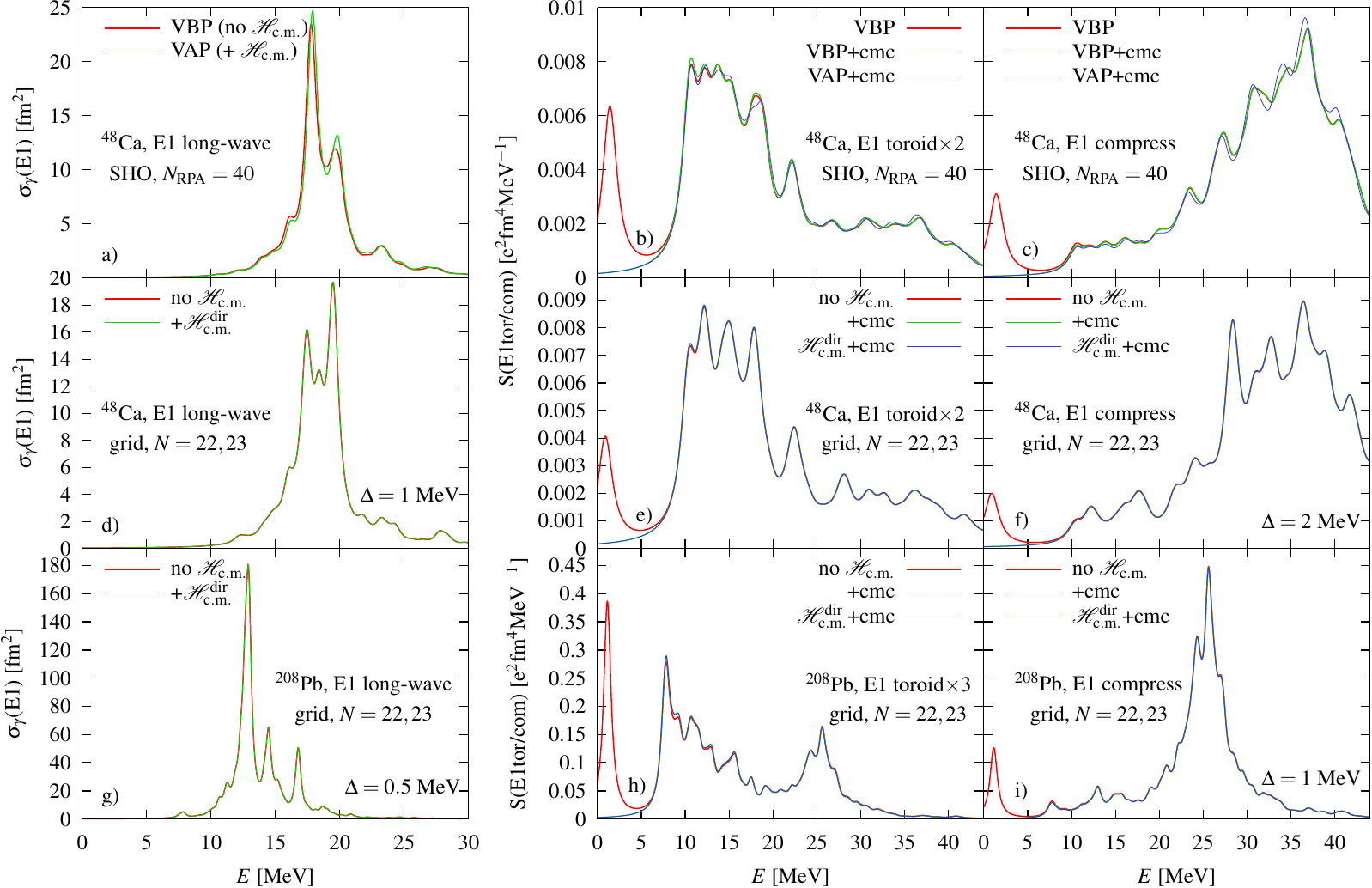}
\caption{Strength functions for long-wave E1 and isoscalar toroidal/compression E1 transitions in $^{48}$Ca and $^{208}$Pb, giving also the effect of center-of-mass correction -- either as $\mathcal{H}_{\mathrm{c.m.}}$ or as a correction in transition operator -- ``cmc'' (\ref{E1vtccm}). Strength of the toroidal transition was increased 2- or 3-times to get a reasonable scaling.}\label{fig_vtccm}
\end{figure}
Table \ref{tab_E1RPA} shows also a significant decrease of the energy of the first E1 state (after the spurious one), which has isoscalar character, with the increasing basis. Low-lying states are an important component of the so-called pygmy resonance \cite{Savran2013}, although in the case of lead, the most of strength is concentrated in the second lowest state, whose downshift is not so dramatic (going as 7.913, 7.798, 7.697, 7.641, 7.637, 7.636 for VAP approach with $N_\mathrm{RPA}=20\!\!-\!\!120$; or as 7.901, 7.788, 7.691, 7.636, 7.633, 7.633 for VBP approach). Accurate determination of the energy of low-lying pygmy mode is probably guaranteed only with continuum RPA \cite{Daoutidis2011} (although still on the one-phonon level, which underestimates the fragmentation).

The influence of the kinetic center-of-mass term $\mathcal{H}_{\mathrm{c.m.}}$ (only direct term was used in the grid-based calculation) and the cmc correction in the transition operators (\ref{E1vtccm}) is depicted in fig.~\ref{fig_vtccm} for E1 transitions. Smaller basis ($N_\mathrm{RPA} = 40$) was used to demonstrate the effect. Effective charges were $z_p=N/A,\,z_n=-Z/A$ for long-wave E1 and $z_p=z_n=0.5,\,g_p=g_n=0.88\times0.7$ for toroidal/compression E1. It is clear that VAP approach has certain influence on the overall shape of the strength function, but it doesn't seem to be very important (Fig.~\ref{fig_vtccm}a-c). Term $\mathcal{H}_\mathrm{c.m.}$ in RPA has one interesting property: It removes the isoscalar center-of-mass strength in the transition operators with time-even densities (see the exhaustion of EWSR by spurious state in Table \ref{tab_E1RPA}), but doesn't have such effect on the time-odd current. In fact, the strength of the spurious state for toroidal/compression transition is almost 100-times larger than for VBP (with no $\mathcal{H}_\mathrm{c.m.}$ in RPA), and the same behavior is found for VBP+$\mathcal{H}_\mathrm{c.m.}^\mathrm{dir}$ (so the ``cmc'' in transition operator must be included in such cases). The reason can be traced down to the structure coefficients: coefficients $c_p^{(\nu{+})}$ acquire a sign opposite to $c_p^{(\nu{-})}$, so only the time-even quantities are reduced. When $\mathcal{H}_\mathrm{c.m.}$ is omitted (VBP approach), the coefficients $c_p^{(\nu{+})}$ and $c_p^{(\nu{-})}$ have the same sign, and the opposite effect is observed: spurious time-odd strength is reduced as $\sim1/E$, while the time-even strength keeps the sum-rule. Again, disappearance of the isoscalar E1 EWSR contribution of the spurious state with VAP can be related to cancellation of the mass constant (coming from the double commutator of kinetic term and $rY_{1\mu}$) by $\mathcal{H}_{\mathrm{c.m.}}$, which was not included in the ground-state EWSR estimate, so the total relative EWSR of E1 RPA states goes down to 0\% in Table \ref{tab_E1RPA}.

Finally, a comparison of the calculation times for particular RPA procedures is given in table \ref{tab_cpu}. Calculation of the matrix elements scales like $O(N^2)$. Matrix algorithms scale like $O(N^3)$ and consist of the following steps: square root of the matrix $P$ (\ref{half_RPA}), matrix multiplication to calculate $C^TQC$ (\ref{CQC}) and its diagonalization in two steps -- Householder transformation (bringing the matrix to tri-diagonal form) and Householder-like iterations (gradually decreasing the off-diagonal elements) -- and finally, the conversion of eigenvectors $\vec{R}_\nu$ to structure constants $c_p^{(\nu\pm)}$.
\begin{table}[t]
\caption{Duration of the most time-consuming parts of a single-threaded RPA calculation (E1, VBP) on 2.5 GHz Intel i5 (Sandy Bridge) processor.}\label{tab_cpu}
\centering
\begin{tabular}{|c|r|r|r|}
\hline
\rule{0pt}{12pt} & $^{40}$Ca & $^{48}$Ca & $^{208}$Pb \\
\hline
$N_\mathrm{RPA}$ & $100\ $ & $120\ $ & $120\ \ $ \\
\hline
$A_{pp'},B_{pp'}$ & 22 s & 41 s & 317 s \\
$\sqrt{P}$ & 1 s & 1 s & 35 s \\
$C^T QC$ & 9 s & 23 s & 944 s \\
Householder & 13 s & 33 s & 1214 s \\
3-diag.~iter. & 6 s & 14 s & 365 s \\
$c_p^{(\nu\pm)}$ & 8 s & 18 s & 557 s \\
\hline
\end{tabular}
\end{table}

\section{Influence of tensor and spin terms}\label{sec_spin-tens}
Full Skyrme functional (\ref{Skyrme_DFT}) contains also the time-odd terms which are not active in the calculation of the ground state of an even-even nucleus. Especially the spin terms ($\tilde{b}_0,\,\tilde{b}_0',\,\tilde{b}_2,\,\tilde{b}_2',\,\tilde{b}_3,\,\tilde{b}_3'$) are difficult to estimate experimentally, since they are not coupled to other time-even terms by Galilean invariance \cite{Dobaczewski1995}. We can fix these terms by a condition that the functional is fully equivalent to the density-dependent two-body interaction (\ref{V_skyrme}). For this reason, the present work is restricted mainly to parametrizations, which include the $\mathcal{J}^2$ (tensor) term (parameters $\tilde{b}_1,\,\tilde{b}_1'$, containing both time-even and time-odd parts) -- SLy7 \cite{SLy6} and SGII \cite{SGII} -- and which don't apply tweaking of the individual parameters, as is done for $b_4'$ in SkI3 and SkI4 \cite{SkI3}, although the tweaked functionals sometimes better describe the M1 resonance \cite{Vesely2009,Nesterenko2010}.

\begin{figure}[h]
\includegraphics[width=\textwidth]{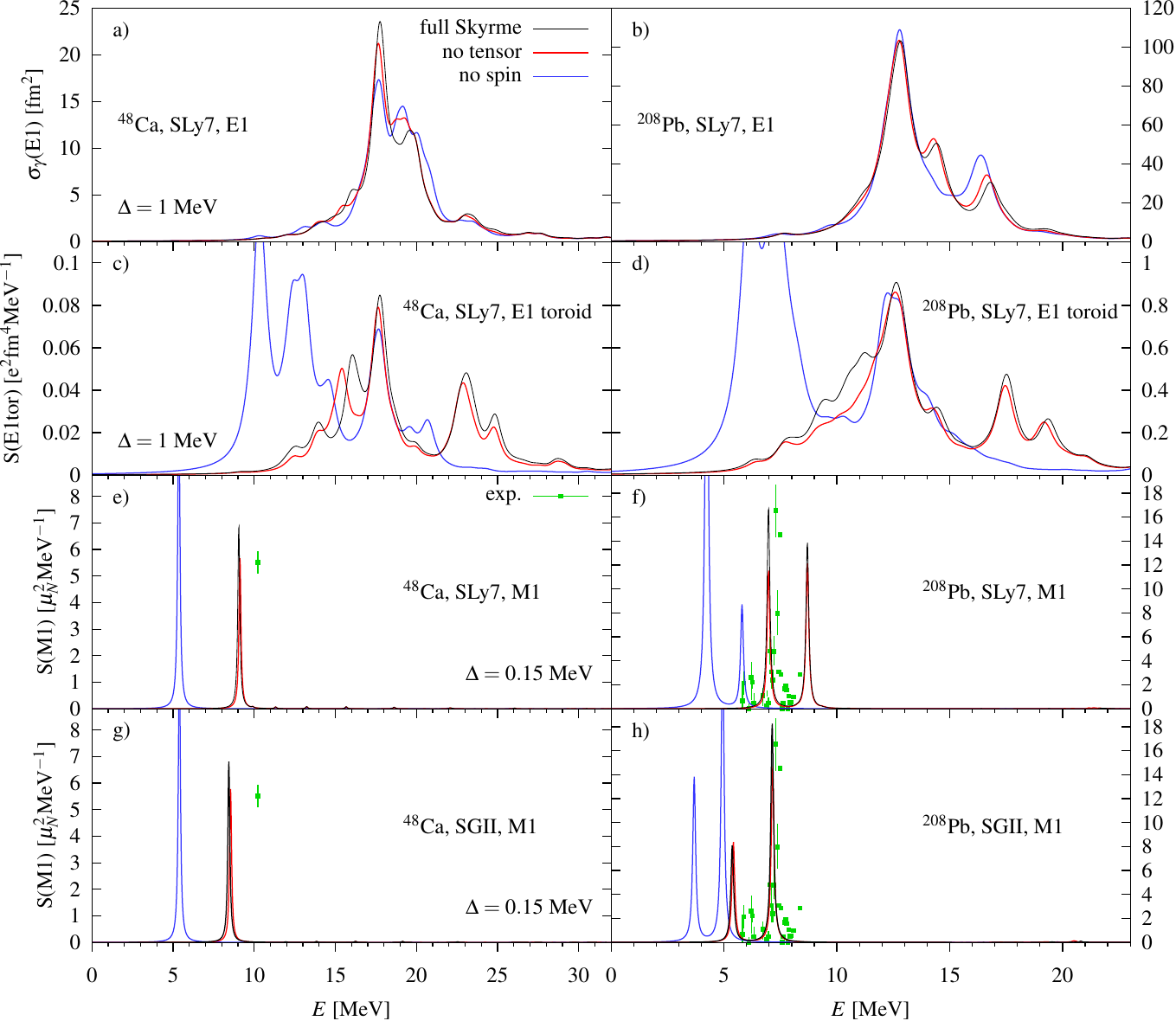}
\caption{Strength functions for E1, toroidal E1 (with natural charges and cmc) and M1 transitions in $^{48}$Ca and $^{208}$Pb, giving also the cases with omitted spin or tensor terms in the residual interaction, and the experimental data for $^{48}$Ca M1 \cite{Steffen1983} and $^{208}$Pb M1 \cite{Laszewski1988}}\label{fig_spin-tens}
\end{figure}
The importance of spin and tensor terms is demonstrated in Fig.~\ref{fig_spin-tens}, which shows the calculations with omitted spin or tensor terms. In electric dipole transitions, the impact is clearly visible only in the higher order term, represented here by toroidal strength function, calculated with natural charges. Unfortunately, this quantity (being mostly isovector) is probably not measurable. However, magnetic dipole transitions are experimentally accessible, and confirm the necessity of inclusion of the spin terms, as was also found previously \cite{VeselyPhD}. With regard to accuracy of M1 for the given parametrizations, SLy7 appears to provide slightly better agreement with experiment, although the second peak in $^{208}$Pb is beyond the experimental range \cite{Laszewski1988}. On the other hand, the first peak according to SGII may be identified with the $1^+$ state at 5.8445 MeV, having isoscalar nature with $B(\mathrm{M1})\!\!\uparrow\,=1.0(4)\,\mu_N^2$ \cite{Muller1985}, which is however too weak to explain the calculated $B(\mathrm{M1})\!\uparrow\,=5.7\,\mu_N^2$. Tensor term was found to have a minor influence in all cases.

\section{Comparison of exact and long-wave E1 s.f.}
The transition probabilities and strength functions are usually evaluated with long-wave versions of the transition operators (\ref{tran}). It is therefore instructive to give a comparison to the results obtained with the exact transition operators (\ref{exactM}), which can be calculated easily in the case of full RPA.
\begin{figure}[h]
\includegraphics[width=\textwidth]{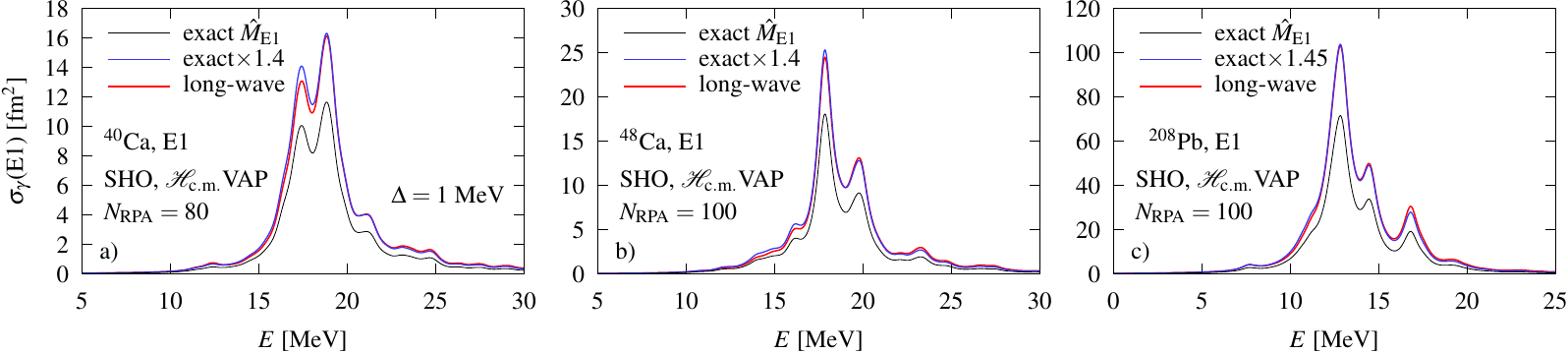}
\caption{Comparison of electric dipole strength functions for $^{40,48}$Ca and $^{208}$Pb obtained with usual long-wave and exact transition operators (with natural charges). The overall strength of the exact s.f.~is found to be uniformly reduced due to usage of the bare mass.}\label{fig_E1exact}
\end{figure}

As can be seen in Fig.~\ref{fig_E1exact} (calculated here with inclusion of the kinetic center-of-mass term), ``exact'' strength function has significanly reduced amplitude (1.4-times). This fact can be explained by the inadequate use of bare mass in the nuclear current (\ref{j_nuc}). Effective mass is smaller, but only in isovector transitions, as was mentioned also in the explanation of EWSR (\ref{EWSR-wf}), and can be demonstrated for E1 compression transitions. These can be calculated either by current-based transition operator $\hat{M}_\mathrm{com}$ or by the density-based $\hat{M}_\mathrm{com'}$, which are related by the continuity equation. Isoscalar transition ($z_p=z_n=1$) shows nearly equal results for both choices, while isovector transition ($z_p=N/A,\,z_n=-Z/A$) gives reduced strength for current-based operator.
\begin{figure}[!b]
\includegraphics[width=\textwidth]{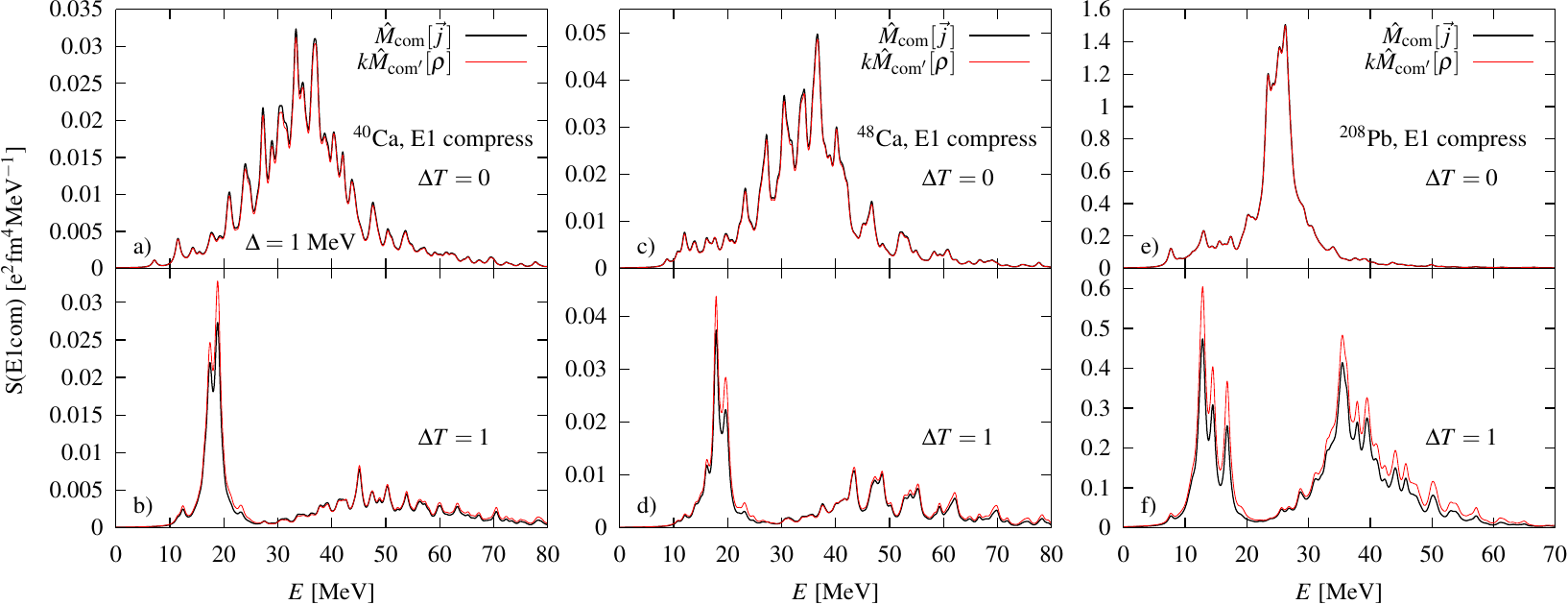}
\caption{Comparison of current- and density-based compression strength for $^{40,48}$Ca and $^{208}$Pb. Better agreement in scale is found for isoscalar s.f.}\label{fig_E1vtcT0}
\end{figure}

\begin{figure}[h]
\includegraphics[width=\textwidth]{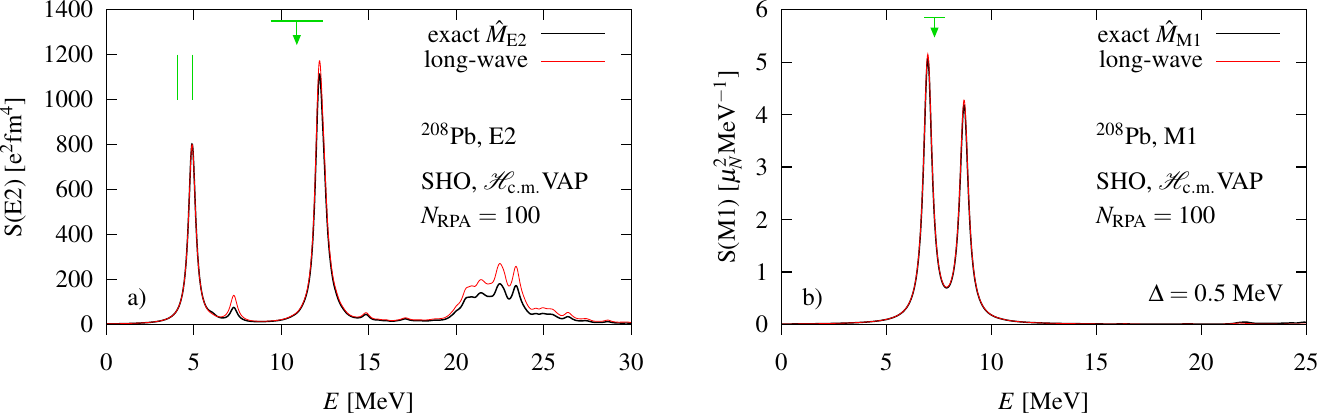}
\caption{Comparison of ``exact'' and long-wave E2 and M1 strength for $^{208}$Pb (with natural charges). Isoscalar transitions of E2 (two largest peaks) are not reduced, in contrast with the isovector residue. Green lines in a) show the position of the first two $2^+$ states \cite{A208} and the centroid and width of isoscalar giant quadrupole resonance \cite{Youngblood2004}, same for M1 spin-flip resonance \cite{Laszewski1988}.}\label{fig_E2M1exact}
\end{figure}

Comparison with exact operators is also done for M1 and E2 transitions of $^{208}$Pb in Fig.~\ref{fig_E2M1exact}. Quadrupole resonance was calculated with the natural charges ($z_p=1,\,z_n=0$), so the resulting strength function is a superposition of both isoscalar and isovector component.

A closer look on the presented results also shows that the effective mass is not a constant, but depends on multipolarity and also on energy (except long-wave E1). This point was not further studied here, and the remaining sections present only the results involving long-wave or toroidal/compression operators.

\section{Full RPA versus SRPA}
Faster calculation of the strength function can be achieved through separable RPA \cite{Nesterenko2002,Nesterenko2006}, whose spherical formulation is described in appendix \ref{app_SRPA}, and the main features will be summarized also here. SRPA requires a set of input operators which provide generating fields for nuclear excitation, and the resulting responses are giving rise to a separable form of the residual interaction. The accuracy of the separable interaction is proportional to the number of input operators, which have to be chosen by a clever way to cover the most important aspects of the full interaction. The following time-even operators were utilized for electric transitions:
\begin{equation}
\label{Q1234}
\begin{split}
\hat{Q}_1 &= \int \mathrm{d}^3r \, \hat{\rho}(\vec{r}) \, r^\lambda Y_{\lambda\mu},\quad
\hat{Q}_2 = \int \mathrm{d}^3r \, \hat{\rho}(\vec{r}) \, r^{\lambda+2}  Y_{\lambda\mu}, \\
\hat{Q}_3 &= \int \mathrm{d}^3r \, \hat{\rho}(\vec{r}) \,
j_\lambda(0.9x_\lambda r/r_\mathrm{diff}),\quad
\hat{Q}_4 = \int \mathrm{d}^3r \, \hat{\rho}(\vec{r}) \, r^\lambda Y_{\lambda\mu}
j_\lambda(1.2x_\lambda r/r_\mathrm{diff})
\end{split}
\end{equation}
The operators were used in consecutive order, so, e.g., the 2-operator SRPA means that $\hat{Q}_1$ and $\hat{Q}_2$ were used. Accurate description of toroidal transitions required also additional operators containing spin
\begin{equation}
\label{Q567}
\hat{Q}_{5,6} = \int\mathrm{d}^3 r\: [\vec{\nabla}\cdot\hat{\vec{\mathcal{J}}}(\vec{r})]
\cdot\Big\{\begin{array}{l} r Y_{1\mu} \\ r^3 Y_{1\mu} \end{array}, \quad
\hat{Q}_7 = \int\mathrm{d}^3 r\: \hat{\vec{\mathcal{J}}}(\vec{r})\cdot
\vec{\nabla}\times r^3 \vec{Y}_{1\mu}^1
\end{equation} 
and time-odd operators
\begin{equation}
\label{P_add}
\hat{P}_{8} = \int\mathrm{d}^3 r\: [\vec{\nabla}\times\hat{\vec{s}}(\vec{r})]
\cdot\vec{\nabla}\times r^3\vec{Y}_{1\mu}^1, \quad
\hat{P}_{9,10} = \int\mathrm{d}^3 r\: \hat{\vec{j}}(\vec{r})
\cdot\vec{\nabla}\times \Big\{\begin{array}{l} r\vec{Y}_{1\mu}^1 \\
r^3 \vec{Y}_{1\mu}^1\end{array}
\end{equation}
making use of familiar Skyrme currents (\ref{Jd_op}).

Separate operators were used for protons and neutrons. Moreover, another time-conjugate operators were created as
\begin{equation}
\hat{P}_{k} = \mathrm{i}[\hat{H},\hat{Q}_{k}]\qquad \textrm{or}\qquad
\hat{Q}_{k} = \mathrm{i}[\hat{H},\hat{P}_{k}]
\end{equation}
so the total number of input operators and the dimension of separable interaction is 4-times larger than the numbers given here.

\begin{figure}[t]
\includegraphics[width=\textwidth]{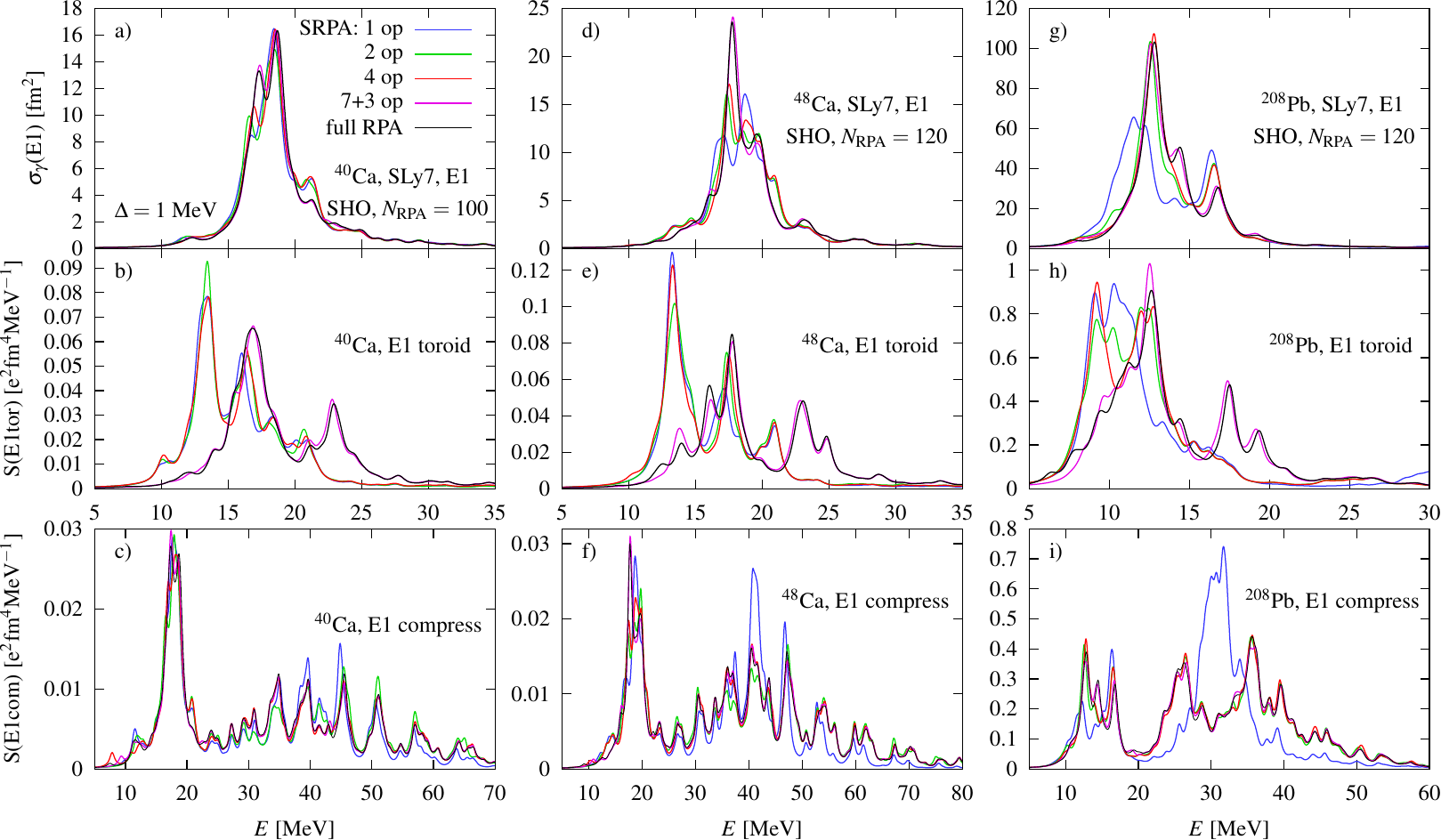}
\caption{Comparison of E1 strength functions calculated by full RPA and SRPA with increasing number of input operators.}\label{fig_E1srpa}
\end{figure}
Figure \ref{fig_E1srpa} shows the results for electric dipole strength functions. Toroidal and compression transitions were calculated with natural charges and center-of-mass correction in operators (\ref{cmc-generic}). As can be seen, one operator already gives correct position of giant dipole resonance (GDR). The second operator corrects also the compression strength function, but much more operators (containing also spin) are needed for accurate reproduction of the toroidal s.f. Even in that case, the calculation time for $^{208}$Pb is reduced from one hour (full RPA) to around 2 minutes (SRPA with 4000-point strength function).

\begin{figure}
\centering
\includegraphics[angle=90,width=0.6\textwidth]{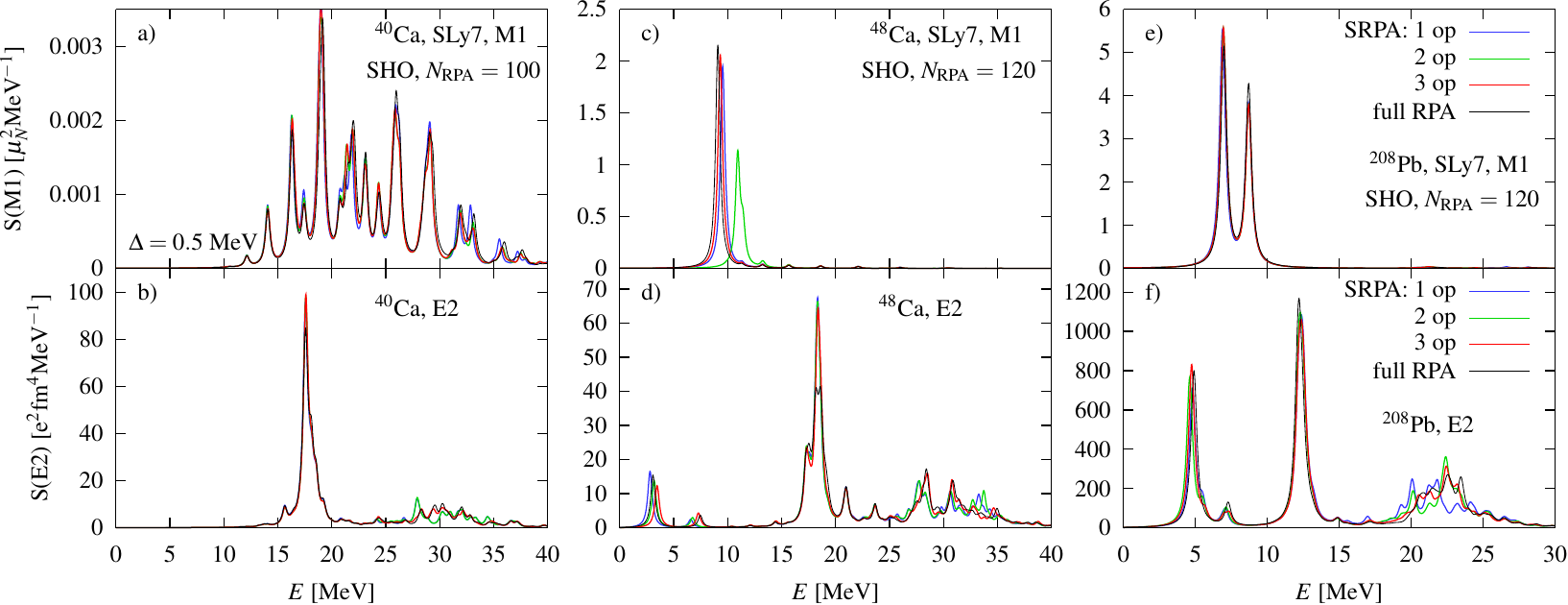}
\caption{Comparison of E1 strength functions calculated by full RPA and SRPA with increasing number of input operators.}\label{fig_M1E2srpa}
\end{figure}
Other multipolarities are satisfactorily described already with fewer operators. Figure \ref{fig_M1E2srpa} shows results for electric quadrupole (with natural charges; using $\hat{Q}_1,\,\hat{Q}_2,\,\hat{Q}_5$) and M1 transitions. Magnetic transition were calculated with up to three operators:
\begin{equation}
\label{P_magn}
\hat{P}_1 = \vec{\sigma}\cdot r^0\vec{Y}_{1\mu}^0,\quad
\hat{P}_2 = \vec{\sigma}\cdot r^2\vec{Y}_{1\mu}^0,\quad
\hat{P}_3 = \vec{l}\cdot r^2\vec{Y}_{1\mu}^0.
\end{equation}
The $^{48}$Ca M1 s.f.~experienced a large shift for 2 input operators, demonstrating that SRPA is prone to instabilities. Such occasional deviations show that a perfect description by means of SRPA cannot be guaranteed a priori, and this method remains mainly an interesting mathematical tool to provide a first estimate of the strength function, when full RPA is numerically not feasible. 

\chapter{Physical results}\label{ch_results}
This chapter will present selected results, either published or submitted for publication, calculated mostly with full RPA in the axial (cylindrical) symmetry, with exception of the first part dealing with pygmy resonance in spherical nuclei. The programs described in this work were utilized in the following areas of research:
\begin{itemize}
\item toroidal nature of the low-energy (pygmy) E1 mode \cite{Repko2013,Reinhard2014,Nesterenko2015}
\item vortical, toroidal and compression transitions by spherical SRPA in tin isotopes \cite{Kvasil2013} (not given here)
\item low lying $2^+$ states in rare earths \cite{Nesterenko-lowE2} (not given here)
\item monopole transitions in spherical nuclei \cite{Kvasil2015-ischia,Kvasil2015} and in deformed $^{24}$Mg \cite{Nesterenko-Mg24} (not given here)
\item description of axial full RPA with a sample calculation of $^{154}$Sm E1, E2, M1  \cite{Repko-istros}
\item magnetic (M1) transitions in deformed $^{50}$Cr \cite{Pai2016}
\end{itemize}

Strength functions displayed in this chapter employ a double-folding procedure, which makes the Lorentz-smoothing parameter $\Delta$ energy-dependent. Formula (\ref{sf}) is then replaced by
\begin{align}
\label{sf_df}
S_n(\mathrm{E/M}\lambda\mu; E) &= \sum_\nu E^n
B(\mathrm{E/M}\lambda\mu,0\rightarrow\nu)\delta_{\Delta_\nu}(E_\nu-E) \\
\delta_{\Delta_\nu}(E_\nu-E) &= \frac{\Delta_\nu}{2\pi[(E_\nu-E)^2+(\Delta_\nu/2)^2]},  \\[-8pt]
&\qquad\qquad\qquad\qquad\textrm{where }\Delta_\nu = \begin{cases}
\Delta_0\quad\textrm{for }E_\nu < E_0 \\
\Delta_0+a(E_\nu-E_0)\quad\textrm{for }E_\nu > E_0 \end{cases} \nonumber
\end{align}
and $E_0$ is the nucleon emission thershold (the smaller of $p/n$ separation energies). Other parameters are chosen as
\begin{equation}
\label{df_param}
\Delta_0 = 0.15\ \mathrm{MeV},\qquad a = 0.15
\end{equation}

Transition densities and currents (\ref{trans_rho},\ref{trans_cur}) are calculated either for one state, or are averaged over multiple states in the given energy interval. In this averaging, there is an ambiguity in the overall phase factors of the structure constants. For this reason, the transition currents are weighted by transition matrix elements \cite{Repko2013}, which also somewhat suppress the non-collective states, so that the resulting density better expresses the nature of the excitations in a given energy region.
\begin{subequations}
\label{cur_avg}
\begin{align}
\delta\rho_q^\mathrm{(E\lambda)}(\vec{r}) &= \sum_{\nu\in(E_1,E_2)}
\langle[\hat{C}_\nu^{\phantom{|}},\hat{M}_{\lambda\mu}^\mathrm{E}]\rangle^*\,
\langle[\hat{C}_\nu^{\phantom{|}},\hat{\rho}_q(\vec{r})]\rangle \\
\delta\vec{j}_q^\mathrm{(\lambda)}(\vec{r}) &= \sum_{\nu\in(E_1,E_2)}
\langle[\hat{C}_\nu^{\phantom{|}},\hat{M}_{\lambda\mu}]\rangle^*\,
\langle[\hat{C}_\nu^{\phantom{|}},\hat{\vec{j}}_q(\vec{r})]\rangle
\end{align}
\end{subequations}

The multipolar $\mu$-components of the strength functions and other quantities (for given $\lambda$) are denoted also by $K^\pi$, and then they are understood as a sum of $\mu=\pm K$ components, while $\pi=(-1)^\lambda$ for electric transitions and $\pi=(-1)^{\lambda+1}$ for magnetic transitions.

\section{Toroidal nature of low-energy E1 mode}
\begin{figure}[h]
\includegraphics[width=\textwidth]{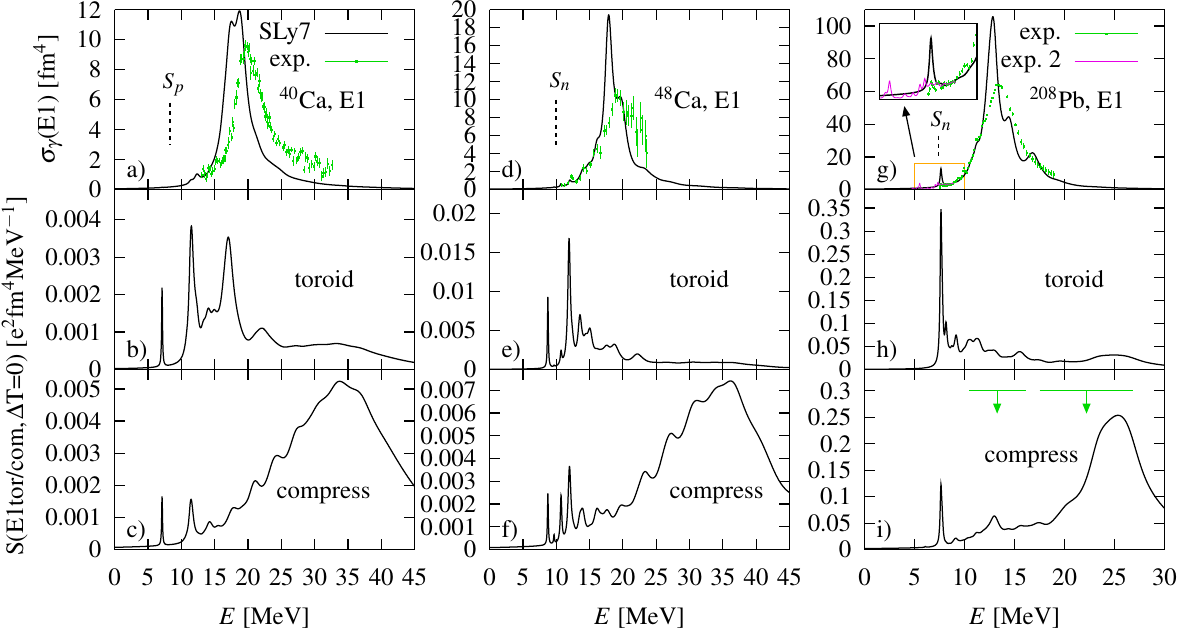}
\caption{Double-folded strength functions of $^{40,48}$Ca, $^{208}$Pb for E1 transitions: GDR, and isoscalar toroidal and compression. Experimental data are given for photoabsorption cross section: $^{40}$Ca \cite{Ahrens1975}, $^{48}$Ca \cite{OKeefe1987}, $^{208}$Pb \cite{Veyssiere1970}, for $(p,p')$-deduced $B(\mathrm{E1})$ in $^{208}$Pb (exp.~2) \cite{Poltoratska2012}; and for isoscalar E1 \cite{Youngblood2004}. The transitions were calculated with full RPA with Skyrme parametrization SLy7 and including $\mathcal{H}_\mathrm{c.m.}$ (with HF VAP). Isoscalar transition operators were taken with $z_p=z_n=0.5,\,g_p=g_n=0.88\times0.7$}\label{fig_pygmy_sf}
\end{figure}

Figure \ref{fig_pygmy_sf} shows the E1 strength functions, calculated with full RPA for spherical nuclei with large SHO basis ($N_\mathrm{RPA}=80$ for $^{40}$Ca and $N_\mathrm{RPA}=100$ for $^{48}$Ca and $^{208}$Pb) and with all center-of-mass corrections (kinetic + operator). Double folding was done with (\ref{df_param}). The inset in Fig.~\ref{fig_pygmy_sf}g shows that the (one-phonon) RPA cannot reproduce the experimentally observed fragmentation in $^{208}$Pb \cite{Poltoratska2012}.

\begin{figure}[t]
\includegraphics[width=\textwidth]{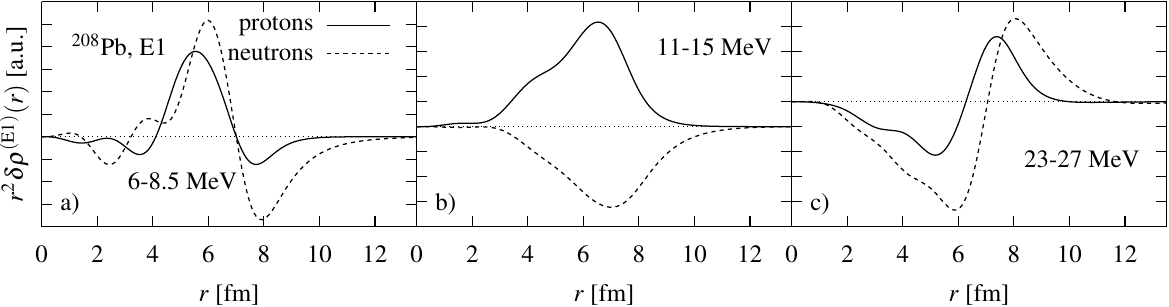}
\caption{Transition densities of $^{208}$Pb weighted by long-wave E1 operator in given energy intervals. Calculated by full RPA, Skyrme SLy7, including $\mathcal{H}_\mathrm{c.m.}$ (with HF VAP).}\label{fig_Pb208rho}
\end{figure}

\begin{figure}[t]
\centering
\includegraphics[width=0.9\textwidth]{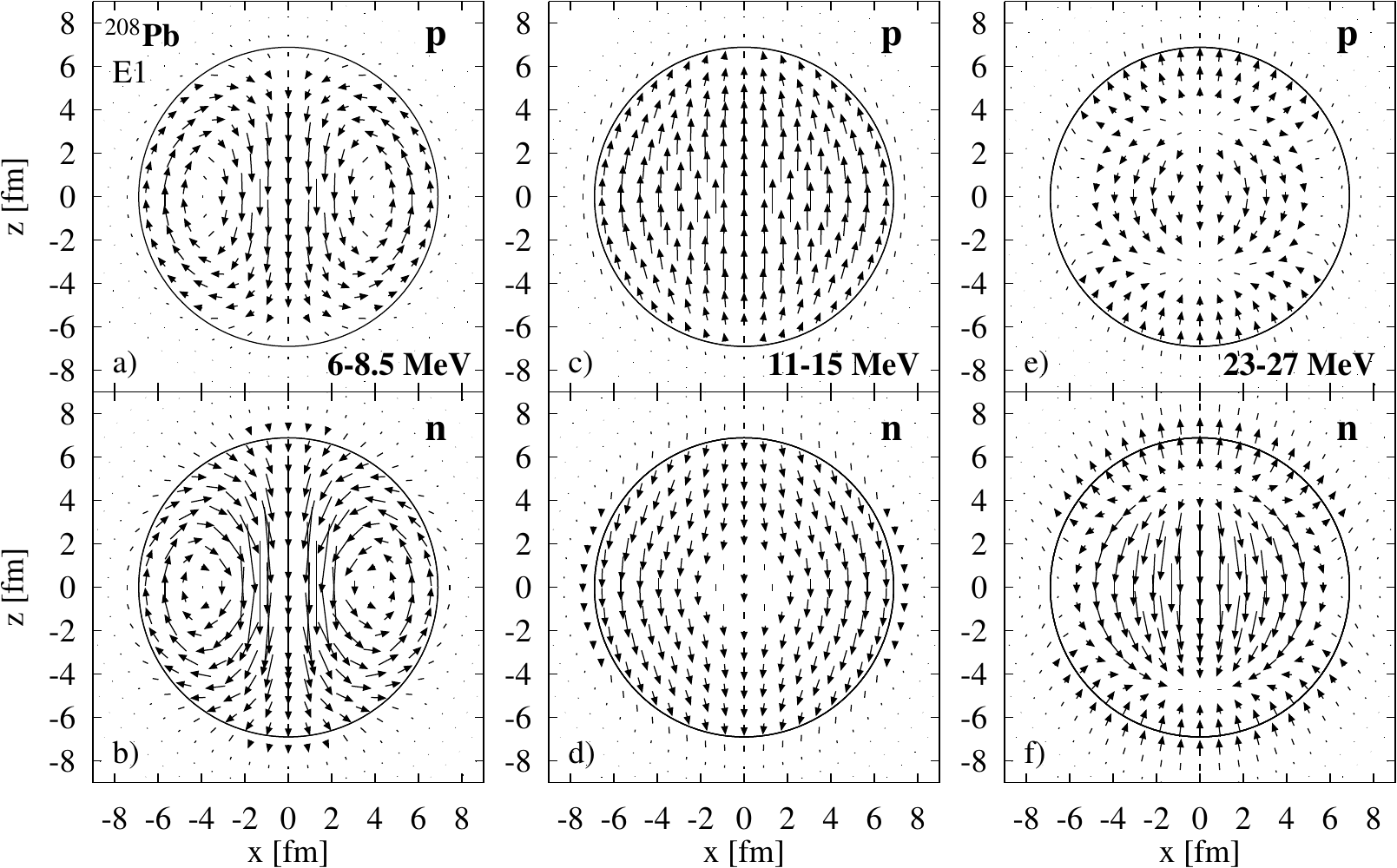}
\caption{Transition currents of $^{208}$Pb weighted by long-wave E1 operator in given energy intervals.}\label{fig_Pb_cur}
\end{figure}

Giant dipole resonance (GDR) of heavy neutron-rich nuclei contains a low-energy branch, around the nucleon emission threshold (Fig.~\ref{fig_pygmy_sf}g), which is called ``pygmy'' mode and is mostly interpreted as oscillation of the neutron skin with respect to proton-neutron core \cite{Savran2013}. To confirm this assumption, I calculated transition densities (Fig.~\ref{fig_Pb208rho}) and transition currents (Fig.~\ref{fig_Pb_cur}) of $^{208}$Pb in the pygmy region (here chosen as 6-8.5 MeV). Weighting operator in (\ref{cur_avg}) was long-wave (isovector) E1 operator to avoid any bias in the interpretation by ``forcing'' certain type of motion due to the operator choice.

\begin{figure}[t]
\centering
\includegraphics[width=0.7\textwidth]{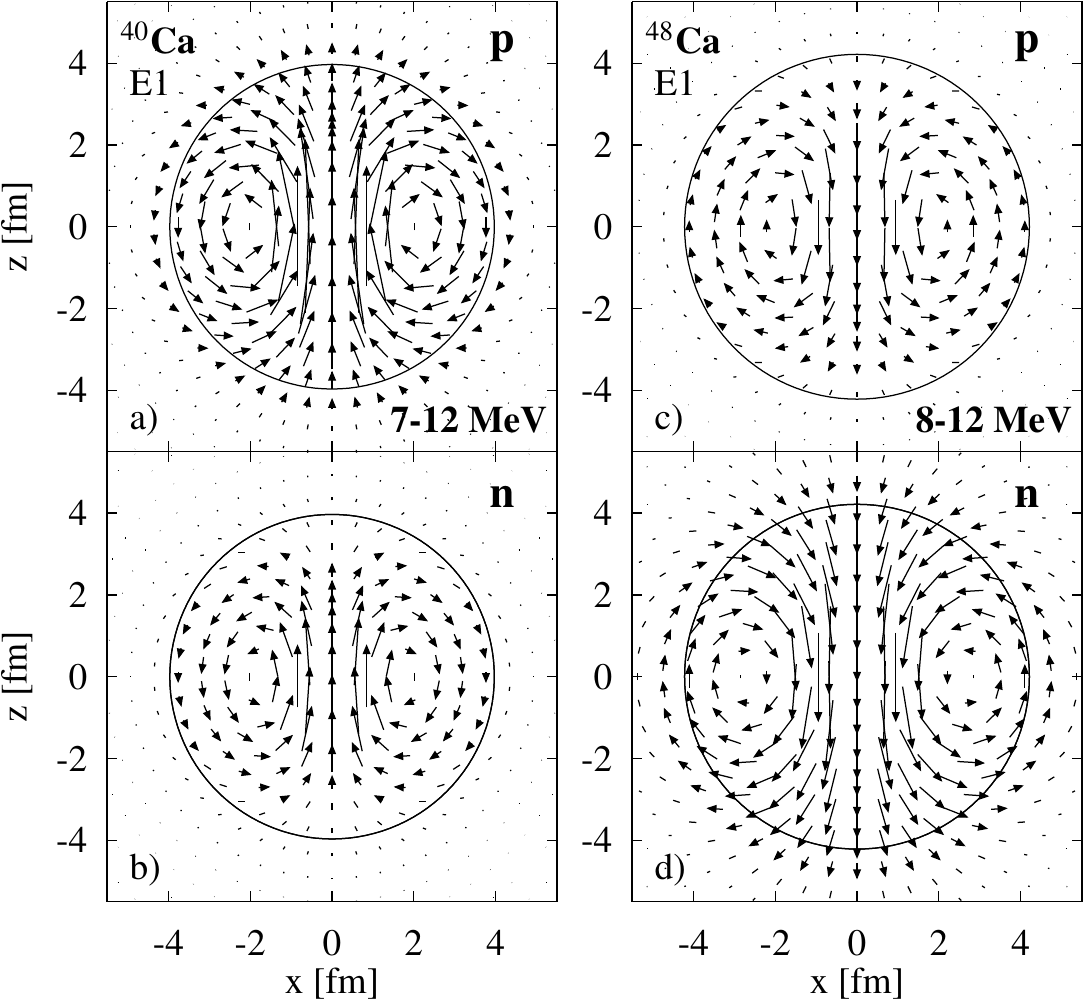}
\caption{Transition currents of $^{40}$Ca and $^{48}$Ca weighted by long-wave E1 operator in given energy intervals.}\label{fig_Ca_cur}
\end{figure}

Although the transition densities appear to confirm the ``pygmy'' picture of oscillating neutron skin, we should be careful at this interpretation due to the fact that the transition density is not sensitive to vortical motion according to the continuity equation (divergence of curl is zero):
\begin{equation}
-\mathrm{i}kc\delta\rho = -\partial_t\delta\rho = \vec{\nabla}\cdot\delta\vec{j}
\end{equation}
And indeed, Figures \ref{fig_Pb_cur}ab and \ref{fig_Ca_cur} indicate toroidal flow, together with the larger amplitude of the toroidal s.f.~in comparison with compression s.f. In fact, a motion reminiscent of skin vibration appears to be present in the high-energy area of the compression resonance (Fig.~\ref{fig_pygmy_sf}i, Fig.~\ref{fig_Pb_cur}ef).

We can therefore assume the following interpretation of the increased low-energy strength of the photoabsorption cross section with increasing neutron excess: The low-energy E1 states have mostly isoscalar toroidal nature with some compression component, and the electromagnetic strength is related to the compensating center-of-mass motion of the protons in response to the compressional component of the neutron skin. Isoscalar nature of the low-lying E1 states was also experimentally confirmed in $^{40}$Ca \cite{Papakonstantinou2011} and $^{48}$Ca \cite{Derya2014}. Calcium isotopes were theoretically analyzed also with second-RPA \cite{Gambacurta2011}, which includes two-phonon configurations, increasing the strength and fragmentation of the low-lying E1 strength, and the characteristic neutron-skin vibration was not confirmed.

The EWSR of the long-wave isovector E1 is exhausted by 0.35\%, 0.24\% and 1.08\% in the selected intervals for $^{40,48}$Ca (Fig.~\ref{fig_Ca_cur}) and $^{208}$Pb (Fig.~\ref{fig_Pb_cur}ab), respectively.

\section{Deformed $^{154}$Sm: E0, E1, E2}
\begin{figure}[h]
\includegraphics[width=\textwidth]{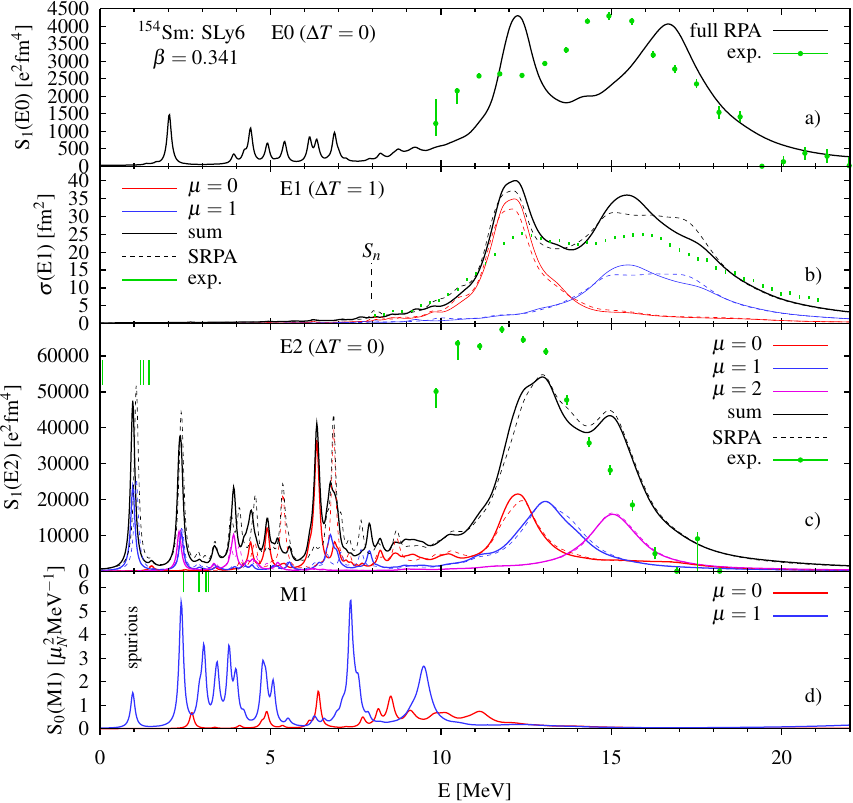}
\caption{\!\![revised] Double-folded strength functions of $^{154}$Sm for electric monopole, dipole and quadrupole resonance, compared to the experimental data (isoscalar \cite{Youngblood2004}, isovector \cite{Carlos1974}), and magnetic dipole transitions. Experimental distributions are given with respective errorbars, and individual states are given by vertical lines.}\label{fig_Sm154sf}
\end{figure}
Calculation of giant resonances was performed also for deformed nucleus $^{154}$Sm, which has experimental data available for isovector E1 (by photoabsorption) \cite{Carlos1974} as well as isoscalar E0 and E2 (by $\alpha$-scattering) \cite{Youngblood2004}. Parametrization SLy6 \cite{SLy6} was utilized, which was fitted without the tensor term (tensor term was found to cause certain problems in axial E1 RPA -- too high spurious state and broken rotational symmetry for spherical nuclei -- related probably to the Hartree-Fock). Equilibrium deformation $\beta=0.341$ was determined by HF. Strength functions of E0, E1, E2, as well as M1, are depicted in Fig.~\ref{fig_Sm154sf}. They were calculated by full RPA and SRPA with 5 input operators
\begin{equation}
\begin{split}
\hat{Q}_1 &= r^\lambda Y_{\lambda\mu},\quad
\hat{Q}_2 = r^{\lambda+2} Y_{\lambda+2,\mu},\quad
\hat{Q}_3 = j_\lambda(0.6r\ldots)Y_{\lambda\mu},\\
&\quad\hat{Q}_4 = j_\lambda(0.9r\ldots)Y_{\lambda\mu},\quad
\hat{Q}_5 = j_\lambda(1.2r\ldots)Y_{\lambda\mu}
\end{split}
\end{equation}
and the single-particle levels were taken up to 40 MeV. In the case of E2 ($\mu=0$), the second operator was replaced by $r^2Y_{00}$. The calculation time was around 24 hours for the most demanding full RPA cases (E1, $\mu=1$, 22570 $2qp$ pairs, using 24 GB of RAM and 8 threads on 12-core 3.46 GHz Intel Xeon Westmere workstation, spurious state was at 2.102 MeV; same for E2, $\mu=1$, 22558 $2qp$, with spurious state at 0.962 MeV; later optimizations led to over 2-fold speedup), while SRPA calculation took 1 hour for E1, and 2 hours for E2, respectively, on 2.5 GHz Intel i5 Sandy Bridge laptop (using one thread). However, the results of full RPA calculation for E2 ($\mu=1$) could be reused to evaluate also M1 at almost no cost ($\lambda$ is not a good quantum number in the body-fixed system, so the multipolarity is provided only by the transition operator). Double folding parameters were chosen as (\ref{df_param}), isoscalar transition used $z_p=z_n=1$, isovector charges were $z_p=N/A,\,z_n=-Z/A$ and magnetic transition was calculated with natural charges. The first state of E2 ($\mu=1$) and M1 ($\mu=1$) is spurious and corresponds to rotation of the nucleus.

Isoscalar giant quadrupole resonance (IS GQR) is clearly up-shifted in energy, compared to the experimental data \cite{Youngblood2004}. This shift is caused by effective mass smaller than 1 ($m^*/m\approx0.7$ for SLy-forces), while GQR can be accurately reproduced with parametrizations having $m^*/m\approx1$ \cite{Nesterenko2006} (e.g. SkT6 \cite{SkT6}) -- however, these forces fail to describe the GDR. There was an attempt to resolve such tensions by a new parametrization SV-bas ($m^*/m\approx0.9$) \cite{SVbas}, which however fails in the calculation of M1 transition (due to non-positive-definite matrix $P$). For this reason, and also due to omission of the tensor term and due to the tuning of $b_4'$ term -- which is inconsistent with 2-body Skyrme interaction (\ref{V_skyrme}) -- the SV-bas force was not used in the present work.

\begin{figure}[t]
\centering
\includegraphics[width=0.9\textwidth]{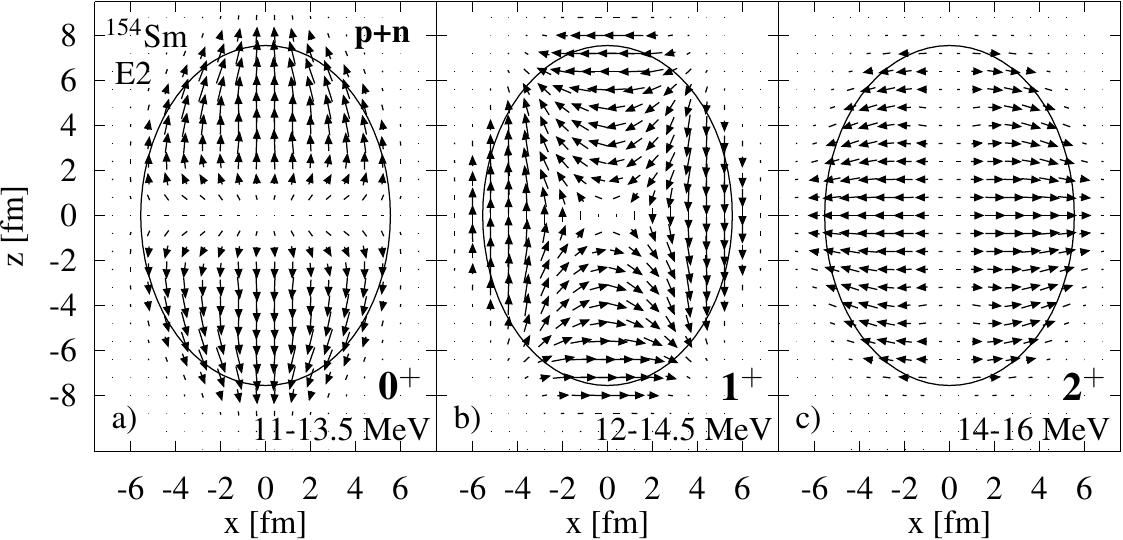}
\caption{\![revised $0^+$] Transition currents of $^{154}$Sm, evaluated in the area of giant quadrupole resonance, weighted by isoscalar E2 operator.}\label{fig_Sm154E2}
\end{figure}
It is instructive to show the transition currents corresponding to individual branches of the GQR (Fig.~\ref{fig_Sm154E2}). $K^\pi=0^+$ branch has a character of $\beta$-vibration and the lowest energy, and mixes with E0 resonance. On the contrary, $K^\pi=2^+$ branch has the highest energy and a character of $\gamma$-vibration, and is relative pure, so that also separable RPA describes it accurately (Fig.~\ref{fig_Sm154sf}c).

\begin{figure}[t]
\centering
\includegraphics[width=0.7\textwidth]{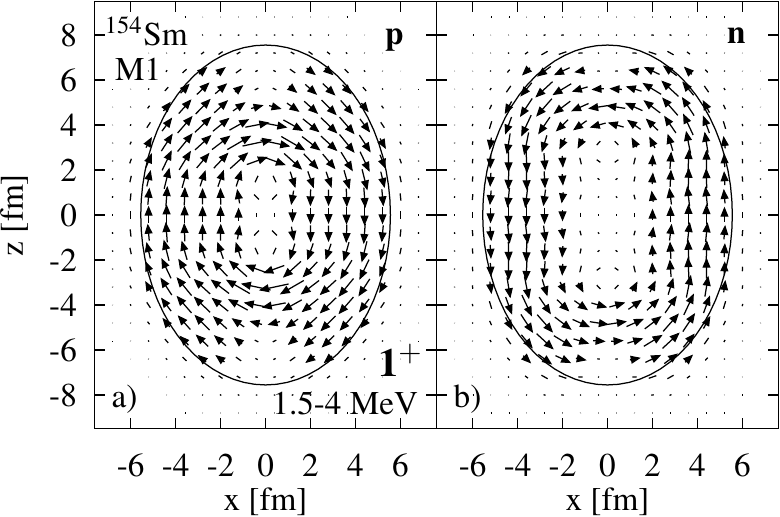}
\caption{Transition currents of $^{154}$Sm, evaluated in the area of scissor mode, weighted by the M1 operator.}\label{fig_Sm154M1}
\end{figure}
\begin{figure}[t]
\centering
\includegraphics[width=0.8\textwidth]{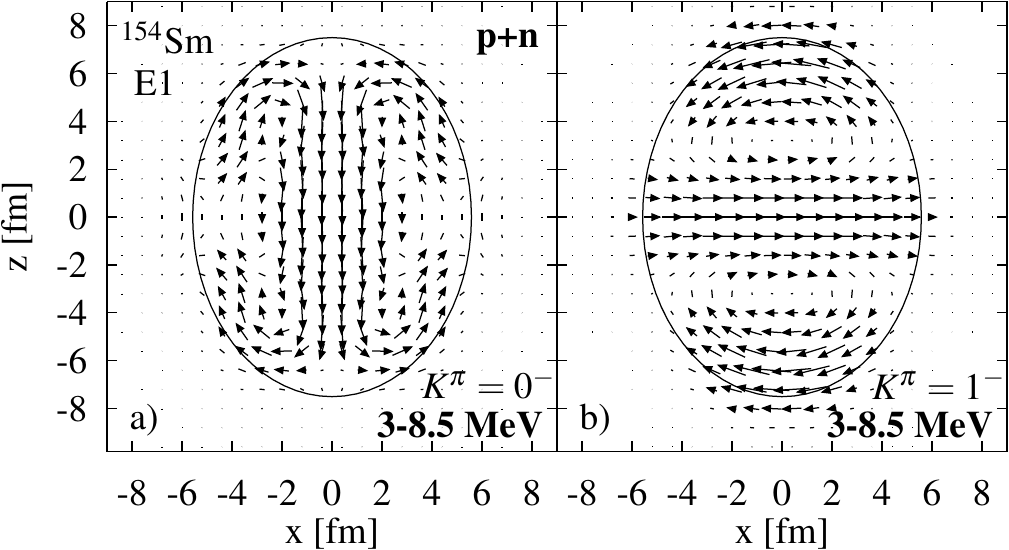}
\caption{\![revised] Transition currents of $^{154}$Sm, evaluated in the area of ``pygmy'' mode, weighted by the isovector E1 operator.}\label{fig_Sm154pygmy}
\end{figure}

Finally, two more pictures are given: for the scissor mode of M1 resonance (Fig.~\ref{fig_Sm154M1}) and for low-energy mode of the E1 resonance (Fig.~\ref{fig_Sm154pygmy}). The first case is not present in spherical nuclei, and the second case shows a clear difference between the two components ($\mu=0,1$) of toroidal resonance caused by broken spherical symmetry.

\section{Scissor and spin-flip parts of M1 resonance in $^{50}$Cr}
\begin{figure}[!b]
\centering
\includegraphics[width=0.7\textwidth]{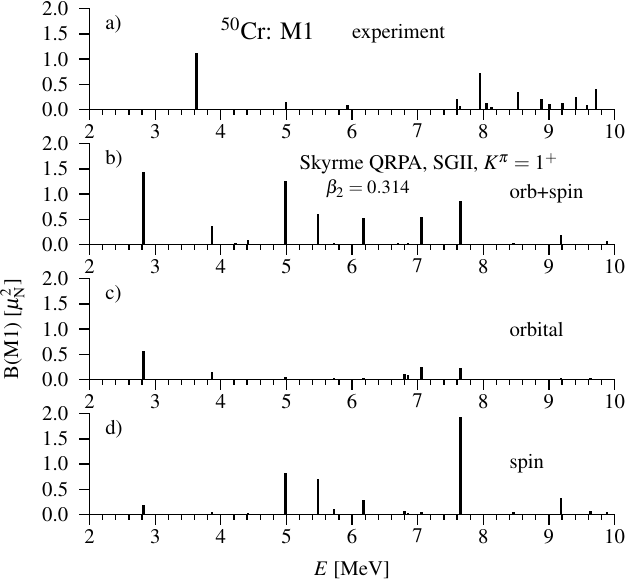}
\caption{Comparison of experimental and theoretical M1 transitions with energy up to 10 MeV in $^{50}$Cr. Experimental values of B(M1)$\uparrow$ have an uncertainity of about 10\% \cite{Pai2016}. RPA results are given for component $\mu=\pm1$. Additional plots show the transition strengths calculated only with either c) orbital, or d) spin part of the M1 transition operator.}\label{fig_Cr50_M11}
\end{figure}

\begin{figure}[t]
\centering
\includegraphics[width=0.7\textwidth]{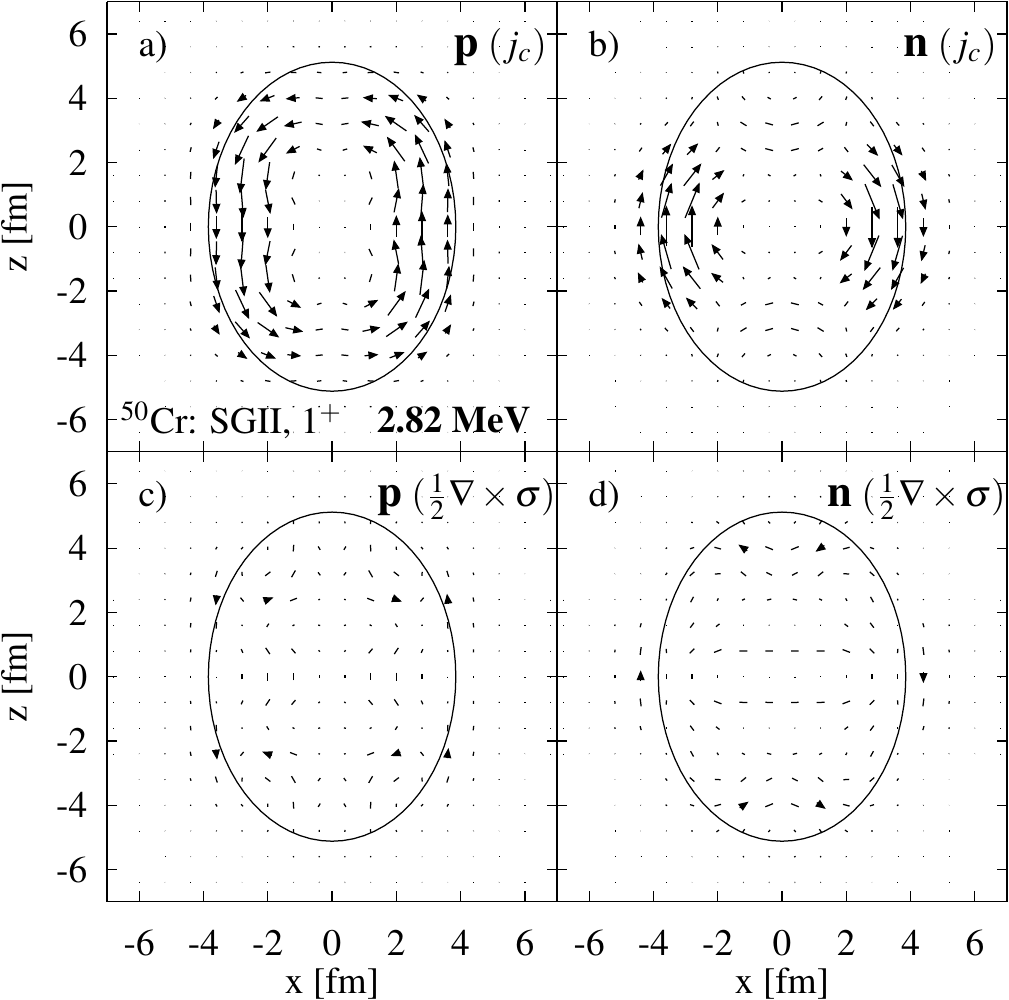}
\caption{Transition currents (convective and magnetization) of the 2.82 MeV scissor mode.}\label{fig_Cr50_scissor}
\end{figure}

Recent experimental data on M1 transitions in deformed nucleus $^{50}$Cr \cite{Pai2016} allow to investigate accuracy of theoretical predictions for this light deformed nucleus. Full quasiparticle RPA calculation utilized Skyrme force SGII \cite{SGII}, which was developed with aim to reproduce the Gamow-Teller transitions. Equilibrium deformation was $\beta=0.314$. Figure \ref{fig_Cr50_M11} gives the comparison of M1 ($\mu=1$) transition probabilities, calculated by full RPA (s.p.~basis up to 50 MeV), with the experiment. The scissor mode (see also Fig.~\ref{fig_Cr50_scissor}) is concentrated in one state, in agreement with the experiment, and the spin-flip mode is spread in multiple states of higher energy. Skyrme calculation underestimates the energy of both modes, while the weak fragmentation of calculated spin-flip mode is caused by one-phonon nature of RPA.

\begin{figure}[h]
\centering
\includegraphics[width=0.7\textwidth]{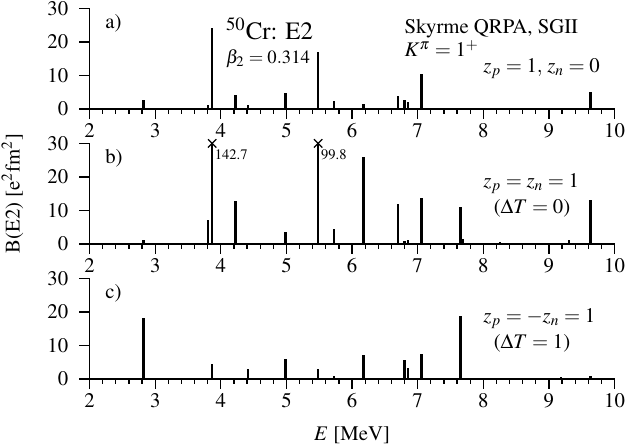}
\caption{Transitions $K^\pi=1^+$ in $^{50}$Cr evaluated with E2 operator ($\mu=\pm1$). Most of the states have isoscalar character, except isovector scissor mode at 2.82 MeV and spin-flip mode at 7.65 MeV with mixed character. Two strongest isoscalar transitions are truncated, giving their B(E2).}\label{fig_Cr50_E21}
\end{figure}

\begin{figure}[h]
\centering
\includegraphics[width=0.7\textwidth]{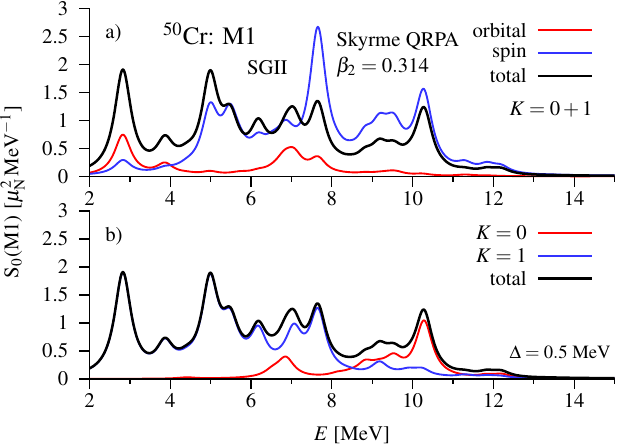}
\caption{\![revised] Total calculated $^{50}$Cr M1 strength given as a strength function. Orbital and spin part show constructive interference up to 7 MeV, and destructive interference beyond that.}\label{fig_Cr50_M1sf}
\end{figure}

Orbital motion of the states beyond 3 MeV is mostly isoscalar, except the strongest 7.65 MeV spin-flip state, as can be demonstrated by evaluating the electric quadrupole transition probabilities (Fig.~\ref{fig_Cr50_E21}). The $\mu=0$ branch of M1 transitions gives minor contribution, and becomes important only in the higher-energy region (9-12 MeV; see Fig.~\ref{fig_Cr50_M1sf}).

\chapter{Summary}
Skyrme Random Phase Approximation was successfully implemented for spherical and axially symmetric nuclei, including the Coulomb and pairing interaction. Spherical formalism involved advanced angular-momentum-coupling techniques and enabled to formulate the results in a convenient and rotationally-invariant way of density/current reduced matrix elements, in contrast with previous formulations, which were either not manifestly invariant \cite{Reinhard1992} or too cumbersome \cite{Terasaki2005,Colo2013}.

Kinetic center-of-mass term was implemented in Hartree-Fock before variation, as well as in RPA, and led to an interesting behavior with respect to removal of the spurious motion. Nevertheless, this term can be safely omitted due to relativelly small influence on the giant resonances. In the axial case, the logarithmic singularity of the Coulomb integral was correctly removed by a procedure inspired by the cartesian case, which can be treated analytically (to a certain degree).

Large-basis calculation on the spherical closed-shell nuclei $^{40,48}$Ca and $^{208}$Pb was used to demonstrate the influence of the oscillator length, tensor ($\mathcal{J}^2$) and spin terms in Skyrme functional, accuracy of the long-wave transition operators and separable RPA method. Last chapter gives an analysis of selected physical topis, such as isoscalar toroidal character of the low-energy (``pygmy'') E1 mode, electric multipolar resonances in deformed $^{154}$Sm and demonstration of scissor and spin-flip M1 transitions in $^{50}$Cr. All these calculations are restricted to one-phonon excitations with a discrete basis of s.p.~states, so the fragmentation due to coupling with complex configurations and escape width are simulated only by Lorentz smoothing of the strength functions. Better description should include multi-phonon configurations, however, it is not clear whether such methods are physically valid for density functionals. For example, the zero-range interaction leads to divergent correlation energy \cite{Moghrabi2010}. If we stay in the present one-phonon RPA framework, there is still a possibility of extension to $\beta$-transitions.

The results of calculations with the methods described in this work were published in \cite{Repko2013,Reinhard2014,Nesterenko2015,Kvasil2013,Kvasil2015-ischia,Kvasil2015} and submitted for publication in \cite{Nesterenko-lowE2,Nesterenko-Mg24,Repko-istros,Pai2016}.

\appendix

\chapter{Detailed derivation of Skyrme functional from two-body interaction}\label{app_skyr-dft}
Skyrme interaction (\ref{V_skyrme}) can be completely rewritten (including its exchange term) into a density functional (\ref{Skyrme_DFT}) in terms of generalized one-body densities and currents (\ref{Jd_gs}).

Indices $j,k,l$ will denote cartesian coordinates, so the summations run over $\{x,y,z\}$. Index $s$ in $\psi_{\alpha s}(\vec{r})$ denotes spin projection of a given wave function.
\begin{small}
\begin{equation*}
\begin{array}{ll}
\langle \vec{r}_1 s_1,\vec{r}_2 s_2|\alpha\beta\rangle = \psi_{\alpha s_1}(\vec{r}_1) \psi_{\beta s_2}(\vec{r}_2), \ \ & \displaystyle
\langle \alpha\beta| \delta(\vec{r}_1-\vec{r}_2)|\alpha\beta\rangle =
\int \psi_\alpha^\dagger(\vec{r})\psi_\alpha^{\phantom{|}}(\vec{r})
\psi_\beta^\dagger(\vec{r})\psi_\beta^{\phantom{|}}(\vec{r})\mathrm{d}^3 r \\
\langle \vec{r}_1 s_1,\vec{r}_2 s_2|\beta\alpha\rangle = \psi_{\beta s_1}(\vec{r}_1) \psi_{\alpha s_2}(\vec{r}_2), \ \ &
\langle \vec{r}_1 s_1,\vec{r}_2 s_2|\hat{P}_\sigma|\beta\alpha\rangle = \psi_{\beta s_2}(\vec{r}_1) \psi_{\alpha s_1}(\vec{r}_2)\phantom{\Big{|}}
\end{array}
\end{equation*}
\end{small} \\[-6pt]
Spin-exchange term $\hat{P}_\sigma=\frac{1}{2}(1+\vec{\sigma}_1\cdot\vec{\sigma}_2)$ (\ref{P_sigma}) can either act to reverse the effect of HF exchange term (with later setting $\vec{r}_1=\vec{r}_2$), or is taken simply from its definition and introduces $\vec{\sigma}$ matrices into integrals:
\begin{align*}
\langle \alpha\beta|\hat{P}_\sigma\delta(\vec{r}_1-\vec{r}_2)|\beta\alpha\rangle &=
\int \psi_\alpha^\dagger(\vec{r})\psi_\alpha^{\phantom{|}}(\vec{r})
\psi_\beta^\dagger(\vec{r})\psi_\beta^{\phantom{|}}(\vec{r})\mathrm{d}^3 r \\
\langle \alpha\beta|\hat{P}_\sigma\delta(\vec{r}_1-\vec{r}_2)|\alpha\beta\rangle &=
\frac{1}{2}\int \psi_\alpha^\dagger(\vec{r})\psi_\alpha^{\phantom{|}}(\vec{r})
\psi_\beta^\dagger(\vec{r})\psi_\beta^{\phantom{|}}(\vec{r})\mathrm{d}^3 r \\
&\quad {}+
\frac{1}{2}\int [\psi_\alpha^\dagger(\vec{r})\vec{\sigma}\psi_\alpha^{\phantom{|}}(\vec{r})]\cdot
[\psi_\beta^\dagger(\vec{r})\vec{\sigma}\psi_\beta^{\phantom{|}}(\vec{r})]\mathrm{d}^3 r \\
&= \langle\alpha\beta|\hat{P}_\sigma\delta(\vec{r}_1-\vec{r}_2)\hat{P}_\sigma|\beta\alpha\rangle = \langle\alpha\beta|\delta(\vec{r}_1-\vec{r}_2)|\beta\alpha\rangle
\end{align*}
Term with $t_0$ and $x_0$ can be now obtained easily:
\begin{equation}
\label{t0-term}
\begin{split}
\sum_{\alpha\beta}\langle\alpha\beta|t_0(1+x_0\hat{P}_\sigma)\delta(\vec{r}_1-\vec{r}_2)|\alpha\beta\rangle &= \int \bigg[
t_0\Big(1+\frac{x_0}{2}\Big)\rho^2 + \frac{t_0 x_0}{2}\vec{s}^2 \bigg]\mathrm{d}^3 r \\
\sum_{\alpha\beta\in q}\langle\alpha\beta|t_0(1+x_0\hat{P}_\sigma)\delta(\vec{r}_1-\vec{r}_2)|\beta\alpha\rangle &= \int \bigg[
t_0\Big(\frac{1}{2}+x_0\Big)\rho_q^2 + \frac{t_0}{2}\vec{s}_q^2 \bigg]\mathrm{d}^3 r
\end{split}
\end{equation}
Parts of $t_1$ term:
\begin{footnotesize}
\begin{align*}
\langle\alpha\beta|(1+x_1\hat{P}_\sigma)&\delta(\vec{r}_1-\vec{r}_2)(\vec{\nabla}_1-\vec{\nabla}_2)^2|\alpha\beta\rangle = \\
&= \Big(1+\frac{x_1}{2}\Big)\int \psi_\alpha^\dagger \psi_\beta^\dagger
\big\{[\Delta\psi_\alpha^{\phantom{|}}]\psi_\beta^{\phantom{|}} -
2[\vec{\nabla}\psi_\alpha^{\phantom{|}}]\cdot[\vec{\nabla}\psi_\beta^{\phantom{|}}]
+ \psi_\alpha^{\phantom{|}}[\Delta\psi_\beta^{\phantom{|}}]\big\}
\mathrm{d}^3 r \\
&\ \ {}+ \frac{x_1}{2} \int \psi_\alpha^\dagger \psi_\beta^\dagger
\bigg\{[\vec{\sigma}\Delta\psi_\alpha^{\phantom{|}}]\cdot[\vec{\sigma}\psi_\beta^{\phantom{|}}]
-2\sum_{j,k}[\partial_j\sigma_k\psi_\alpha^{\phantom{|}}][\partial_j\sigma_k\psi_\beta^{\phantom{|}}]
+[\vec{\sigma}\psi_\alpha^{\phantom{|}}]\cdot[\vec{\sigma}\Delta\psi_\beta^{\phantom{|}}]\bigg\} \mathrm{d}^3 r \\
\langle\alpha\beta|(1+x_1\hat{P}_\sigma)&\delta(\vec{r}_1-\vec{r}_2)(\vec{\nabla}_1-\vec{\nabla}_2)^2|\beta\alpha\rangle = \\
=& \Big(\frac{1}{2}+x_1\Big)\int \psi_\alpha^\dagger \psi_\beta^\dagger
\big\{[\Delta\psi_\alpha^{\phantom{|}}]\psi_\beta^{\phantom{|}} -
2[\vec{\nabla}\psi_\alpha^{\phantom{|}}]\cdot[\vec{\nabla}\psi_\beta^{\phantom{|}}]
+ \psi_\alpha^{\phantom{|}}[\Delta\psi_\beta^{\phantom{|}}]\big\}
\mathrm{d}^3 r \phantom{\bigg{|}}\\
&\ \ {}+ \frac{1}{2} \int \psi_\alpha^\dagger \psi_\beta^\dagger
\bigg\{[\vec{\sigma}\Delta\psi_\alpha^{\phantom{|}}]\cdot[\vec{\sigma}\psi_\beta^{\phantom{|}}]
-2\sum_{j,k}[\partial_j\sigma_k\psi_\alpha^{\phantom{|}}][\partial_j\sigma_k\psi_\beta^{\phantom{|}}]
+[\vec{\sigma}\psi_\alpha^{\phantom{|}}]\cdot[\vec{\sigma}\Delta\psi_\beta^{\phantom{|}}]\bigg\} \mathrm{d}^3 r
\end{align*}
\end{footnotesize}\\[-2pt]
I prepare derivatives of the previously defined densities:
\begin{small}
\begin{align*}
\Delta\rho_q(\vec{r}) &= 2\tau_q(\vec{r}) + \sum_{\alpha\in q}\big\{
[\Delta\psi_\alpha^{\phantom{|}}(\vec{r})]^\dagger \psi_\alpha^{\phantom{|}}(\vec{r}) +
\psi_\alpha^\dagger(\vec{r}) [\Delta\psi_\alpha^{\phantom{|}}(\vec{r})]\big\} \\
\Delta\vec{s}_q(\vec{r}) &= 2\vec{T}_q(\vec{r}) + \sum_{\alpha\in q}\big\{
[\vec{\sigma}\Delta\psi_\alpha^{\phantom{|}}(\vec{r})]^\dagger \psi_\alpha^{\phantom{|}}(\vec{r}) +
\psi_\alpha^\dagger(\vec{r}) [\vec{\sigma}\Delta\psi_\alpha^{\phantom{|}}(\vec{r})]\big\} \\
\ [\vec{\nabla}\rho(\vec{r})]^2 - 4[\vec{j}(\vec{r})]^2 &=
2\sum_{\alpha\beta}\big\{
[\vec{\nabla}\psi_\alpha^{\phantom{|}}(\vec{r})]^\dagger \cdot
[\vec{\nabla}\psi_\beta^{\phantom{|}}(\vec{r})]^\dagger \psi_\alpha^{\phantom{|}}(\vec{r})\psi_\beta^{\phantom{|}}(\vec{r}) \\[-8pt]
& \qquad\qquad{}+
\psi_\alpha^\dagger(\vec{r})\psi_\beta^\dagger(\vec{r})
[\vec{\nabla}\psi_\alpha^{\phantom{|}}(\vec{r})]\cdot
[\vec{\nabla}\psi_\beta^{\phantom{|}}(\vec{r})] \big\} \\
\ [\partial_j s_k(\vec{r})]^2-4[\mathcal{J}_{jk}(\vec{r})]^2 &=
2\sum_{\alpha\beta}\big\{
[\partial_j\sigma_k\psi_\alpha^{\phantom{|}}(\vec{r})]^\dagger
[\partial_j\sigma_k\psi_\beta^{\phantom{|}}(\vec{r})]^\dagger \psi_\alpha^{\phantom{|}}(\vec{r})\psi_\beta^{\phantom{|}}(\vec{r}) \\[-8pt]
& \qquad\qquad {}+
\psi_\alpha^\dagger(\vec{r})\psi_\beta^\dagger(\vec{r})
[\partial_j\sigma_k\psi_\alpha^{\phantom{|}}(\vec{r})]
[\partial_j\sigma_k\psi_\beta^{\phantom{|}}(\vec{r})] \big\} \\
\int_V [2\rho\Delta\rho + 2(\vec{\nabla}\rho)^2] \mathrm{d}^3 r &=
\int_V \Delta(\rho^2) \mathrm{d}^3 r =
\oint_{\partial V} \vec{\nabla}(\rho^2)\cdot \mathrm{d}\vec{S} = 0
\end{align*}
\end{small}\\[-2pt]
The whole $t_1$ term:
\begin{small}
\begin{equation}
\begin{split}
-\frac{1}{8}t_1 \sum_{\alpha\beta}&\,\langle\alpha\beta|(1+x_1\hat{P}_\sigma)[(\overleftarrow{\nabla}_1-\overleftarrow{\nabla}_2)^2\delta(\vec{r}_1-\vec{r}_2) + \delta(\vec{r}_1-\vec{r}_2)(\overrightarrow{\nabla}_1-\overrightarrow{\nabla}_2)^2]|\alpha\beta\rangle = \\
&= \int 
\bigg\{ {-}\frac{t_1(2+x_1)}{16}[2(\Delta\rho-2\tau)\rho-(\vec{\nabla}\rho)^2+4\vec{j}^2] \\
&\qquad\qquad{}-\frac{t_1 x_1}{16} \bigg[2(\Delta\vec{s}-2\vec{T})\cdot\vec{s}-\sum_{j,k}[(\partial_j s_k)^2-4(\mathcal{J}_{jk})^2]\bigg]
\bigg\}\mathrm{d}^3 r \\
&= \int
\bigg\{ \frac{t_1(2+x_1)}{16}[3(\vec{\nabla}\rho)^2+4\rho\tau-4\vec{j}^2]\\
&\qquad\qquad{}+
\frac{t_1 x_1}{16} \bigg[4\vec{s}\cdot\vec{T}+\!\!\sum_{j,k=x,y,z}\!\![3(\partial_j s_k)^2-4(\mathcal{J}_{jk})^2]\bigg]
\bigg\}\mathrm{d}^3 r \\
-\frac{1}{8}t_1 \sum_{\alpha\beta\in q}&\,\langle\alpha\beta|(1+x_1\hat{P}_\sigma)[(\overleftarrow{\nabla}_1-\overleftarrow{\nabla}_2)^2\delta(\vec{r}_1-\vec{r}_2) + \delta(\vec{r}_1-\vec{r}_2)(\overrightarrow{\nabla}_1-\overrightarrow{\nabla}_2)^2]|\beta\alpha\rangle = \phantom{\Bigg{|}}\\
&= \int
\bigg\{ \frac{t_1(1+2x_1)}{16}[3(\vec{\nabla}\rho_q)^2+4\rho_q\tau_q-4\vec{j}_q^2]\\
&\qquad\qquad{}+
\frac{t_1}{16} \bigg[4\vec{s}_q\cdot\vec{T}_q+\sum_{j,k}[3(\partial_j s_{q;k})^2-4(\mathcal{J}_{q;jk})^2]\bigg]
\bigg\}\mathrm{d}^3 r
\end{split}
\end{equation}
\end{small}\\[-6pt]
Parts of $t_2$ term:
\begin{footnotesize}
\begin{align*}
\langle\alpha\beta|(1+x_2\hat{P}_\sigma)\overleftarrow{\nabla}_1\cdot
\delta(\vec{r}_1-\vec{r}_2)(\vec{\nabla}_1-\vec{\nabla}_2)|\alpha\beta\rangle =
\Big(1+\frac{x_2}{2}\Big)\int \psi_\beta^\dagger (\vec{\nabla}\psi_\alpha^\dagger)\cdot
[\psi_\beta^{\phantom{|}}(\vec{\nabla}\psi_\alpha^{\phantom{|}}) -
(\vec{\nabla}\psi_\beta^{\phantom{|}})\psi_\alpha^{\phantom{|}}]
\mathrm{d}^3 r \\
{}+ \frac{x_2}{2} \int \sum_{j} \bigg\{
[(\partial_j\psi_\alpha^\dagger)\vec{\sigma}(\partial_j\psi_\alpha^{\phantom{|}})]
\cdot[\psi_\beta^\dagger\vec{\sigma}\psi_\beta^{\phantom{|}}]
-[(\partial_j\vec{\sigma}\psi_\alpha^{\phantom{|}})^\dagger\psi_\alpha^{\phantom{|}}] \cdot
[\psi_\beta^\dagger(\partial_j\vec{\sigma}\psi_\beta^{\phantom{|}})]
\bigg\} \mathrm{d}^3 r \\
\langle\alpha\beta|(1+x_2\hat{P}_\sigma)\overleftarrow{\nabla}_1\cdot
\delta(\vec{r}_1-\vec{r}_2)(\vec{\nabla}_1-\vec{\nabla}_2)|\beta\alpha\rangle =
\Big(\frac{1}{2}+x_2\Big)\int \psi_\beta^\dagger (\vec{\nabla}\psi_\alpha^\dagger)\cdot
[(\vec{\nabla}\psi_\beta^{\phantom{|}})\psi_\alpha^{\phantom{|}} -
\psi_\beta^{\phantom{|}}(\vec{\nabla}\psi_\alpha^{\phantom{|}})]
\mathrm{d}^3 r \\
{}- \frac{1}{2} \int \sum_{j} \bigg\{
[(\partial_j\psi_\alpha^\dagger)\vec{\sigma}(\partial_j\psi_\alpha^{\phantom{|}})]
\cdot[\psi_\beta^\dagger\vec{\sigma}\psi_\beta^{\phantom{|}}]
-[(\partial_j\vec{\sigma}\psi_\alpha^{\phantom{|}})^\dagger\psi_\alpha^{\phantom{|}}] \cdot
[\psi_\beta^\dagger(\partial_j\vec{\sigma}\psi_\beta^{\phantom{|}})]
\bigg\} \mathrm{d}^3 r
\end{align*}
\end{footnotesize}\\[-2pt]
Useful derivatives:
\begin{small}
\begin{align*}
[\vec{\nabla}\rho(\vec{r})]^2 + 4[\vec{j}(\vec{r})]^2 &=
2\sum_{\alpha\beta}\big\{
\psi_\beta^\dagger(\vec{r})[\vec{\nabla}\psi_\alpha^{\phantom{|}}(\vec{r})]^\dagger \cdot
[\vec{\nabla}\psi_\beta^{\phantom{|}}(\vec{r})] \psi_\alpha^{\phantom{|}}(\vec{r})\\[-8pt]
&\qquad\qquad{}+
\psi_\alpha^\dagger(\vec{r})[\vec{\nabla}\psi_\beta^{\phantom{|}}(\vec{r})]^\dagger\cdot
[\vec{\nabla}\psi_\alpha^{\phantom{|}}(\vec{r})]\psi_\beta^{\phantom{|}}(\vec{r}) \big\} \\
\ [\partial_j s_k(\vec{r})]^2+4[\mathcal{J}_{jk}(\vec{r})]^2 &=
2\sum_{\alpha\beta}\big\{
[\partial_j\sigma_k\psi_\alpha^{\phantom{|}}(\vec{r})]^\dagger \psi_\alpha^{\phantom{|}}(\vec{r})
\psi_\beta^\dagger(\vec{r}) [\partial_j\sigma_k\psi_\beta^{\phantom{|}}(\vec{r})] \\[-8pt]
&\qquad\qquad{}+
\psi_\alpha^\dagger(\vec{r})[\partial_j\sigma_k\psi_\alpha^{\phantom{|}}(\vec{r})]
[\partial_j\sigma_k\psi_\beta^{\phantom{|}}(\vec{r})]^\dagger\psi_\beta^{\phantom{|}}(\vec{r})
 \big\}
\end{align*}
\end{small}\\[-2pt]
The whole $t_2$ term:
\begin{small}
\begin{equation}
\begin{split}
\frac{1}{4}t_2\sum_{\alpha\beta}&\,\langle\alpha\beta|(1+x_2\hat{P}_\sigma)(\overleftarrow{\nabla}_1-\overleftarrow{\nabla}_2)\cdot
\delta(\vec{r}_1-\vec{r}_2)(\overrightarrow{\nabla}_1-\overrightarrow{\nabla}_2)|\alpha\beta\rangle =\\
&= \int \bigg\{
\frac{t_2(2+x_2)}{16}[4\rho\tau-(\vec{\nabla}\rho)^2-4\vec{j}^2] \\
&\qquad\qquad{}+
\frac{t_2 x_2}{16}\bigg[4\vec{s}\cdot\vec{T}-\!\!\sum_{j,k=x,y,z}\!\![(\partial_j s_k)^2+4(\mathcal{J}_{jk})^2]\bigg]\bigg\} \mathrm{d}^3 r \\
&= \int \bigg\{
{-}\frac{t_2(1+2x_2)}{16}[4\rho_q\tau_q-(\vec{\nabla}\rho_q)^2-4\vec{j}_q^2] \\
&\qquad\qquad{}-
\frac{t_2}{16} \bigg[4\vec{s}_q\cdot\vec{T}_q-\!\!\sum_{j,k=x,y,z}\!\![(\partial_j s_{q;k})^2+4(\mathcal{J}_{q;jk})^2]\bigg]\bigg\} \mathrm{d}^3 r
\end{split}
\end{equation}
\end{small}\\[-6pt]
Density-dependent $t_3$ term is just a simple variation of $t_0$ term:
\begin{small}
\begin{equation}
\begin{split}
\frac{1}{6}t_3\sum_{\beta\gamma}\langle\beta\gamma|(1+x_3\hat{P}_\sigma)\delta(\vec{r}_1-\vec{r}_2)\rho^\alpha\Big(\frac{\vec{r_1}+\vec{r}_2}{2}\Big)|\beta\gamma\rangle &= \int \bigg[
\frac{t_3(2+x_3)}{12}\rho^{\alpha+2} + \frac{t_3 x_3}{12}\rho^\alpha\vec{s}^2
\bigg]\mathrm{d}^3 r \\
\frac{1}{6}t_3\sum_{\beta\gamma\in q}\langle\beta\gamma|(1+x_3\hat{P}_\sigma)\delta(\vec{r}_1-\vec{r}_2)\rho^\alpha\Big(\frac{\vec{r_1}+\vec{r}_2}{2}\Big)|\gamma\beta\rangle &= \int \bigg[
\frac{t_3(1+2x_3)}{12}\rho^\alpha\rho_q^2 + \frac{t_3}{12}\rho^\alpha\vec{s}_q^2
\bigg]\mathrm{d}^3 r
\end{split}
\end{equation}
\end{small}\\[-6pt]
Parts of $t_4$ term:
\begin{small}
\begin{align*}
\langle\alpha\beta|&\,(\vec{\sigma}_1+\vec{\sigma}_2)\cdot[\overleftarrow{\nabla}_1\times
\delta(\vec{r}_1-\vec{r}_2)(\overrightarrow{\nabla}_1-\overrightarrow{\nabla}_2)]|\alpha\beta\rangle = \\
&= \sum_{ijk}\varepsilon_{ijk}\int \big[(\sigma_i\partial_j\psi_\alpha^{\phantom{|}})^\dagger \psi_\beta^\dagger +
(\partial_j\psi_\alpha^{\phantom{|}})^\dagger (\sigma_i\psi_\beta^{\phantom{|}})^\dagger\big]
\big[(\partial_k\psi_\alpha^{\phantom{|}})\psi_\beta^{\phantom{|}} -
\psi_\alpha^{\phantom{|}}(\partial_k\psi_\beta^{\phantom{|}})\big]
\mathrm{d}^3 r \\
\langle\alpha\beta|&\,(\vec{\sigma}_1+\vec{\sigma}_2)\cdot[\overleftarrow{\nabla}_1\times
\delta(\vec{r}_1-\vec{r}_2)(\overrightarrow{\nabla}_1-\overrightarrow{\nabla}_2)]|\beta\alpha\rangle = \phantom{\bigg{|}}\\
&= \frac{1}{2}\sum_{ijk}\varepsilon_{ijk}\int \big[(\sigma_i\partial_j\psi_\alpha^{\phantom{|}})^\dagger \psi_\beta^\dagger +
(\partial_j\psi_\alpha^{\phantom{|}})^\dagger (\sigma_i\psi_\beta^{\phantom{|}})^\dagger\big]
\big[\psi_\alpha^{\phantom{|}}(\partial_k\psi_\beta^{\phantom{|}}) -
(\partial_k\psi_\alpha^{\phantom{|}})\psi_\beta^{\phantom{|}}\big]
\mathrm{d}^3 r \\
&\quad {}+
\frac{1}{2}\sum_{ijkn}\varepsilon_{ijk}\int \big[(\sigma_n\sigma_i\partial_j\psi_\alpha^{\phantom{|}})^\dagger (\sigma_n\psi_\beta^{\phantom{|}})^\dagger +
(\sigma_n\partial_j\psi_\alpha^{\phantom{|}})^\dagger (\sigma_n\sigma_i\psi_\beta^{\phantom{|}})^\dagger\big] \\[-8pt]
&\qquad\qquad\qquad\qquad{}\times
\big[\psi_\alpha^{\phantom{|}}(\partial_k\psi_\beta^{\phantom{|}}) -
(\partial_k\psi_\alpha^{\phantom{|}})\psi_\beta^{\phantom{|}}\big]
\mathrm{d}^3 r
\end{align*}
\end{small}\\[-6pt]
I will substitute following relations into the exchange term:
\[ \sigma_i\sigma_n = \delta_{in}+\mathrm{i}\sum_p\varepsilon_{inp}\sigma_p
\qquad\textrm{and}\qquad
\sum_i\varepsilon_{ijk}\varepsilon_{inp} = \delta_{jn}\delta_{kp} - \delta_{jp}\delta_{kn} \]
\begin{footnotesize}
\begin{align*}
\langle\alpha\beta|&\,(\vec{\sigma}_1+\vec{\sigma}_2)\cdot[\overleftarrow{\nabla}_1\times
\delta(\vec{r}_1-\vec{r}_2)(\overrightarrow{\nabla}_1-\overrightarrow{\nabla}_2)]|\beta\alpha\rangle = \phantom{\bigg{|}}\\
&= \sum_{ijk}\varepsilon_{ijk}\int \big[(\sigma_i\partial_j\psi_\alpha^{\phantom{|}})^\dagger \psi_\beta^\dagger +
(\partial_j\psi_\alpha^{\phantom{|}})^\dagger (\sigma_i\psi_\beta^{\phantom{|}})^\dagger\big]
\big[\psi_\alpha^{\phantom{|}}(\partial_k\psi_\beta^{\phantom{|}}) -
(\partial_k\psi_\alpha^{\phantom{|}})\psi_\beta^{\phantom{|}}\big]
\mathrm{d}^3 r \\
&\quad {}+
\frac{\mathrm{i}}{2}\sum_{jk}\varepsilon_{ijk}\int
\big[(\partial_j\psi_\alpha^{\phantom{|}})^\dagger \psi_\beta^\dagger -
(\partial_j\psi_\alpha^{\phantom{|}})^\dagger \psi_\beta^\dagger\big]
(\sigma_{1k}\sigma_{2j}-\sigma_{1j}\sigma_{2k})
\big[\psi_\alpha^{\phantom{|}}(\partial_k\psi_\beta^{\phantom{|}}) -
(\partial_k\psi_\alpha^{\phantom{|}})\psi_\beta^{\phantom{|}}\big]
\mathrm{d}^3 r
\end{align*}
\end{footnotesize}\\[-2pt]
The first part is equal to $(-1)\times$direct term and the second part is zero. Then I use following relations:
\begin{small}
\begin{align*}
\sum_{ijk}\varepsilon_{ijk} (\partial_j\psi)^\dagger\sigma_i(\partial_k\psi)
&= \sum_{ijk}^{xyz}\varepsilon_{ijk} \big[ \partial_j(\psi^\dagger\sigma_i\partial_k\psi)
- (\psi^\dagger\sigma_i\partial_j\partial_k\psi)\big] =
\vec{\nabla}\cdot\big[\psi^\dagger(\vec{\nabla}\times\vec{\sigma})\psi\big] - 0\\
&= \sum_{ijk}\varepsilon_{ijk} \partial_k [(\sigma_i\partial_j\psi)^\dagger\psi] =
-\vec{\nabla}\cdot\big\{[(\vec{\nabla}\times\vec{\sigma})\psi]^\dagger\psi\big\}\\
\sum_{jk}\varepsilon_{ijk} (\partial_j\psi)^\dagger(\partial_k\psi)
&=
\sum_{jk}\varepsilon_{ijk} \big[ \partial_j(\psi^\dagger\partial_k\psi)
- (\psi^\dagger\partial_j\partial_k\psi)\big] =
\big[\vec{\nabla}\times(\psi^\dagger\vec{\nabla}\psi)\big]_i - 0 \\
&=
\sum_{jk}\varepsilon_{ijk} \partial_k[(\partial_j\psi)^\dagger\psi] =
-\big\{\vec{\nabla}\times[(\vec{\nabla}\psi)^\dagger\psi]\big\}_i
\end{align*}
\begin{align*}
\langle\alpha\beta|&(\vec{\sigma}_1+\vec{\sigma}_2)\cdot[\overleftarrow{\nabla}_1\times
\delta(\vec{r}_1-\vec{r}_2)(\overrightarrow{\nabla}_1-\overrightarrow{\nabla}_2)]|\alpha\beta\rangle = \\
&\qquad=\int \Big\{
\vec{\nabla}\cdot\big[\psi_\alpha^\dagger(\vec{\nabla}\times\vec{\sigma})\psi_\alpha^{\phantom{|}}\big]\psi_\beta^\dagger\psi_\beta^{\phantom{|}}
+\big[\vec{\nabla}\times(\psi^\dagger\vec{\nabla}\psi)\big]\cdot
(\psi_\beta^\dagger\vec{\sigma}\psi_\beta^{\phantom{|}}) \\[-8pt]
&\qquad\qquad\qquad{}-\sum_{ijk}\varepsilon_{ijk}\Big[
[(\partial_j\psi_\alpha^{\phantom{|}})^\dagger\sigma_i\psi_\alpha^{\phantom{|}}]
(\psi_\beta^\dagger\partial_k\psi_\beta^{\phantom{|}})
+[(\partial_j\psi_\alpha^{\phantom{|}})^\dagger\psi_\alpha^{\phantom{|}}]
(\psi_\beta^\dagger\sigma_i\partial_k\psi_\beta^{\phantom{|}})\Big]\Big\}\mathrm{d}^3 r \phantom{\bigg{|}}
\end{align*}
\end{small}\\[-2pt]
Remaining indexed terms (after addition of the $\overleftarrow{\nabla}_2$ term, i.e., symmetrization in $\alpha\leftrightarrow\beta$) can be obtained by subtraction of the following two lines:
\begin{align*}
\langle\alpha|\hat{\mathcal{J}}_{ji}|\alpha\rangle
\partial_k(\psi_\beta^\dagger\psi_\beta^{\phantom{|}}) &=
\frac{\mathrm{i}}{2}
\big[(\partial_j\psi_\alpha^{\phantom{|}})^\dagger\sigma_i\psi_\alpha^{\phantom{|}}
-\psi_\alpha^\dagger\sigma_i(\partial_j\psi_\alpha^{\phantom{|}})\big]
\big[(\partial_k\psi_\beta^{\phantom{|}})^\dagger\psi_\beta^{\phantom{|}}
+\psi_\beta^\dagger(\partial_k\psi_\beta^{\phantom{|}})\big] \\
\partial_j(\psi_\alpha^\dagger\sigma_i\psi_\alpha^{\phantom{|}})
\langle\beta|\hat{j}_k|\beta\rangle &=
\frac{\mathrm{i}}{2}
\big[(\partial_j\psi_\alpha^{\phantom{|}})^\dagger\sigma_i\psi_\alpha^{\phantom{|}}
+\psi_\alpha^\dagger\sigma_i(\partial_j\psi_\alpha^{\phantom{|}})\big]
\big[(\partial_k\psi_\beta^{\phantom{|}})^\dagger\psi_\beta^{\phantom{|}}
-\psi_\beta^\dagger(\partial_k\psi_\beta^{\phantom{|}})\big]
\end{align*}
The whole $t_4$ term:
\begin{small}
\begin{align}
\frac{\mathrm{i}}{4}t_4\sum_{\alpha\beta}\langle\alpha\beta|&\,(\vec{\sigma}_1+\vec{\sigma}_2)\cdot[(\overleftarrow{\nabla}_1-\overleftarrow{\nabla}_2)\times
\delta(\vec{r}_1-\vec{r}_2)(\overrightarrow{\nabla}_1-\overrightarrow{\nabla}_2)]|\alpha\beta\rangle = \nonumber\\
&= \frac{t_4}{2}\int \big[ {-}\rho\vec{\nabla}\!\cdot\!\vec{\mathcal{J}}-\vec{s}\cdot(\vec{\nabla}\!\times\!\vec{j}) + \vec{\mathcal{J}}\cdot\vec{\nabla}\rho - \vec{j}\cdot(\vec{\nabla}\!\times\!\vec{s})
\big]\mathrm{d}^3 r = \\
&= t_4\int \big[ {-}\rho\vec{\nabla}\!\cdot\!\vec{\mathcal{J}}-\vec{s}\cdot(\vec{\nabla}\!\times\!\vec{j}) \big]\mathrm{d}^3 r \nonumber\\
\frac{\mathrm{i}}{4}t_4\sum_{\alpha\beta\in q}\langle\alpha\beta|&\,(\vec{\sigma}_1+\vec{\sigma}_2)\cdot[(\overleftarrow{\nabla}_1-\overleftarrow{\nabla}_2)\times
\delta(\vec{r}_1-\vec{r}_2)(\overrightarrow{\nabla}_1-\overrightarrow{\nabla}_2)]|\beta\alpha\rangle = \phantom{\Bigg{|}} \nonumber\\
&= \frac{t_4}{2}\int \big[ \rho_q\vec{\nabla}\!\cdot\!\vec{\mathcal{J}}_q+\vec{s}_q\cdot(\vec{\nabla}\!\times\!\vec{j}_q) - \vec{\mathcal{J}}_q\cdot\vec{\nabla}\rho_q + \vec{j}_q\cdot(\vec{\nabla}\!\times\!\vec{s}_q)
\big]\mathrm{d}^3 r \nonumber\\
\label{t4-term}
&= t_4\int \big[ \rho_q\vec{\nabla}\!\cdot\!\vec{\mathcal{J}}_q+\vec{s}_q\cdot(\vec{\nabla}\!\times\!\vec{j}_q) \big]\mathrm{d}^3 r
\end{align}
\end{small}

\chapter{Derivation of the matrix element of spin-orbital current}\label{app_Jab}
The steps of the derivation are here presented in the form of references, (V.~\emph{number}), to the formulae from the book of Varshalovich \cite{Varshalovich1988}, which are given here after application of necessary substitutions and other transformations.

Spherical components of the single particle matrix element are
\begin{equation}
\label{Jab_me}
\big[\langle\alpha|\hat{\vec{\mathcal{J}}}(\vec{r})|\beta\rangle\big]_M =
-\frac{\mathrm{i}}{2}
\big\{ \psi_\alpha^\dagger[(\vec{\nabla}\times\vec{\sigma})_M
\psi_\beta^{\phantom{\dagger}}] -
(-1)^M[(\vec{\nabla}\times\vec{\sigma})_{-M}
\psi_\alpha^{\phantom{\dagger}}]^\dagger\psi_\beta^{\phantom{\dagger}}
\big\}
\end{equation}
\begin{equation}
{[\vec{\nabla}\times\vec{\sigma}]}_M = -\mathrm{i}\sqrt{2}\sum_{\mu\nu}
C_{1\mu1\nu}^{1M} \sigma_{\nu} \nabla_{\mu}\qquad
(\sigma_\nu = 2\hat{s}_\nu)
\tag{V.~1.2.28}
\end{equation}
\begin{equation}
\nabla_\mu \psi_\beta^{\phantom{+}} = \sum_{l_2=l_\beta\pm1} R_\beta^{(\pm)}
\frac{(-1)^{j_\beta+l_\beta-\frac{1}{2}}}{\sqrt{2l_2+1}}
\sum_K
\begin{Bmatrix} j_\beta & K & 1 \\ l_2 & l_\beta & \frac{1}{2}\end{Bmatrix}
C_{j_\beta,m_\beta,1,\mu}^{K,m_\beta+\mu}
\Omega_{K,m_\beta+\mu}^{l_2}
\tag{V.~7.1.24}
\end{equation}
\begin{equation}
\sigma_\nu \Omega_{K,m_\beta+\mu}^{l_2} = \sum_{K'}
(-1)^{l_2+K'-\frac{1}{2}}
\sqrt{6(2K+1)}\,
\begin{Bmatrix} K & K' & 1 \\ \frac{1}{2} & \frac{1}{2} & l_2 \end{Bmatrix}
C_{K,m_\beta+\mu,1,\nu}^{K',m_\beta+\mu+\nu}
\Omega_{K',m_\beta+\mu+\nu}^{l_2}
\tag{V.~7.1.28}
\end{equation}
\begin{equation}
\sum_{\mu\nu}
C_{1\mu1\nu}^{1M}
C_{K,m_\beta+\mu,1,\nu}^{K',m_\beta+M}
C_{j_\beta,m_\beta,1,\mu}^{K,m_\beta+\mu} =
\sqrt{3(2K+1)}\,
\begin{Bmatrix} 1 & 1 & 1 \\ K' & j_\beta & K \end{Bmatrix}
C_{1,M,j_\beta,m_\beta}^{K',m_\beta+M}
\tag{V.~8.7.12}
\end{equation}
\begin{equation}
\sum_K (2K+1)
\begin{Bmatrix} j_\beta & 1 & K \\ l_2 & \frac{1}{2} & l_\beta \end{Bmatrix}
\begin{Bmatrix} l_2 & \frac{1}{2} & K \\ 1 & K' & \frac{1}{2} \end{Bmatrix}
\begin{Bmatrix} 1 & K' & K \\ j_\beta & 1 & 1 \end{Bmatrix} =
(-1)^{j_\beta+K'}
\begin{Bmatrix} j_\beta & K' & 1 \\ l_\beta & l_2 & 1 \\ \frac{1}{2} & \frac{1}{2} & 1 \end{Bmatrix}
\tag{V.~9.8.5}
\end{equation}
Together:
\begin{equation}
{[\vec{\nabla}\times\vec{\sigma}]}_M \psi_\beta^{\phantom{+}} =
-6\mathrm{i} \sum_{l_2=l_\beta\pm1}
\frac{R_\beta^{(\pm)}}{\sqrt{2l_2+1}}
\sum_{K'}
\begin{Bmatrix} j_\beta & K' & 1 \\ l_\beta & l_2 & 1 \\ \frac{1}{2} & \frac{1}{2} & 1 \end{Bmatrix}
C_{1,M,j_\beta,m_\beta}^{K',m_\beta+M}
\Omega_{K',m_\beta+M}^{l_2}
\end{equation}
\begin{align}
\frac{\Omega_{j_\alpha m_\alpha}^{l_\alpha\dagger} \Omega_{K',m_\beta+M}^{l_2}}{\sqrt{(2j_\alpha+1)(2l_\alpha+1)(2l_2+1)}} =
& \sum_L
\frac{(-1)^{j_\alpha+m_\alpha+K'+L+\frac{1}{2}}\sqrt{2K'+1}}{\sqrt{4\pi(2L+1)}}
\begin{Bmatrix} l_\alpha & l_2 & L \\ K' & j_\alpha & \frac{1}{2} \end{Bmatrix}
\nonumber\\
&C_{l_\alpha 0 l_2 0}^{L 0}
C_{j_\alpha,-m_\alpha,K',m_\beta+M}^{L,m_\beta-m_\alpha+M}
Y_{L,m_\beta-m_\alpha+M}
\tag{V.~7.2.40}
\end{align}
\begin{align}
C_{j_\alpha,-m_\alpha,K',m_\beta+M}^{L,m_\beta-m_\alpha+M}
C_{1,M,j_\beta,m_\beta}^{K',m_\beta+M} &=
\sum_J (-1)^{j_\beta+K'} \sqrt{(2J+1)(2K'+1)}\,
\begin{Bmatrix} j_\beta & j_\alpha & J \\ L & 1 & K' \end{Bmatrix} \nonumber\\
&\quad\times C_{j_\beta,m_\beta,j_\alpha,-m_\alpha}^{J,m_\beta-m_\alpha}
C_{1,M,J,m_\beta-m_\alpha}^{L,m_\beta-m_\alpha+M} \tag{V.~8.7.35}\\
&= \sum_J (-1)^{K'+J+j_\alpha+M} \sqrt{(2L+1)(2K'+1)} \nonumber\\[-6pt]
&\quad\times\begin{Bmatrix} j_\beta & 1 & K' \\ L & j_\alpha & J \end{Bmatrix}
C_{j_\alpha,-m_\alpha,j_\beta,m_\beta}^{J,m_\beta-m_\alpha}
C_{L,m_\beta-m_\alpha+M,1,-M}^{J,m_\beta-m_\alpha} \nonumber
\end{align}
Together with (\ref{sph_vectors}) it gives
\begin{align}
\psi_\alpha^\dagger [(\vec{\nabla}\times\vec{\sigma}) \psi_\beta^{\phantom{\dagger}}]
= &
6\mathrm{i} \sum_{l_2=l_\beta\pm1}
\frac{R_\alpha^{(0)} R_\beta^{(\pm)}}{\sqrt{4\pi}}
\sum_{LJ} (-1)^{m_\alpha+L+J-\frac{1}{2}}\,
C_{l_\alpha 0 l_2 0}^{L 0}
C_{j_\alpha,-m_\alpha,j_\beta,m_\beta}^{J,m_\beta-m_\alpha}
\vec{Y}_{J,m_\beta-m_\alpha}^L \nonumber\\
\label{Jab_part1}
& \times \sum_{K'} (2K'+1)
\begin{Bmatrix} j_\beta & K' & 1 \\ l_\beta & l_2 & 1 \\ \frac{1}{2} & \frac{1}{2} & 1 \end{Bmatrix}
\begin{Bmatrix} \frac{1}{2} & l_2 & K' \\ L & j_\alpha & l_\alpha \end{Bmatrix}
\begin{Bmatrix} j_\beta & 1 & K' \\ L & j_\alpha & J \end{Bmatrix}
\end{align}
Second term of (\ref{Jab_me}) is evaluated analogously:
\begin{equation}
(-1)^M\big\{{[\vec{\nabla}\times\vec{\sigma}]}_{-M} \psi_\alpha^{\phantom{\dagger}}\big\}^\dagger =
6\mathrm{i}(-1)^M\!\! \sum_{l_1=l_\alpha\pm1}
\frac{R_\alpha^{(\pm)}}{\sqrt{2l_1+1}}
\sum_{K'}
\begin{Bmatrix} j_\alpha & K' & 1 \\ l_\alpha & l_1 & 1 \\ \frac{1}{2} & \frac{1}{2} & 1 \end{Bmatrix}
C_{1,-M,j_\alpha,m_\alpha}^{K',m_\alpha-M}
\Omega_{K',m_\alpha-M}^{l_1\dagger}
\end{equation}
\begin{align}
\frac{\Omega_{K',m_\alpha-M}^{l_1\dagger} \Omega_{j_\beta m_\beta}^{l_\beta}}
{\sqrt{(2l_1+1)(2j_\beta+1)(2l_\beta+1)}} = &
\sum_L
\frac{(-1)^{K'+m_\alpha-M+j_\beta+L+\frac{1}{2}}\sqrt{2K'+1}}{\sqrt{4\pi(2L+1)}}
\begin{Bmatrix} l_1 & l_\beta & L \\ j_\beta & K' & \frac{1}{2} \end{Bmatrix}
\nonumber\\
& \times C_{l_1 0 l_\beta 0}^{L 0}
C_{K',M-m_\alpha,j_\beta,m_\beta}^{L,m_\beta-m_\alpha+M}
Y_{L,m_\beta-m_\alpha+M}
\tag{V.~7.2.40}
\end{align}
\begin{align}
C_{K',M-m_\alpha,j_\beta,m_\beta}^{L,m_\beta-m_\alpha+M}
C_{j_\alpha,-m_\alpha,1,M}^{K',M-m_\alpha} &=
\sum_J
(-1)^{K'+j_\beta-L}
\sqrt{(2J+1)(2K'+1)}
\begin{Bmatrix} j_\alpha & j_\beta & J \\ L & 1 & K' \end{Bmatrix} \nonumber\\
&\quad\times C_{j_\alpha,-m_\alpha,j_\beta,m_\beta}^{J,m_\beta-m_\alpha}
C_{1,M,J,m_\beta-m_\alpha}^{L,m_\beta-m_\alpha+M} \tag{V.~8.7.35}\\
&= \sum_J (-1)^{K'+j_\beta+L+M+1}
\sqrt{(2L+1)(2K'+1)} \nonumber\\[-6pt]
&\quad\times\begin{Bmatrix} j_\alpha & 1 & K' \\ L & j_\beta & J \end{Bmatrix}
C_{j_\alpha,-m_\alpha,j_\beta,m_\beta}^{J,m_\beta-m_\alpha}
C_{L,m_\beta-m_\alpha+M,1,-M}^{J,m_\beta-m_\alpha}\nonumber
\end{align}
Together:
\begin{align}
[(\vec{\nabla}\times\vec{\sigma}) \psi_\alpha^{\phantom{\dagger}}]^\dagger \psi_\beta^{\phantom{|}} &=
6\mathrm{i} \sum_{l_1=l_\alpha\pm1}
\frac{R_\alpha^{(\pm)} R_\beta^{(0)}}{\sqrt{4\pi}}
\sum_{LJ}
(-1)^{m_\alpha-\frac{1}{2}}\,
C_{l_1 0 l_\beta 0}^{L 0}
C_{j_\alpha,-m_\alpha,j_\beta,m_\beta}^{J,m_\beta-m_\alpha}
\vec{Y}_{J,m_\beta-m_\alpha}^L \nonumber\\
\label{Jab_part2}
& \qquad\times \sum_{K'} (2K'+1)
\begin{Bmatrix} j_\alpha & K' & 1 \\ l_\alpha & l_1 & 1 \\ \frac{1}{2} & \frac{1}{2} & 1 \end{Bmatrix}
\begin{Bmatrix} \frac{1}{2} & l_1 & K' \\ L & j_\beta & l_\beta \end{Bmatrix}
\begin{Bmatrix} j_\alpha & 1 & K' \\ L & j_\beta & J \end{Bmatrix}
\end{align}

The sums over $K'$ in (\ref{Jab_part1}, \ref{Jab_part2}) can be evaluated after decomposition of $9j$ symbol into $6j$ symbols (I take $g=1/2$ in the cited formula).
\begin{align}
\begin{Bmatrix} j_\beta & K' & 1 \\ l_\beta & l_\beta\pm1 & 1 \\ \frac{1}{2} & \frac{1}{2} & 1 \end{Bmatrix}
\begin{Bmatrix} 1 & 1 & 1 \\ \frac{1}{2} & \frac{1}{2} & \frac{1}{2} \end{Bmatrix}
= &-\frac{1}{3}
\begin{Bmatrix} j_\beta & K' & 1 \\ \frac{1}{2} & \frac{1}{2} & l_\beta\pm1 \end{Bmatrix}
\begin{Bmatrix} l_\beta\pm1 & l_\beta & 1 \\ \frac{1}{2} & \frac{1}{2} & j_\beta \end{Bmatrix} \nonumber\\
&-\frac{(-1)^{K'+l_\beta-\frac{1}{2}}}{18}
\begin{Bmatrix} j_\beta & K' & 1 \\ l_\beta\pm1 & l_\beta & \frac{1}{2} \end{Bmatrix}
\tag{V.~10.9.9}
\end{align}
I then evaluate some $6j$ symbols using tables 9.1 and 9.10 from \cite{Varshalovich1988} (given $6j$ symbols are non-zero only for the given $j_\beta=l_\beta\pm\frac{1}{2}$).
\begin{small}
\begin{equation*}
\begin{Bmatrix} 1 & 1 & 1 \\ \frac{1}{2} & \frac{1}{2} & \frac{1}{2} \end{Bmatrix}
= -\frac{1}{3}, \quad
\begin{Bmatrix} l_\beta & \!l_\beta+1\!\! & \!1 \\ \frac{1}{2} & \frac{1}{2} & \!\!l_\beta+\frac{1}{2} \end{Bmatrix} = \frac{1}{\sqrt{6(l_\beta+1)}}, \quad
\begin{Bmatrix} l_\beta & l_\beta-1\!\! & \!1 \\ \frac{1}{2} & \frac{1}{2} & \!\!l_\beta-\frac{1}{2} \end{Bmatrix} = \frac{1}{\sqrt{6l_\beta}}
\end{equation*}
\end{small} \\[-6pt]
\begin{equation}
\sum_{K'} (2K'+1)
\begin{Bmatrix} j_\beta & 1 & K' \\ \frac{1}{2} & l_\beta^\pm & \frac{1}{2} \end{Bmatrix}
\begin{Bmatrix} \frac{1}{2} & l_\beta^\pm & K' \\ L & j_\alpha & l_\alpha \end{Bmatrix}
\begin{Bmatrix} L & j_\alpha & K' \\ j_\beta & 1 & J \end{Bmatrix} = -
\begin{Bmatrix} j_\beta & j_\alpha & J \\ l_\beta^\pm & l_\alpha & L \\ \frac{1}{2} & \frac{1}{2} & 1 \end{Bmatrix}
\tag{V.~9.8.5}
\end{equation}
\begin{align}
\sum_{K'} (-1)^{K'-\frac{1}{2}} (2K'+1)
\begin{Bmatrix} 1 & j_\beta & K' \\ \frac{1}{2} & l_\beta^\pm & l_\beta \end{Bmatrix}
\begin{Bmatrix} \frac{1}{2} & l_\beta^\pm & K' \\ L & j_\alpha & l_\alpha \end{Bmatrix}
\begin{Bmatrix} L & j_\alpha & K' \\ j_\beta & 1 & J \end{Bmatrix}& \nonumber\\
\displaystyle = (-1)^{j_\alpha+j_\beta+l_\alpha+L+J+1}
\begin{Bmatrix} l_\beta & l_\alpha & J \\ L & 1 & l_\beta^\pm \end{Bmatrix}
\begin{Bmatrix} l_\beta & l_\alpha & J \\ j_\alpha & j_\beta & \frac{1}{2} \end{Bmatrix}&
\tag{V.~9.8.6}
\end{align}
Final result is then
\begin{equation}
\langle\alpha|\vec{\mathcal{J}}_q(\vec{r})|\beta\rangle =
\sum_{LJ} \bigg[\sum_{ss'}^{0\pm,\pm0}
\mathcal{A}_{\alpha\beta LJ}^{\vec{J},ss'} R_\alpha^{(s)} R_\beta^{(s')} \bigg]
\frac{(-1)^{j_\alpha+j_\beta+L+m_\alpha-\frac{1}{2}}}{2\sqrt{4\pi}}\,
C_{j_\alpha,-m_\alpha,j_\beta,m_\beta}^{J,m_\beta-m_\alpha}
\vec{Y}_{J,m_\beta-m_\alpha}^L
\end{equation}
where coefficients $\mathcal{A}_{\alpha\beta LJ}^{\vec{J},ss'}$ are given at (\ref{spin-orb_me}).

\chapter{Separable RPA in the spherical symmetry}\label{app_SRPA}
Calculation demands of the full RPA can grow rapidly as one increases the $2qp$ basis. Although in the spherical case the computional load is not so dramatic, I present here also the separable RPA \cite{Nesterenko2002,Nesterenko2006}, as a more efficient calculation scheme, useful mainly for a quick calculation of the strength functions. SRPA is outlined here according to my master thesis \cite{RepkoMgr}, where it was derived for general wavefunctions, whereas here I assume spherical symmetry, and the final formulae are given in a fully rotationally invariant form.

Residual interaction is first approximated by a sum of separable terms
\begin{equation}
\label{SRPA_hamilt}
\hat{H} = \hat{H}_0 + \hat{V}_\mathrm{res} =
\sum_\gamma \varepsilon_\gamma^{\phantom{|}}
\hat{\alpha}_\gamma^+\hat{\alpha}_\gamma^{\phantom{|}}
- \frac{1}{2}\sum_{qk,q'k'}^{\mu_{k'}=-\mu_k^{\phantom{|}}}
(-1)^{\mu_k}{:}\big[\kappa_{qk,q'k'}^{\phantom{|}}
\hat{X}_{qk}^{\phantom{+}}\hat{X}_{q'k'}
+ \eta_{qk,q'k'}^{\phantom{+}}
\hat{Y}_{qk}^{\phantom{+}}\hat{Y}_{q'k'}\big]{:}
\end{equation}
where $k$ labels one of $K$ separable one-body operators $\hat{X}$ (time-even) and $\hat{Y}$ (time-odd). These operators will be obtained by means of the linear response theory from given input operators $\hat{Q}_{qk}$ (time-even) and $\hat{P}_{qk}$ (time-odd) acting on nucleons $q$. I choose a perturbed ground state:
\begin{equation}
|q_{qk}^{\phantom{+}},p_{qk}^{\phantom{+}}\rangle = \prod_q\prod_{k=1}^K \mathrm{e}^{-\mathrm{i}q_{qk}\hat{P}_{qk}}\mathrm{e}^{-\mathrm{i}p_{qk}\hat{Q}_{qk}} |\textrm{HF+BCS}\rangle
\end{equation}
This form was chosen on the basis of Thouless thorem \cite{Ring1980}, so that it remains a Slater state, and therefore it makes sense to use mean-field density functional. Effective one-body part of (\ref{SRPA_hamilt}) in the perturbed ground state becomes
\begin{equation}
\hat{h} = \hat{H}_0 + \mathrm{i}\sum_{\tilde{q}\tilde{k}}
\sum_{qk,q'k'}^{\mu_{k'}=-\mu_k^{\phantom{|}}}(-1)^{\mu_k}\Big(
q_{\tilde{q}\tilde{k}} \kappa_{qk,q'k'}^{\phantom{+}} \hat{X}_{qk}^{\phantom{+}}\langle[\hat{X}_{q'k'}^{\phantom{+}},\hat{P}_{\tilde{q}\tilde{k}}]\rangle +
p_{\tilde{q}\tilde{k}} \eta_{qk,q'k'}^{\phantom{+}} \hat{Y}_{qk}^{\phantom{+}}\langle[\hat{Y}_{q'k'}^{\phantom{+}},\hat{Q}_{\tilde{q}\tilde{k}}]\rangle \Big)
\end{equation}
I define a basis of operators $\hat{X}_{qk}$, $\hat{Y}_{qk}$ (to get a simple result) by setting
\begin{subequations}\label{str_mtrx}
\begin{align}
\kappa_{q'k',qk}^{-1} &= \mathrm{i}(-1)^{\mu_k}
\langle[\hat{X}_{q'k'},\hat{P}_{qk}]\rangle =
\sum_{\alpha\beta\in q}^{\alpha\geq\beta} \frac{2\mathrm{i}}{2\lambda_k+1}
X^*_{q'k';\alpha\beta} P_{qk;\alpha\beta} \\
\eta_{q'k',qk}^{-1} &= \mathrm{i}(-1)^{\mu_k}
\langle[\hat{Y}_{q'k'},\hat{Q}_{qk}]\rangle =
\sum_{\alpha\beta\in q}^{\alpha\geq\beta} \frac{2\mathrm{i}}{2\lambda_k+1}
Y^*_{q'k';\alpha\beta} Q_{qk;\alpha\beta}
\end{align}
\end{subequations}
where I used (\ref{comm}) and $\gamma_T^A(-1)^{l_\alpha+l_\beta+\lambda}A_{\alpha\beta} = A_{\alpha\beta}^*$ (\ref{rme_hermit}). The effective one-body Hamiltonian is then, together with conditions on hermiticity
\begin{equation}
\label{SRPA_effH}
\hat{h} = \hat{H}_0 + \sum_{qk}(q_{qk}\hat{X}_{qk} + p_{qk}\hat{Y}_{qk}),\quad
q_{qk} = (-1)^{\mu_k} q_{q\bar{k}},\quad p_{qk} = (-1)^{\mu_k} p_{q\bar{k}}
\end{equation}
where $\bar{k}$ labels an operator with opposite projection, $\mu_{q\bar{k}} = -\mu_{qk}$. By equating (\ref{SRPA_effH}) with Skyrme effective Hamiltonian in the perturbed ground state, I obtain
\begin{subequations}
\begin{align}
\hat{X}_{qk} &= \mathrm{i}\sum_{dd'}^\textrm{even}\int\mathrm{d}^3 r
\frac{\delta^2\mathcal{H}}{\delta J_d \delta J_{d'}}
\langle[\hat{P}_{qk},\hat{J}_d(\vec{r})]\rangle \hat{J}_{d'}(\vec{r}) \\
\hat{Y}_{qk} &= \mathrm{i}\sum_{dd'}^\textrm{odd}\int\mathrm{d}^3 r
\frac{\delta^2\mathcal{H}}{\delta J_d \delta J_{d'}}
\langle[\hat{Q}_{qk},\hat{J}_d(\vec{r})]\rangle \hat{J}_{d'}(\vec{r})
\end{align}
\end{subequations}
In terms of reduced matrix elements:
\begin{subequations}\label{XY_op}
\begin{align}
X_{qk;\gamma\delta} & =
\frac{-2\mathrm{i}}{2\lambda+1}\sum_{dd'}^\textrm{even}\int\mathrm{d}^3 r
\bigg(\frac{\delta^2\mathcal{H}}{\delta J_d \delta J_{d'}}
\sum_{\alpha\beta\in q}^{\alpha\geq\beta} P_{qk;\alpha\beta} J_{d;\alpha\beta}^*(r)
\bigg) J_{d';\gamma\delta}(r) \\
Y_{qk;\gamma\delta} & =
\frac{-2\mathrm{i}}{2\lambda+1}\sum_{dd'}^\textrm{odd}\int\mathrm{d}^3 r
\bigg(\frac{\delta^2\mathcal{H}}{\delta J_d \delta J_{d'}}
\sum_{\alpha\beta\in q}^{\alpha\geq\beta} Q_{qk;\alpha\beta} J_{d;\alpha\beta}^*(r)
\bigg) J_{d';\gamma\delta}(r)
\end{align}
\end{subequations}
where large parentheses indicate \emph{responses} of the operators $\hat{Q}_{qk}$ and $\hat{P}_{qk}$ and they need to be calculated only once (this is one of the numerical advantages of SRPA; the second one is the reduction of matrix dimension, see (\ref{SRPA_eq3})). It should be emphasized that the index $q$ in $\hat{X}_{qk}$ and $\hat{Y}_{qk}$ denotes the \emph{origin} of of these operators (i.e.~it labels the corresponding generating operators $\hat{Q}_{qk},\hat{P}_{qk}$), and not the type of the nucleons on which they act, as it was used in the operators $\hat{J}_{d;q},\hat{Q}_{qk},\hat{P}_{qk}$; the operators $\hat{X}_{qk},\hat{Y}_{qk}$ act on both protons and neutrons.

To better approximate the residual interaction, input operators $\hat{Q}_{qk},\hat{P}_{qk}$ come in pairs, where only one of them is given a priori, and the second one is defined by relations
\begin{equation}
\hat{P}_{qk} = \mathrm{i}[\hat{H},\hat{Q}_{qk}] \qquad \textrm{or} \qquad
\hat{Q}_{qk} = \mathrm{i}[\hat{H},\hat{P}_{qk}],
\end{equation}
which in terms of reduced matrix elements turn into
\begin{equation}
\label{second-QP}
P_{qk;\alpha\beta} = \mathrm{i}\varepsilon_{\alpha\beta}
Q_{qk;\alpha\beta} - Y_{qk;\alpha\beta} \qquad\textrm{or}\qquad
Q_{qk;\alpha\beta} = \mathrm{i}\varepsilon_{\alpha\beta}
P_{qk;\alpha\beta} - X_{qk;\alpha\beta}.
\end{equation}
The calculation of the matrix elements then proceeds as
\[  Q\rightarrow Y\rightarrow P\rightarrow X \qquad\textrm{or}\qquad
P\rightarrow X\rightarrow Q\rightarrow Y \]

Evaluation of the RPA equation (\ref{RPA_eq}) using separable Hamiltonian (\ref{SRPA_hamilt}) leads to
\begin{small}
\begin{subequations}\label{SRPA_eq1}
\begin{align}
(\varepsilon_{\alpha\beta}-E_\nu)c_{\alpha\beta}^{(\nu-)} & =
\!\!\sum_{qk,q'k'}\!\! \frac{(-1)^{l_\beta+\mu_k}}{\sqrt{2\lambda+1}}
\big(\kappa_{qk,q'k'}^{\phantom{+}}\langle[\hat{C}_\nu^+,\hat{X}_{q'k'}^{\phantom{+}}]\rangle X_{qk;\alpha\beta}
+ \eta_{qk,q'k'}^{\phantom{+}}\langle[\hat{C}_\nu^+,\hat{Y}_{q'k'}^{\phantom{+}}]\rangle Y_{qk;\alpha\beta}\big) \\
(\varepsilon_{\alpha\beta}+E_\nu)c_{\alpha\beta}^{(\nu+)} & =
\!\!\sum_{qk,q'k'}\!\!\frac{(-1)^{l_\beta+\mu_k}}{\sqrt{2\lambda+1}}
\big(\kappa_{qk,q'k'}^{\phantom{+}}\langle[\hat{C}_\nu^+,\hat{X}_{q'k'}^{\phantom{+}}]\rangle X_{qk;\alpha\beta}
- \eta_{qk,q'k'}^{\phantom{+}}\langle[\hat{C}_\nu^+,\hat{Y}_{q'k'}^{\phantom{+}}]\rangle Y_{qk;\alpha\beta} \big)
\end{align}
\end{subequations}
\end{small} \\[-6pt]
To reduce the number of equations, I introduce coefficients $\bar{q}_{qk}^\nu,\bar{p}_{qk}^\nu$, whose notation was inspired by the correspondence $[\hat{H},\hat{C}_\nu^+] \leftrightarrow q_k[\hat{V},\hat{P}_k]$
\begin{subequations}\label{comm-CXY}
\begin{align}
\sum_{qk}\kappa_{q'k',qk}^{-1}\bar{q}_{qk}^\nu & =
(-1)^{\mu_{k'}}\langle[\hat{C}_\nu^+,\hat{X}_{q'k'}^{\phantom{+}}]\rangle = 
\frac{(-1)^{l_\alpha+\lambda}}{\sqrt{2\lambda_\nu+1}}
\sum_{\alpha>\beta}
(c_{\alpha\beta}^{(\nu-)}+c_{\alpha\beta}^{(\nu+)})X_{q'k';\alpha\beta}^{\phantom{+}} \\
\sum_{qk}\eta_{q'k',qk}^{-1}\bar{p}_{qk}^\nu & =
(-1)^{\mu_{k'}}\langle[\hat{C}_\nu^+,\hat{Y}_{q'k'}^{\phantom{+}}]\rangle =
\frac{(-1)^{l_\alpha+\lambda+1}}{\sqrt{2\lambda+1}}
\sum_{\alpha>\beta}
(c_{\alpha\beta}^{(\nu-)}-c_{\alpha\beta}^{(\nu+)})Y_{q'k';\alpha\beta}^{\phantom{+}}
\end{align}
\end{subequations}
Equations (\ref{SRPA_eq1}) then become
\begin{subequations}\label{SRPA_eq2}
\begin{align}
(\varepsilon_{\alpha\beta}-E_\nu)c_{\alpha\beta}^{(\nu-)} & =
\sum_{qk,q'k'} \frac{(-1)^{l_\beta}}{\sqrt{2\lambda+1}}
\big(X_{qk;\alpha\beta}\bar{q}_{qk}^\nu
+ Y_{qk;\alpha\beta}\bar{p}_{qk}^\nu\big) \\
(\varepsilon_{\alpha\beta}+E_\nu)c_{\alpha\beta}^{(\nu+)} & =
\sum_{qk,q'k'}\frac{(-1)^{l_\beta}}{\sqrt{2\lambda+1}}
\big(X_{qk;\alpha\beta}\bar{q}_{qk}^\nu
- Y_{qk;\alpha\beta}\bar{p}_{qk}^\nu\big)
\end{align}
\end{subequations}
After elimination of $c_{\alpha\beta}^{(\nu\pm)}$ from (\ref{comm-CXY}) and (\ref{SRPA_eq2}), I am left with a matrix equation
\begin{equation}
\label{SRPA_eq3}
D\vec{R} =
\begin{pmatrix} F^{(XX)}-\kappa^{-1} & F^{(XY)} \\ F^{(YX)} & F^{(YY)}-\eta^{-1} \end{pmatrix} \binom{\bar{q}^\nu}{\bar{p}^\nu} = \binom{0}{0}
\end{equation}
where I defined matrix $D$, vector $\vec{R}$, and the matrices $F$ as
\begin{equation}
\label{F_me}
\begin{array}{ll}
\displaystyle
F_{q'k',qk}^{(XX)} = \frac{1}{2\lambda+1}
\sum_{\alpha\geq\beta}
\frac{2\varepsilon_{\alpha\beta}X^*_{q'k';\alpha\beta}X_{qk;\alpha\beta}}{\varepsilon_{\alpha\beta}^2-E_\nu^2}, &
\displaystyle
F_{q'k',qk}^{(XY)} = \frac{1}{2\lambda+1}
\sum_{\alpha\geq\beta}
\frac{2E_\nu X^*_{q'k';\alpha\beta}Y_{qk;\alpha\beta}}{\varepsilon_{\alpha\beta}^2-E_\nu^2}, \\
\displaystyle
F_{q'k',qk}^{(YX)} = \frac{1}{2\lambda+1}
\sum_{\alpha\geq\beta}
\frac{2E_\nu Y^*_{q'k';\alpha\beta}X_{qk;\alpha\beta}}{\varepsilon_{\alpha\beta}^2-E_\nu^2}, \phantom{\Bigg|} &
\displaystyle
F_{q'k',qk}^{(YY)} = \frac{1}{2\lambda+1}
\sum_{\alpha\geq\beta}
\frac{2\varepsilon_{\alpha\beta}Y^*_{q'k';\alpha\beta}Y_{qk;\alpha\beta}}{\varepsilon_{\alpha\beta}^2-E_\nu^2}
\end{array}
\end{equation}
Reduced matrix elements $X_{qk;\alpha\beta}$ and $Y_{qk;\alpha\beta}$ are either real or imaginary depending on the $\hat{M}_{\lambda\mu}^{\mathrm{E/M}}$ (see also (\ref{EM_sel_rules})) and the $\bar{q}^\nu$ and $\bar{p}^\nu$ are chosen in a way that $c_{\alpha\beta}^{(\nu\pm)}$ remains real
\begin{equation}
\label{XY-cc}
X_{qk;\alpha\beta}^* = \gamma_T^M X_{qk;\alpha\beta},\quad
Y_{qk;\alpha\beta}^* = -\gamma_T^M Y_{qk;\alpha\beta},\quad
\bar{q}_{qk}^* = \gamma_T^M\bar{q}_{qk},\quad
\bar{p}_{qk}^* = -\gamma_T^M\bar{p}_{qk}
\end{equation}

Matrices $D$ and $F$, and the vector $\vec{R}$ are not constant, but depend on the chosen RPA state $\nu$ (or its energy, $E_\nu$, respectively), nevertheless, I omit the index $\nu$, not to increase clutter. SRPA equations are therefore not an usual eigenvalue problem, since the number of their solutions can be much higher than the matrix dimension (number of solutions is equal to the number of $\alpha\beta$ pairs). Moreover, during the calculation of the strength function, the matrix $D$ becomes a continuous function of energy, $D(E)$, so the persistence of the index $\nu$ would cause confusion.

Normalization condition $\sum_{\alpha\geq\beta} (|c_{\alpha\beta}^{(\nu-)}|^2-|c_{\alpha\beta}^{(\nu+)}|^2) = 1$ (\ref{RPA_norm}) becomes \cite{Kvasil1998}
\begin{equation}
\label{R-norm}
\vec{R}^\dagger \frac{\partial D}{\partial E_\nu} \vec{R} = 1
\end{equation}

Transition probability is obtained by combining (\ref{trans_me}) with (\ref{SRPA_eq2})
\begin{subequations}
\begin{align}
\langle[\hat{C}_\nu,\hat{M}_{\lambda\mu}^\mathrm{E}]\rangle &=
\frac{-1}{2\lambda+1}\sum_{\alpha\geq\beta}
\frac{M_{\lambda;\alpha\beta}^\mathrm{E}}{\varepsilon_{\alpha\beta}^2-E_\nu^2}
\Big[ 2\varepsilon_{\alpha\beta}
X_{qk;\alpha\beta} \,\bar{q}_{qk}^\nu +
2\hbar\omega_\nu Y_{qk;\alpha\beta}\,\bar{p}_{qk}^\nu
\Big]^* = \vec{R}^\dagger \vec{A} \\
\langle[\hat{C}_\nu,\hat{M}_{\lambda\mu}^\mathrm{M}]\rangle &=
\frac{-1}{2\lambda+1}\sum_{\alpha\geq\beta}
\frac{M_{\lambda;\alpha\beta}^\mathrm{M}}{\varepsilon_{\alpha\beta}^2-E_\nu^2}
\Big[ 2\hbar\omega_\nu X_{qk;\alpha\beta}\,\bar{q}_{qk}^\nu
+ 2\varepsilon_{\alpha\beta}
Y_{qk;\alpha\beta} \,\bar{p}_{qk}^\nu
\Big]^* = \vec{R}^\dagger \vec{A}
\end{align}
\end{subequations}
where I defined vector $\vec{A}$ (dependent on the energy)
\begin{subequations}
\begin{align}
A_{qk}^{(X)} &= \frac{-2}{2\lambda+1}\sum_{\alpha\geq\beta} \frac{M_{\lambda;\alpha\beta}X_{qk;\alpha\beta}^*}{\varepsilon_{\alpha\beta}^2-E_\nu^2}\times
\Big\{\!\!\begin{array}{l} \varepsilon_{\alpha\beta} \\
E_\nu \end{array} \\
A_{qk}^{(Y)} &= \frac{-2}{2\lambda+1}\sum_{\alpha\geq\beta} \frac{M_{\lambda;\alpha\beta}Y_{qk;\alpha\beta}^*}{\varepsilon_{\alpha\beta}^2-E_\nu^2}\times
\Big\{\!\!\begin{array}{ll} E_\nu & (\mathrm{E}\lambda) \\
\varepsilon_{\alpha\beta}\! & (\mathrm{M}\lambda) \end{array}
\end{align}
\end{subequations}
Reduced transition probability is then
\begin{equation}
\label{BE_0}
B(\textrm{E/M}\lambda\mu;0\rightarrow\nu) = |\langle\nu|\hat{M}_{\lambda\mu}|\textrm{RPA}\rangle|^2 = \vec{A}^\dagger \vec{R}\vec{R}^\dagger\vec{A}
\end{equation}
I will evaluate matrix $\vec{R}\vec{R}^\dagger$ using (\ref{R-norm}). I will use singularity of the matrix $D$ ($\det D(E_\nu) = 0$) and expand it by $j$-row into algebraic supplements $d_{jk}$ ($D^{(jk)}$ is a submatrix of $D$ with omitted $j$-th row and $k$-th column)
\begin{equation}
0 = \det D = \sum_k (-1)^{j+k}D_{jk}\det D^{(jk)} = \sum_k D_{jk} d_{jk} =
\sum_k (D_{jk} + D_{j'k})d_{jk} = \sum_k D_{j'k}d_{jk}
\end{equation}
where the penultimate equality follows from invariance of $\det D$ against addition of $j'\neq j$-th row to the $j$-th row. The previous equation says that $d_{jk}$ is a solution of the equation $D\vec{R} = 0$ where vector $\vec{R}$ was created from $d_{jk}$ using $k$ as a vector index and any fixed $j$. So vector components $R_k$ are proportional to $d_{jk}$:
\begin{equation}
\frac{R_k}{R_{k'}} = \frac{d_{jk}}{d_{jk'}}
\end{equation}
The matrix $D$ is hermitian (\ref{F_me}, \ref{XY-cc}) as well as its algebraic supplement
\begin{equation}
d_{jk}^* = (-1)^{j+k}(\det D^{(jk)})^* = (-1)^{j+k}\det D^{(jk)\dagger} = (-1)^{j+k}\det D^{(kj)} = d_{kj}
\end{equation}
and the derivative of $\det D$ can be calculated by a chain rule applied to its matrix elements
\begin{equation}
\frac{\partial\det D}{\partial E_\nu} = \sum_{ij}\frac{\partial\det D}{\partial D_{ij}} \frac{\partial D_{ij}}{\partial E_\nu} = \sum_{ij} d_{ij}\frac{\partial D_{ij}}{\partial E_\nu}
\end{equation}
Normalization condition (\ref{R-norm}) can be now written as \cite{Kvasil1998}
\begin{align}
1 &= \sum_{kk'} R_k^* \frac{\partial D_{kk'}}{\partial E_\nu} R_{k'} =
R_i^* R_j^{\phantom{*}} \sum_{kk'} \frac{R_k^*}{R_i^*} \frac{\partial D_{kk'}}{\partial E_\nu}\frac{R_{k'}}{R_j} \\
&= R_i^* R_j^{\phantom{*}} \sum_{kk'} \frac{d_{jk}^*}{d_{ji}^*} \frac{\partial D_{kk'}}{\partial E_\nu}\frac{d_{kk'}}{d_{kj}} =
\frac{R_i^* R_j^{\phantom{*}}}{d_{ij}} \frac{\partial\det D}{\partial E_\nu}
\end{align}
Reduced transition probability (\ref{BE_0}) is then
\begin{equation}
B(\lambda\mu;0\rightarrow\nu) = |\langle\nu|\hat{M}_{\lambda\mu}|\textrm{RPA}\rangle|^2 = \sum_{ij}A_i R_i^* R_j A_j^* =
\sum_{ij}\frac{A_i d_{ij} A_j^*}{\frac{\partial\det D}{\partial E_\nu}} = -\frac{\det B}{\ \frac{\partial\det D}{\partial E_\nu}\ }
\end{equation}
where the expanded matrix $B$ was defined by
\begin{equation}
\sum_{ij} A_i d_{ij} A_j^* = -\det \begin{pmatrix} D_{ij} & A_i \\ A_j^* & 0 \end{pmatrix} = -\det B
\end{equation}
The star at $A_j$ means complex conjugation of matrix elements of $\hat{X}_{qk}$ and $\hat{Y}_{qk}$ only (matrix elements of $M_\lambda$ are real), but not the complex conjugation of $E_\nu$ that becomes complex during the evaluation of the strength function.

Strength function of $n$-th order
\begin{align}
\label{sf_def}
S_n(\lambda\mu;E) &= \sum_\nu E_\nu^n B(\lambda\mu;0\rightarrow\nu)\delta_\Delta(E-E_\nu), \\[-8pt]
&\qquad\qquad\qquad\qquad\textrm{where}\ 
\delta_\Delta(E-E_\nu) = \frac{\Delta/2\pi}{(E-E_\nu)^2+(\Delta/2)^2} \nonumber
\end{align}
can be evaluated direcly from determinants of $B$ and $D$ employing their complex-analytic properties with respect to the parameter $E_\nu$. Let's define function $f(z)$ that vanishes at infinity (for $n\leq 2$)
\[ f(E_\nu) = -E_\nu^n\frac{\det B(E_\nu)}{\det D(E_\nu)},\quad
\mathop{\mathrm{Res}}_{\ z=E_\nu} f(z) = -E_\nu^n\frac{\det B(E_\nu)}{\ \frac{\partial\det D}{\partial E_\nu}\ } = E_\nu^n B(\lambda\mu;0\rightarrow\nu) \]
Lorentz smoothing can be obtained directly by shifting the energy by an imaginary constant
\[ f\Big(x+\mathrm{i}\frac{\Delta}{2}\Big) =
\sum_j \frac{1}{x-x_j+\mathrm{i}\Delta/2}\!\mathop{\mathrm{Res}}_{\ z=x_j}\! f(z) =
\sum_j \frac{x-x_j-\mathrm{i}\Delta/2}{(x-x_j)^2+(\Delta/2)^2}\!\mathop{\mathrm{Res}}_{\ z=x_j}\! f(z) \]
\[ -\frac{1}{\pi}\Im\Big[f\Big(x+\mathrm{i}\frac{\Delta}{2}\Big)\Big] =
\sum_j \delta_\Delta(x-x_j)\!\mathop{\mathrm{Res}}_{\ z=x_j}\! f(z) \]
Besides the poles in $E_\nu$, the function $f(z)$ contains poles also in $\pm\varepsilon_{\alpha\beta},-E_\nu$. Negative poles will be neglected, due to their small contribution for positive $E$. The contribution of positive poles ($+\varepsilon_{\alpha\beta}$) is evaluated and removed using
\begin{align*}
\mathop{\mathrm{lim}}_{\ z\rightarrow\varepsilon_{\alpha\beta}}\!(z-\varepsilon_{\alpha\beta})^2 A_{qk}^{(X)}(z) A_{q'k'}^{(Y)*}(z)
&= \frac{|M_{\lambda;\alpha\beta}|^2}{(2\lambda+1)^2} \frac{4\varepsilon_{\alpha\beta}\varepsilon_{\alpha\beta}X_{qk;\alpha\beta}^* Y_{q'k';\alpha\beta}}{(\varepsilon_{\alpha\beta}+\varepsilon_{\alpha\beta})^2} \\
&= \frac{|M_{\lambda;\alpha\beta}|^2}{2\lambda+1}
\mathop{\mathrm{lim}}_{\ z\rightarrow\varepsilon_{\alpha\beta}}\!(\varepsilon_{\alpha\beta}-z)F_{qk,q'k'}^{(XY)}(z)
\end{align*}
\begin{equation}
-\!\!\mathop{\mathrm{Res}}_{\ z=\varepsilon_{\alpha\beta}}\! f(z) =
\!\!\mathop{\mathrm{lim}}_{\ z\rightarrow\varepsilon_{\alpha\beta}}\!\!(z-\varepsilon_{\alpha\beta})\frac{\det B(z)}{\det D(z)} =
\frac{|M_{\lambda;\alpha\beta}|^2}{2\lambda+1}
\!\mathop{\mathrm{lim}}_{\ z\rightarrow\varepsilon_{\alpha\beta}}\!\sum_{ij}\frac{D_{ij}(z)d_{ij}(z)}{\det D(z)}
= \frac{|M_{\lambda;\alpha\beta}|^2}{2\lambda+1}
\end{equation}
The final strength function (for $n\in\{0,1,2\}$) is then
\begin{equation}
\label{SRPA_sf}
S_n(\lambda\mu;E) = \frac{1}{\pi}\Im\bigg[z^n\frac{\det B(z)}{\det D(z)}\bigg]_{z=E+\mathrm{i}\frac{\Delta}{2}} + \sum_{\alpha\geq\beta} \varepsilon_{\alpha\beta}^n
\frac{|M_{\lambda;\alpha\beta}|^2}{2\lambda+1}
\delta_\Delta(E-\varepsilon_{\alpha\beta})
\end{equation}



\bibliography{bibl2b}

\begin{thebibliography}{10}
\addcontentsline{toc}{chapter}{Bibliography}
\providecommand{\doi}[1]{\href{http://dx.doi.org/#1}{\path{#1}}}

\bibitem{Ring1980}
P.~Ring and P.~Schuck:
\newblock {\em {The Nuclear Many-Body Problem}}.
\newblock Springer-Verlag, Berlin Heidelberg, 1980.

\bibitem{Ring2015}
P.~Ring:
\newblock {On the way to a microscopic derivation of covariant density
  functional theory}.
\newblock {\em {Journal of Physics: Conference Series}} {\bf {580}} ({2015})
  {012005}.
\newblock doi:\doi{10.1088/1742-6596/580/1/012005}

\bibitem{Muther1988}
H.~Müther, R.~Machleidt, and R.~Brockmann:
\newblock {Dirac-Brueckner-Hartree-Fock approach in finite nuclei}.
\newblock {\em {Physics Letters B}} {\bf {202}} ({1988}) {483--488}.
\newblock doi:\doi{10.1016/0370-2693(88)91848-5}

\bibitem{Caurier2005}
E.~Caurier, G.~Martínez-Pinedo, F.~Nowacki, A.~Poves, and A.~P. Zuker:
\newblock {The shell model as a unified view of nuclear structure}.
\newblock {\em {Reviews of Modern Physics}} {\bf {77}} ({2005}) {427--488}.
\newblock doi:\doi{10.1103/RevModPhys.77.427}

\bibitem{Barett2013}
B.~R. Barrett, P.~Navrátil, and J.~P. Vary:
\newblock {Ab initio no core shell model}.
\newblock {\em {Progress in Particle and Nuclear Physics}} {\bf {69}} ({2013})
  {131--181}.
\newblock doi:\doi{10.1016/j.ppnp.2012.10.003}

\bibitem{Dytrych2013}
T.~Dytrych, K.~D. Launey, J.~P. Draayer, P.~Maris, J.~P. Vary, E.~Saule,
  U.~Catalyurek, M.~Sosonkina, D.~Langr, and M.~A. Caprio:
\newblock {Collective Modes in Light Nuclei from First Principles}.
\newblock {\em {Physical Review Letters}} {\bf {111}} ({2013}) {252501}.
\newblock doi:\doi{10.1103/PhysRevLett.111.252501}

\bibitem{Bender2003}
M.~Bender, P.-H. Heenen, and P.-G. Reinhard:
\newblock {Self-consistent mean-field models for nuclear structure}.
\newblock {\em {Reviews of Modern Physics}} {\bf {75}} ({2003}) {121--180}.
\newblock doi:\doi{10.1103/RevModPhys.75.121}

\bibitem{Skyrme1959}
T.~H.~R. Skyrme:
\newblock {The effective nuclear potential}.
\newblock {\em {Nuclear Physics}} {\bf {9}} ({1959}) {615--634}.
\newblock doi:\doi{10.1016/0029-5582(58)90345-6}

\bibitem{Vautherin1972}
D.~Vautherin and D.~M. Brink:
\newblock {Hartree-Fock Calculations with Skyrme's Interaction. I. Spherical
  Nuclei}.
\newblock {\em {Physical Review C}} {\bf {5}} ({1972}) {626--647}.
\newblock doi:\doi{10.1103/PhysRevC.5.626}

\bibitem{Reinhard2011}
J.~Erler, P.~Klüpfel, and P.-G. Reinhard:
\newblock {Self-consistent nuclear mean-field models: example
  Skyrme-Hartree-Fock}.
\newblock {\em {Journal of Physics G: Nuclear and Particle Physics}} {\bf {38}}
  ({2011}) {033101}.
\newblock doi:\doi{10.1088/0954-3899/38/3/033101}

\bibitem{Gogny1980}
J.~Dechargé and D.~Gogny:
\newblock {Hartree-Fock-Bogolyubov calculations with the $D1$ effective
  interaction on spherical nuclei}.
\newblock {\em {Physical Review C}} {\bf {21}} ({1980}) {1568--1593}.
\newblock doi:\doi{10.1103/PhysRevC.21.1568}

\bibitem{Gogny2009}
W.~Younes and D.~Gogny:
\newblock {Microscopic calculation of $^{240}$Pu scission with a finite-range
  effective force}.
\newblock {\em {Physical Review C}} {\bf {80}} ({2009}) {054313}.
\newblock doi:\doi{10.1103/PhysRevC.80.054313}

\bibitem{Walecka1986}
B.~D. Serot and J.~D. Walecka:
\newblock {The Relativistic Nuclear Many-Body Problem}.
\newblock {\em {Advances in Nuclear Physics}} {\bf {16}} ({1986}) {1--320}.

\bibitem{Vretenar2005}
D.~Vretenar, A.~V. Afanasjev, G.~A. Lalazissis, and P.~Ring:
\newblock {Relativistic Hartree-Bogoliubov theory: static and dynamic aspects
  of exotic nuclear structure}.
\newblock {\em {Physics Reports (Review Section of Physics Letters)}} {\bf
  {409}} ({2005}) {101--259}.
\newblock doi:\doi{10.1016/j.physrep.2004.10.001}

\bibitem{Niksic2014}
T.~Nikšić, N.~Paar, D.~Vretenar, and P.~Ring:
\newblock {DIRHB-A relativistic self-consistent mean-field framework for atomic
  nuclei}.
\newblock {\em {Computer Physics Communications}} {\bf {185}} ({2014})
  {1808--1821}.
\newblock doi:\doi{10.1016/j.cpc.2014.02.027}

\bibitem{Rodriguez2010}
T.~R. Rodríguez and J.~L. Egido:
\newblock {Triaxial angular momentum projection and configuration mixing
  calculations with the Gogny force}.
\newblock {\em {Physical Review C}} {\bf {81}} ({2010}) {064323}.
\newblock doi:\doi{10.1103/PhysRevC.81.064323}

\bibitem{Yao2011}
J.~M. Yao, H.~Mei, H.~Chen, J.~Meng, P.~Ring, and D.~Vretenar:
\newblock {Configuration mixing of angular-momentum-projected triaxial
  relativistic mean-field wave functions. II. Microscopic analysis of low-lying
  states in magnesium isotopes}.
\newblock {\em {Physical Review C}} {\bf {83}} ({2011}) {014308}.
\newblock doi:\doi{10.1103/PhysRevC.83.014308}

\bibitem{Reinhard1992}
P.-G. Reinhard:
\newblock {From sum rules to RPA: 1. Nuclei}.
\newblock {\em Annalen der Physik} {\bf 504} (1992) 632--661.
\newblock doi:\doi{10.1002/andp.19925040805}

\bibitem{Terasaki2005}
J.~Terasaki, J.~Engel, M.~Bender, J.~Dobaczewski, W.~Nazarewicz, and
  M.~Stoitsov:
\newblock {Self-consistent description of multipole strength in exotic nuclei:
  Method}.
\newblock {\em {Physical Review C}} {\bf {71}} ({2005}) {034310}.
\newblock doi:\doi{10.1103/PhysRevC.71.034310}

\bibitem{Colo2013}
G.~Colò, L.~Cao, N.~Van~Giai, and L.~Capelli:
\newblock {Self-consistent RPA calculations with Skyrme-type interactions: The
  skyrme\_rpa program}.
\newblock {\em {Computer Physics Communications}} {\bf {184}} ({2013})
  {142--161}.
\newblock doi:\doi{10.1016/j.cpc.2012.07.016}

\bibitem{Terasaki2010}
J.~Terasaki and J.~Engel:
\newblock {Self-consistent Skyrme quasiparticle random-phase approximation for
  use in axially symmetric nuclei of arbitrary mass}.
\newblock {\em {Physical Review C}} {\bf {82}} ({2010}) {034326}.
\newblock doi:\doi{10.1103/PhysRevC.82.034326}.

\bibitem{Yoshida2013}
K.~Yoshida and T.~Nakatsukasa:
\newblock {Shape evolution of giant resonances in Nd and Sm isotopes}.
\newblock {\em {Physical Review C}} {\bf {88}} ({2013}) {034309}.
\newblock doi:\doi{10.1103/PhysRevC.88.034309}

\bibitem{Nesterenko2002}
V.~O. Nesterenko, J.~Kvasil, and P.-G. Reinhard:
\newblock {Separable random phase approximation for self-consistent nuclear
  models}.
\newblock {\em {Physical Review C}} {\bf {66}} ({2002}) {044307}.
\newblock doi:\doi{10.1103/PhysRevC.66.044307}

\bibitem{Nesterenko2006}
V.~O. Nesterenko, W.~Kleinig, J.~Kvasil, P.~Vesely, P.-G. Reinhard, and D.~S.
  Dolci:
\newblock {Self-consistent separable random-phase approximation for Skyrme
  forces: Giant resonances in axial nuclei}.
\newblock {\em {Physical Review C}} {\bf {74}} ({2006}) {064306},
  \href{http://arxiv.org/abs/nucl-th/0512045}{\path{arXiv:nucl-th/0512045}}.
\newblock doi:\doi{10.1103/PhysRevC.74.064306}

\bibitem{Repko2013}
A.~Repko, P.-G. Reinhard, V.~O. Nesterenko, and J.~Kvasil:
\newblock {Toroidal nature of the low-energy E1 mode}.
\newblock {\em {Physical Review C}} {\bf {87}} ({2013}) {024305}.
\newblock doi:\doi{10.1103/PhysRevC.87.024305}

\bibitem{Reinhard2014}
P.-G. Reinhard, V.~O. Nesterenko, A.~Repko, and J.~Kvasil:
\newblock {Nuclear vorticity in isoscalar E1 modes: Skyrme-random-phase
  approximation analysis}.
\newblock {\em {Physical Review C}} {\bf {89}} ({2014}) {024321}.
\newblock doi:\doi{10.1103/PhysRevC.89.024321}

\bibitem{Slater1951}
J.~C. Slater:
\newblock {A Simplification of the Hartree-Fock Method}.
\newblock {\em {Physical Review}} {\bf {81}} ({1951}) {385--390}.
\newblock doi:\doi{10.1103/PhysRev.81.385}

\bibitem{Varshalovich1988}
D.~A. Varshalovich, A.~N. Moskalev, and V.~K. Khersonskii:
\newblock {\em {Quantum Theory of Angular Momentum}}.
\newblock World Scientific, Singapore, 1988.

\bibitem{Repko-specf_qm}
A.~Repko:
\newblock {\em {Špeciálne funkcie v kvantovej mechanike}}, {27.10.2015}.
\newblock
  \href{http://a-repko.sk/knihy/specf_qm.pdf}{\path{http://a-repko.sk/knihy/specf_qm.pdf}}

\bibitem{SGII}
N.~Van~Giai and H.~Sagawa:
\newblock {Spin-isospin and pairing properties of modified Skyrme
  interactions}.
\newblock {\em {Physics Letters B}} {\bf {106}} ({1981}) {379--382}.
\newblock doi:\doi{10.1016/0370-2693(81)90646-8}

\bibitem{SLy6}
E.~Chabanat, P.~Bonche, P.~Haensel, J.~Meyer, and R.~Schaeffer:
\newblock {A Skyrme parametrization from subnuclear to neutron star densities -
  Part II. Nuclei far from stabilities}.
\newblock {\em {Nuclear Physics A}} {\bf {635}} ({1998}) {231--256}.
\newblock doi:\doi{10.1016/S0375-9474(98)00180-8}

\bibitem{SkT6}
F.~Tondeur, M.~Brack, M.~Farine, and J.~M. Pearson:
\newblock {Static nuclear properties and the parametrisation of Skyrme forces}.
\newblock {\em {Nuclear Physics A}} {\bf {420}} ({1984}) {297--319}.
\newblock doi:\doi{10.1016/0375-9474(84)90444-5}

\bibitem{Kvasil2011}
J.~Kvasil, V.~O. Nesterenko, W.~Kleinig, P.-G. Reinhard, and P.~Vesely:
\newblock {General treatment of vortical, toroidal, and compression modes}.
\newblock {\em {Physical Review C}} {\bf {84}} ({2011}) {034303}.
\newblock doi:\doi{10.1103/PhysRevC.84.034303}

\bibitem{Greiner1996}
W.~Greiner and J.~A. Maruhn:
\newblock {\em {Nuclear Models}}.
\newblock Springer-Verlag, Berlin Heidelberg, 1996.

\bibitem{VeselyPhD}
P.~Veselý:
\newblock {\em {Collective Nuclear Excitations within Skyrme Separable RPA}}.
\newblock PhD thesis, {Charles University in Prague}, {2009}.

\bibitem{Lipparini1989}
E.~Lipparini and S.~Stringari:
\newblock {Sum rules and giant resonances in nuclei}.
\newblock {\em {Physics Reports (Review Section of Physics Letters)}} {\bf
  {175}} ({1989}) {103--261}.
\newblock doi:\doi{10.1016/0370-1573(89)90029-X}

\bibitem{Hassan1980}
M.~H. Hassan:
\newblock {Coefficients of Talmi and of Moshinsky and Smirnov for the harmonic
  oscillator basis}.
\newblock {\em {Journal of Physics A: Mathematical and General}} {\bf {13}}
  ({1980}) {1903--1924}.
\newblock doi:\doi{10.1088/0305-4470/13/6/014}

\bibitem{Edwards1974}
H.~M. Edwards:
\newblock {\em {Riemann's Zeta Function}}.
\newblock {Academic Press}, New York, 1974.

\bibitem{Dobaczewski1997}
J.~Dobaczewski and J.~Dudek:
\newblock {Solution of the Skyrme-Hartree-Fock equations in the Cartesian
  deformed harmonic oscillator basis. I. The method}.
\newblock {\em {Computer Physics Communications}} {\bf {102}} ({1997})
  {166--182}.
\newblock doi:\doi{10.1016/S0010-4655(97)00004-0}

\bibitem{Dobaczewski2005}
M.~V. Stoitsov, J.~Dobaczewski, W.~Nazarewicz, and P.~Ring:
\newblock {Axially deformed solution of the Skyrme-Hartree-Fock-Bogolyubov
  equations using the transformed harmonic oscillator basis. The program HFBTHO
  (v1.66p)}.
\newblock {\em {Computer Physics Communications}} {\bf {167}} ({2005})
  {43--63}.
\newblock doi:\doi{10.1016/j.cpc.2005.01.001}

\bibitem{Abramowitz1972}
M.~Abramowitz and I.~A. Stegun:
\newblock {\em {Handbook of Mathematical Functions With Formulas, Graphs, and
  Mathematical Tables}}.
\newblock {National Bureau of Standards}, Washington, D.C., 10th edition, 1972.

\bibitem{Carroll2011}
J.~D. Carroll, A.~W. Thomas, J.~Rafelski, and G.~A. Miller:
\newblock {\em {P}roton form-factor dependence of the finite-size correction to
  the {L}amb shift in muonic hydrogen}, 2011.
\newblock \href{http://arxiv.org/abs/1108.2541v1}{\path{arXiv:1108.2541v1
  [physics.atom-ph]}}

\bibitem{Bender2000}
M.~Bender, K.~Rutz, P.-G. Reinhard, and J.~A. Maruhn:
\newblock {Pairing gaps from nuclear mean-field models}.
\newblock {\em {European Physical Journal A}} {\bf {8}} ({2000}) {59--75}.
\newblock doi:\doi{10.1007/s10050-000-4504-z}

\bibitem{Reinhard1999}
P.-G. Reinhard, D.~J. Dean, W.~Nazarewicz, J.~Dobaczewski, J.~A. Maruhn, and
  M.~R. Strayer:
\newblock {Shape coexistence and the effective nucleon-nucleon interaction}.
\newblock {\em {Physical Review C}} {\bf {60}} ({1999}) {014316}.
\newblock doi:\doi{10.1103/PhysRevC.60.014316}

\bibitem{Guo2007}
L.~Guo, J.~A. Maruhn, and P.-G. Reinhard:
\newblock {Triaxiality and shape coexistence in germanium isotopes}.
\newblock {\em {Physical Review C}} {\bf {76}} ({2007}) {034317}.
\newblock doi:\doi{10.1103/PhysRevC.76.034317}

\bibitem{Wang2012}
M.~Wang, G.~Audi, A.~H. Wapstra, F.~G. Kondev, M.~MacCormick, X.~Xu, and
  B.~Pfeiffer:
\newblock {The AME2012 atomic mass evaluation (II). Tables, graphs and
  references}.
\newblock {\em {Chinese Physics C}} {\bf {36}} ({2012}) {1603--2014}.
\newblock doi:\doi{10.1088/1674-1137/36/12/003}

\bibitem{Savran2013}
D.~Savran, T.~Aumann, and A.~Zilges:
\newblock {Experimental studies of the Pygmy Dipole Resonance}.
\newblock {\em {Progress in Particle and Nuclear Physics}} {\bf {70}} ({2013})
  {210--245}.
\newblock doi:\doi{10.1016/j.ppnp.2013.02.003}

\bibitem{Daoutidis2011}
I.~Daoutidis and P.~Ring:
\newblock {Relativistic continuum quasiparticle random-phase approximation in
  spherical nuclei}.
\newblock {\em {Physical Review C}} {\bf {83}} ({2011}) {044303}.
\newblock doi:\doi{10.1103/PhysRevC.83.044303}

\bibitem{Dobaczewski1995}
J.~Dobaczewski and J.~Dudek:
\newblock {Time-odd components in the mean field of rotating superdeformed
  nuclei}.
\newblock {\em {Physical Review C}} {\bf {52}} ({1995}) {1827--1839}.
\newblock doi:\doi{10.1103/PhysRevC.52.1827}

\bibitem{SkI3}
P.-G. Reinhard and H.~Flocard:
\newblock {Nuclear effective forces and isotope shifts}.
\newblock {\em {Nuclear Physics A}} {\bf {584}} ({1995}) {467--488}.
\newblock doi:\doi{10.1016/0375-9474(94)00770-N}

\bibitem{Vesely2009}
P.~Vesely, J.~Kvasil, V.~O. Nesterenko, W.~Kleinig, P.-G. Reinhard, and V.~Y.
  Ponomarev:
\newblock {Skyrme random-phase-approximation description of spin-flip M1 giant
  resonance}.
\newblock {\em {Physical Review C}} {\bf {80}} ({2009}) {031302(R)}.
\newblock doi:\doi{10.1103/PhysRevC.80.031302}

\bibitem{Nesterenko2010}
V.~O. Nesterenko, J.~Kvasil, P.~Vesely, W.~Kleinig, and P.-G. Reinhard:
\newblock {Skyrme-random-phase-approximation description of spin-flip and
  orbital M1 giant resonances}.
\newblock {\em {International Journal of Modern Physics E}} {\bf {19}} ({2010})
  {558--567}.
\newblock doi:\doi{10.1142/S0218301310014972}

\bibitem{Steffen1983}
W.~Steffen, H.-D. Gräf, A.~Richter, A.~Härting, W.~Weise, U.~Deutschmann,
  G.~Lahm, and R.~Neuhausen:
\newblock {Form factor of the M1 transition to the 10.23 MeV state in $^{48}$Ca
  and the role of the $\Delta$(1232)}.
\newblock {\em {Nuclear Physics A}} {\bf {404}} ({1983}) {413--427}.
\newblock doi:\doi{10.1016/0375-9474(83)90267-1}

\bibitem{Laszewski1988}
R.~M. Laszewski, R.~Alarcon, D.~S. Dale, and S.~D. Hoblit:
\newblock {Distribution of M1 Transitions in $^{208}$Pb}.
\newblock {\em {Physical Review Letters}} {\bf {61}} ({1988}) {1710--1712}.
\newblock doi:\doi{10.1103/PhysRevLett.61.1710}

\bibitem{Muller1985}
S.~Müller, G.~Küchler, A.~Richter, H.~P. Blok, H.~Blok, C.~W. de~Jager,
  H.~de~Vries, and J.~Wambach:
\newblock {High-Resolution Inelastic Electron Scattering and the Isoscalar
  Nature of the M1 Transitions to the $J^\pi=1^+$ State at $E_x$ = 5.846 MeV in
  $^{208}$Pb}.
\newblock {\em {Physical Review Letters}} {\bf {54}} ({1985}) {293--296}.
\newblock doi:\doi{10.1103/PhysRevLett.54.293}

\bibitem{A208}
M.~J. Martin:
\newblock {Nuclear Data Sheets for A=208}.
\newblock {\em {Nuclear Data Sheets}} {\bf {108}} ({2007}) {1583--1806}.
\newblock doi:\doi{10.1016/j.nds.2007.07.001}

\bibitem{Youngblood2004}
D.~H. Youngblood, Y.-W. Lui, H.~L. Clark, B.~John, Y.~Tokimoto, and X.~Chen:
\newblock {Isoscalar E0--E3 strength in $^{116}$Sn, $^{144}$Sm, $^{154}$Sm, and
  $^{208}$Pb}.
\newblock {\em {Physical Review C}} {\bf {69}} ({2004}) {034315}.
\newblock doi:\doi{10.1103/PhysRevC.69.034315}

\bibitem{Nesterenko2015}
V.~O. Nesterenko, A.~Repko, P.-G. Reinhard, and J.~Kvasil:
\newblock {Relation of E1 pygmy and toroidal resonances}.
\newblock {\em {EPJ Web of Conferences (CGS15, Dresden, Germany, 2014)}} {\bf
  {93}} ({2015}) {01020}.
\newblock doi:\doi{10.1051/epjconf/20159301020}

\bibitem{Kvasil2013}
J.~Kvasil, A.~Repko, V.~O. Nesterenko, W.~Kleinig, P.-G. Reinhard, and
  N.~Lo~Iudice:
\newblock {Toroidal, compression and vortical dipole strengths in $^{124}$Sn}.
\newblock {\em {Physica Scripta}} {\bf {T154}} ({2013}) {014019}.
\newblock doi:\doi{10.1088/0031-8949/2013/T154/014019}

\bibitem{Nesterenko-lowE2}
V.~O. Nesterenko, V.~G. Kartavenko, W.~Kleinig, J.~Kvasil, A.~Repko, R.~V.
  Jolos, and P.-G. Reinhard:
\newblock {Skyrme random-phase-approximation description of lowest
  $K^\pi=2_\gamma^+$ states in axially deformed nuclei}.
\newblock {\em {Physical Review C}} {\bf {93}} ({2016}) {034301}.
\newblock doi:\doi{10.1103/PhysRevC.93.034301}

\bibitem{Kvasil2015-ischia}
J.~Kvasil, V.~O. Nesterenko, A.~Repko, D.~Božík, W.~Kleinig, and P.-G.
  Reinhard:
\newblock {Deformation effects in Giant Monopole Resonance}.
\newblock {\em {Journal of Physics: Conference Series}} {\bf {580}} ({2015})
  {012053}.
\newblock doi:\doi{10.1088/1742-6596/580/1/012053}

\bibitem{Kvasil2015}
J.~Kvasil, D.~Božík, A.~Repko, P.-G. Reinhard, V.~O. Nesterenko, and
  W.~Kleinig:
\newblock {Monopole giant resonance in $^{100-132}$Sn, $^{144}$Sm and
  $^{208}$Pb}.
\newblock {\em {Physica Scripta}} {\bf {90}} ({2015}) {114007}.
\newblock doi:\doi{10.1088/0031-8949/90/11/114007}

\bibitem{Nesterenko-Mg24}
J.~Kvasil, V.~O. Nesterenko, A.~Repko, P.-G. Reinhard, and W.~Kleinig:
\newblock {Deformation-induced splitting of the monopole giant resonance in
  $^{24}$Mg}.
\newblock {\em {EPJ Web of Conferences (NSRT15, Dubna, Russia, 2015)}} {\bf
  {107}} ({2016}) {05003}.
\newblock doi:\doi{10.1051/epjconf/201610705003}

\bibitem{Repko-istros}
A.~Repko, J.~Kvasil, V.~O. Nesterenko, and P.-G. Reinhard:
\newblock {Skyrme RPA for spherical and axially symmetric nuclei}.
\newblock \emph{Submitted to proceedings of ISTROS (Slovakia, 2015)},
  \href{http://arxiv.org/abs/1510.01248}{\path{arXiv:1510.01248 [nucl-th]}}

\bibitem{Pai2016}
H.~Pai, T.~Beck, J.~Beller, R.~Beyer, M.~Bhike, V.~Derya, U.~Gayer, J.~Isaak,
  Krishichayan, J.~Kvasil, B.~Löher, V.~O. Nesterenko, N.~Pietralla,
  G.~Martínez-Pinedo, L.~Mertes, V.~Y. Ponomarev, P.-G. Reinhard, A.~Repko,
  P.~C. Ries, C.~Romig, D.~Savran, R.~Schwengner, W.~Tornow, V.~Werner,
  J.~Wilhelmy, A.~Zilges, and M.~Zweidinger:
\newblock {Magnetic dipole excitations of $^{50}$Cr}.
\newblock {\em {Physical Review C}} {\bf {93}} ({2016}) {014318}.
\newblock doi:\doi{10.1103/PhysRevC.93.014318}

\bibitem{Poltoratska2012}
I.~Poltoratska, P.~von Neumann-Cosel, A.~Tamii, T.~Adachi, C.~A. Bertulani,
  J.~Carter, M.~Dozono, H.~Fujita, K.~Fujita, Y.~Fujita, K.~Hatanaka, M.~Itoh,
  T.~Kawabata, Y.~Kalmykov, A.~M. Krumbholz, E.~Litvinova, H.~Matsubara,
  K.~Nakanishi, R.~Neveling, H.~Okamura, H.~J. Ong, B.~Özel Tashenov, V.~Y.
  Ponomarev, A.~Richter, B.~Rubio, H.~Sakaguchi, Y.~Sakemi, Y.~Sasamoto,
  Y.~Shimbara, Y.~Shimizu, F.~D. Smit, T.~Suzuki, Y.~Tameshige, J.~Wambach,
  M.~Yosoi, and J.~Zenihiro:
\newblock {Pygmy dipole resonance in $^{208}$Pb}.
\newblock {\em {Physical Review C}} {\bf {85}} ({2012}) {041304(R)}.
\newblock doi:\doi{10.1103/PhysRevC.85.041304}

\bibitem{Ahrens1975}
J.~Ahrens, H.~Borchert, K.~H. Czock, H.~B. Eppler, H.~Gimm, H.~Gundrum,
  M.~Kröning, P.~Riehn, G.~Sita~Ram, A.~Zieger, and B.~Ziegler:
\newblock {Total nuclear photon absorption cross sections for some light
  elements}.
\newblock {\em {Nuclear Physics A}} {\bf {251}} ({1975}) {479--492}.
\newblock doi:\doi{10.1016/0375-9474(75)90543-6}

\bibitem{OKeefe1987}
G.~J. O'Keefe, M.~N. Thompson, Y.~I. Assafiri, R.~E. Pywell, and K.~Shoda:
\newblock {The photonuclear cross sections of $^{48}$Ca}.
\newblock {\em {Nuclear Physics A}} {\bf {469}} ({1987}) {239--252}.
\newblock doi:\doi{10.1016/0375-9474(87)90108-4}

\bibitem{Veyssiere1970}
A.~Veyssiere, H.~Beil, R.~Bergere, P.~Carlos, and A.~Lepretre:
\newblock {Photoneutron cross sections of $^{208}$Pb and $^{197}$Au}.
\newblock {\em {Nuclear Physics A}} {\bf {159}} ({1970}) {561--576}.
\newblock doi:\doi{10.1016/0375-9474(70)90727-X}

\bibitem{Papakonstantinou2011}
P.~Papakonstantinou, V.~Y. Ponomarev, R.~Roth, and J.~Wambach:
\newblock {Isoscalar dipole coherence at low energies and forbidden E1
  strength}.
\newblock {\em {European Physical Journal A}} {\bf {47}} ({2011}) {14}.
\newblock doi:\doi{10.1140/epja/i2011-11014-7}

\bibitem{Derya2014}
V.~Derya, D.~Savran, J.~Endres, M.~N. Harakeh, H.~Hergert, J.~H. Kelley,
  P.~Papakonstantinou, N.~Pietralla, V.~Y. Ponomarev, R.~Roth, G.~Rusev, A.~P.
  Tonchev, W.~Tornow, H.~J. Wörtche, and A.~Zilges:
\newblock {Isospin properties of electric dipole excitations in $^{48}$Ca}.
\newblock {\em {Physics Letters B}} {\bf {730}} ({2014}) {288--292}.
\newblock doi:\doi{10.1016/j.physletb.2014.01.050}

\bibitem{Gambacurta2011}
D.~Gambacurta, M.~Grasso, and F.~Catara:
\newblock {Low-lying dipole response in the stable $^{40,48}$Ca nuclei with the
  second random-phase approximation}.
\newblock {\em {Physical Review C}} {\bf {84}} ({2011}) {034301}.
\newblock doi:\doi{10.1103/PhysRevC.84.034301}

\bibitem{Carlos1974}
P.~Carlos, H.~Beil, R.~Bergère, A.~Leprêtre, A.~de~Miniac, and A.~Veyssière:
\newblock {The giant dipole resonance in transition region of the samarium
  isotopes}.
\newblock {\em {Nuclear Physics A}} {\bf {225}} ({1974}) {171--188}.
\newblock doi:\doi{10.1016/0375-9474(74)90373-X}

\bibitem{SVbas}
P.~Klüpfel, P.-G. Reinhard, T.~J. Bürvenich, and J.~A. Maruhn:
\newblock {Variations on a theme by Skyrme: A systematic study of adjustments
  of model parameters}.
\newblock {\em {Physical Review C}} {\bf {79}} ({2009}) {034310}.
\newblock doi:\doi{10.1103/PhysRevC.79.034310}

\bibitem{Moghrabi2010}
K.~Moghrabi, M.~Grasso, G.~Colò, and N.~Van~Giai:
\newblock {Beyond Mean-Field Theories with Zero-Range Effective Interactions: A
  Way to Handle the Ultraviolet Divergence}.
\newblock {\em {Physical Review Letters}} {\bf {105}} ({2010}) {262501}.
\newblock doi:\doi{10.1103/PhysRevLett.105.262501}

\bibitem{RepkoMgr}
A.~Repko:
\newblock {\em {Giant Resonances in Atomic Nuclei}}.
\newblock Master thesis, {Charles University in Prague}, {2011}.

\bibitem{Kvasil1998}
J.~Kvasil, N.~Lo~Iudice, V.~O. Nesterenko, and M.~Kopál:
\newblock {Strength functions for collective excitations in deformed nuclei}.
\newblock {\em {Physical Review C}} {\bf {58}} ({1998}) {209--219}.
\newblock doi:\doi{10.1103/PhysRevC.58.209}

\end{thebibliography}


\chapwithtoc{List of Abbreviations}
\begin{tabular}{ll}
$2qp$ & two-quasiparticle pairs (like $\hat{\alpha}\hat{\alpha}$ or $\hat{\alpha}^+\hat{\alpha}^+$) \\
BCS & Bardeen-Cooper-Schrieffer (theory of pairing) \\
BHF & Br\"uckner-Hartree-Fock \\
cmc & center-of-mass correction \\
DFT & density functional theory \\
E-M & Euler-Maclaurin (formula, summation, correction) \\
EWSR & energy-weighted sum rule \\
GDR & giant dipole resonance (E1) \\
GQR & giant quadrupole resonance (E2) \\
GMR & giant monopole resonance (E0) \\
HF & Hartree-Fock \\
HFB & Hartree-Fock-Bogoliubov \\
QRPA & quasiparticle random phase approximation \\
r.m.e. & reduced matrix elements \\
RPA & random phase approximation \\
s.f. & strength function \\
SHO & spherical harmonic oscillator \\
s.p. & single-particle/one-body (basis, matrix elements) \\
SRPA & separable random phase approximation \\
VAP & variation after projection \\
VBP & variation before projection \\
w.f. & wavefunction \\
\end{tabular}

\vspace{15pt}
Symbol $\Delta$ is used for two different purposes which may cause confusion: either as an energy-smoothing parameter (in the units of MeV), or as a grid spacing (lattice parameter) in the units of fm. There are also some other possible collisions, such as $\alpha,\,b,\,\delta,\,J,\,j,\,Q,\,P,\,p$, but hopefully all of them should be clear by the context or by the attributes (index, hat).


\openright
\end{document}